\documentclass[pra,onecolumn,nofootinbib,10pt,longbibliography]{revtex4-1}

\renewcommand{\thesection}{\arabic{section}}
\renewcommand{\thesubsection}{\thesection.\arabic{subsection}}
\renewcommand{\thesubsubsection}{\thesubsection.\arabic{subsubsection}}

\makeatletter
\renewcommand{\p@subsection}{}
\renewcommand{\p@subsubsection}{}
\makeatother

\usepackage{etoolbox}
\patchcmd{\section}
  {\centering}
  {\raggedright}
  {}
  {}
\patchcmd{\subsection}
  {\centering}
  {\raggedright}
  {}
  {}
\patchcmd{\subsubsection}
  {\centering}
  {\raggedright}
  {}
  {}

  \usepackage{graphicx}
\usepackage{amsmath} 		
\usepackage{amssymb}		
\usepackage{amsthm}   		

\usepackage[mathscr]{eucal}	
\usepackage{mathrsfs}		
\DeclareMathAlphabet{\mathpzc}{OT1}{pzc}{m}{it}  

\usepackage{mathtools}

\usepackage{bm}		
\usepackage{verbatim}     
\usepackage{xspace}	
\usepackage{enumerate}  
\usepackage{url}

\usepackage{epigraph}    		

\providecommand{\textquote}[1]{\textsl{#1}}      
\providecommand{\textperson}[1]{\textsc{#1}}   

\bibpunct[,]{[}{]}{,}{n}{}{,}
\setcitestyle{square,numbers}

\def\be{\begin{equation}}      
\def\ee{\end{equation}}

\def\beu{\begin{equation*}}   
\def\eeu{\end{equation*}}

\def\bsub{\begin{subequations}}  
\def\esub{\end{subequations}}

\def\ie{i.e.}  

\providecommand{\stext}[1]{\text{\tiny{#1}}}  
\def\realsymbol{\mathbb{R}}
\def\rationalsymbol{\mathbb{Q}}

\def\integersymbol{\mathbb{Z}}
\def\naturalsymbol{\mathbb{N}}

\providecommand{\abs}[1]{\left\lvert#1\right\rvert}   

\DeclareMathOperator*{\argmin}{arg\,min}	
\DeclareMathOperator{\sgn}{sgn}			

\providecommand{\bv}[1]{\boldsymbol{#1}}       

\providecommand{\opdet}[1]{\abs}  

\def\transsymbol{\text{\tiny T}}  		
\providecommand{\trans}{^{\transsymbol}}  
\providecommand{\inv}{^{-1}}			

\DeclareMathSymbol{\bigtimes}{\mathop}{symbols}{"02}   

\def\binfo{\mathcal{I}}  				
\def\given{\,|\,}  						

\providecommand{\del}{\partial}

\def\K{\mathcal{K}}
\def\S{\mathcal{S}}
\def\D{\mathcal{D}}

\def\C{\mathcal{C}}

\def\J{\mathcal{J}}

\def\Z{\mathbb{Z}}

\def\p{p}
\def\P{\mathcal{P}}
\def\p{\rho}
\def\m{m}

\def\X{\mathcal{X}}
\def\Y{\mathcal{Y}}

\def\Dmin{\EuScript{D}_{\stext{min}}}
 \def\Dbar{\bar{\EuScript{D}}}
\def\f{\mathcal{F}}

\def\U{\EuScript{U}}

\def\fQ{\lfloor Q_s \rfloor}
\def\cQ{\lceil Q_s \rceil}

\def\A{\mathcal{A}}
\def\H{\mathcal{H}}
\def\G{\mathcal{G}}
\def\Q{\mathcal{Q}}
\def\I{\mathcal{I}}
\def\L{\mathcal{L}}

\def\share{\EuScript{S}}
\def\surplus{\mathscr{S}}
\def\deficiency{\mathscr{D}}
\def\round{\xi} 

\def\Gini{\EuScript{G}}
\def\LAC{\text{LAC}}

\begin{document} 


\title{The Census and the Second Law: An Entropic Approach\\ to Optimal Apportionment for the U.S.\ House of Representatives}
\author{A.E.~Charman}
\thanks{This work is dedicated to the memory of an ever inquisitive and congenial teacher, Professor Harold Lecar, who took time from studying how the brain works in order to try to ensure that elections work they way they should.  I shall certainly miss what undoubtedly would have been his insights (and outrage) regarding the 2016 Presidential contest, and the unprecedented events surrounding the certification of the 2020 election results.}
\affiliation{Department of Physics, University of California, Berkeley\\Berkeley, CA 94720, USA}
\email{acharman@physics.berkeley.edu}
\date{July 4, 2021 [version 2.1]}

\begin{abstract}
The Constitutionally mandated task of assigning Congressional seats to the various U.S.\ States proportional to their represented populations (``according to their numbers'') has engendered much contention, but rather less consensus.  Using the same principles of entropic inference that underlie the foundations of information theory and statistical thermodynamics, and also enjoy fruitful application in image processing, spectral analysis, machine learning, econometrics, bioinformatics, and a growing number of other fields, we motivate and explore a method for Congressional apportionment based on minimizing relative entropy (also known as Kullback-Leibler divergence), or, equivalently, maximizing Shannon entropy.\\

 The ideal apportionment of seats to states is that which minimizes (subject to prescribed constraints on the total number of representatives, as well as on the minimum and maximum number of representatives per state) the relative entropy or discrimination information, arising as the natural measure of divergence between two probability distributions---in this case a uniform distribution of representational weight, or polling probability, across all represented individuals, as in an ideal or direct democracy, and a Congress-mediated distribution of representational weight, where individuals receive political representation indirectly, through their elected representatives.  Equivalently, the optimal apportionment maximizes (subject to said constraints) the Shannon entropy, which is the natural measure of uncertainty or missing information associated with the indirect distribution. Statistically speaking, the optimal apportionment maximizes the uniformity of the sampling distribution induced by the division of seats, consistent with states receiving a whole number of representatives (between allowed lower and upper bounds).  In terms of communication theory, we might say that the entropic apportionment gives each constituent as equal a voice as possible.  If we view representational weight as a finite resource to be distributed amongst the represented population, the entropic measure is identical with the Theil index long employed in economics to measure inequality in the distribution of wealth or income, or in ecology to measure the distribution of biomass, species abundance, genetic information, or reproductive fitness.\\

In actual application to U.S.\ Census data,  entropic apportionment tends to perform similarly to the well-known Huntington-Hill and Webster-Willcox methods, but enjoys a more fundamental motivation and more natural mathematical properties than either. In fact, both of these traditional methods can be viewed as approximations to the entropic apportionment method.  Besides Congressional apportionment, the method is also directly applicable to other multi-regional or multi-constituency legislatures, to party-list proportional voting systems used in various parliamentary elections, and similar settings, where the task is to allocate a finite, discrete number of seats or other resources, and the primary goal is one of maximal proportionality or equity.  In addition, the same entropic figure-of-merit can be used in parallel to compare different choices for the total number of representatives, and then subsequently to assess different Congressional district sizes, after seats are assigned and proposed district boundaries drawn.  
\end{abstract}


\maketitle



\begin{epigraphs}
\qitem{\textquote{It has been said that democracy is the worst form of government except all the others that have been tried.}}{\textperson{Winston Churchill} }
\vspace{2 ex}
\qitem{\textquote{``The world isn't fair, Calvin.''\\``I know Dad, but why isn't it ever unfair in my favor?''}} 
{ \textperson{Bill Watterson} }
\vspace{2 ex}
\qitem{\textquote{You should call it entropy, for two reasons. In the first place your uncertainty function has been used in statistical mechanics under that name, so it already has a name.  In the second place, and more important, no one really knows what entropy really is, so in a debate you will always have the advantage.}}{ \textperson{John von Neumann}, to \textperson{Claude Shannon} } 
\end{epigraphs}
 
 %
\setlength{\parindent}{0em}
\setlength{\parskip}{1.75 ex plus 0.5ex minus 0.2ex}
%

%
%

\section{Introduction and Motivation}\label{section:introduction}

With each U.S.\ Census arrives a burst of renewed interest in the associated political, legal, philosophical, and mathematical issues surrounding the subsequent apportionment of Representatives to the various states ``according to their respective numbers,'' as required by the U.S.\ Constitution.  With much at stake in terms of legislative influence, the loss or gain of districts, and as we have recently seen, votes in the Electoral College, debate periodically percolates through courtrooms and the Capitol, in law reviews and even scientific journals \cite{seaton:1900,willcox:1916,huntington:1921,chafee:1929,bliss:1929,willcox:1941,schmeckebie:1941,morse:1948,jewell:1962,appel:1965,mckay:1965,dixon:1968,cortner:1970,pitkin:1972,papayanopoulos:1973,birkhoff:1976,adams:1977,balinski_young:1979, balinski_young:1980,orourke:1980, eig:1981,grofman_scarrow:1981,balinski_young:1982,cain:1984,michigan:1984,young:1985,grillidicortona:1987,young:1994,ernst:1994,saari:1995,park:2000,huckabee:2000,benoit:2000,mills:2001,young:2001,huckabee:2001,balinski_young:2001,edelman_sherry:2002,neubauer:2003,baumle_poston:2004,balinski:2005,edelman:2006,balinski_laraki:2010,crocker:2010,caufield:2010,cipra:2010,szpiro:2010,taylor_pacelli:2010,thirtythousand:2010,burnett:2011,maltz:2011,devilliers_nielsen:2012,krabill_fielding:2012,suzuki:2015,robinson:2016}.  In intervening years, attention tends to wane, but such lacunae actually provide us with opportunities to reconsider more dispassionately goals and methods for this consistently contentious Constitutional obligation.  

Both the periodic apportionment of representatives across states, and subsequent redistricting within states, have struck many commentators as curious inversions in the usual democratic dynamic:  instead of voters choosing their representatives, elected representatives in effect choose their constituencies.  Certainly circumspection is called for as we confront questions of some complexity, controversy, and consequence.  But as suggested by ``I cut/you choose'' fair-division principles familiar to game theorists and suspicious siblings everywhere, evaluating apportionment algorithms \textit{before} knowing actual census counts might serve to encourage better debate, and possibly even better legislation, based on methodological merits, and not particular political consequences.\footnote{After all, one could hardly expect Representatives to be so broad-minded as to vote their own districts out of existence.}

Here we explore, from an information-theoretic perspective, the fundamental questions posed by apportionment:  what does it mean, or should it mean, to distribute seats proportionally?  How should we round rational numbers to integers, and thereby \cite{cipra:2010} ``convert census data into congressional seats?''  Also, we touch on certain related questions: how many representatives should represent the people, and exactly which people\footnote{We will end up addressing here the question of the \textit{size} of districts, or variations in the sizes of districts, but not the nature of their geographic \textit{boundaries} or other determinants of their \textit{constituencies}  That is, we will speak to the question of \textit{how many} people each Representative represents, but not the thornier political questions of exactly \textit{who} is to be represented by \textit{whom}, a problem receiving much deserved attention of late, but left for another day.} should each of these representatives represent?  

Tentative answers to the questions raised may be considered from various perspectives, including the legal, the political, the moral, and the mathematical, leading to still more questions.  What do we understand to be the ultimate goals and evaluative criteria for census, choice of House size, apportionment, or districting, and what technical procedures or algorithms best achieve these goals?  Are the intent and application of these various methods consistent on their face with the U.S.\ Constitution, and, if so, would they otherwise require modification of relevant court decisions or federal laws and regulations?  Finally, would they stand any reasonable chance of being accepted by a majority of Congress, or would they otherwise be likely to be enacted if within the discretion of the existing executive branch, and could they withstand any subsequent challenges in federal courts or the Supreme Court?

After exploring the Constitutional ground rules and basic features of the problem, we shall here focus primarily on technical issues at the level of allocating representatives to states, along with some subsequent discussion of the optimal choice of the overall House size and determination of sizes for districts, if not their exact make-up.  We will mostly set aside various issues of political feasibility\footnote{Particularly because no major changes to the \textit{status quo} seem likely any time soon, given the current level of polarization and dysfunction in Congress....}\,\footnote{Parts of this document date prior to the 2016 U.S.\ election. Subsequent events have only made it  more unlikely that we as a nation will, at least in any short term, take positive steps to improve fairness in representative governance, since the one political party that now controls all three branches of the federal government has been the main beneficiary of existing inefficiencies and inequities.  Still, one may naively contend that it is important to know what ``fair'' apportionment looks like, and how to measure it, in the hope that rational debate might some day resume....}, and attempt to avoid wading too deeply into the related if tangential quagmires such as districting and gerrymandering.\footnote{However, we note in passing that entropy and information theory can also ``inform'' issues such as optimal district-drawing, and detection or prevention of gerrymandering.  This topic will be pursued in a subsequent paper.}

Despite the plethora of suggestions and algorithms for apportionment made over the last 225 years or so, the criterion argued for here---which might immediately come to the mind of many modern physicists, chemists, electrical engineers, and statisticians---has received surprisingly little serious attention or consideration.\footnote{Along with some other methods, entropic apportionment has been explored by Agnew \cite{agnew:2008}, and independently by Ossipoff \cite{ossipoff:2013}, although its truly fundamental justification, unique properties, and multiple advantages were not emphasized.}  Our aim is therefore to argue for the merits of a method for apportioning representatives by borrowing key ideas about \textit{entropy} from the fields of statistical mechanics and information theory.

The actual algorithm advocated can determine the ``optimal'' apportionment of representatives to states for a fixed total House size.  If desired it can also be used concurrently to choose the optimal size of the House of Representatives within some allowed range, as well as the optimal size (but not actual demographic make-up) of each Congressional district within individual U.S.\ states.  Importantly, it not only generates optimal apportionments, House sizes, and district sizes, but can objectively compare or rank different proposals.

The method amounts to minimizing a certain measure of discrepancy between actual and ideal polling probabilities, or equivalently weights of representation, held by represented individuals under a proposed apportionment.  In communication theory, this measure is known as the \textit{relative entropy}, and in statistics, as the \textit{discrimination information}, \textit{information gain}, or \textit{Kullback-Leibler divergence}, and is closely related to notions of \textit{entropy}, \textit{availability}, or \textit{exergy} in statistical thermodynamics. 

In a certain sense, we describe how allocations of Congressional seats may assume their ``fairest'' values in the same way that in electrical engineering we describe how better codes lead to shorter messages, or in physics why temperatures tend to equalize---essentially via a principle of \textit{maximum entropy}.  While such parallels between physics and politics might initially seem far-fetched, we will endeavor to carefully motivate our method from plausible desiderata, guided by what little the U.S.\ Constitution has to say on the matter.

The modern notion of information-theoretic entropy was introduced by Claude Shannon in the mid-20th century \cite{shannon:1948a,shannon:1948b,shannon:1949} as a measure of uncertainty or missing information, building on some earlier ideas of Hartley and Nyquist.  But Shannon's measure has deep connections to the physical concept of thermodynamic entropy as explored earlier by Carnot, Clausius, Kelvin, Maxwell, Boltzmann, Gibbs, Planck, Einstein, and other pioneers of thermal physics and statistical mechanics \cite{jaynes:1983,jaynes:2003}.  The closely related notion of \textit{relative entropy} provides the most natural information-theoretic measure of similarity between two probability distributions over the same sample space, or more generally between any two sets of nonnegative proportions or additive measures that might not even have an obvious  probabilistic interpretation but which can be defined on a distributive lattice of possibilities.  We use relative entropy as the most natural measure of \textit{discrepancy} or \textit{inequity} in the effective \textit{indirect} representational weight of individuals as represented by Congress, as compared to the equal weights that would hold in a pure direct democracy.  This relative entropy itself was employed at least as far back as World War II, in the code-breaking work of Alan Turing and I.J.~Good in Great Britain \cite{good:1983}, and popularized within the statistics community by researchers Solomon Kullback and Richard Leibler \cite{kullback:1959} while working for the U.S.\ government.\footnote{They did, however, work in one of the least transparent parts of the executive branch, the National Security Agency, so perhaps it is not altogether unexpected that their ideas did not make its way directly to Congress or the Census Bureau.}

\newpage
And while we focus on the apportionment problem for the U.S.\ House of Representatives, the algorithm advocated here can be easily adapted for other sorts of apportionment, including other federal or regional systems involving political subdivisions of unequal size, or party-list proportional representation systems used in many parliamentary democracies.  In fact, general applicability of a single underlying apportionment rule will be a guiding desideratum.

Indeed, it turns out that the same entropic figure of merit was introduced decades ago by the econometrician Henri Theil \cite{theil:1967} as a measure of distributional inequality of some finite asset or resource, and has been used to quantify inequality of income or wealth distributions in economics, as well as species, fitness, genomic, or biomass diversity in ecology.
 
One can only approach with some measure of humility any question of mathematical principle that compelled the attention of Edward Huntington, Marston Morse, John von Neumann, and Garrett Birkhoff, or a question of democratic practice addressed by the likes of Alexander Hamilton, Thomas Jefferson, John Quincy Adams, and Daniel Webster.  While well aware that there will almost certainly never be any one method of apportionment universally agreed upon as ``best'' by all interested parties in all situations for all times, we nevertheless contend that the entropic framework proposed here possesses a more natural interpretation and more objective justification than methods used previously, and provides a unified perspective and figure-of-merit from which we can generate optimal apportionments, compare or rank alternative apportionments, quantify the effects of different choices of House size, and assess choices regarding the sizes of Congressional districts, all while in actual practice leading to sensible apportionments that when applied to real census data tend to agree with one or both of the most commonly recommended methods, known as the Huntington-Hill method and the Webster-Willcox method---but enjoying a more fundamental motivation and interpretation than either.

\section{The Congressional Apportionment Problem}

What requirements the U.S.\ Constitution \cite{constitution:1789} does prescribe are spelled out in Article I, Section 2, later modified by the 14th Amendment:\footnote{In order to enforce voting rights of newly enfranchised (male) African American citizens, the 14th Amendment stipulated that in the event that voting rights in federal elections were denied to any adult males entitled to vote, the ``basis of representation'' (i.e., population for apportionment) should be reduced in proportion to the fraction of those denied the vote relative to the total number of adult male citizens.  Neither the later 19th nor 26th Amendments explicitly expanded this penalty clause to include women, or 18--20 year-old voters.}
\begin{quote}
Representatives ... shall be apportioned among the several States which may be included in this Union, according to their respective Numbers.... The actual Enumeration [of which] shall be made within three Years after the first Meeting of the Congress of the United States, and within every subsequent Term of ten Years, in such Manner as they shall by Law direct....  The Number of Representatives shall not exceed one for every thirty Thousand, but each State shall have at Least one Representative....
\end{quote}
The phrase ``according to their respective Numbers'' clearly suggests an intent of \textit{proportional} representation, in some sense and to whatever extent possible, but the Constitution is otherwise reticent on precise definitions or methodology, hence the talk of an apportionment ``problem,'' because in general the assigned number of Representatives cannot all be in exact common proportion to the respective represented populations as revealed by the ``actual Enumeration.''  More details of the long legal and historical context may be found in Appendix~\ref{background} (starting on page \pageref{background}).

\subsection{Inputs and Outputs}

In determining the apportionment of the $R$ representatives (current statue specifying that $R = 435$) amongst the $S$ states (where currently of course $S = 50$ in the U.S.), a reasonable reading of  the ``according to their respective numbers'' clause denies use of any other demographic information \textit{except} the state-level represented population counts $p_s$, for $s = 1 , \dotsc, S$.  These so-called ``populations for apportionment,'' or \textit{represented populations}, are mathematically speaking, specified as nonnegative whole numbers, practically speaking, subject to some uncertainty or measurement error, but, legally speaking, accepted as exact for the purpose of apportionment, as determined from the most recent official decennial census (now conducted by the Census Bureau, part of the Department of Commerce).

Given officially accepted values for the populations $\bv{p} = (p_1, \dotsc, p_S)$, a \textit{feasible} apportionment $\bv{a} = (a_1, \dotsc, a_S)$ is an assignment of a nonnegative integer number of representatives or seats to each state, summing to a prescribed total of $R = \sum\limits_s a_s$ voting representatives in the House, for which in addition, the number of seats awarded to each state must satisfy
\be
\lambda_ s \le a_s \le u_s,
\ee
where the U.S.\ Constitution requires $\lambda_s = 1$ as a lower bound, and $u_s =  \max\bigl[1, \lfloor p_s/\Dmin  \rfloor \bigr]$ as an upper bound, for all states $s = 1, \dotsc, S$, $\Dmin = 30\hspace{0.88pt}000$ being the mandated minimum number of represented person per Representative.\footnote{But the meaning of ``but'' is not entirely unambiguous, so it is not entirely clear whether the Constitution strictly rules out a state with less than $\Dmin$ inhabitants, or whether the requirement for at least one representative overrules the requirement on a minimum number of persons per representative.  However, the issue has never needed adjudication, and in the event of some future demographic drop due to major natural or sociopolitical disaster, if the population of some U.S.\ state ever were to fall below $\Dmin$, we imagine that the Country would have more important things to worry about.  For mathematical convenience, we have expressed the upper bounds $u_s$ in a manner that always remains consistent with the lower bounds $\lambda_s$.}  The represented populations, the apportionment bounds, and related quantities are defined and discussed in more detail in Appendix~\ref{parameters} (starting on page \pageref{parameters}).

\subsection{Fractional Dessert but Integer Rewards}

\vspace{-4pt}
The over-arching goal of \textit{proportionality} (``according to their respective numbers'')  suggests that the apportionments $\bv{a}$ are to be chosen such that the ratios of delegation sizes ${a_{s}}/{a_{s'}}$ between pairs of states are, \textit{in some sense to be made precise}, close to the ratios of represented populations ${p_{s}}/{p_{s'}}$ of those same states, but the Constitution appears otherwise silent on what exactly should be meant by closeness to this ideal, or how it should be measured or maximized.  As noted by Balinski and Young \cite{balinski_young:1977r}, two of the most prominent scholars of apportionment, the overarching ``issue is to find an operational method for interpreting this mandate, and to identify the essential properties that any fair and reasonable method ought to have.''

Congressional apportionment therefore requires converting ``census data to Congressional seats'' \cite{cipra:2010} according to some definite mathematical procedure, hopefully one viewed as reasonable by all of the various stakeholders.  But analogous problems arise in other political or economic contexts, so such apportionment problems are by no means unique to the U.S.\ Congress.  For example, in party-list proportional representation systems,\footnote{Proportional representation systems are currently used, in some form, in Algeria, Argentina, Belgium, Brazil, Bulgaria, Cambodia, Colombia, Costa Rica, Croatia, Czech Republic, Denmark, Estonia, Finland, Germany, Guatemala, Hong Kong, Indonesia, Israel, Italy, Latvia, Morocco, the Netherlands, New Zealand, Norway, Peru, Poland, Portugal, Slovakia, Spain, Sri Lanka, Sweden, Turkey, Uruguay, and other countries and regions.  Countries like Mexico and the U.K.~use mixed systems combining proportional representation and first-past-the-post voting.  Currently, the United Kingdom uses no fewer than six different voting methods, with members of the House of Commons elected by first-past-the-post plurality voting within districts (periodically proposed by the U.K.~Boundary Commission and approved by the House of Commons, based on preference for ``geographically naturally'' boundaries over population equality, as well as legislation specifying separate quotas for England, Wales, Scotland, and Northern Island).  For elections to the European Parliament (soon to be irrelevant), party-list systems were instead used (except in Northern Ireland).} the number of seats earned by a given political party is intended to be proportional to the number of valid votes cast for that party.  Mathematically, the problem is virtually the same as Congressional apportionment, only with the $p_s$ representing votes cast for the $s$th party rather than persons residing within the $s$th state, the major difference being that the minimum allowed apportionment is usually zero, so parties must reach some threshold of popularity before earning a first representative in the parliament.  Apportionment of delegates from member countries to the European Parliament is set by treaty rather than mathematical formula, but perhaps (quasi-)proportional representation might be a more satisfactory approach.  At regional or local levels, various executive or legislative bodies, councils, etc., may also involve assigning representatives from unequally-sized constituencies.

More generally, our basic problem is that of distributing or assigning a finite number $R$ of unit \textit{resources} that are discrete, indivisible, but essentially identical and interchangeable, to $S$ \textit{categories} which are discrete and exclusive, when the apportionments $\bv{a}$ are intended to be, to whatever extent possible, proportional to certain populations, votes, magnitudes, counts, weights, or other measures $\bv{p}$ of merit or dessert, possibly subject to additional constraints on the minimum or maximum number of resources to be allocated to each category.\footnote{Elections to the Scottish Parliament use an \textit{additional member system}, where voters cast separate ballots for a local representative within single-member districts, and a party-list vote for representation within larger regions, but a second vote determines the total number of representatives for a party including those elected to the single-member constituencies after taking into account the seats gained in each region by each party in the first ballot.  Scandinavian countries also include so-called ``leveling'' or ``adjustment'' seats, which supplement members directly elected by each constituency to better ensure that each party's share of the total seats in the national legislature is proportional to the party's overall share of votes.  Bi-proportional or other schemes are also employed.  For instance, proportional representation with respect to \textit{both} party and region might be sought---a particular apportionment scheme developed by mathematician F.~Pukelsheim was used within Switzerland for cantonal and municipal elections with precisely this goal.  Increasingly, single transferable vote (STV) selection is also combined with proportional representation, which adds another layer of complication.  But with appropriate modifications, the entropic method advocated here should be applicable to these more complex systems of representation.}  So in our subsequent discussion, we may think of  ``seats'' more broadly as the discrete items to be apportioned, the ``House size'' to be the number of items to be apportioned, the ``States'' to be the regions, parties, political units, economic agents or other categories to which the items are apportioned, and ``populations'' to be the counts, votes, sizes, or other measures of dessert in proportion to which the items are to be distributed.  The goal is to find an apportionment method that does the fairest job of apportioning the discrete items, in some well-defined sense of ``fairest'' that ought to be motivated, quantified, and justified, particularly when, generally speaking, no perfect distribution of rewards can be said to exist.

\section{Towards an Optimal Solution to Apportionment}


Given the impossibility of perfect proportionality in general, the real problem is not one of failure to find a workable method, but rather a surfeit of methods that can and have been suggested, all using different but ostensibly reasonable-looking rules, and all offering some semblance of approximate proportionality.  To avoid \textit{ad hocness}, argument must turn to why one approach is any better, or more fair, or less arbitrary, than any other.

\subsection{Guiding Principles: Uniformity, Universality, and Uniqueness}

We  therefore seek a principled way to winnow down the plethora of plausible-looking apportionment methods.  At the highest level of abstraction, we will be guided by three underlying principles based upon what we will call \textit{the Three U's}: \textit{Uniformity} of individual representation, \textit{Universality} of apportionment methodology, and \textit{Uniqueness} of the resulting algorithm and its output.  While admittedly sounding both a bit vague and high-handed, these \textit{desiderata} will lead us toward more concrete quantitative criteria and, with the help of some mathematical theorems from information theory, finally to a single, well-motivated, well-behaved, and easily executed apportionment algorithm.

\subsubsection{Uniformity of Representation}

First and foremost, we take seriously the idea of the House of Representatives as ``the People's House,'' and the associated goal of maximum equality or \textit{Uniformity} of representation between and among individuals, rather than other political units or actors.  Many apportionment methods and metrics are instead overtly (and overly) state-centric, focusing on the number of representatives to which states are somehow entitled, or on the fairness of certain states (hypothetically) gaining marginal seats from, or losing seats to, other states.  In our view, states do not deserve representatives; people of the various states deserve an equal say in their governance.  Only, in the U.S.\ system based on a (sometimes strange) mix of representative democracy and federalism, Congressional representation must be granted to the people through their respective states of residence, and representatives cannot be shared or split across states.\footnote{Similarly, in party-list systems, it is voters who deserve proportional representation, not political parties who deserve seats, but ultimately, finite, whole numbers of seats must be granted to discrete parties for whom the voters voted.}

Undergirding this outlook is what we call the \textit{One Person, One Voice} (OPOV) principle.  In essentially all cases involving proportional representation or other forms of political apportionment, the underlying goal would seem to be to enjoy the advantages and efficiencies of representative government, but in a way that embodies, as nearly as possible, the ideal of equality inherent in direct democracy.  In the context of U.S.\ legislative apportionment, a fundamental corollary of the Equal Protection clause is commonly referred to as the ``One Person, One Vote'' principle.\footnote{Traditionally, reference was made to the ``one man, one vote'' principle.  Perhaps more historically accurate would have been the ``one sufficiently wealthy white man, one vote'' principle.}  In Australia, they instead speak of the concept of ``one vote, one value.''  But the U.S.\ Constitution is very clear that Congressional representatives are to be allocated based on the full represented populations of the states, not their respective numbers of voters, or eligible voters, or even just citizens\footnote{See Appendix~\ref{background} for some discussion of the legal history.}, so \textit{One Person, One Voice} would seem a more accurate catchphrase to embody the goal of uniformity of representation in this context.

\subsubsection{Universality of Methodology}

A second natural principle is \textit{Universality}.  Although we focus here on the case of U.S.\ Congressional apportionment, we view this task as just one of many closely related political or economic apportionment problems, to which one and the same methodology or strategy should be applicable whenever the goal is equality or proportionality of representation or, more broadly, fairness in the distribution or division of some limited, discrete resource.  In particular, in the case of the U.S.\ Congress, the Constitution requires that $a_s \ge 1$ for each state, but this is not necessarily or even typically the case for, say, party-list proportional representation, where parties with few votes may not receive any seats, and in principle there could even be more parties than seats ($S > R$).  Universality demands that in the face of any these tasks, essentially the same goals should be pursued and same methods adopted, just with different interpretations of the variables and with different choices for the auxiliary constraints, rather than relying on some \textit{ad hoc} or artificial adjustment to the optimization method itself or to its output.  This alone seems to rule out about half of the traditional apportionment algorithms, namely any that \textit{automatically} assign at least one seat to each party even in the absence of explicit lower bound constraints.

A second aspect of Universality arises as we recall that the congressional apportionment problem in the narrow sense, i.e., awarding of legislative seats, is part of a hierarchical sequence of related political tasks, namely:
\begin{enumerate}
\item choice of House size $R$ (total number of representatives);
\item inter-state apportionment of representatives $\bv{a}$;
\item intra-state choice of congressional district sizes $d_{sk}$, when district-based representation is employed (which is currently required by law for all states with $a_s > 1$),
\end{enumerate}
all based on the represented populations $\bv{p}$ obtained from the census, and consistent with Constitutional lower and upper bounds.  It seems to us that the \textit{very same} principles of fair allocation or measures of proportionality which guide the apportionment of seats between states, should also be able to inform better choices of overall House size, or to assess inequities in Congressional district sizes within states.\footnote{A fourth task involves choosing the districts within states, i.e., the people whom Representatives are mean to represent.  Although also of fundamental concern, problems of re-districting, and of detecting or defending against gerrymandering, are largely distinct from the problem of apportionment itself.}

While it is evident that with fixed population counts, certain seat totals $R$ divide up more reasonably (with fewer ``leftover'' seats to award) than others, and while it is equally clear that differences in district sizes within states are, from the point of view of constituents, as undemocratic and unfair as differences between states,\footnote{Here we are thinking of inequalities in legislative representation.  Because of the structure of the electoral college, inter-state inequities in apportionment also introduce an additional layer of unfairness in presidential elections.} few methods of apportionment purport to quantify these sorts of mismatches using the \textit{same type of measure} employed to judge the apportionments themselves.  We consider it a major selling point that our proposed method achieves this unified perspective in a completely natural way.

\subsubsection{Uniqueness of Algorithms and Outcomes}

Altogether, desiderata for the apportionment methodology should entail \textit{Uniqueness} of the specific algorithm, and (near)-uniqueness of the output.  As to method, we ought to be able to pinpoint one best algorithm, eliminate---or at the very least minimize---\textit{ad hoc} assumptions or arbitrary choices, and supply fundamental justifications why this one method is preferred to all others, instead of just pointing to a few ostensibly desirable operational features, plausible-sounding properties, or empirical claims of performance.  This algorithm in turn should produce a unique apportionment, except in the unlikely event of exact ties arising from symmetries or accidental commensurability between the populations counts, in which cases (but only in such cases) the only fair way to choose amongst the remaining possibilities would be by random lottery.

Below, we shall attempt to argue why, out of all the different \textit{types} of methods that have been suggested, these three guiding principles single out \textit{global optimization} as best\.footnote{Although subsequently, such global optimization methods can be reinterpreted or implemented as so-called divisor methods or ranking methods.}  Then, in seeking an optimand, objective function, or figure-of-merit to optimize that embodies these principles, we will argue why the goal should be to equalize the representational shares or weight of individuals, rather than matching apportionments to state-level quotas, or balancing district sizes.\footnote{Although subsequently, we can re-express the figure-of-merit in terms of either quotas, or average district sizes, if desired.}  Finally, we will invoke some ideas from information theory to arrive at the most natural way to measure (and then minimize) inequality in the individual weights of representation.

\subsection{Quota, Divisor, Ranking, or Optimization Method?}

For a summary of various properties that might be expected of an apportionment method, and of the ``paradoxes'' that arise when certain of these properties fail to hold, see Appendix~\ref{paradoxes} (page \pageref{paradoxes} {ff}.).  For a survey describing and comparing various types of apportionment methodologies used or suggested, see Appendix~\ref{typologies} (page \pageref{typologies} {ff}.).  Specific methods are discussed in Appendix~\ref{historical_methods} (page \pageref{historical_methods} {ff}.).

In evaluating any apportionment methods, we ought to at least try to focus first on democratic principle before politics or pragmatics.  In the Federal Courts, legal claims often come down to one state arguing why it deserves a marginal representative more than does another state.  In Congress itself, debates and votes on apportionment are usually (if not always overtly) strategic, with keen attention paid to which apportionments or apportionment methods favor which states or which sorts of states---for instance, large or small, urban or rural, and/or predominately ``red'' (Republican-leaning) or ``blue'' (Democratic-leaning).  But, really we should endeavor to prioritize rights of the people rather than interests of States, or of political parties or particular demographics.

That is, a fair apportionment method should above all highlight the goal of equality of representation of individuals, within the confines of federalism, rather than any claims to seats by the states \textit{per se}.\footnote{Or in a party-list system, apportionment should emphasize the rights of voters over claims by the parties.}  For these reasons, we are extremely skeptical of \textit{quota} methods that demand $\abs{a_s - q_s} < 1$ for every $s = 1, \dotsc, S$, where the exact state quotas $q_s = \tfrac{p_s}{P} R = \tfrac{p_s}{\Dbar}$ are regarded as the ideal number of seats to which each state is entitled based on the goal of purely proportional representation.  (See Appendix~\ref{parameters} for more details on state quotas and related quantities).

Whilst with almost any method, the exact state quotas $q_s$ emerge as \textit{ideal targets} for apportionment,\footnote{After all, they are just proportional to the respective state populations, or can even be identified with the populations, measured in multiples of the overall average district size $\Dbar$.} there is no compelling reason to insist \textit{a priori} that the apportionments must remain within one unit of the corresponding quotas for all states, rather than, say, trying to match representatives to quotas at the level of counties or boroughs or precincts or other sub-units within states, or else larger geographic regions or groups of states, or any other sub-divisions of the populace.

In our reading, the prescribed Constitutional lower bounds (specifically, requirements of at least one representative per state) are not at all intended to deliberately nudge Congressional representation away from proportionality or to purposefully over-represent smaller states\footnote{Given the so-called ``Great Compromise,'' exaggerating the influence of smaller states would seem to be the job for the U.S.\ Senate.  Indeed, as of the 2010 U.S.\ Census, $18$ Senators represented just over $51\%$ of the population, with the remaining $49\%$ of the population represented by $82$ Senators.  As of 2017, it is estimated that the population of just Los Angeles County exceeds that of $43$ U.S.\ States.  This sort of is mismatch is only expected to get worse given projected demographic trends.  As we will see, relative entropy could also be used to characterize this inequality, but there is no Constitutional remedy short of a fundamental and entirely implausible amendment process.}, but merely reflect a desire to ensure that no individuals entitled in principle to a national voice are left without any Congressional representation whatsoever.  Likewise, the upper bounds (implicit in the prescribed minimum number $\Dmin$ of persons per Congressional representative) are not intended to deliberately under-represent larger states, but merely reflect a sense of the scale of workable district sizes or representational ratios that might contribute to efficient operation of the House (providing a diversity of ideas, but not an unwieldy number of members), or an effective democratic dynamic (wherein each member represents a sufficiently large sampling of people and viewpoints, balancing opportunity for healthy debate with likelihood of common ground and compromise).\footnote{In early drafts of the Constitution, $\Dmin$ was apparently set at $40\hspace{1pt}000$, but George Washington thought that number too high, and asked the convention to reconsider.  In the \textit{Federalist Papers} 55 \cite{hamilton:1788}, Madison warned of the democratic dangers of not expanding the House as the population grew, and in fact later proposed (along with what became the Bill of Rights) another amendment (never ratified) that would have mandated maximum district sizes: 
\begin{quote}
After the first enumeration ..., there shall be one representative for every 30,000, until the number shall amount to 100, after which the proportion shall be so regulated by Congress, that there shall be not less than 100 representatives, nor less than one representative for every 40,000 persons, until the number of representatives shall amount to 200; after which the proportion shall be so regulated by Congress, that there shall not be less than 200 representatives, nor more than one representative for every 50,000 persons.
\end{quote}}

Thus, we are also led to reject methods based on rounding the states' so-called \textit{fair shares},\footnote{Again, see Appendix~\ref{parameters} for a precise definition of these quantities.} because the very notion of ``fair'' share as used in the literature overemphasizes the claims of states rather than focusing on individuals, and otherwise exaggerates what should be an altogether auxiliary role played by the lower and upper bound constraints.

Regarding so-called \textit{divisor} methods: amongst all of the academic literature, and committee reports, and Congressional debates, we can see no principled way to select definitively amongst the myriad quotient \textit{rounding rules} in and of themselves.  (See Appendix~\ref{typologies} for more details).  Disagreement at this level has been ongoing for a couple of centuries, and though Huntington's method of so-called \textit{Equal Proportions} \cite{huntington:1921,huntington:1928,huntington:1941} has earned approval from the National Academy of Sciences and has been entrenched in Congressional statute for over $75$ years, the various justifications offered have proven less than compelling, the method is of dubious universality, being inapplicable without \textit{ex post} modification to party-list systems, and little actual consensus has ever emerged amongst experts on either the political or mathematical sides.

Choice of \textit{pairwise comparison} tests would appear equally arbitrary, despite Huntington's advocacy to the contrary,\footnote{And anyhow, workable pairwise comparison criteria (which lead to transitive rankings) all turn out to be mathematically equivalent to divisor rules.} and criteria involving mere \textit{local} optimality or \textit{Pareto} optimality lack compelling rationale when there exist well-behaved rules guaranteeing globally optimal apportionments that maximize some overall measure of fairness.  Additionally, a global optimization method induces a total ordering of all feasible apportionments, so can be used not only to select an optimal apportionment but to compare or rank any and all proposed apportionments in a meaningful way.

So we narrow the methodological search to apportionment via \textit{variational principle}, based on \textit{constrained global optimization} (i.e., minimization of some measure of unfairness, inequity, inequality, or disproportionality, or maximization of some measure of fairness, equity, uniformity, or proportionality), subject to auxiliary inequality constraints on lower or upper bounds for each state, and to equality or inequality constraints on the total number of seats to be apportioned.

\subsection{Match Quotas, Equalize District Sizes, or Balance Representational Weights?}


We have decided that the apportionment problem is one of mapping a set of rational numbers to a set of nearby integers in some \textit{optimally fair} way.  But what precisely should be near to what, and how should closeness or fairness be measured?

\newpage
Obviously  the exact state quotas $q_s = \tfrac{p_s}{P}R$ are not generically integers, so the $a_s$ cannot be chosen exactly equal to the $q_s$, without violating Constitutional (and common-sense) mandates dictating that each state receive at least one representative but always a whole number of representatives.

As a corollary, given the proposed apportionments, the state-level average district sizes $\bar{d}_{s} = \tfrac{p_s}{a_s}$, cannot all be all chosen to be equal to the national average $\bar{D} =\tfrac{P}{R}$, and likewise the representational weights (or equivalently, polling probabilities)  $\bar{w}_{sn} = \tfrac{a_s}{R} \tfrac{1}{p_s} = \tfrac{a_s}{q_s}\tfrac{1}{P}$ cannot all assume the value associated with an ideal direct democracy, namely $\bar{w} = \tfrac{1}{P}$ for all represented individuals in all districts of all states.  Almost always, there must be some cost in terms of democratic fairness in order to gain the presumed practical efficiencies of indirect or representative government.  This is of why we speak of apportionment as a \textit{problem}.

(Again, all of these various quantities, including the represented populations $p_s$, proposed apportionments $a_s$, and various related quantities such as quotas $q_s$, average district sizes $\bar{d}_{s}$, and weights of representation $\bar{w}_{sn}$, are discussed in more detail in Appendix~\ref{parameters}).

From the perspective of a single state,\footnote{Or rather, from the point of view of the state's own governor, legislature, attorney general, or other political leadership.} the most natural assessment might compare $a_s$ to $q_s$, indicating whether the state received as many Congressional seats as to which it is ``entitled'' based on its population.  Or else perhaps the state will look at its average district size $\bar{d}_s$ in comparison to some other state's average $\bar{d}_{s'}$, in attempting to justify a claim that it marginally ``deserves'' an additional seat more than the other state.

From the perspective of any one incumbent Congressional Representative, the key comparison would be between his or her district size $d_{sk}$ and  the average size $\Dbar$, i.e., whether his or her district is larger than it ``should'' be, which effects the difficulty and cost of campaigning and provisioning of constituent services.  (After apportionment, but before Congressional districts are assigned, the state-level average district size $\bar{d}_k$ would again become the natural quantity to consider).
 
But to reiterate, we maintain that the relevant perspective is not that of a state (or party, or a congressional representative), but of the represented individuals themselves.  And the only natural comparison that relates directly to individual representation will involve the weights of representation, or equivalently, polling probabilities.

While virtually any figure-of-merit to be optimized may be re-expressible in terms of state-level variables, fundamentally it should be interpretable as a measure of disproportionality in representation across all individuals, not just discrepancies between seats deserved and seats allotted to states.  But how do we choose a measure of fairness to maximize, or of unfairness to minimize?  Any number of functions might be imagined.  To avoid the same sort of criticisms of \textit{ad hockery} that we have leveled against other approaches, we had better try to construct this function from compelling principles.


%

\section{Entropic Apportionment}\label{entropic_apportionment}

To recap: we seek to motivate a unique and universal mathematical procedure that allots discrete seats to states by optimizing a natural global measure of equity of individual weights of representation across all represented inhabitants of those states, subject to constraints on the total number of representatives, a minimum number of representatives per state, and a minimum number of persons per representative.

We must tolerate some cost in terms of imperfect equality in order to gain the presumed advantages and efficiencies of a representative government.  But the task of fair apportionment is to contain this cost by rounding the allotments of seats to whole numbers, so as to minimize some principled measure of discrepancy between the actual and ideal weights of representation, or equivalently (since the ideal weights are all equal), to maximize a measure of the uniformity of the realized weights across all represented individuals.

Shifting to an equivalent interpretation in terms of polling probabilities, we can equivalently speak of making the indirect polling probability distribution $\bar{\pi}_{sk} =  \tfrac{a_s}{R} \tfrac{1}{p_s}$ as close as possible to the direct polling distribution, $\bar{\pi} = \tfrac{1}{P}$.
But once we accept that weights of representation can be interpreted as probabilities (or even just as nonnegative, additive proportions or measures), we contend that the choice of a figure-of-merit becomes clear, as then various arguments all point to a single functional form used to measure the discrepancy between two probability distributions over the same space, namely the relative entropy, otherwise known as the Kullback-Leibler divergence, or discrimination information \cite{kullback:1959,cover_thomas:1991,mackay:2003,jaynes:2003}.

And since the direct distribution is by design completely uniform across all represented individuals, choosing the indirect probability distribution to be as close as possible to the direct probability distribution amounts to choosing the former to be as uniform as possible, and similar arguments single out the Shannon entropy as the natural measure of uniformity of any probability distribution.

The entropy and relative entropy are closely related quantities, and in the current context, the relative entropy can just be expressed in terms of a difference of Shannon entropies, so entropy maximization and relative entropy minimization lead to equivalent variational principles.  More generally, the relative entropy offers somewhat more flexibility than the entropy itself, and can allow for weighted voting or representational schemes.\footnote{For example, in corporate governance, individual votes might be weighted by the number of stock shares.}


\subsection{Some Characterizations of and Motivations for Relative Entropy}

For a more thorough discussion of entropy and relative entropy and an explication of a variety of information-theoretic arguments all leading to relative entropy minimization or entropy maximization, see Appendix~\ref{entropy} (starting on page \pageref{entropy}).  For a discussion of (relative) entropy as a measure of distributional inequality in the context of economics and ecology, see Appendix~\ref{inequality} (starting on page \pageref{inequality}).  Here we will summarize one fundamental characterization theorem, and suggest some additional heuristic interpretations or motivations.

We seek a quantified measure of  discrepancy, departure, or \textit{divergence}, $\K(\bv{\omega} ; \bv{\mu})$ between one ``trial'' distribution $\bv{\omega}$ and an ideal, reference or target distribution $\bv{\mu}$.  These can be any probability distributions---or really any nonnegative, normalized weights over the same set of possibilities---but for our task, ultimately we can interpret the $\omega_{sn} = \bar{\pi}_{sn}$ as the  indirect or Congress-mediated weights of representation across all represented individuals assuming uniform intra-state representation, while $\mu_{sn} = \bar{\pi} = \tfrac{1}{P}$ will represent the direct or democratically ideal distribution of weights of representation.  But we can proceed somewhat more generally.

First, the most basic notions of fairness demand that only the numbers of people in the various states should matter, not the identities of particular people, the identities of particular states, nor which particular people reside in which particular states.  Beyond how many people live in each state, the measure of divergence should ignore any other demographic features or labels.  How we arbitrarily choose to name individuals or their states, or the order in which we happen to list states, or individuals within states, should not affect our judgements of similarity between different distributions of representational weight.  This means that the divergence measure $\K(\bv{\omega}; \bv{\mu})$ must be numerically \textit{invariant under permutation} of the arbitrary labels we use to refer to states and individuals within states.

Second, adding ``irrelevant''\footnote{Irrelevant, that is, only in the legal context of Congressional apportionment.  Obviously their interests, rights, and dignity should matter in other contexts.} individuals (for example, foreign nationals, or unrepresented residents of Washington, D.C.) to the list of individuals, but with zero direct and zero indirect weight of representation, should not change the measure.  This says that the divergence $\K$ must be \textit{extensible}, or  invariant under the addition or removal of superfluous alternatives with \textit{no} weight under \text{either} the trial or target distributions.
 
Third, if we aggregate some of the individuals into a group and consider only their aggregated weight of representation in the measure of divergence, then the corresponding \textit{change} $\Delta \K$ to the divergence $\K$ should depend only on the weights of representation within the grouped individuals, and not the weights of other individuals.  We then say that $\K(\bv{\omega}; \bv{\mu})$ satisfies a \textit{branching} property.

Together, these imply that the divergence must assume an \textit{additive} form,
\be
\K(\bv{\omega}; \bv{\mu},S) = \sum\limits_{s = 1}^{S} \sum\limits_{n = 1}^{p_s}  \omega_{sn} \, \psi( \omega_{sn}, \mu_{sn}),
\ee
where $\psi(\omega_{sn}, \mu_{sn})$ should assume the same functional form  for all individuals in all states.   Therefore, the divergence $\K$ must be some \textit{average} (with respect to the trial distribution) of some function $\psi(\omega, \mu)$ of the trial and target weights assigned to each person.

Fourth, making small changes to the weights of representation should lead to small changes in the measure of discrepancy.  That is, we demand that $\K$ is a \textit{continuous} function of all of its inputs (except perhaps at certain boundary points where the weights vanish).

Fifth, if the two distributions of weights $\bv{\omega}$ and $\bv{\mu}$ are exactly equal, then obviously there is no discrepancy, and so a measure of discrepancy $\K(\bv{\mu}; \bv{\mu})$  should vanish. Otherwise, there must be some discrepancy, so the measure should be non-negative.  We say that $\K(\bv{\omega}; \bv{\mu})$ is \textit{positive semidefinite}.

Sixth, the divergence between the best-case uniform distribution of representational weights and a worst-case ``dictatorial'' distribution of weights should be positive, but finite:
\be
0 < \K\bigl(1,0, \dotsc, 0; \tfrac{1}{P}, \dotsc, \tfrac{1}{P}, S\bigr) < \infty, 
\ee
for any populations size $P$ satisfying $1 < P < \infty$.  The lower bound flows from positive semidefiniteness as described above.  The upper bound is imposed because giving no weight to individuals deserving some weight may be highly undesirable, but is not logically contradictory.  On the other hand, it would be nonsensical to assign positive weight of representation to some individual who is not even entitled to it---otherwise, where would never even know when to stop including people in the summation---so the divergence can be infinite in such cases.  We refer to this as an \textit{ordering} assumption.

Seventh, and finally, it is natural to demand that divergence should be \textit{additive} under conditions of independence.  So for example, in order to measure the degree of disproportionality of representation in the U.S.A.\ via the House of Representatives, \textit{and} in Mexico via its Chamber of Deputies, the overall divergence should just be the sum of the divergences associated with the two nations.  Or, if we consider conducting multiple \textit{independent} polls of individuals within one country, then the overall divergence between sampling distributions based on the indirect versus direct weights of representation should be the sum of divergences for each poll separately.

It turns out \cite{aczel:1975} that the only functional form for the divergence $\K(\bv{\omega}; \bv{\mu},S)$ satisfying all seven of these properties is
\be
\K(\bv{\omega}; \bv{\mu}, S) =  \kappa \sum\limits_{s = 1}^{S} \sum\limits_{n = 1}^{p_s} \omega_{sn} \, \log \Bigl[ \frac{\omega_{sn}}{\mu_{sn}} \Bigr],
\ee
which is the Kullback-Leibler divergence, or relative entropy, of the distribution $\bv{\omega}$ relative to the distribution $\bv{\mu}$.  Here $\kappa$ is some positive constant reflecting the choice of units and/or the base of the logarithms.  Except where otherwise noted, we will take $\kappa = 1$ and $\log x = \log_2 x$, thereby measuring (relative) entropies in \textit{bits} (short for \textit{binary digits}).

The relative entropy sees widespread application in physics, communication theory, statistics, economics, machine learning, image processing, bioinformatics, ecology, and other fields, both as a measure of divergence between probability distributions, and as the basis of variational principles or other algorithms used to: assign probability distributions based on limited information; approximate one probability distribution by a simpler model; or test statistical hypotheses predicting different probability distributions.  

Relative entropy then naturally extends to the problem of assessing representational inequality, and the associated task of minimizing such inequality in the optimal apportionment of Congressional seats to states, or parliamentary seats to parties.  As mentioned, entropy has already been used in closely related distributional, social choice, and social welfare contexts. For example, Sewell, \textit{et al}. \cite{sewell:2009} have developed a maximum-entropy stochastic voting scheme, and Theil \cite{theil:1967} introduced an entropic measure of inequality for income or wealth distributions.  In fact, entropy reflects inequality in the share of the total income (in a country, or state, or city, etc.) controlled by given shares of the population, in the same way that it measures inequality in the share of the total votes in the U.S.\ House of representatives controlled by given shares of the population.

No other measure besides relative entropy can possess all of the defining characteristics listed above, nor several other natural mathematical properties enjoyed by relative entropy.  This alone is enough to recommend it.  Broadly brushed, the following interpretations of or rationales for entropic apportionment may be further offered:
\begin{itemize}
\item the overarching objective is to give an equal voice to equal numbers of people.  The entropic approach takes very seriously, indeed almost literally, the goal of giving individuals an \textit{equal say} in their governance, because communication theory provides an explicit way to compare the democratic signal associated with their ``voices'' and to quantify the extent to which individuals do have an equal say in Congress;
\item  institutional and cultural norms effectively \textit{presume} that everyone gets equal representation, which means that, implicitly, democratic ``messages'' are encoded according to a the direct polling distribution.  But messages are effectively sampled based on the indirect, Congress-mediated polling distribution.  Minimizing the relative entropy minimizes the number of spurious or extraneous bits of information associated with this mismatch;
\item we choose that apportionment which would make it most difficult to distinguish direct from indirect polling (if one is given the results of repeated polling, and asked to guess which sampling procedure was used);\footnote{In the social choice literature, one reads of the ``random dictator'' model which obviates certain impossibility theorems---at the expense of leaving a decision in the hands of  one randomly chosen person.  For what it is worth, we could also interpret entropic apportionment in terms of a hypothetical choice of a  random dictator, made either directly or indirectly.  The entropic apportionment is that which is predicted to yield a particular individual as randomly or uniformly as possible if we would have to choose indirectly.}
\item we seek the most \textit{uniform} distribution of representational weights across individuals, and Shannon entropy is the natural measure of uniformity, and so maximizing entropy maximizes uniformity;
\item we seek the least informative or least biased distribution of representational weights across individuals, such that no more information is put into the indirect polling probability distribution than is needed to satisfy the constraints, and no individual is arbitrarily given more weight under indirect sampling than is justified by the prior information;
\item the more unequal are the representational weights, the less surprise we experience, less uncertainty we resolve, and less information we receive in learning of the outcome of indirect (Congress-mediated) polling;
\item  even if one is reluctant to attach a probabilistic interpretation to the  weights of representation, they are by necessity nonnegative (the worse that can happen is individuals lack any representation) and additive (the net weight of representation of any group is the sum of the weights of representations for all individuals within the group).  Compatible with this structure, the most natural variational principle measuring departure  of one such measure from another is once again the relative entropy.
\end{itemize}
As to the last point: we do continue to maintain that it is both meaningful and natural to interpret the weights of representation \textit{probabilistically}, namely as polling probabilities, and so we may apply without apology the full apparatus of information theory.  But  even if we were to avoid attaching probabilistic meanings to the representational weights, we would still be led to the Shannon entropy as the natural measure of the uniformity of these weights, and the relative entropy as a measure of divergence or discrepancy between two distributions of weights, in particular those distributions associated with representative (Congress-mediated) versus direct (unmediated) democracy, and therefore to the principle of relative entropy minimization or entropy maximization as the most justifiable variational principle for apportionment.

Next, we turn to some specific features of this entropic figure-of-merit and its properties in the context of apportionment.

\subsection{Apportionment via Entropic Optimization}

Given the  state populations $\bv{p}$ and a proposed House size $R$, entropic apportionment selects $\bv{a}$ by minimizing what is called the relative entropy or Kullback-Leibler divergence between the indirect and direct polling distributions:
\be
\K(\bv{a}, \bv{p},R,S) = \sum\limits_{s=1}^{S} \sum\limits_{n=1}^{p_s}  \bar{\pi}_{sn} \log \bigl[  \tfrac{\bar{\pi}_{sn} }{\bar{w}} \bigr] =   \sum\limits_{s=1}^{S} \sum\limits_{n=1}^{p_s}   \tfrac{a_s}{p_s R} \log \Bigl[\tfrac{\frac{a_s}{p_s R}}{\frac{1}{P}} \Bigr] =  \sum\limits_{s=1}^{S}  \tfrac{a_s}{R} \log \Bigl[ \tfrac{a_s/R}{p_s/P} \Bigr],
\ee
subject to the constraint $\sum\limits_{s=1}^{S} a_s = R$ on the total number of seats,\footnote{This equality constraint can be replaced with inequality constraints to find the optimal apportionment over some range of House sizes.  This is actually our recommended procedure, as discussed below.} and to lower and upper bound constraints on individual states, $\lambda_s \le a_s \le u_s$.  That is, we minimize the discrimination information between the probability of selecting constituents directly and at random in a pure democracy, versus selecting constituents indirectly via their congressional representatives.  Refer again to Appendix~\ref{entropy} for more details.

This relative entropy further simplifies to
\be
\K(\bv{a}, \bv{q},R,S) = \sum\limits_{s=1}^{S} \tfrac{a_s}{R} \log \bigl[\tfrac{a_s/R}{q_s/R}\bigr] = \tfrac{1}{R}\sum\limits_{s=1}^{S} a_s \log \bigl[\tfrac{a_s}{q_s}\bigr],  
\ee
which is the form typically most convenient for numerical optimization, using the incremental ``greedy'' algorithm outlined in Appendix~\ref{optimization} (beginning on page \pageref{optimization}).  Notice that while this quantity does start off fundamentally as a measure of divergence between the direct and indirect polling probabilities across all represented individuals, it can be expressed as a sum over the state-level variables $\bv{a}$ and $\bv{q}$, and can also be interpreted as the discrimination information between choosing a state at random, with probability proportional to the size of its congressional delegation, and choosing a state with probability proportional to its population.  Crucially, this  discrimination information is derived and motivated at the level of represented persons, but because all represented persons within a state are treated equivalently, the relative entropy over the space of represented person reduces to a relative entropy over the space of States.\footnote{This is possible because (i) the sub-sampling within states was implicitly assumed to be uniform, and (ii) the relative entropy is an $f$-divergence, as defined and discussed in Appendix~\ref{entropy}.}

Because the ideal democratic polling distribution is completely uniform, we can also write the relative entropy as the \textit{deficit} in the Shannon entropy compared to the democratic ideal,
\be
\K(\bv{a}, \bv{q},R,S) = \S_{\stext{max}}(P) - \S(\bv{a}, \bv{q},R)    ,
\ee
where
\be
S_{\stext{max}} = S_{\stext{max}}(P) = \log P
\ee
represents the maximum possible Shannon entropy, associated with choosing uniformly at random amongst the total population of $P$ persons,
and 
\be
\S(\bv{a}, \bv{p},R,S) =  - \sum\limits_{s=1}^{S} \sum\limits_{n=1}^{p_s} \bar{\pi}_{sn} \log \bar{\pi}_{sn} =
- \sum\limits_{s=1}^{S} \sum\limits_{n=1}^{p_s}  \tfrac{a_s}{p_s R} \log \bigl[ \tfrac{a_s}{p_s R} \bigr] = -\tfrac{1}{R} \sum\limits_{s=1}^{S}  a_s \log \bigl[ \tfrac{a_s}{q_s}\tfrac{1}{P} \bigr] 
\ee
is the Shannon entropy when polling persons indirectly via their representatives, and quantifies the degree of uniformity in the representational weights across all represented individuals.  So for given populations $\bv{p}$ and a fixed house size $R$, minimizing the relative entropy $\K$ is equivalent to maximizing the Shannon entropy $\S$, both subject to suitable constraints on the allowed values of $\bv{a}$.

Expressing in this way a relative entropy as the difference between the maximum possible entropy and actual entropy is possible if and only if the relative entropy compares one probability distribution to the completely uniform distribution.  But more generally, the entropy deficit $(\S_{\stext{max}} - \S)$ is known as the \textit{absolute redundancy} and plays an important role in information theory, as it measures the inefficiency of the coding relative to the total capacity of the communication channel.\footnote{The normalized difference $\bigl(1 - \tfrac{\S}{\S_{\stext{max}}}\bigr)$ is known as the \textit{relative redundancy}, or \textit{data compression ratio}.}

From the various expressions for the relative entropy, we may infer that it always lies in the range
\be\label{kbounds2}
0 \le \K(\bv{a}, \bv{q},R,S) \le \log P.
\ee 
The so-called Gibbs inequality (which undergirds the mathematical foundations of statistical mechanics) guarantees  that the relative entropy $\K(\bv{a}, \bv{q},R,S)$ is nonnegative, and vanishes if and only if  $a_s = q_S$ for all states $s = 1, \dots S$.  Of course, for this to happen, it must be the case that all state-level quotas are exact integers, which would be highly unlikely in real-world examples.\footnote{But in such a fortuitous case, virtually any apportionment method worth the name would recognize the assignment $\bv{a} = \bv{q}$ as optimal.}  The lower bound on Shannon entropy entails the upper bound on $\K$, but this bound can only be achieved if exactly one individual out of the population is assigned any positive polling probability, which is not mathematically possible under our assumptions,\footnote{Additionally, this would also violate lower bound and other Constitutional constraints for U.S.\ Congressional apportionment.  The actual largest achievable value of $\K$ consistent with both upper and lower bund constraints depends on the distributional details, but generally speaking, the worst case will leave the largest states having only one seat, and give the smallest state or states the most seats possible without exceeding the upper bounds.}
unless there happens to be a state for which $p_s = 1$.  So generically, both bounds in Equation \eqref{kbounds2} tend to be strict, with the equity of the proposed apportionment decreasing with increasing $\K$ or, equivalently, decreasing $\S$.

Yet another way to re-write the relative entropy is in terms of the resulting (average) district sizes across states:
\be\begin{split}
\K( \bv{\bar{d}}; \bv{a}, \bar{D},S) &= \sum\limits_{s=1}^{S} \sum\limits_{n=1}^{p_s}   \tfrac{a_s}{p_s R} \log \Bigl[\tfrac{\frac{a_s}{p_s R}}{\frac{1}{P}} \Bigr] =  \sum\limits_{s=1}^{S} \sum\limits_{n=1}^{p_s}  \tfrac{1}{R} \tfrac{1}{\bar{d}_s}  \log \bigl[\tfrac{P}{R} \tfrac{1}{\bar{d}_s}  \bigr]
= \log\bigl[ \tfrac{P}{R}\bigr]   - \sum\limits_{s=1}^{S} \tfrac{a_s}{R}  \log [ \bar{d}_s ]  \\
&= \log [\bar{D}] -   \log \Bigl[ \bigl\{ \prod\limits_{s=1}^{S} \bar{d}_s{}^{a_s}  \bigr\}^{\frac{1}{R}} \Bigr]
= \log \biggl[  \tfrac{\bar{D}}{ \bigl\{ \prod\limits_{s=1}^{S} \bar{d}_s{}^{a_s}  \bigr\}^{\frac{1}{R}}   } \biggr].
\end{split}
\ee
Information theory aside, we see that for given population counts $\bv{p}$ and a fixed house size $R$, optimization of inter-state apportionments  according to this entropic rule will have a very simple interpretation.  Note that the argument of the first logarithm is the \textit{arithmetic mean} of the district sizes across all states, $\Dbar = \tfrac{P}{R} = \tfrac{1}{R} \sum\limits_{s=1}^{S} a_s \bar{d}_s$,  whereas the argument of the second logarithm is the \textit{geometric mean}  of the (intra-state average\footnote{Note that the intra-state average district sizes here still refer to arithmetic averages, $\bar{d}_s = \frac{1}{a_s} \sum\limits_{k=1}^{a_s} d_{sk} = \tfrac{p_s}{a_s}$.}) district sizes $\bar{d}_s$.  Recalling that the logarithm function is strictly monotone, we see that minimizing the relative entropy is equivalent to maximizing the geometric mean of all district sizes for a fixed arithmetic mean of district sizes.  The arithmetic-geometric-mean inequality\footnote{Recall that the \textit{Harmonic-Geometric-Arithmetic-Quadratic Mean Inequality} says that if $x_1, \dotsc, x_n$ are all positive real numbers, then their harmonic mean, geometric mean, arithmetic mean, and quadratic mean (root-mean square) satisfy:
$0 < \Bigl( \frac{1}{n} \sum\limits_{j = 1}^{n} \frac{1}{x_j}  \Bigr)^{-1} \le \Bigl( \prod\limits_{j = 1}^{n}  x_j \Bigr)^{1/n} \le \Bigl( \frac{1}{n} \sum\limits_{j = 1}^{n} x_j  \Bigr) \le \Bigl( \frac{1}{n} \sum\limits_{j = 1}^{n} x_j^2  \Bigr)^{1/2}$, with equality between the means if and only if $x_1 =  \dotsb = x_n$.  The inequality actually generalizes to any sequence of power means.} guarantees that the geometric mean is no greater than the arithmetic mean, with equality if and only if all intra-state averaged district sizes are equal, $\bar{d}_s = \Dbar$ for $s = 1, \dotsc, S$, consistent with the positive definiteness of $\K$.  More will be said on this formulation below, when we discuss evaluation of district sizes themselves.

%

\subsection{Comparison of Huntington, Webster, and Entropic Apportionments}\label{comparison_HWE}

The Huntington (or Huntington-Hill) \cite{huntington:1921,huntington:1928} and Webster (or Webster-Willcox) \cite{willcox:1916,willcox:1941} schemes are perhaps the most widely advocated and discussed of the historic apportionment methods, so provide obvious foils for entropic apportionment.  As mentioned, the former is the methodology long mandated by Congressional statute.

In actual application to modern U.S.\ Census data, the Webster and Huntington methods tend to produce either identical apportionments, or similar apportionments, differing perhaps in the allocation of the last one or two representatives between pairs of states, although more profound disagreements are definitely possible.  Applied to the same real-world Census data, the entropic apportionment method tends to agree with both the Webster and Huntington methods when the latter both agree, or more often than not with the Huntington method otherwise, although differences with both can occur.

In artificial examples where the populations can be chosen arbitrarily, so as to intentionally create difficult boundary cases, entropic apportionment tends to accord with both the Webster and Huntington methods when the latter two coincide, to agree with the more plausible of their answers when the differences between the output of the Huntington and Webster apportionments are non-zero but small, or effectively to interpolate between them when the differences become more pronounced. 

See Appendix~\ref{tables} (beginning on page \pageref{tables}) for some numerical examples.

\subsubsection{Taylor Expansions}

Given that the Huntington and Webster methods are the most widely advocated historical approaches, this similar performance on the part of entropic apportionment offers some reassurance of reasonable output for the entropic method, beyond its theoretical justification.  It is not difficult to understand why these three methods tend to produce similar results for assignment of seats, because, apart from an overall pre-factor, their respective figures-of-merit all agree up to second order when Taylor expanded with respect to the apportionments $a_s$, about the exact state quotas $q_s$.

Recall that the Webster objective function to be minimized, usually motivated as the mean-squared deviation in best-case shares of representatives (over all individuals), may be defined as
\be\label{w4}
\U_{\stext{W}}(\bv{a}, \bv{q}; R,S,P) = \tfrac{1}{P} \sum\limits_s p_s \bigl( \tfrac{1}{\bar{d}_s} - \tfrac{1}{\Dbar} \bigr)^2 = \tfrac{R}{P^2} \sum\limits_s  \tfrac{ (a_s - q_s  )^2}{q_s},
\ee
while the Huntington objective function arises as the mean-squared deviation of (best-case) district sizes, and can be expanded as
\be
\begin{split}
\U_{\stext{H}}(\bv{a}, \bv{q}; R,S,P) &= \tfrac{1}{R} \sum\limits_s a_s \bigl( \bar{d}_s - \Dbar \bigr)^2 
 = \tfrac{P^2}{R^3} \sum\limits_s \tfrac{(a_s-q_s)^2}{a_s}  \\
 &=  \tfrac{P^2}{R^3} \sum\limits_s  \tfrac{q_s}{a_s}  \tfrac{(a_s-q_s)^2}{q_s} 
 \approx  \tfrac{P^2}{R^3} \sum\limits_s \tfrac{(a_s-q_s)^2}{q_s} + \dotsc;
\end{split}
\ee
while the Kullback-Leibler divergence is 
\be\label{k_expansion_2}
\K(\bv{a}, \bv{q}; R,S,P) = \sum\limits_s \tfrac{a_s}{R} \log_2 \bigl(  \tfrac{a_s}{q_s} \bigr) \approx \tfrac{1}{2 \ln 2 } \tfrac{1}{R} \sum\limits_s \tfrac{(a_s-q_s)^2}{q_s} + \dotsc,
\ee
(For that matter, the \textit{dual} Kullback-Leibler divergence also has the same leading-order Taylor expansion:
\be
\tilde{\K}(\bv{a}, \bv{q}; R,S,P) = \sum\limits_s \tfrac{q_s}{R} \log_2 \bigl(\tfrac{q_s}{a_s} \bigr) \approx \tfrac{1}{2 \ln 2 } \tfrac{1}{R} \sum\limits_s \tfrac{(a_s-q_s)^2}{q_s} + \dotsc,
\ee
but this functional is disfavored on other grounds compared to primal relative entropy $\K$).

Of course, we are inclined to turn this argument around and assert that the Webster and Huntington optimands are mere rational-function approximations to a more fundamental measure, namely the relative entropy.  This connection is an important if often under-appreciated reason why the chi-squared statistic $\chi^2 = \sum\limits_s  \tfrac{ (a_s - q_s  )^2}{q_s}$ appearing on the right-hand side of these expansions has been so widely used in statistical goodness-of-fit tests: $\chi^2$ turns out to be an \textit{approximation} to the Kullback-Leibler divergence $\K$, or, equivalently, to the \textit{log-likelihood statistic} under multinomial sampling.

But in addition to its more compelling theoretical motivation, the entropic method enjoys an important additional advantage over either the Webster or Huntington methods, by virtue of supplying a definitive \textit{normalization} for the global measure of representational proportionality, which, given the populations $\bv{p}$ across the $S$ states, allows us not only to choose an optimal apportionment $\bv{a}$ given a \textit{fixed} choice for $R$, but also to objectively compare apportionments across different numbers of total representatives, and therefore to assess simultaneously the choice of house size $R$ along with the proposed distribution of these seats.

That is, for a prescribed value of $R$, the entropic, Webster, and Huntington methods often lead to the same or similar constrained optimum $\bv{a}$ when  a ``good'' apportionment exists for which $\bv{a} \approx \bv{q}$, but notice that their respective objective functions, as conventionally defined, all impose very different scaling behavior with respect to $R$ itself.  Only the relative entropy $\K$ provides a well-justified absolute measure of the total inequity in representation across all $P$ {individuals}, one that can be used to compare the overall quality of proposed apportionments as $R$ is varied.  More will be said about this below, in Section~\ref{house_size}.

\subsubsection{Sequential Optimization and Recursive Optimization}\label{recursive_HWE}

All three of these apportionment methods also share additional fundamental properties.\footnote{However, these three methods are not unique in this respect.  Various other methods, such as the dual entropic method, can also enjoy these same properties.}  Notice that their respective objective functions can be re-written as
\bsub
\begin{align}
\U_{\stext{H}} &= \tfrac{1}{R} \sum\limits_s a_s\bar{d}_s^2  - \Dbar^2 = \tfrac{1}{R} \sum\limits_s \tfrac{p_s^2}{a_s} - \Dbar^2,\\
\U_{\stext{W}} &= \tfrac{1}{P} \sum\limits_s p_s \tfrac{1}{\bar{d}_s^2} - \tfrac{1}{\Dbar^2} = \tfrac{1}{P} \sum\limits_s \tfrac{a_s^2}{p_s} - \tfrac{1}{\Dbar^2},\\
\K &= \log \Dbar - \tfrac{1}{R} \sum\limits_s a_s \log \bar{d}_s = \log \Dbar - \tfrac{1}{R} \sum\limits_s a_s \log \tfrac{p_s}{a_s},
\end{align}
\esub
which in each case is of the general form
\be\label{u2}
\U(\bv{a}, \bv{p}, R, S) =  \zeta(R,S,P) \,\sum\limits_{s=1}^{S} \psi(a_s, p_s) + \eta(R,S,P),
\ee
where $P = \sum\limits_s p_s$ and $R = \sum\limits_s a_a$ as usual, but, once the common pre-factor $\zeta(R,S,P)$ is accounted for, any one contribution $\psi(a_s, p_s)$ for the $s$th state does not depend explicitly on $R$, $S$, or $P$, nor on the values of $a_{s'}$ or $p_{s'}$ for $s' \neq s$.

This decomposability has two very appealing consequences.  It is shown in Appendix~\ref{optimization} that any apportionment method that can be formulated via constrained optimization of an additive,  discretely-convex function may be implemented via a ``greedy'' algorithm, where individual seats are allotted in succession by local optimization.  But in general, the $q_s$ that appear in such an objective function must be frozen at their ``final'' values determined by $\bv{p}$ and the total number $R$ of seats to be apportioned.  It follows that intermediate stages of the greedy optimization procedure would \textit{not} necessarily correspond to optimal apportionments of fewer than $R$ seats.  However, if the objective function assumes the form \eqref{u2}, for the purposes of determining the optimal apportionments $\bv{a}$ that minimize $\U$ given fixed populations $\bv{p}$, the pre-factor $\zeta(R,S,P)$ and the offset $\eta(R,S,P)$ are irrelevant, and can be dropped.  But then the function $\sum_{s=1}^{S} \psi(a_s, p_s)$ being optimized has no explicit dependence on $R$, so during the greedy allocation, the intermediate stages {do} yield the optimal apportionments for successive house sizes up to final total $R$. 

In other words, this structural feature of the objective functions (together with convexity) implies that the optimal apportionments, according to any of these three methods, can be generated \textit{sequentially}.  That is, the optimal apportionment of $(R + \Delta R)$ seats can be found first by finding the optimal apportionment of $R$ seats, then, continuing from this assignment, allotting the remaining $\Delta R$ seats by locally optimizing the same measure.  

Second, the apportionments resulting from optimization of objective function of this form will satisfy a certain sort of \textit{recursive self-consistency}: suppose the optimal apportionment $\bv{a}$ for $R$ seats across all $S$ states is found.  If we arbitrarily split the states into two (or more) subsets, and re-perform the optimization within each subset separately, while constraining the sub-total of seats within each subset to be equal to the sum of seats assigned altogether to those states in the aforementioned global optimization, then the resulting optimal apportionment will be unchanged.

Neither of these properties necessarily holds for objective functions that cannot be written in the form \eqref{u2}, for example with the minimax criteria of the Adams or Jefferson methods, or the distance-from-quota function associated with the Hamilton-Vinton method, or, say, with any optimand of the form \mbox{$\U_{\alpha\beta} = \tfrac{1}{R} \sum\limits_s a_s \bigl\lvert\bar{d}_s^\beta - \Dbar^\beta \rvert^{\alpha}$} for $\alpha \neq 2$ and $\beta \neq 0$.

%

\subsection{Equivalent Formulations of Entropic Apportionment}

Although we have argued that global (constrained) optimization is the most natural mathematical framework for apportionment, and have verified an efficient algorithm to find the exact optimum in the case of entropic apportionment, it may prove reassuring that the entropic apportionment method can also be formulated and implemented explicitly either as a ranking method, divisor method, or else a pairwise comparison method.

Actually, in our greedy optimization algorithm we already have the makings of an equivalent ranking method.  For any specified house size $R$, minimizing $\K = \sum\limits_s \tfrac{a_s}{R} \log \tfrac{a_s}{q_s}  = \sum\limits_s \tfrac{a_s}{R} \log \tfrac{a_s/R}{p_s/P}$ is equivalent to maximizing $\sum\limits_s a_s \log \tfrac{p_s}{a_s}$.  As shown in Appendix~\ref{optimization}, the associated optimal apportionment can be generated by assigning the seats one by one, to the state which will lead to the largest increase in $\sum\limits_s a'_s \log(p_s/a'_s)$ compared to the previous seat counts.  That is, we assign the next seat to the state with the largest increment to
 \be
(a'_s +1) \log \tfrac{p_s}{a'_s +1} - a'_s \log \tfrac{p_s}{a_s} = 
\log \Bigl[  \bigl(\tfrac{p_s}{a'_s+1} \bigr)^{a'_s+1}  \bigr( \tfrac{p_s}{a_s}\bigr)^{-a'_s}  \Bigr] = \log \tfrac{ p_s^{a'_s+1}  {a'_s}^{a'_s}   }{ p_s^{a'_s} (a'_s + 1)^{a'_s+1} }
= \log \tfrac{ p_s {a'_s}^{a'_s}   }{ (a'_s + 1)^{a'_s+1} },
\ee
and since the logarithm is monotonic, this is equivalent to a standard ranking method that awards seats in decreasing order of the ranking index 
\be
\rho(a'_s, p_s, P, R) \propto  \tfrac{ e \, p_s {a'_s}^{a'_s} }{(a'_s + 1)^{a'_s+1} },
\ee
where we have included an overall factor of the Napier-Euler number $e = 2.71828\dotsc$ anticipating a connection to the quotient rounding rule to be introduced momentarily.  As we have just seen, along the way to the final target house size $R$, this algorithm automatically generates optimal apportionments for all house sizes ranging from $\sum\limits_{s=1}^{S}  \lambda_s$ (just equal to $S$ for the U.S.\ Congressional case) up to $R$, inclusive.

These ranking indices can also be re-written as
\be
\rho(a_s, p_s) = \frac{p_s}{\I(a_s, a_s+1)},
\ee
where $\I(a_s, a_s+1)$ is the so-called \textit{identric mean} of the successive integers $a_s$ and $a_s+1$, namely
\be
\I(a_s,a_s+1) = \tfrac{1}{e} \tfrac{(a_s+1)^{a_s+1}}{a_s^{a_s}}.
\ee
Then, by interpreting the ranking index $\rho(a_s, p_s)$ as the largest value $D^{\star}$ for the divisor $D$ at which the quotient $Q_s = \tfrac{p_s}{D}$ would be rounded up to exactly $(a_s+1)$, we see that the entropic method is also equivalent to a divisor method using the identric mean $\theta(Q_s) = \I(\lfloor Q_s \rfloor, \lceil Q_s \rceil)$ as the rounding threshold for quotients.\footnote{Agnew \cite{agnew:2008} also offers a proof of this equivalence, but his argument does not mention the convex nature of the optimization, and so strictly speaking can only demonstrate that the identric-mean divisor and ranking methods lead to a \textit{local} minimum of the relative entropy.  But in fact the constrained optimum so obtained will be global.}

Alternatively, the entropic apportionment can be characterized in terms of pairwise stability, or Pareto optimality, with respect to a Huntington-type comparison test, where the comparison function
\be
{T}_{\K}(a_0, q_o; a_u; q_u) =  \sgn( q_u a_o  - q_o a_u ) \, \bigl[  a_o \ln( \tfrac{a_o}{q_o} ) + a_u \ln( \tfrac{a_u}{q_u}) + (q_o - a_o) + (q_u - a_u) \bigr]
\ee
may be taken to define the extent to which the state $o$ is relatively over-represented relative to the state $u$.  Starting with any initial apportionment consistent with the constraints, pairs of states can be compared, and seats shifted one at a time from a relatively over-represented to a relatively under-represented state if the exchange would decrease $T_{\K}$ while leaving all allotments between prescribed lower and upper bounds; and when no further feasible exchanges are possible that would improve matters, an entropically optimal apportionment will have been achieved.  Of course, there is really no need to introduce the function ${T}_{\K}(a_0, q_o; a_u; q_u)$, since the relative entropy can be explored directly, and the global nature of any (constrained) local optima follows from its additivity and convexity properties.


\subsection{More Maximum Entropy}

Actually, once we embrace the maximum entropy principle for assigning prior probability distributions, we can apply it successively at various levels in order to better illuminate the thinking underlying the proposed apportionment rule.

The basic idea is in effect to work hierarchically (and backwards), inferring maximum entropy distributions conditional on the problem parameters, some of which are as yet undetermined, then inferring these parameters by further entropy maximization.\footnote{Finding parameterized probability distributions by maximizing entropy conditional on information not yet known is a common tactic in Bayesian statistical theory \cite{jaynes:1983,jaynes:2003}.}

State populations for apportionment $\bv{p}$ are assumed fixed and known throughout any one apportionment task.  We then presume, counterfactually at this stage, that also $(a_1, \dotsc, a_S)$, and hence $R$, are somehow known.  Consider first the space of $R$ Congressional seats:  to assign the most equitable probability distribution over representatives, we can maximize the entropy, subject to no constraints other than the  assumptions of normalization of the probabilities over the $R$ possibilities.\footnote{It turns out that non-negativity of probabilities will be satisfied automatically upon optimization, and need not be imposed as an explicit constraint.}  Unsurprisingly, the result is just a uniform distribution assigning equal probabilities
\be
\rho_{sk} = \tfrac{1}{R}
\ee
to all representatives, indexed by a $(s,k)$ pair for $k = 1, \dotsc, a_s$ and $s = 1, \dotsc, k$.

Next consider distributions over the $P$ represented persons.  The unconstrained maximum entropy  distribution $\mu_{sn}$ ( over the set of all represented individuals for $n = 1, \dots, p_s$, and $s = 1, \dots, S$) will also be a uniform distribution, with probability $\tfrac{1}{P}$ per person, but this does not account for the fact that individuals are actually to be represented through their Congressional delegates.  If we add the minimal additional constraint that the total weight of representation of the population within any state must equal the representational share in Congress of that state's delegation, 
\be
\sum\limits_{s=1}^{p_s} \mu_{sn} = \sum\limits_{k=1}^{a_s} \p_{sk} =  \sum\limits_{k=1}^{a_s} \tfrac{1}{R} = \tfrac{a_s}{R},
\ee
then the result is, again not surprisingly, a piecewise uniform distribution,
\be
\mu_{sn} = \tfrac{1}{p_s} \tfrac{a_s}{\sum\limits_{s'=1}^{S} a_{s'}} =  \tfrac{1}{p_s} \tfrac{a_s}{R}  = \tfrac{1}{R} \tfrac{1}{\bar{d}_s},
\ee
equal for all persons (indexed by $n$) within the $s$th state.\footnote{Here we are using \textit{rho} for the ``representatives,''  and \textit{mu} for the ``masses.''}

But now we can maximize the entropy once more, this time with respect to the $(a_1, \dotsc, a_s)$  parameters, in order to find the optimal apportionment assuming a fixed value of $R$.

Extending this idea, we could then also maximize entropy with respect to $R$ itself, and/or with respect to possible district partitioning, ideas to be explored in the next section. 


\section{Beyond Mere Apportionment:  Evaluating the Fairness of House Sizes and District Sizes}

A great advantage of our entropic approach is that it allows use of the very same information-theoretic measure of uniformity to assess or compare different proposed apportionments of a fixed total of $R$ seats, but also to evaluate different choices for the size $R$ of the House itself, or the sizes of districts within states, and can even compare apportionments across time or across different populations. 

\subsection{Choosing the House Size}\label{house_size}

Besides generating optimal apportionments $\bv{a}$ for any given $R$, the output of the entropic apportionment algorithm also provides a principled and meaningful way to compare possible apportionments across different house sizes (all for the same populations $\bv{p}$).  As $R$ is varied, and $\bv{a}$ is chosen optimally at each value of $R$ under consideration, a smaller value of the relative entropy $\K$, or equivalently, a larger value of the Shannon entropy $\S$, indicates objectively a more fair apportionment overall.

Equivalently, according to our entropic criterion, different values of $R$ are to be preferred (at least in terms of representational equity, if not necessarily institutional efficiency) to the extent that the \textit{ratio} between the arithmetic and geometric means of district sizes across states is made closer to unity.

While other apportionment optimands such as the Huntington-Hill or Webster-Willcox figures-of-merit arguably could be deployed to select $R$, we would seem to lack any principle way to fix the $R$-dependence in the pre-factor, which would not effect the apportionment $\bv{a}$ for given $R$, but would effect the choice of $R$ itself.

Obviously, practical and institutional concerns are paramount when addressing the question of  House size, but conventionally, a variety of mathematical suggestions for selection or optimization of $R$ have been considered in the past.  For instance:
\begin{enumerate}
\item When using a quota method, one can choose a nominal House size $R_0$, calculate quotas, and then round all quotas based on the chosen rule, leading to a final House size $R$ close but not necessarily exactly equal to $R$.  However, this begs the question of how to choose $R_0$ in the first place. 

\item When using a divisor rule, one can adopt a divisor $D$ representing a nominal target for the nominal district size, then calculate $\bv{a$} without further iteration, resulting in an average district size $\bar{D}$ close but not not necessarily equal to $D$.  But we are left wondering how to adopt an initial choice of $D$.
 
\item Invoke the ``Wyoming'' rule, or \textit{smallest-entitled-unit rule}: since every state is required to receive at least one representative, it might seem plausible to try to choose $R$ so that the $\bar{D} = \tfrac{P}{R}$ is as close as possible to the population $p_{\stext{min}} = \min\limits_{s} [ p_s ]$ of the least populated state \cite{shugart:2013}.  While not a terrible idea, this has a minimax feel, and does not actually attempt to account in any sense for disproportionality summed or averaged over all state populations.

\item  Use a ``Rhode Island,'' or \textit{most-over-represented rule}:  sometimes a small-ish state with more than one representative can be  over-represented to a greater extent than the smallest state itself (as measured, say, by the differences between $\bar{D}_s$ and $\bar{D}$).  So we could try to choose $R$ iteratively, in order to minimize the maximum degree of over-representation.  But this minimax strategy emphasizes over-representation over under-representation, and ignores variations in average district size which do not affect the most-over represented state. 

\item Use the \textit{cube root rule}: Taagepera \cite{taagepera:1972,shugart:2014} found an empirical power-law relationship between the overall size of the population of a country and the size of its ``lower-house'' national assembly, where  approximately $R \sim P^{1/3}$.
Using a very simple model based on the ideas of representatives as ``interest aggregators,'' and efficient balancing of representative-constituent versus inter-representative communication channels, he was able to motivate a scaling relation of the form
\be
R \approx (2 \alpha P)^{\frac{1}{3}}
\ee
in the typical limit for which $P \gg R$, and where $\alpha$ is supposed to represent the ``politically active'' fraction of the population, which might be taken to be something like the average proportion of the population which are literate adults.\footnote{Based on U.S.\ Census estimates, about $77\%$ of the US population is aged $18$ or older and about $86\%$ is estimated to be literate.  So as of 2017, this would suggest an optimal House size of about $R \approx 754$.}  But before we get too excited, observe that there is appreciable scatter in the empirical record, and that almost all data look somewhat linear on the sort of  log-log plot used to motivate this power law.
\end{enumerate}

\begin{figure}[t!]
\includegraphics[scale=0.70]{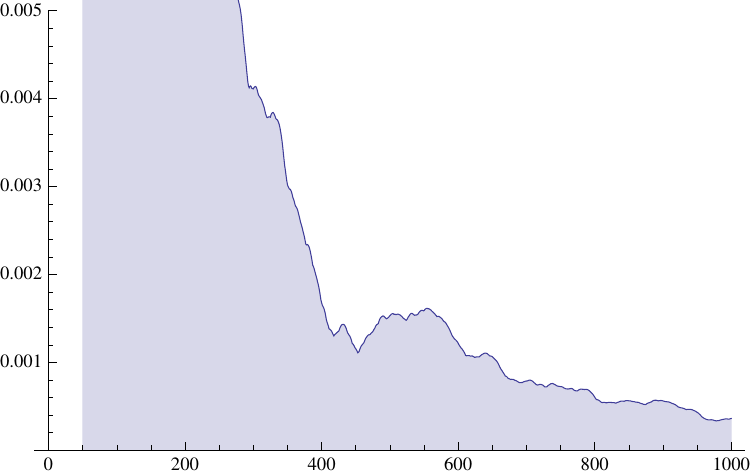}\\
\caption{Trend in discrimination information (in bits) as a function of House size, from $R = 50$ to $R = 1\,000$, for the 1990 U.S\ census data.  The vertical scale is cut off to highlight behavior at larger $R$.  At the prescribed house size of $R = 435$, the Kullback-Leibler divergence is $\K = 0.00140531$, but nearby local optima happen to do better, particularly $R = 418$ (corresponding to $\K = 0.00130027$) or $R = 453$ ($\K = 0.00110856$), as well as significantly larger Houses such as $R = 624$ ($\K = 0.00105713$), and $R = 978$  ($\K = 0.000333297$).}
\label{kl_trend_1990}
\end{figure}

But instead, relative entropy offers a more straightforward and principled way to choose $R$ \textit{a posteriori} (within some range deemed practical \textit{a priori}) without introducing additional \textit{ad hoc} assumptions.  This use of relative entropy is illustrated in Figure~\ref{kl_trend_1990}, which shows the Kullback-Leibler information $\K$ as a function of House size for the 1990 U.S.\ Census data.  There is evident a general downward trend as a function of increasing $R$, for the simple reason that it is typically easier to approximate rational ratios when we have bigger denominators with which to work.\footnote{A branch of number theory known as \textit{Diophantine approximation} concerns the approximation of real numbers (such as quotas or representational weights) by rational numbers with bounds on the denominator.  If $\alpha = \frac{a}{b}$  is some \textit{rational} number to be approximated by another rational number $\tfrac{m}{n}$, then obviously the approximation would be perfect if $m = i a$ and $n = i b$ for some non-zero integer $i$, but if $\tfrac{m}{n} \neq \alpha$, then $\lvert \frac{m}{n} - \alpha \rvert = \lvert \frac{m}{n} - \frac{a}{b} \rvert  \ge \frac{1}{nb}$, so achieving smaller approximation errors generally involves larger denominators.  A  rational number is perfectly approximated by itself, but is in a sense badly approximated by any other rational number.  And somewhat counterintuitively, this sort of lower bound for approximation by rational numbers of other rational numbers is larger than the lower bound for approximating algebraic numbers, which is  larger than the lower bound for transcendental numbers.  Bounds for \textit{simultaneous} approximation of multiple numbers by rationals all with the same denominator can also be derived.} 
But the improvement is far from monotonic, and certain house sizes perform better according to the entropic measure of fairness than do nearby choices.  For example, a somewhat smaller house size of $R = 418$ or a somewhat larger House size $R = 453$ does better than the currently mandated House size of $R = 435$, while still larger House sizes such as $R = 635$ or $R = 978$ do better still.\footnote{A further advantage of a larger House size would be less distortion from democratic ``one person-one vote'' principles in the Electoral College, since a larger number of Electoral College votes earned by House membership would dilute those granted by Senate membership.  Because the current arrangement greatly favors some rural, Republican-leaning states, we do not anticipate great interest in expanding the House from the presently constituted Congress.  There is also currently afoot a popular movement involving the \textit{National Popular Vote Interstate Compact} which sets forth rules for awarding electoral college votes that, if adopted by the legislatures of sufficient states, would guarantee victory for the popular vote winner regardless of how other states would assign their electoral college votes.  Signed by several moderate to large ``bluer'' (Democratic-leaning) states, this movement also appears largely stalled as of 2021, since in order to take effect, some smaller states and traditional ``swing'' states would have to voluntarily sacrifice their exaggerated influence enjoyed under the current system.} 

\begin{figure}[t!]
\includegraphics[scale=0.70]{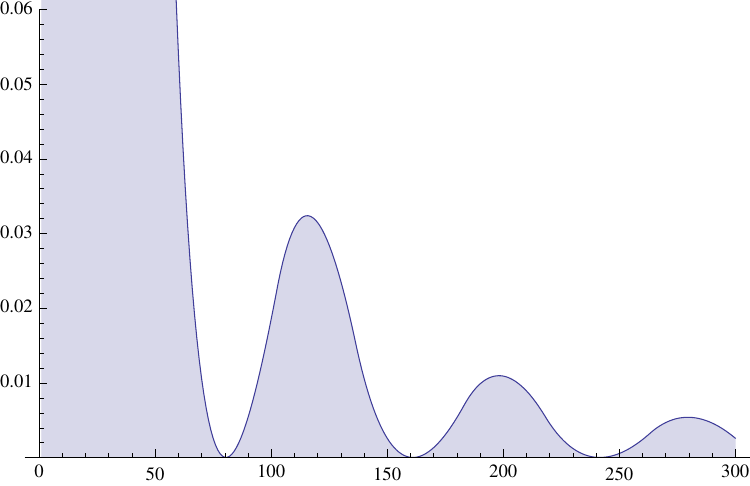}\\
\caption{Trend in discrimination information (in bits) as a function of House size  $R$, for the artificial populations in the example of Table~\ref{table:above_quota} in Appendix~\ref{tables}, which can be found on page \pageref{table:above_quota}.  Notice the ``magic numbers'' leading to especially low values of relative entropy at values of $R$ where all of the low-population states are assigned the same number of seats.  For example, while $\K = 0.0216786$ at the specified house size $R = 102$, the relative entropy drops to $\K = 0.0000160278$ at $R = 161$.}
\label{above_quota_trend}
\end{figure}

\begin{figure}[t!]
\includegraphics[scale=0.70]{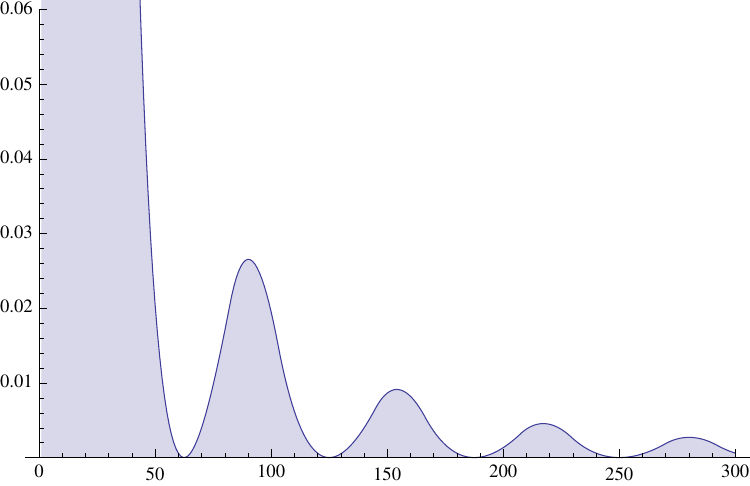}\\
\caption{Trend in discrimination information (in bits) as a function of House size $R$, for the artificial populations in the example summarized in Table~\ref{table:below_quota} on page \pageref{table:below_quota} in Appendix~\ref{tables}.  Again notice the ``magic numbers'' leading to especially low values of relative entropy at values of $R$ for which all of the low-population states are assigned the same number of seats.  For example, while $\K = 0.0219509$ at the specified house size $R = 98$, the relative entropy drops to $\K = .00000303316$ at $R = 125$.}
\label{below_quota_trend}
\end{figure}

Differences can be even more dramatic in artificial examples, as shown in Figures \ref{above_quota_trend} and \ref{below_quota_trend}.  We see that certain house sizes can lead to especially good apportionments, because for fixed populations, certain whole numbers divide more proportionally, with fewer ``leftover'' seats that lead to a mismatch.

\newpage
Therefore, in principle the most sensible approach to apportionment would \textit{not} fix $R$ in advance, but explore apportionments over some range of $R$ values acceptable to stakeholders and consistent with upper and lower bound constraints,\footnote{This was in fact not uncommon practice in early Congressional apportionments in the United States.} and adopt \textit{ex post} the choice leading to the smallest value of $\K$ within the agreed-upon range.  Arbitrarily fixing the House size at $R = 435$ will almost always be sub-optimal, particularly as the overall national population grows and shifts, and is widely recognized as introducing significant distortion---both malrepresentation across states, and average under-representation (i.e., over-sized districts) compared to world-wide democratic norms.\footnote{Again, we do not anticipate that politicians in the Republican party will take action any time soon to correct the exaggerated influence of states like Wyoming and North Dakota.  On the other hand, Democratic-leaning Rhode Island is very over-represented as of 2010, and is projected to be so in 2020, so perhaps there remains some political room to maneuver on this issue. See Appendix~\ref{apportioned_representatives} for further discussion.}

\subsection{Entropic Quantities in Context}

In choosing amongst different possible apportionments for given values of the house size $R$ and fixed population counts $\bv{p}$, the sign of the difference in $\K$ is all-important, but the {magnitude} of the difference in $\K$ is really irrelevant; if a better apportionment with a smaller value of  $\K$ (and consistent with all constraints) is achievable, then there is no justifiable reason not to adopt it, no matter how small the improvement in the value of $\K$.\footnote{Of course, below some level, small differences in the value of $\K$ across different apportionments might be comparable to those attributable to the probable uncertainties or likely temporal variations in the values for the population counts themselves.  But Constitutionally speaking, the certified census totals are to be taken as exact for the purposes of apportionment.  But see Section~\ref{sensitivity}.}

But if using the relative entropy $\K$ to compare, for instance, apportionments across different values of $R$, questions inevitably arise as to the meaning or importance of some given difference in the values of the respective relative entropies $\K$, since increasing or decreasing the total number of seats will inevitably impose some exogenous economic or political costs, that must be balanced against the goal of maximal proportionality.

Admittedly, no definitive answers can be offered, since the issues will involve not just the equity of representation as quantified in the entropy, but also expected utility incorporating other sorts of logistical, administrative, institutional, financial, and political costs and benefits.  But we can say a bit more about what numerical values for relative entropy or entropy are typical, and what a given difference might mean, while highlighting that what might look like a small absolute difference can have a clear information-theoretic consequence.

\subsubsection{Some Orders of Magnitude and Expected Scaling Relations for Relative Entropies of Apportionment}\label{oom1}

To get a rough idea of typical numerical scales, we can ignore disparities between very large and small states, and just assume that $q_s \sim O(\tfrac{R}{S})$ for most every $s$.

Using the fundamental logarithm inequality, but recalling we are measuring entropies in bits, we can write
\be\label{kub2}
0 \le \K \le \tfrac{1}{\ln 2} \tfrac{1}{R} \sum\limits_{s=1}^{S} a_s \bigl( \tfrac{a_s}{q_s} - 1\bigr) =  \tfrac{1}{\ln 2} \Bigl\{  \sum\limits_{s=1}^{S} \tfrac{1}{R}  \tfrac{a_s^2}{q_s} - 1 \Bigr\}.
\ee
For an \textit{arbitrary} (non-optimized) apportionment, we might estimate $a_s^2$ by assuming a multinomial distribution (as if the seats are randomly assigned to the states, regardless of their populations), so that
\be
a_s^2 \sim R \tfrac{1}{S}(1 - \tfrac{1}{S}) + \tfrac{R^2}{S^2},
\ee
and so
\be\label{ubu}
0 \le \K \lesssim \tfrac{1}{\ln 2}\bigl\{  \tfrac{1}{R} S \tfrac{ R \frac{1}{S}(1 - \frac{1}{S}) + \frac{R^2}{S^2}   }{ \frac{R}{S} }  -1 \bigr\}
= \tfrac{1}{\ln 2}\bigl\{ \tfrac{S-1}{R}  \bigr\},
\ee
which is independent of the total population $P$, but approximately scales inversely with the average number of seats per state.

Instead, for an \textit{optimized} or nearly optimized apportionment, we would of course anticipate a smaller value for $\K$, which we may attempt to estimate using the Taylor expansion\footnote{Actually, the Taylor expansion \eqref{k_expansion_2} and the upper bound \eqref{kub2} would lead to similar estimates (within a factor of two or so), since we can use $\sum\limits_s a_s = \sum\limits_s q_s = R$ to write $\tfrac{1}{R} \sum\limits_s a_s \bigl( \tfrac{a_s}{q_s} - 1 \bigr) = \sum\limits_s \tfrac{ (a_s - q_s)^2}{q_s}$.} of Equation \eqref{k_expansion_2}.  In the multi-representative (\ie,  $R \gg S > 1$) limit, we can assume $| a_s - q_s | \sim O(\tfrac{1}{2})$ on average, so that:
\be\label{kest2}
\K \approx \tfrac{1}{2 \ln 2 } \tfrac{1}{R} \sum\limits_{s=1}^{S} \tfrac{(a_s-q_s)^2}{q_s} \sim \tfrac{1}{2 \ln 2 } \tfrac{1}{R}  S \tfrac{( \frac{1}{2} )^2}{\frac{R}{S}}  \sim \tfrac{1}{8 \ln 2 } \tfrac{S^2}{R^2}, 
\ee
which is again independent of the overall population $P$, and small compared to unity when $R \gg S > 1$, scaling inversely with the square of the average number of representatives per state.

In the opposite limit, where $1 \le R \ll S$ (which could happen in party-list type voting systems), so that generically $q_s \sim O(\tfrac{R}{S}) \ll 1$, we would expect $O(R)$ out of the $S$ states to have $O(1)$ seats and $O(S-R)$ to have $0$ seats, and hence
\be
\K \sim \tfrac{1}{2 \ln 2 } \tfrac{1}{R} \bigl\{ R \tfrac{(1)^2}{\frac{R}{S}} + (S-R) \tfrac{ (\frac{R}{S})^2 }{\frac{R}{S}}  \bigr\}
\approx  \tfrac{1}{2 \ln 2 }  \bigl\{ 1 + \tfrac{S}{R} - \tfrac{R}{S} \bigr\} \approx \tfrac{1}{2 \ln 2 } \tfrac{S}{R},
\ee
 suggesting that in the limit of very few seats but states of comparable size, the optimal apportionment cannot do much better than an arbitrary apportionment---perhaps a factor of two or so lower.
 
In between the large-$R$ and small-$R$ limits, approximate scaling laws are a bit harder to deduce, but we can roughly extrapolate.

Another approximation for the scaling of $\K$ may be derived from \textit{Pinsker's inequality}, which (again recalling we are measuring entropies in bits) says that
\be
\sqrt{ \ln 2\, \K} \ge \tfrac{\,1}{\sqrt{2}} \sum\limits_{s=1}^{S} \sum\limits_{n=1}^{p_s} \lvert   \tfrac{a_s}{R} \tfrac{1}{p_s} - \tfrac{1}{P} \rvert  = \tfrac{\,1}{\sqrt{2}} \sum\limits_{s=1}^{S} \sum\limits_{n=1}^{p_s} \tfrac{1}{P}  \lvert \tfrac{a_s}{q_s}  - 1 \rvert.
\ee
When $R \gg S > 1$, the right-hand side is expected to be about
\be
 \tfrac{\,1}{\sqrt{2}} \sum\limits_{s=1}^{S} \sum\limits_{n=1}^{p_s} \tfrac{1}{P}  \lvert \tfrac{a_s}{q_s}  - 1 \rvert
 = \tfrac{\,1}{\sqrt{2}} \sum\limits_{s=1}^{S} \sum\limits_{n=1}^{p_s} \tfrac{1}{P}  \tfrac{\lvert a_s - q_s \rvert}{q_s}
 \sim  \tfrac{\,1}{\sqrt{2}} S \tfrac{P}{S} \tfrac{1}{P} \tfrac{1/2}{R/S} =  \tfrac{\,1}{2\sqrt{2}} \tfrac{S}{R},
 \ee
 from which we can estimate
\be
\K \gtrsim \tfrac{1}{8 \ln 2} \tfrac{S^2}{R^2},
\ee
identical to the estimate based on the Taylor expansion.

In the opposite limit where $1 \le  R \ll S$, we have
\be
 \tfrac{\,1}{\sqrt{2}} \sum\limits_{s=1}^{S} \sum\limits_{n=1}^{p_s} \tfrac{1}{P}  \lvert \tfrac{a_s}{q_s}  - 1 \rvert \sim \tfrac{\,1}{\sqrt{2}}  \bigl[    R \tfrac{1}{S}  \lvert \tfrac{S}{R}  - 1 \rvert + (S-R) \tfrac{1}{S} \lvert -1 \rvert| \bigr]
= \tfrac{\,1}{\sqrt{2}} \bigl[ 1 - \tfrac{R}{S} + 1 - \tfrac{R}{S} \bigr],
\ee
so that instead
\be
\K \gtrsim \tfrac{2}{\ln 2} \bigl[ 1 - \tfrac{R}{S} \bigr]^2,
\ee
which is substantially smaller than the corresponding Taylor series estimate, around $O(1)$ rather than $O(\tfrac{S}{R})$, but keep in mind that this is only a  rough estimate for what is itself only a lower bound.

We may also be interested in the typical magnitude of the \textit{change} $\Delta \K$ in $\K$  under the smallest possible shift in an apportionment, namely the exchange of one representative between one  pair of states.  In the case where $1 < S \ll R$ and the apportionments are not too far from optimal, a Taylor expansion suggests that
\be
\lvert \Delta \K \rvert  \gtrsim \tfrac{1}{\ln 2} \tfrac{S}{R^2},
\ee
which is smaller by a factor of $O(S\inv)$ than the value of $\K$ itself.  In the opposite limit where $1 \le R \ll S$, and most of the apportionments will be either $0$ or $1$, the typical difference would be instead
\be
\lvert \Delta \K \rvert  \gtrsim \tfrac{1}{\ln 2} \tfrac{\sqrt{S}}{R} \tfrac{1}{\sqrt{P}},
\ee
supposing that the population is divided randomly between states.

Under an increment $\Delta R$ in the House size (where we assume $| \Delta R | \lesssim R$), we can roughly estimate a characteristic magnitude in the change in $\K$ from the estimate for $\K$ itself, leading to
\be 
\lvert \Delta \K \rvert \sim \tfrac{1}{4\, \ln 2} \, O\bigl( \tfrac{S^2}{R^2} \tfrac{\Delta R}{R} \bigr)
\ee
in the case where $1 < S \ll R$, and the apportionments are both near optimal, or else
\be 
\lvert \Delta \K \rvert \sim \tfrac{1}{2\, \ln 2} \, O\bigl( \tfrac{S}{R} \tfrac{\Delta R}{R} \bigr)
\ee
when instead $1 \le R \ll S$.

\subsubsection{Small Differences Can Still Be Important Differences}


As an illustrative example, consider the case of the 2000 U.S.\ Census, under different hypothetical values for $R$.  The relative entropy is $\K = 0.00136783$ for the prescribed House size of $R = 435$, but $\K = 0.00108120$ at a moderately larger size $R = 488$, for a difference of only $\Delta \S = -\Delta \K \approx 2.87 \cdot 10^{-4} \; \mbox{bits}$, which is certainly small compared to unity in any conventional sense.\footnote{Note also that these entropies are about a factor of two smaller than the estimate supplied by equation \eqref{kest2}, which turns out to be $\K \sim 0.002383 \; \mbox{bits}$.}  But should this numerical difference $\Delta \K$ actually matter to us?  

We might first think to look at lower and upper bounds, since the magnitude of the difference between relative entropies can be no larger than the difference between their upper and lower bounds.  All relative entropies are bounded below by zero, but available upper bounds (associated with \textit{least-uniform} distributions) are so weak as to be next to useless in assessing or comparing near-optimal (most uniform) distributions,  In the current example, we know that the relative entropy $\K$ cannot exceed $\log_2 P =  28.0682 \; \mbox{bits}$, and can only approach this bound to the extent that all of the probability is concentrated on a single (dictatorial) individual.  Recognizing that all individuals within a state must be treated equally for the purposes of apportionment, the upper bound is reduced to $\K \le 9.1502 \; \mbox{bits},$\footnote{Compare this to our estimate from Equation \eqref{ubu}, which is $\K \lesssim 12.5226 \; \mbox{bits}$.} and imposing the Constitutional limits on each state's apportionment further reduces this to  $\K \le 3.7241\; \mbox{bits}$.\footnote{This tighter bound can be calculated by giving each state one seat, then assigning as many seats as possible to the smallest states, in order, while ensuring that $\bar{d}_s \ge \Dmin$.}  But this bound remains orders of magnitude greater than the relative entropies and relative entropy differences in which we are interested, so is not particularly useful.

Another point of comparison might be the U.S.\ Senate.  With California and Wyoming each getting two Senators each, the resulting level of representational disproportionality in the Senate ought to register as egregious by almost any democratic measure, so should definitely provide an upper bound on what  counts as significantly unequal.  Calculating the relative entropy between direct polling of individuals and \textit{Senate}-mediated polling, we find using the same 1990 U.S.\ Census data that $\K_{S} = 0.69424 \; \mbox{bits}$.  This is less than one bit but still still orders of magnitude larger than the differences in relative entropy the emerge for different House sizes, so does not provide much in the way of a meaningful bound.
 
A different approach will be needed.  Remember that the relative entropy $\K$ represents an informational inefficiency per polling of one individual from the population.  But in, say, the 2004 U.S.\ election (towards the middle of the census/reapportionment cycle), on the order of  $V \approx 1.2 \cdot 10^{8}$  voters cast ballots, so we take that as a conservative estimate for the number of politically interested citizens whose opinions we imagine are to be effectively polled (say under independent sampling with replacement) on important issues.  Suppose we also conservatively estimate that Congress votes on about $5$ issues per year of broad and substantial concern to the American populace,\footnote{Some could regard this as a significant underestimate.  Others, perhaps looking at the productivity in recent Congresses, might think this number is, if anything, too high.  Also note that for every ``primary'' vote on major legislation, there will most likely be several auxiliary, preliminary, or procedural votes, but these will be highly correlated with the primary vote, and so need not be counted separately for purposes of approximation.} or some $C_{10} \approx (5)(10) = 50$ major decisions over a $10$-year census/reapportionment cycle.

So the total amount of discrimination information between polling in a direct democracy and polling mediated through Congressional representatives would be something like 
\be
H_{\stext{D}} \sim  V \, C_{10}\, \Delta S  = 1.15 \cdot 10^6 \;\mbox{bits}.
\ee
But what should we make of this, or to what should it be compared?

Relative to the total of around $V C_{10} \sim 6.1 \cdot 10^{9} \; \mbox{bits}$ of polling information (assuming binary opinions on yes-or-no-votes), this is a tiny amount of ``misinformation'' due to distorted polling probabilities.  But perhaps a better comparison is to the actual amount of information required to specify the votes of members of the House, no larger than $C_{10} R \sim 2.2 \cdot 10^3 \; \mbox{bits}$,  or even the mere $C_{10} \sim 50 \; \mbox{bits}$ of information regarding actual legislative outcomes in up-or-down votes.

Another way to think about these numbers is in terms of the combinatorics of the imagined sampling frequency distributions over the decadal span.  The actual difference $\Delta \K$ means that for every way one can generate a polling sample agreeing with the frequency distribution expected for indirect, Congress-mediated sampling, there are about $2^{1.15 \cdot 10^6} \approx 10^{3.46 \cdot 10^5}$ more ways to generate a sample agreeing with the frequency distribution expected from direct, uniform sampling. 

A third approach involves thinking about statistical discrimination information.  Suppose a statistician were shown only the frequency distribution (\ie, who was sampled, how many times) from (what would only later be revealed to be) a ten-year, indirect (Congress-mediated) sample, and asked if this sample originated from the direct or indirect sampling method.  The expected weight of evidence in favor of the indirect sampling hypothesis would be about $3.46 \cdot 10^6 \; \mbox{decibels}$, which is enormous.  In comparison, only about $30 \; \mbox{dB}$ would typically be considered reasonable evidence in favor of a hypothesis, and $100 \; \mbox{dB}$ as extremely strong evidence.\footnote{For non-Bayesian readers, note that $30 \; \mbox{dB}$ in weight of evidence would, \textit{very} roughly speaking, correspond in a typical frequentist null hypothesis significance test (NHST) to a $p$-value of about $p = 0.0001$, where in many disciplines $p = 0.01$ or even $p = 0.05$ is often held as the threshold for reporting a ``statistically significant'' result.  In elementary particle physics, with more stringent conventions, the threshold for reporting  ``evidence for'' a new particle is typically $p = 0.003$, or about $25\; \mbox{dB}$, while the standard for ``discovery'' of a particle is $p = 0.0000003$, or about $65\; \mbox{dB}$.  A weight of evidence of $3.46 \cdot 10^6 \; \mbox{dB}$ would correspond to a $p$-value of about $p \approx 10^{-3.46 \cdot 10^5}$, which is ridiculously small by any reasonable statistical standard.}

How small would the difference need to be before it might  be disregarded, statistically speaking?  Conventionally, a change of about $3 \ \mbox{dB}$ is considered a just-noticeable difference in log-odds, while a change of about $30 \ \mbox{dB}$  in weight of evidence is typically considered strong evidence.  taking the threshold somewhere between these limits, in the current example, a difference in $\K$ might be considered ignorable somewhere between $\lvert \Delta \K \rvert \sim 1.6 \cdot 10^{-10} \; \mbox{bits}$ and $\lvert \Delta \K \rvert \sim 1.6 \cdot 10^{-9} \; \mbox{bits}$, much smaller than the observed difference quoted above, and smaller even than the smallest change in $\K$ that could be anticipated given the scale of the populations, so effectively \textit{any} observable change in $\K$ consistent with the constraints could be regarded as statistically meaningful.\footnote{It is an oft-repeated truism that statistical significance is not the same thing as practical significance.  But here, we are primarily concerned with inequities in sampling information and concomitant distortion of political influence, so in a sense messaging and information are precisely what are at issue.  But additionally, the probability that even the smallest change in an apportionment (the addition or removal of a seat, or a shift in a seat form one state to another) would lead to actual substantial differences in who gets represented by whom, and by which party, or what constituent services are provided to what individuals, or what legislation is passed, is difficult to calculate, but will clearly be close to unity.  Minimal changes in apportionment can produce large changes in expected utility amongst represented individuals, so in this sense essentially any difference in our figure-of-merit that could motivate a shift in apportionment ought to be considered as potentially significant.}

\subsection{Entropy and Districting}\label{districting}

In  cases involving apportionment and districting coming before the Federal courts, and in other legal debates \cite{chafee:1929,bliss:1929,schmeckebie:1941,jewell:1962,dixon:1968,cortner:1970,pitkin:1972,adams:1977,orourke:1980,grofman_scarrow:1981,eig:1981,balinski_young:1982,cain:1984,michigan:1984,kromkowski:1991,young:1994,balinski_young:2001,edelman_sherry:2002,balinski:2005,edelman:2006,rosenberg:2009,cipra:2010,caufield:2010,maltz:2011,krabill_fielding:2012,suzuki:2015,robinson:2016}, scholars, plaintiffs, and judges have all invoked various quantitative measures of inequality regarding district sizes (either within states or across states), such as the variance, or absolute or relative differences between largest and smallest districts.\footnote{Litigation regarding gerrymandering has also addressed many other demographic factors besides just the sizes of districts, but these issues will be discussed elsewhere.}  But it remains difficult to justify any one of these traditional statistics over the various alternatives, at the level of districts themselves.  Instead, we should be thinking, not about variation in district sizes \textit{per se}, but about inequalities in the weight of representation afforded individuals within the districts, as a consequence of these differences in district sizes, and once again the natural way to measure these differences is via relative entropy.  Only at this later stage of assessment, where apportionments have been made and specific district boundaries have been proposed, such that we can unambiguously attribute individuals to specific districts, we should use polling probabilities based on the actually proposed district partition rather than the within-state averages.

\newpage
\subsubsection{Optimization of District Sizes}

The Kullback-Leibler divergence between the direct (uniform) and indirect (representative-mediated) polling probability distributions \textit{given proposed districting information} would be
\be
\K_d =  \sum\limits_{s=1}^{S}\sum\limits_{k=1}^{a_s} \sum\limits_{i =1}^{d_{sk}} \pi_{sk} \log\bigl[\tfrac{\pi_{sk}}{\frac{1}{P}} \bigr]
= \sum\limits_{s=1}^{S}\sum\limits_{k=1}^{a_s} \sum\limits_{i =1}^{d_{sk}} \tfrac{1}{R} \tfrac{1}{d_{sk}} \log\Bigl[\tfrac{ \frac{1}{R} \frac{1}{d_{sk}}}{\frac{1}{P}} \Bigr]
= \log P + \sum\limits_{s=1}^{S}\sum\limits_{k=1}^{a_s} \sum\limits_{i =1}^{d_{sk}} \tfrac{1}{R} \tfrac{1}{d_{sk}} \log\bigl[ \tfrac{1}{R} \tfrac{1}{d_{sk}} \bigr],
\ee
where the sum is taken over all represented individuals with each (proposed) district of each state.

Since all individuals within a given district are still presumed to have equal weights of representation, this simplifies to 
\be
\begin{split}
\K_d &= \sum\limits_{s=1}^{S}\sum\limits_{k=1}^{a_s} \tfrac{1}{R} \log\bigl[\tfrac{\Dbar}{d_{sk}} \bigr]
= \log[\Dbar]  - \tfrac{1}{R} \sum\limits_{s=1}^{S}\sum\limits_{k=1}^{a_s} \log[d_{sk}] \\
&= \log[\Dbar]  - \log \,\Bigl[ \prod\limits_{s=1}^{S} \prod\limits_{k=1}^{a_s} d_{sk}^{\frac{1}{R}} \Bigr]
= \log \,\biggl[  \tfrac{\Dbar}{ \bigl( \prod\limits_{s=1}^{S} \prod\limits_{k=1}^{a_s} d_{sk} \bigr)^{\frac{1}{R}}} \biggr].
\end{split}
\ee
For a given apportionment $\bv{a}$, this means that uniformity in the distribution of weight of representation is improved to the extent that the \textit{geometric mean} of the district sizes can be made as large as possible.

Not surprisingly, this constrained optimum is achieved when the districts are chosen to be nearly as equal in size as possible, with the constraint that each encompasses a whole number of individuals.  See Appendix~\ref{district_sizes} (page \pageref{district_sizes} ff.) for an elementary proof.  That is, in the very best case realizable, the $s$th state would have $\eta_s = p_s \!\! \! \mod a_s =  a_s( \bar{d}_s - \lfloor \bar{d}_s \rfloor \bigr)$ districts of size $\lceil \bar{d}_s \rceil$, and $(a_s - \eta_s)$ districts of size $\lfloor \bar{d}_s\rfloor$.  But the more important feature of relative entropy is not that it judges equal or nearly equal sized districts to be best---presumably every advocate of democracy already believed that---but rather that it provides a unique, principled way to quantify deviations from this ideal, for the purposes of assessing possible district boundaries proposed within particular states, or for comparing within-state inequities in districting to cross-state inequities due to the apportionment itself, for example in order to better inform the debate on optimal House size.

\subsubsection{Entropy Chains Using Increasingly Detailed State and District-Level Information}

It may be useful to consider a sequence of Shannon entropies associated with a hierarchy of polling probability distributions, all defined similarly, except that each uses successively more complete or accurate information about the districting, and therefore each successive entropy is constrained to be smaller than its predecessor.

At the most uniform end of the hierarchy, $\S_{\stext{max}}$ still corresponds to the maximum possible entropy, associated with uniform polling probabilities $\bar{\pi} = \tfrac{1}{P} = \tfrac{1}{R} \tfrac{1}{\Dbar}$ across all individuals:
\be
\S_{\stext{max}} = - \sum\limits_{s=1}^{S} \sum\limits_{n=1}^{p_s} \tfrac{1}{P} \log[ \tfrac{1}{P} ] 
= \sum\limits_{s=1}^{S}  \tfrac{p_s}{P} \log[P]
= - \sum\limits_{s=1}^{S} \tfrac{a_s}{R} \log\bigl[ \tfrac{1}{R} \tfrac{1}{\Dbar}\bigr]  = \log P.
\ee
Mathematically, this is as if each individual acts as his or her own representative, like in a direct democracy.

Next we add constraints demanding that each state receive some whole number of representatives (possibly between specified lower and upper bounds), while the total number of apportioned representatives sums to some specified value $R$.  Let $\bv{a}^*$ be the apportionment maximizing the entropy (or equivalently minimizing the relative entropy) subject to these constraints.  District partitions within states are yet to be determined, so polling probabilities are still to be based on state-level average district sizes $\bar{d}_s^{*} = \tfrac{p_s}{a_{s}^*}$. Then the corresponding polling probabilities are $\pi_{sn}^* = \tfrac{a_s^*}{R}\tfrac{1}{P} = \tfrac{1}{R} \tfrac{1}{\bar{d}_s^*}$, where $d_s^* = \tfrac{p_s}{a_s^*}$, and the Shannon entropy becomes 
\be
\S_{a^*} = - \sum\limits_{s=1}^{S} \sum\limits_{n=1}^{p_s} \tfrac{a_s^{*}}{R} \tfrac{1}{p_s} \log\bigl[ \tfrac{a_s^{*}}{R} \tfrac{1}{p_s} \bigr] 
=- \sum\limits_{s=1}^{S} \tfrac{a_s^*}{R} \log\bigl[ \tfrac{1}{R} \tfrac{1}{\bar{d}_s^*} \bigr] = \S_{\stext{max}} - \K_{a^*}, 
\ee
and must satisfy $S_{a^*} \le S_{\stext{max}}$, where the inequality will be strict unless every state's apportionment $a_s^{*}$ is exactly equal to the corresponding state quota $q_s$, which is wildly unlikely when $P \gg R$, except in factitious examples where values for $R$ and all of the $p_{s'}$ are carefully chosen to produce integral quotas.

Any other feasible apportionment $\bv{a}$ satisfying the same constraints yields instead an entropy
\be
\S_a = - \sum\limits_{s=1}^{S} \sum\limits_{n=1}^{p_s} \tfrac{a_s}{R} \tfrac{1}{p_s} \log\bigl[ \tfrac{a_s}{R} \tfrac{1}{p_s} \bigr] 
=- \sum\limits_{s=1}^{S} \tfrac{a_s}{R} \log\bigl[ \tfrac{1}{R} \tfrac{1}{\bar{d}_s} \bigr] = S_{\stext{max}} - \K_{a}, 
\ee
where $\bar{d}_s = \frac{p_s}{a_s}$ is the state-level average district size based on the chosen (but possibly sub-optimal) apportionment $\bv{a}$.  Because $\bv{a}^*$ was chosen optimally, it must be the case that $\S_a \le \S_{a*}$, where the equality will be strict unless $\bv{a} = \bv{a}^*$, or else a tie arose, so that $\bv{a}$ is just as good as $\bv{a}^*$.

Next, once a feasible (and possibly but not necessarily, optimal) apportionment $\bv{a}$ is adopted, state populations can be partitioned into districts.  Suppose the districts are chosen to be of optimal size $d_{sk}^{\dag}$, in the sense of maximizing the entropy subject to the additional constraints that the $s$th state population is divided into exactly $a_s$ mutually exclusive and exhaustive districts each containing a whole number of represented individuals from that state.  With polling stratified by these districts, the polling probabilities become $\pi_{ski}^{\dag} = \tfrac{1}{R} \tfrac{1}{d_{sk}^{\dag}}$, and the entropy is
\be
\S_{d^{\dag}} = -\sum\limits_{s=1}^{S}\sum\limits_{k=1}^{a_s} \sum\limits_{i =1}^{d^{\dag}_{sk}} \tfrac{1}{R} \tfrac{1}{d^{\dag}_{sk}} \log\bigl[ \tfrac{1}{R} \tfrac{1}{d^{\dag}_{sk}} \bigr]  =  -\sum\limits_{s=1}^{S} \Bigl\{ \tfrac{ \eta_s}{R} \log\bigl[ \tfrac{1}{R} \tfrac{1}{\lceil \bar{d}_s \rceil} \bigr]   +  \tfrac{(a_s-\eta_s)}{R} \log\bigl[ \tfrac{1}{R} \tfrac{1}{\lfloor \bar{d}_s \rfloor} \bigr]  \Bigr\},
\ee
where it must be the case that $\S_d^{\dag} \le \S_{a}$ necessarily, and the inequality will be strict unless every state population $p_s$ is exactly divisible by the corresponding $a_s$ (\ie, $\eta_s = 0$ for all $s = 1, \dotsc, S$), so that districts within each state can be chosen to be exactly equal in size, which is also very unlikely to happen in actual practice.

If a different (but still mutually exclusive and exhaustive) district partitioning is used within the states, possibly of sub-optimal sizes $d_{sk}$, then the polling probabilities are instead $\pi_{ski} = \tfrac{1}{R} \tfrac{1}{d_{sk}}$, and the corresponding entropy is
\be\label{Sd}
\S_d = -\sum\limits_{s=1}^{S}\sum\limits_{k=1}^{a_s} \sum\limits_{i =1}^{d_{sk}} \tfrac{1}{R} \tfrac{1}{d_{sk}} \log\bigl[ \tfrac{1}{R} \tfrac{1}{d_{sk}} \bigr]  = -\sum\limits_{s=1}^{S}\sum\limits_{k=1}^{a_s}  \tfrac{1}{R} \log\bigl[ \tfrac{1}{R} \tfrac{1}{d_{sk}} \bigr], 
\ee
where $\S_d \le \S_d^{\dag}$ necessarily, and the inequality will be strict unless the district sizes are chosen to be as nearly equal as possible (\ie, $d_{sk} = d_{sk}^{\dag}$ under some consistent choice\footnote{In order  to meaningfully compare district sizes, we might, for example, demand that for each state $s$, the size of the districts are sequenced in non-decreasing order with respect to the index $k$.} of the labeling $k$ for districts).  In practice, the districts need not be chosen quite so fastidiously, and variations in district sizes substantially larger than $\pm 1$ person are typically tolerated by the Courts, so maximally equal district sizes tends to happen only in states that have a single (``at large'') district.  Since this last entropy \eqref{Sd} is a well-defined Shannon entropy over a finite number of possibilities, it also satisfies $0 \le \S_d$, with equality if and only if the population consists of only one inhabitant in one state, in which case there would be nothing to apportion.

\newpage
Altogether, these various Shannon entropies will satisfy the chain of inequalities
\be
0 \le \S_d \le S_{{d}^{\dag}} \le S_a \le S_a^* \le S_{\stext{max}}
\ee
always, and
\be
0 < \S_d < S_{{d}^{\dag}} < S_a \le S_a^* < S_{\stext{max}}
\ee
typically.

So when faced with a question as to whether a variation in the sizes of a certain state's districts should be acceptable, Courts might look at that state's contribution to the entropy deficit $(\S_{{d}^{\dag}} - S_{d})$, i.e., the part actually under the control of the states, compared to, say, that state's contribution to the deficit $(\S_{\stext{max}} - \S_{{d}^{\dag}}) = \K_{{d}^{\dag}}$ arising from the nature of the apportionment itself.  Some conventional standard might be adopted specifying that the former could not introduce significantly more entropy loss than the latter, for instance.  Otherwise it could be reasonably argued that the sub-optimal districting is squandering whatever potential level of equity was achieved by the apportionment itself.

\subsubsection{Intra-State Entropies}

Actually, if we are focusing on districts within individual states, it perhaps makes better sense to effectively \textit{re-normalize} the entropies by conditioning on the separate state-level polling outcomes, rather than quantifying each contribution to the overall national-level entropy.  That is, we can define the sequence of \textit{state-conditional} entropies:
\bsub
\begin{align}
\S_{s} &= -\sum\limits_{n = 1}^{p_s}  \tfrac{1}{p_s} \log \tfrac{1}{p_s} = \log p_s ,\\
\S_{d_s^{\dag}} &= -\sum\limits_{k = 1}^{a_s} \sum\limits_{i = 1}^{d_{sk}^{\dag}}  \tfrac{1}{d_{sk}^{\dag}} \log \tfrac{1}{d_{sk}^{\dag}}  
=   \log[ a_s] +  \tfrac{1}{a_s} \sum\limits_{k = 1}^{a_s} \log d_{sk}^{\dag} = \log[ a_s] +  \tfrac{\eta_s}{a_s}\log \lceil \bar{d}_s \rceil  + \bigl(1 - \tfrac{\eta_s}{a_s} \bigr)  \log \lfloor \bar{d}_s \rfloor  ,\\
\S_{d_s} &= -\sum\limits_{k = 1}^{a_s}\sum\limits_{i = 1}^{d_{sk}}  \tfrac{1}{a_s} \tfrac{1}{d_{sk}} \log\bigl[\tfrac{1}{a_s} \tfrac{1}{d_{sk}} \bigr]
=  \log[ a_s] +  \tfrac{1}{a_s} \sum\limits_{k = 1}^{a_s} \log d_{sk},
\end{align}
\esub
which can be related to intra-state relative entropies (relative that is, to an intra-state uniform distribution),
 \bsub
\begin{align}
\K_{d_s^{\dag}} &=  \S_{s} - \S_{d_s^{\dag}} = \sum\limits_{k=1}^{a_s} \sum\limits_{i=1}^{d_{sk}^{\dag}} \tfrac{1}{a_s d_{sk}^{\dag}} \log\bigl[ \tfrac{p_s}{a_s d_{sk}^{\dag}} \bigr] 
=  \tfrac{1}{a_s}  \sum\limits_{k=1}^{a_s} \log\bigl[ \tfrac{\bar{d}_s}{d_{sk}^{\dag}} \bigr]
= \tfrac{\eta_s}{a_s} \log\bigl[ \tfrac{\bar{d}_s}{\lceil  \bar{d}_s  \rceil} \bigr]  + (1 - \tfrac{\eta_s}{a_s}) \log\bigl[ \tfrac{\bar{d}_s}{\lfloor  \bar{d}_s \rfloor} \bigr], \\
\K_{d_s}  &=  \S_{s} - \S_{d_s} = \sum\limits_{k=1}^{a_s} \sum\limits_{i=1}^{d_{sk}} \tfrac{1}{a_s d_{sk}} \log\bigl[ \tfrac{p_s}{a_s d_{sk}} \bigr] 
=  \tfrac{1}{a_s}  \sum\limits_{k=1}^{a_s} \log\bigl[ \tfrac{\bar{d}_s}{d_{sk}} \bigr],
\end{align}
\esub
but also may be related to the national-level entropies via the recursivity property of Shannon entropy:
\bsub
\begin{align}
\S_{\stext{max}} &=  -\sum\limits_{s=1}^{S} \tfrac{p_s}{P} \log \tfrac{p_s}{P} + \sum\limits_{s=1}^{S} \tfrac{p_s}{P}\, \S_{s},\\
\S_{a^*} &=  -\sum\limits_{s=1}^{S} \tfrac{\,a_s^*}{R} \log \tfrac{\,a_s^*}{R} + \sum\limits_{s=1}^{S} \tfrac{\,a_s^*}{R}\, \S_{s},\\
\S_{a} &=  -\sum\limits_{s=1}^{S} \tfrac{a_s}{R} \log \tfrac{a_s}{R} + \sum\limits_{s=1}^{S} \tfrac{a_s}{R}\, \S_{s},\\
\S_{d^{\dag}} &= -\sum\limits_{s=1}^{S} \tfrac{a_s}{R} \log \tfrac{a_s}{R} + \sum\limits_{s=1}^{S} \tfrac{a_s}{R}\, \S_{d^{\dag}_s},\\
\S_{d} &= -\sum\limits_{s=1}^{S} \tfrac{a_s}{R} \log \tfrac{a_s}{R} + \sum\limits_{s=1}^{S} \tfrac{a_s}{R}\, \S_{d_s},
\end{align}
\esub
which in each case corresponds to an entropy associated with choosing amongst states, plus the average across states of the conditional entropy regarding the sampling of individuals within each state.

So a state's adoption of district sizes could then be judged based on whether the entropy deficit $(\S_{d^{\dag}_s} - \S_{d_s}) = (\K_{d_s} - \K_{d^{\dag}_s})$ is sufficiently small (or at least not too large) compared to  $(\S_{s} - \S_{d^{\dag}_s}) = \K_{d_s^{\dag}}$ itself, by some adopted convention.

We could also ``grade on a curve,'' and look at the equality of representation afforded within say, California compared to Colorado or Connecticut.  But here it would still seem to make sense to remove from the relative entropy the component beyond the control of states themselves, resulting from the overall apportionment.  In addition, there will be another ambiguity which we address below, as to whether we would then compare on an overall or  per-capita basis.



\subsubsection{Some Orders of Magnitude for Intra-State Entropies}\label{oom2}

At the intra-state level, what sort of values for $\K_{d_s}$ might we expect?  Using again the Pinsker inequality and the fundamental logarithm inequality, we may infer
\be
\tfrac{1}{2 \ln 2}  \Bigl[ \sum\limits_{k = 1}^{a_s} \sum\limits_{i=1}^{d_{sk}}   \bigl\lvert  \tfrac{1}{a_s d_{sk}} - \tfrac{1}{p_s} \bigr\rvert  \Bigr]^2 \le \K_{d_s} \le \tfrac{1}{\ln 2} \sum\limits_{k = 1}^{a_s} \sum\limits_{i=1}^{d_{sk}}    \tfrac{1}{a_s d_{sk}}  \bigl(  \tfrac{p_s}{a_s d_{sk}} - 1 \bigr), 
\ee
which simplifies to 
\be
\tfrac{1}{2 \ln 2}  \Bigl[ \sum\limits_{k = 1}^{a_s}  \tfrac{1}{a_s} \bigl\lvert  1 - \tfrac{d_{sk}}{\bar{d}_s} \bigr\rvert  \Bigr]^2 \le \K_{d_s} \le \tfrac{1}{\ln 2} \sum\limits_{k = 1}^{a_s} \tfrac{1}{a_s}  \bigl(  \tfrac{\bar{d}_s}{d_{sk}} - 1 \bigr),
\ee
since all inhabitants of the same district share the same weight of representation.  Assuming that the fractional variation in district size is small, \ie, $ \tfrac{ \lvert d_{sk} - \bar{d}_s \rvert}{\bar{d}_s} \sim O(f_s) \ll 1$, we estimate
\be
0 \le \tfrac{1}{2 \ln 2}  \lvert f_s \rvert^2 \lesssim \K_{d_s} \lesssim  \tfrac{1}{\ln 2}  \tfrac{1}{\sqrt{a_s}} \lvert f_s \rvert,
\ee
where on the right-hand side we incorporated the fact that the variations in district size are expected to be small in magnitude, but of opposite sign with more or less equal probability.
 
Note that $f_s = 0$ if $a_s = 1$ (or else if at-large representation were to be employed for states with $a_s > 1$, which is currently disallowed by statute), but otherwise Courts typically tolerate some moderately small values for the \textit{relative} variation $f_s$---small, that is, compared to unity, but perhaps larger than the minimum possible, which would be $O({a_s}/{p_s})$.  Of course from our point of view, imposing an acceptable tolerance on relative variations in the $d_{sk}$, then considering the effects on $\K_{d_s}$, gets things backwards.  Instead, Courts could adopt directly a limit on $\K_{d_s}$.

\subsection{Comparing Across Different Populations}

We have seen how to use the relative entropy to score different proposed apportionments $\bv{a}$ given a prescribed House size $R$ and population counts $\bv{p}$, to compare optimal apportionments under different house sizes $R$ but again for the same fixed population distribution $\bv{p}$, and to assess different possible partitions of a given State's population $p_s$ into $a_s$ districts of sizes $d_{sk}$.

But there are also cases where we might wish to use the entropy or relative relative to facilitate comparison of the equity of apportionments in distinct populations.  For example, in the previous section, we saw how we might want to compare the distributions of district sizes in different states.

Or, we might also seek to compare the overall apportionments with the U.S.\ in different years.  On the one hand, such comparisons lie strictly beyond the scope of the apportionment problem itself, since whether Congressional representation was fairly chosen in the past should not, as a matter of logical or ethical principle, affect the problem of finding the best apportionment in the present.  On the other hand, in, any political debate over whether the House of Representatives should be expanded from its current size, inevitably it will be asked whether a proposed apportionment in 2020 would be more or less fair than was the apportionment made in 1920.

Unlike most types of assessment within a fixed population, these sorts of judgements really can have no definitive answer until we agree further on fair terms of comparison.  Do we care about the absolute number of affected people, weighted by their degree of misrepresentation, since the total ``anti-democratic disutility'' of political decisions made by a (possibly unrepresentative) government will likely scale proportionally to the size of the affected population?  Or do we care most about a typical or average per-capita level of distortion of democratic voice, resulting from an apportionment and/or districting, which might provide a better measure of intrinsic disproportionality amongst individuals making up populations of very different sizes? 

Different situations might involve different goals and different tradeoffs, so no definitive answer seems possible,\footnote{For instance, how might we compare  (dis)utility in a world where some number of individuals suffer from under-representation, to an alternate world in which those individual never existed at all?} yet in either case entropies quantify something important.

in Section~\ref{oom1}, we learned that both typical and optimized values for the relative entropy $\K$ are \textit{expected} to depend (at the national level) on $S$ and $R$ but not explicitly on the total population $P$, while in Section~\ref{oom2}, we saw that realistic choices of district sizes will lead to intra-state relative entropies $\K_{d_s}$ that are either independent of $p_s$ or tend to decrease with $p_s$.

Even though $\K_a$ is based on probability distributions over a space of $P$ polling possibilities, by using a \textit{relative} entropy we have effectively removed the contribution to entropy that scales with $\log P$, to quantify the entropy deficit.  Likewise, $\K_{d_s}$ is based on distributions over $p_s$ possibilities, but, as a relative entropy, on average does not tend to grow with $p_s$.  The relative entropy always measures the net ``democratic distortion'' involved, or difference in information gained, in one instance of polling of one representative individual, either directly or else indirectly via the House of Representatives.
 
That is to say, $\K_a$ and $\K_{d_s}$ in effect \textit{already} represent per-person average measures of representational inequity, and should not be normalized further by any sort of division by $P$ or $\log(P)$. So at the national level, if we are primarily concerned with the quality of individual representation, we should normally focus on $\K_a$ itself.  If we are more concerned with the total amount of inequity, then we might want to look at the quantity $P\, \K_a$.  Similarly, at the state level, we could look at $\K_{d_s}$ or $\Delta \K_{d_s}$ as an individual measure, and $p_s \,\K_{d_s}$ or $p_s \,\Delta \K_{d_s}$ as a cumulative measure.

\subsection{Apportionment Based on District-Conditioned Entropies?}

To make optimal apportionments, we are advocating use of the relative entropy to compare a uniform or ideal polling distribution to an indirect or Congress-mediated distribution.  At the stage of assigning seats to States (and/or assessing the total number of seats), the size and constitution of Congressional districts remain undetermined, so in inferring the indirect distribution, we use for within-state polling the maximum entropy probability distributions consistent with knowing $\bv{a}$ and $\bv{p}$ but not the exact demographic make-up of all districts, which is \textit{mathematically} equivalent to assuming either: (i) effective districts of average within-state size, given the apportionment $\bv{a}$; (ii) weight of representation of each state's Congressional delegation shared equally amongst the states represented population, (iii) at-large representation, or (iv) the use averaging over the possible district partitions compatible with the proposed apportionment.

So in effect we are using the relative entropy of an averaged polling distribution, since we cannot foretell the actual partition of each state's population.  A natural question arises as to whether to use instead the average of the relative entropy rather than the relative entropy of the averaged distribution.\footnote{Convexity ensures that the average of the relative entropies across different intra-state polling distributions will be bounded from below by the entropy of the average of the these distributions, but in general we do not know ahead of time how close these entropies will be.}  In order to determine the former, we would need to choose for each state a probability distribution over possible district partitions, but the Constitution effectively precludes use of any information about the states other than their populations---including information about their propensities for drawing district boundaries.

One hypothetical exception might be entertained---if some federal statute required all states to adopt best-case district boundaries following reapportionment, such that the size of the subsequent districts could be inferred with certainty from $p_s$ and $a_s$.  Even then, we would still not know which particular individuals would be assigned to which districts, and accounting for this uncertainty would still mean averaging out variation and result in uniform intra-state polling probabilities.\footnote{Recall that the entropy or relative entropies involve a sum over $s$ and $n$, where $(s,n)$ is meant to uniquely index a particular (i.e., namable) individual within the $s$th state, so even if the sizes of the districts can be predicted, the assignment of this individual to a district cannot be known before the actual boundaries are drawn.}  It could be argued that even if the assignment of individuals to districts would remain uncertain, we could still predict what the value of the entropy would become at the post-districting stage once the districts were drawn to best-case standards, so perhaps the associated entropy could be used at the pre-districting stage to select the apportionment $\bv{a}$?

We remain dubious of this idea, even in the unlikely event that State-level district sizes would ever be so narrowly constrained by law---when proposing and evaluating different apportionments of seats to states, it would seem most natural and most fair to continue to spread the weight of representation of a Congressional delegate uniformly over the state's population.  But fortunately, this modification would tend to make little difference in practice.  As shown in Appendix~\ref{optimization}, this variant of the relative entropy could be optimized using the same ``greedy'' algorithm as the district-average entropy, while Appendix~\ref{district_conditioning} (beginning on page \pageref{district_conditioning}) verifies that for realistic parameters relevant to U.S.\ demographics, the difference between this prospective entropy presuming best-case district-sizes and that based on average-district sizes would be very small, and unlikely to change judgements as to the optimal apportionment or optimal House size.


\section{Update on the 2020 Census and Reapportionment}\label{Census2021}

As of Spring 2021, new U.S.\ Congressional apportionments have now been proposed based on the 2020 Census numbers, allocating once again a total of $R = 435$ seats, and calculated once again using the Huntington-Hill Method of Equal Proportions, as is still mandated by Congressional statute. 

\subsection{Comparison Between the 2020 Entropic and Huntington Apportionments}

Referring to Table~\ref{Table:2020} on page \pageref{Table:2000} in Appendix~\ref{tables}, notice that, unlike recent decades, the Huntington-Hill and entropic methods actually disagree in 2020, where in comparison to the Huntington assignment,  the entropic method withholds one seat each from two low-population states, Montana and Rhode Island, and awards one extra seat each to two larger states, New York and Ohio.  The relative entropy/discrimination information for the entropic assignment is $\K = 0.00171902\; \mbox{bits}$, compared to a higher $\K = 0.00189995\; \mbox{bits}$ for the Huntington-Hill assignment prescribed by Congressional statute.  This table also lists each State's contribution $\K_s = \tfrac{a_s}{R} \log \tfrac{a_s}{q_s}$ to the overall, national-level discrimination information $\K = \sum\limits_s \K_s = \sum\limits_s \tfrac{a_s}{R} \log \tfrac{a_s}{q_s}$, with the signs reflecting whether a state's allotment is below or above its exact quota.  We can see how the cost of depriving Montana and Rhode Island of second seats is more than offset by awarding these seats instead to New York and Ohio.

As with the officially proposed Huntington apportionment, in the entropic apportionment, California still loses one seat relative to its previous delegation, for the first time since it was granted statehood, confirming that this drop is a robust consequence of the populations as reported, reflecting somewhat slower growth in the Golden State compared to some parts of the West and South.\footnote{Although given the political and societal circumstances during 2020, it is understandable if there is a certain amount of public skepticism regarding these Census numbers, particularly in States with substantial populations of Hispanic or undocumented residents.  In fact, the three states which most underperformed compared to pre-Census demographic projections in absolute counts (and which were also among the most underperforming in relative terms) were Texas, Florida, and Arizona, which also have some of the highest proportions of Hispanic residents in the U.S., around $40\%$ for Texas, $26\%$ for Texas, and $32\%$ for Arizona.  Hispanic residents account for about $38\%$ of California's population, although it came just slightly ahead of its projection.}

But some of the most surprising---even disconcerting---features of the 2020 Huntington apportionment could be avoided in the entropic assignment.  According to the Census Bureau's preliminary reporting, New York missed out on a $27$th representative by a mere $89$ people, while Minnesota held on to its $8$th seat by virtue of only $26$ residents.  Given some of the presumed additional uncertainties surrounding the 2020 Census, conducted as it was during the Covid-19 pandemic,\footnote{We readily imagine that the number of people in New York who neglected to complete forms, were otherwise overlooked, or perhaps moved out of New York City for just a few weeks or months during the height of the outbreak in the Spring of 2020, greatly exceeded an $89$ person margin.  Even under the best circumstances, undercount rates for the U.S.\ Census are predicted to be on the order of a half-percent or percent.} by an administration that did not exactly generate trust in the reliability or transparency of the process (deprecating immigrants, delaying field operations, and even stopping the count prematurely), the  usual sort of ``political fiction'' surrounding the official Census numbers seems especially absurd this time around, wherein what are tiny differences, almost assuredly lurking in the noise of what are at best imperfect counts, made under unprecedented and challenging conditions, by an administration that did not exactly hide its efforts to add a question about citizenship to the survey, nor its goal of ultimately excluding non-citizens from the populations for apportionment, of what is at best merely a momentary snapshot, as of April 2020, of a dynamic population during an exceptional time, will nevertheless determine Congressional apportionments and Electoral College representation for an entire decade.


That being said, applied to the official 2020 data, the entropic apportionment method does not exhibit the same eyebrow-raising sensitivity to these small and surely meaningless shifts in population counts.  With entropic apportionment, both New York and Minnesota would retain all of their seats relative to the size of their 2010 delegations, and in fact, in 2020 the last $3$ seats (i.e., the $433$rd, $434$th, and $435$th) are awarded by the algorithm to New York, Ohio, and Minnesota, respectively, while Texas, Florida, Arizona, and then California, in that order, would be next in line for additional seats.\footnote{In the Huntington method, New York then Ohio were next in line to receive seats, also followed by Texas, Florida, Arizona, and California.}

All other things being equal, the population of Minnesota would have to drop by $28\hspace{0.75pt}752$ before it would lose the last seat according to the entropic method---a change which is more than $3$ orders of magnitude greater than the reported margin in the Huntington method, and about $0.5\%$ of its population.  In the entropic apportionment, New York State is not even next in line to lose a seat, and its population would have to drop by $129\hspace{0.75pt}179$ before losing its last seat, a margin also over $3$ orders of magnitude larger than in the Huntington method, and about $0.64\%$ of the state's  population. 

By awarding second seats to Montana and Rhode Island, the Huntington-Hill Method exacerbates an already undemocratic bias toward small states in the American federal system.  According to the entropic method, Rhode Island should not earn its $2$nd seat until the House size reaches $R = 442$, and Montana would not receive a second seat until $R = 450$, by which point the two largest states, California and Texas, would have earned two additional seats each.  Intuitively, the Huntington method's assignment of these seats to both Montana and Rhode Island instead of New York and Ohio does appear rather dissatisfactory, because: (a) the fractional parts of the quotas for the former two states are both smaller than the fractional parts of the quotas for the latter two states, (b) both smaller states have quotas closer to their lower than to their upper bounds, while both of the larger states have quotas closer to their upper bounds, and (c) because New York and Ohio are both much more populous, a larger number of residents would suffer in the event of under-representation.  Keep in mind, however, that just looking at state-level quotas can sometimes be misleading---consider for instance the example in Table~\ref{table:not_quota_remainders} of Appendix~\ref{tables}.  And on the other hand,  giving single seats to Rhode Island and Montana does leave them with the largest average district sizes of any states.  This just goes to show that we really ought not to rely on intuitive or \textit{post hoc} arguments as to whether the last few seats were distributed justly.  Once we accept Shannon entropy as a uniquely consistent measure of the uniformity of a distribution, then we should also accept as optimal the apportionment produced by maximizing the entropy, 
and there is no need for further hand-wringing or speculation after the fact about alternatives.

Of course, either as the population of one State is incremented or decremented while holding constant other State populations, or else as population is transferred between two or more States, eventually any integer-valued apportionment of a fixed number of House seats must at some point jump.  But intuitively, it may strike readers as unlikely, and even vaguely suspicious in a forensic accounting sense, that in the very same year, two of these transitions, involving a ``near miss'' and a ``just missed,'' turn out to be so small, in both cases less than $0.0005\%$ of the relevant State's population.  Short of performing numerical simulations, probability distributions for such thresholds would appear difficult to ascertain.  But a \textit{very} rough back-of-the-envelope calculation, based on the $\chi^2$ statistic introduced above, suggests that, \textit{ceteris paribus}, a \textit{typical} fractional shift required to lose (or gain) a seat should be more like $\tfrac{| \Delta p_s |}{p_s} \sim \tfrac{2}{R} \approx 0.46\%$ of the state's population.  However, across all $S$ states, the minimum of such thresholds would be expected to be smaller, by approximately a factor of $O(\tfrac{1}{S})$ or so, corresponding to about $\tfrac{| \Delta p_s |}{p_s} \gtrsim \tfrac{2}{S R} \approx 0.01\%$.  Although quite surprising, perhaps the narrow margins observed in the 2020 Huntington apportionment are not quite as shocking as they first appear.  Still, the fact that at least two thresholds in 2020 were no larger than $O(0.0005\%)$ suggests that either:  (i) some unlikely numerical coincidences have occurred, (ii) possibly some mistake has been introduced in the counts or computations, or (iii) perhaps the fixed House size of $R = 435$ is really starting to ``show its age'' with increasing and increasingly asymmetric population growth, as the competition for the last seats is becoming tighter amongst more states.

\subsection{Sensitivity to Undercount}\label{sensitivity}

Although, thankfully, the sensitivities of the apportionment to population changes in New York or Minnesota are substantially less for entropic apportionment in comparison to Huntington apportionment, even in the entropic method, these margins do appear to be somewhat smaller than traditional undercount rates estimated for the U.S.\ Census, so may very well be ``in the noise.''  In the midst of the COVID-19 pandemic, one might anticipate that the undercount rate is not likely to be substantially better than in 2010, and could possibly be worse, although this may be mitigated somewhat by extensive use of the internet.\footnote{In 2010, overall the undercount rate has been estimated to be about $0.6\%$.  The 2010 Census temporarily employed about $635\hspace{0.95pt}000$ workers, and by the summer of 2020, the final self-response rate was estimated at around $66.5\%$.  In 2020, internet response options were widely emphasized for the first time, and only around $500\hspace{0.95pt} 000$ workers were employed, and planned field operations were delayed by the pandemic.  Compared to previous decades, the final \textit{self}-response rate in 2020 fell somewhat, to around $62.1\%$.  However, by October 2020, it was claimed that $99.9\%$ of households had self-responded or had been contacted, but of course making contact with known households is not the same thing as counting people.  In addition, the fact that, on average, the 2020 residential populations fell below their most recent Census Bureau projections (based on birth, death, and migration data) weighs somewhat against larger undercounts.  But it is of course the variations in the undercount rats across states that matter---an exactly uniform undercount fraction across all states would leave the apportionments unchanged. }

Since the political and economic stakes are high, perhaps we ought to consider devoting even more resources to avoid making decade-long apportionment decisions based on measurement errors.  Even more efforts might allow more thorough data collection and validation, or if the Supreme Court were less hostile,\footnote{See the 1999 decision in \textit{Department of Commerce vs.\ U.S.\ House of Representatives}, and also the 1996 decision in \textit{Wisconsin vs.\ City of New York et al.}} we could reconsider statistical adjustment, although statistical imputation will surely continue to be legally and politically fraught.

To analyze these issues further, it is possible, if somewhat tedious, to calculate for each state the \textit{ceteris paribus} thresholds in population shifts needed for gaining or losing a seat. However, such shifts would not provide an especially natural measure of sensitivity, because presumably states are simultaneously susceptible to various amounts of undercount.  Instead, we can randomly perturb the populations (upward, since some amount of undercounting is probably inevitable, while substantial over-counting or double-counting is far less likely in a careful, non-sampling based Census), and make some \textit{Monte Carlo} estimates of the sensitivity, in terms of the statistics of possible seat shifts between states.

\begin{table}[h!]
\begin{small}
\begin{tabular}{r cl c  |r  cl c}
\hline
swapped & (\textit{a}) \;$0.6\%$ & \textit{undercount} & \textit{rate} & swapped  & (\textit{b}) \; $1.2\%$ & \textit{undercount} & \textit{rate} \\
seats & probability & uncertainty & & seats &  probability & uncertainty  \\
\hline\hline
$\phantom{\ge} 0$   & $0.35382$ & $0.002$      &&  $\phantom{\ge} 0$   &  $0.06086$ & $0.001$ \\
$\phantom{\ge} 1$   & $0.50794$  & $0.002$      &&  $\phantom{\ge} 1$   &  $0.36708$ & $0.002$ \\
$\phantom{\ge } 2$  & $0.13072$  & $0.0015$    &&  $\phantom{\ge} 2$   &  $0.41544$ & $0.002$ \\
$\phantom{\ge} 3$   & $0.00750$  & $0.0004$    &&  $\phantom{\ge} 3$   &  $0.13726$ & $0.0015$ \\
$\ge 4$                     & $0.00002$ &  $0.00002$  &&  $\phantom{\ge} 4$   & $0.01758$ & $0.0006$ \\
 &  &    								 &&  $\phantom{\ge} 5$ & 0.00162 & 0.0002 \\
 &  &    								 && $\phantom{\ge} 6$ & 0.00014 & 0.00005 \\
 &  &     								&& 			$\ge 7$ & 0.00002 & 0.00002 \\
\hline
$P(>0)$: & $0.64618$ & $0.002$       &&  $P(>0)$: & $0.93914$ & $0.001$ \\
\text{Avg.:} &  $0.79196$   & $0.003$              && \text{Avg.:} &    $1.68914$   & $0.004$ \\
\text{St.\ Dev.:}  & $0.68661$  & & & \text{St.\ Dev.:}   & $0.85953$  &   \\
 \hline
\hline
\end{tabular}
\end{small}
\caption{Sensitivity of the 2020 entropic apportionments to random undercount rates of (a) $0.6\%$ of each state's population, and (b) $1.2\%$ of each state's population, based on Monte Carlo sampling.  The House size was fixed at $R = 435$.  Undercounts in each state were assumed to be independent and exponentially distributed, and a sample of $n_{\stext{MC}} = 50\hspace{1pt} 000$ random Census results were generated in each of the two cases, based on the assumed undercount rates and the best guess of the true populations from the actual Census.  Recall that each instance of a misapportioned or ``swapped'' seat involves a pair of States, one being awarded an extra seat that another State deserved more.  The probabilities and averages refer to the number of swapped seats amongst all States, relative to the baseline entropic apportionment of the official Census numbers.  In case (a), the odds favor at least one difference---this probability is about $64.6\%$.  In case (b), this probability is substantially higher, slightly over $93.9\%$.}\label{table:sensitivity_summary}
\end{table}

Table~\ref{table:sensitivity_summary}  offers some summary statistics of a preliminary Monte Carlo simulations for the 2020 Census.  See Table~\ref{Table:2020_sensitivity_long} in Appendix~\ref{Table:2020_sensitivity_long} on page \pageref{Table:2020_sensitivity_long} for additional state-by-state results.  We analyzed two cases based on different average undercount rates,\footnote{Although the undercount rates likely varied across the states, we used a common average percentage for these simulations, and drew the undercount values at random from independent exponential distributions.  The choice of an exponential distribution might be criticized as being somewhat arbitrary, although it does enjoy its own maximum entropy rationale.} namely: (a) an average undercount rate of $0.6\%$ of each States' population, corresponding to the estimated undercount in the 2010 Census; and (b) a more pessimistic ``plague-year'' estimate, using a doubled rate of $1.2\%$.

The average discrimination informations were not too different in the scenarios, corresponding to  $\bar{\K}_a = 0.00172634$ in case (a) and $\bar{\K}_b = 0.00173348$ in case (b), although not surprisingly,  the standard deviation is higher in case (b), $\sigma_{\K_b} = 0.00005964$ compared to
$\sigma_{\K_a} = 0.00003370$. (The corresponding histograms have been omitted, but look reassuringly Gaussian).   At $\K_e = 0.00171902$, the discrimination information achieved for the apportionment via the populations as reported is just slightly below both these averages, while the Huntington value $\K_{\stext{H}} = 0.00189995$ is higher---some $5.15$ standard deviations for case (a), and about $2.79$ standard deviations for the noisier case (b), suggesting that the Huntington apportionment would be somewhat to highly atypical for an entropic apportionment for populations near but above the reported values, depending on the assumed undercount rate.


Even in the lower error-rate case (a), the probability that one or more seats have been misapportioned (relative to what would be optimally fair if we knew the ``true'' counts) is substantial, around $64.6\%$, and the modal shift corresponds to $1$ swapped seat in the final apportionment relative to what it otherwise would have been.   In case (b), the probability is even higher, over $93.9\%$, the mode occurs at $2$ swapped seats rather than just $1$, and the expected number of misapportioned seats exceeds $1$ in this case.\footnote{Notice that probability distributions for the number of swapped seats appear to be under-dispersed relative to a Poissonian distribution.} 

Not surprisingly, looking at the state-level statistics, we observe that the highest probabilities of gaining or losing a seat correspond to those states which were close to the  cutoff for the last seat awarded in the entropic apportionment of the unadulterated populations, namely Minnesota, Ohio, New York, Texas, and Florida.  Even in the lower-undercount case (a), Minnesota has over a $31\%$ chance of losing a seat, and this rises to almost $53\%$ in case (b).  Ohio and New York are not too far behind.\footnote{Obviously the joint probabilities amongst States will be stochastically dependent, since (i) one State's loss is another's gain, and (ii) any earlier change in the seat assignments changes the context in which states compete for subsequent seats.  Such correlations were not explored.}  However, do notice that in the simulations, these seats are not really preferentially lost to Montana and Rhode Island, the small states to which the Huntington method has awarded second seats, but more often to large states like Texas, Florida, and (less commonly) California, or with somewhat less frequency, to the medium-sized Arizona or Virginia.  Montana and Rhode Island have only moderately elevated probabilities of gaining seats compared to several other states, slightly higher than Massachusetts or Pennsylvania, but lower than Michigan.

\subsection{Sub-Optimality of Freezing the House Size A Priori}

\begin{figure}[t]
\includegraphics[scale=0.98]{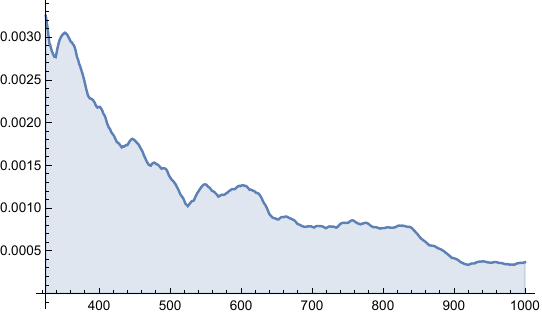}\\
\caption{Trend in optimal discrimination information (in bits) as a function of House size, from $R = 50$ to $R = 1\,000$ total seats, for the 2020 U.S.~Census data.  At the mandated house size of $R = 435$, the Kullback-Leibler divergence is $\K = 0.00171902$, but various local minima could do better, such as $R = 432$ (corresponding to $\K = 0.0017106$), $R = 473$ (corresponding to $\K = 0.00149581$), $R = 525$ (corresponding to $\K = 0.00102247$), $R = 653$ (corresponding to $\K = 0.0008668$), $R = 719$ (corresponding to $\K = 0.000763847$), $R = 796$ (corresponding to $\K = 0.000759089$), or $R = 920$ (corresponding to $\K = 0.000333914$).}
\label{kl_trend_2020}
\end{figure}

\begin{figure}[h]
\includegraphics[scale=0.98]{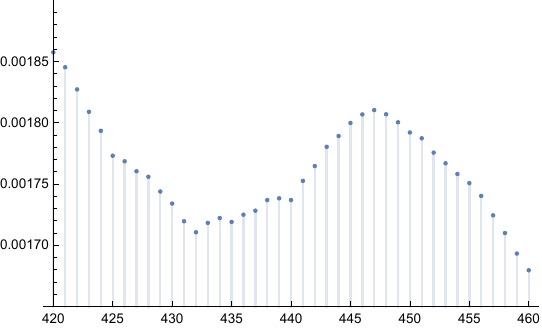}\\
\caption{Trend in optimal discrimination information (in bits) as a function of House size, from $R = 420$ to $R = 460$ total seats, for the 2020 U.S.~Census data.  The apportionment for currently prescribed  House size $R = 435$ is better than for the neighboring totals $R = 434$ or $R = 436$, but worse than for the nearby $R = 432$.}
\label{kl_trend_2020_detail}
\end{figure}

Once again in the 2020 data, we can also discern the sub-optimality introduced by arbitrarily and unnecessarily freezing the House size at the historical value of $R = 435$.  Figure~\ref{kl_trend_2020} plots the optimal discrimination information $\K$ for the entropic apportionment versus possible House sizes from $R = 325$ to $R = 1000$, based on the 2020 Census counts, and Figure~\ref{kl_trend_2020_detail} provides more detail in a smaller interval around the current House size, from $R = 420$ to $R = 460$.  As expected, we see a general (if not monotonic) downward trend reflecting a tendency towards fairer apportionment with larger house sizes, since larger denominators tend to allow more accurate approximation of the exact quotas,\footnote{These mathematical arguments pointing to a tendency toward better proportionality of representation with increasing house size, are largely distinct from purely procedural, parliamentary, or other political arguments as to optimal choice of district sizes, for purposes of representation of community viewpoints, advocacy of democratic preferences or interests, or provision of constituent services.  But we remind readers that a House size of $R = 435$ was first proposed over a century ago, and the U.S.\ population has more than tripled in the interim.  Improvements in communication, transportation, and data processing technologies, and arguably a certain amount of homogenization of popular and political cultures, first with film, then radio and television and now the internet, that has substituted more and more partisan polarization for less and less geographic diversity, have probably made it somewhat easier to represent effectively larger numbers of constituents, but, nevertheless, one can readily make the case that the U.S.\ House of Representatives is overdue for an expansion.} but there are variations on finer scales (of generally decaying amplitude with increasing $R$), with the consequence that some specific House sizes happen to produce local optima that outperform nearby choices in terms of maximizing entropy and minimizing discrimination information.

\newpage
Coincidently, and somewhat unusually, in 2020 there is a local optimum at a House size of $R = 432$, just a bit \textit{smaller} than the current value, in which California would still earn its $52$nd seat, but New York, Ohio, and Minnesota would all receive one less seat.\footnote{However, in 2020, the choice $R = 435$ does happen to lead to a better apportionment than either neighboring size,  $R = 434$ or $R = 436$.  But this is pure happenstance.  If the state-level populations were only somewhat different, the locally optimal House size might be different.}  This apportionment of a slightly smaller House would be more equitable than either the entropic or Huntington-Hill apportionments for $R = 435$ seats.  In addition, there are other local optima at moderately larger values of $R = 473$, with California receiving the last seat, its $57$th, or $R = 525$, with California again receiving the last seat, its $63$rd.\footnote{There is a tendency, if no necessity, for these local optima to occur at points where larger states have just received a seat. Firstly, larger states receive more of the seats overall, so even if we stopped at random, larger states have a higher than average probability of being the recipient of the last seat assigned.  Secondly, when a large state receives an additional seat that is its due, this tends to improve the equality of representation for a large number of people.}

\begin{figure}[b]
\includegraphics[scale=0.98]{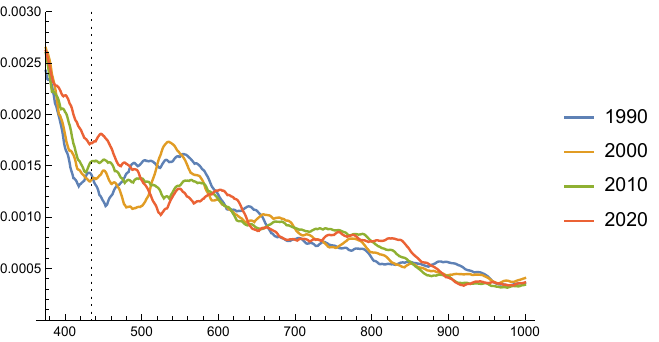}\\
\caption{Trends in optimal discrimination information (in bits) as a function of House size, from $R = 375$ to $R = 1000$ total seats, for the 1990--2020 U.S.~Census data.  The current House size $R = 435$ is indicated as a dotted vertical line.}
\label{trends_recent_4}
\end{figure}

Figure~\ref{trends_recent_4} shows trends in the discrimination information versus hypothetical House size for the four most recent Census data sets.  Although the general contours are clearly correlated, more fine-grained detail, and notably the precise location of local minima and maxima, depend haphazardly on the particulars of the year's counts.  For example, a local minimum for 1990 is near local maxima for 2000, 2010, and 2020, while a local minimum for 2020 is near local maxima for 1990 and 2010, and a local minimum for 2010 is near a pronounced local maximum for 2000 but near an inflection point for 2020.

\newpage
Given that all but the smallest states (with single members) must be prepared to gain or lose a seat after each Census, and in any event must expect to perform re-districting to balance populations within single-member districts based on updated population figures, there is simply no good reason, nor really any discernible partisan advantage, to freezing the House size at a sharp value before calculating apportionments.    Allowing for possible variations in $R$ of just a few percent, or even just a few seats, could lead to demonstrably more equitable apportionments.

\subsection{Fixed House Size in the Context of Continued U.S.\ Population Growth}

For similar reasons, intuition might anticipate that, given any fixed House size $R$, even optimized apportionments would eventually tend to become more disproportionate as the total represented population $P$ grows.

However, in itself, any such effect is not actually expected to be very large,\footnote{This is similar to the way uncertainties in estimates based on random survey sampling depend strongly on the size of the sample, but only very weakly on the total size of the population, when that population is much larger than the sample size.} at least  in a limit where $1 < S \ll R \ll P$.  Given that in the case of apportionment to U.S.\ House of Representatives, $P$ is much, much greater than $R$, and $R$ is greater than $S$ but perhaps not really that much greater, and that each state must receive at least one representative, such effects might still arise, but it is not clear whether any trend is apparent in U.S.\ Congressional apportionments over the last few decades.  If we look at the entropic apportionments for the time period 1960--2020, during which the U.S.\ has assigned $R = 435$ seats to $S = 50$ states while the total population has increased by about $85.4\%$, the overall proportionality as measured by relative entropy has appeared to fluctuate more than decay.  $2020$ has a higher discrimination information than any apportionment in 1980--2010, but lower than in 1960 or 1970.\footnote{When Alaska and Hawaii were admitted as U.S.\ states in 1959, each was given a single representative, and the House size was temporarily increased to $R = 437$ but without changing any other seats.  After the 1960 Census, the total was restored to $R = 435$.}  Of course, in 2020, more individuals will be adversely effected by any given degree of per-capita disproportionality, and the product $P \K$ of population times discrimination information turns out to be higher than in any of the previous six apportionments.  

However, the lack of a clear worsening trend in relative entropy $\K$ with growth in overall population $P$ may have more to do with the fact that, however small,  all states must receive at least one representative, but already by 1960, a House size of only $R = 435$ meant that several of the smallest state populations were substantially below the average district size of the remaining states. 
(In fact, the ratio of largest to smallest average district sizes was actually worse in 1960 than in 2020).

Using the smallest state, Wyoming, as a baseline, the number of representatives in 2020 should be something more like $R \approx 573$ if true proportionality were the goal.\footnote{While Vermont is the next most over-represented state, overall any tendency to over-represent low-population, rural  states tends to favor Republicans, so any changes are likely to face political resistance.}  The best nearby apportionment happens to be for $R = 568$, with $\K =  0.00113373$,  although a smaller House size of $R = 525$ actually does better still, achieving a relative entropy of $\K = 0.00102247$.

Political plausibility aside, overall we recommend entertaining a range of possible House sizes while also allowing for moderately to substantially larger House sizes, committing in advance to accepting the best apportionment within the prior range.

\def\c{\hspace{0.95pt}}
\begin{table}[t!]
\begin{centering}
\begin{scriptsize}
\begin{tabular}{lccclccccccccc}
\hline
   &     &     &      &     &    &    &  & H-H    & entropic    & entropic   & H-H/entropic &    best &   best \\   
Year   &  $S$   & $R$    & $P$     & $\min[\bar{d}_s]$   & $\Dbar$ &    $\max[\bar{d}_s]$    & $\tfrac{\max[\bar{d}_s]}{\min[\bar{d}_s]}$  & $\K$    & $\K$    & $P \times  \K$   & discrepancy &   nearby $R$ & nearby $\K$   \\
\hline\hline
1960 & 50 & 435 & 178\c 559\c 219	& 226\c 167 & 410\c 481 & 484\c 632 &  2.143 &  0.002822  &  0.002822  &  503\c 892	&  0 &  535 &  0.001007  \\
1970	 & 50	 & 435 & 204\c 053\c 325 & 304\c 067 & 469\c 088 & 624\c 181 &  2.053 &  0.002457  &  0.002368  &  483\c 133	&  1 &  535 &  0.001244  \\
1980	& 50	 & 435 & 225\c 907\c 472 & 393\c 345 & 519\c 328 & 690\c 768 & 1.756  &  0.001715  &  0.001715  &  387\c 476	&  0 &  535 &  0.001230  \\
1990	& 50	& 435 & 249\c 022\c 783 & 455\c 975  & 572\c 466 & 803\c 655 & 1.762  &  0.001405  &  0.001405  &  349\c 954	&  0 &  453 &  0.001109  \\
2000	& 50	& 435 & 281\c 424\c 177 & 495\c 304  & 646\c 952 & 905\c 316 & 1.828  &  0.001368  &  0.001368  &  384\c 940	&  0 &  488 &  0.001081  \\
2010	& 50	& 435 & 309\c 183\c 463 & 527\c 624  & 710\c 767 & 994\c 416 & 1.885  &  0.001544  &  0.001544  &  477\c 407	&  0 &  529 &  0.001183  \\
2020	& 50	& 435 & 331\c 108\c 434 & 542\c 704  & 761\c 169 & 990\c 837  & 1.826  &  0.001900  &  0.001719  &  569\c 182	&  2 &  525 &  0.001022  \\
\hline 
\end{tabular}
\end{scriptsize}
\caption{Trends in the official (Huntington-Hill) and entropic apportionments for the U.S.\ House of Representatives following the 1960--2020 decennial censuses, during which time exactly $R = 435$ seats were assigned to $S = 50$ states.  The discrimination information $\K$ tends to be correlated (albeit imperfectly) with other (and more arbitrary) measures of disproportionality, such as the ratio of maximum to minimum average district sizes across states, which, despite being a rather crude metric,  is shown here for reference (and which happen to be the same for both methods). The product $P\K$ of population and discrimination information provides a rough measure of total inequity introduced by the optimal (entropic) apportionment.  The overall discrepancy between the Huntington-Hill (H-H) and entropic recommendations is quantified by the number of swapped pairs of seats.  The ``best nearby''  apportionment refers to the optimal entropic apportionment for any House size within $\pm 100$ seats of the actual value $R = 435$.}\label{Table:recent}
\end{centering}
\end{table}


\subsection{Statehood for Washington, D.C.\ and Puerto Rico?}

During the Presidential election year of 2020 and through at least the first half of 2021, questions of possible Statehood for the District of Columbia and Puerto Rico have also gained some attention.  Although the prospects appear slim in the face of Republican recalcitrance and razor-thin Democratic majorities in the U.S.\ Senate, we have analyzed the hypothetical effects of adding these states.  With a 2020 population of $p_{\stext{DC}} = 691\c533$, Washington, D.C. would become the third smallest state in terms of population, after Wyoming and Vermont, and presumably would be  given $1$ voting representative if it were granted statehood sometime in the next few years.  However, with its population of $p_{\stext{PR}} = 3\c 285\c 874$, Puerto Rico would be a medium sized state (about half the average size, or three-fourths of the median), with a population closest to that of Utah, Iowa, or Nevada, and like them would probably earn a delegation of $4$ representatives based on the current apportionment.  In a hypothetical scenario assuming $S = 50 + 2 = 52$ states, with $R = 435 + 1 + 4 = 440$ seats, the entropic apportionment\footnote{The overseas population of Puerto Rico was not reported by the Census Bureau (although presumably there are some military or other governmental personnel living abroad), so we have used the reported \textit{residential} population rather than what would be its full population for apportionment.} would in fact assign $1$ representative to D.C.\ and $4$ representatives to Puerto Rico, leaving all other state apportionments unchanged,\footnote{With a relative entropy of $\K = 0.001753015$, the overall disproportionality for the optimized $S = 52$, $R = 440$, apportionment including D.C.\ and Puerto Rico is somewhat worse than for the optimized $S = 50$, $R = 435$ apportionment where D.C.\ and Puerto Rico are simply excluded, which yielded $\K = 0.00171902$.  But if we agree that the people of D.C.\ and Puerto Rico are deserving of representation at all, then the relevant comparison is actually to the apportionment for $S = 52$, $R = 435$, but where seats for D.C.\ and Puerto Rico are both held fixed at $a_{\stext{DC}} = a_{\stext{PR}} = 0$, for which the iniquity is of course much worse, corresponding to $\K = 0.018945983$, over an order of magnitude larger.} such that no ``New State Paradox'' would arise under these circumstances.\footnote{With the number of states expanded to $S = 52$ but with a total of only $R = 439$ seats, Minnesota would lose its last seat, while with $R = 441$ seats, Texas would gain an additional seat, relative to the baseline entropic apportionment for $S = 50$, $R = 435$, that excludes D.C.\ and Puerto Rico.}

\section{Conclusion}

Political and philosophical debate as to how best to assign Congressional seats to states based on their respective populations, or how best to assign parliamentary seats to parties based on votes, or how to resolve any number of related problems involving fair allocation of integral rewards based on non-integral desserts, has been ongoing literally for centuries, and we can hardly expect it to subside soon.  Nonetheless, we have endeavored to make the case that the most natural approach is one based on maximum entropy, the same principle used by physicists wanting to make the most justifiable predictions of thermodynamic systems, statisticians wishing  to assign probability distribution in the most unbiased manner, and engineers wishing to make optimal use of a communication channel.  Grounded in information theory, entropic apportionment takes seriously, indeed almost literally, the idea of trying to give an equal say to those represented.

``To apportion,'' as understood by Daniel Webster,\footnote{Webster is quoted in reference \cite{balinski_young:1975}.} ``is to distribute by right measure, to set off in just parts, to assign in due and proper proportion,'' but added, ``that which cannot be done perfectly must be done in a manner as near perfection as can be.''  Entropy provides a way to precisely and meaningfully quantify departures from perfect proportionality.  

\newpage
We began with three underlying principles guiding the choice of apportionment method: \textit{Uniformity} of individual representation, \textit{Universality} of apportionment methodology, and \textit{Uniqueness} of the resulting algorithm, and its output (at least in the absence of ties due to numerical accident or symmetry).  We argued that these \textit{desiderata} restrict viable methodologies to variational principles, involving constrained global optimization of some figure-of-merit which measures the equity or uniformity of the Congress-mediated weights of representation across all represented individuals, or equivalently, the mismatch between these indirect weights of representation $\tfrac{a_s}{R} \tfrac{1}{p_s}$, and the uniform weights of representation, $\tfrac{1}{P}$, associated with an ideal direct democracy.

But such weights of representation must be nonnegative for every individual, and will be additive over any group of individuals, and can be interpreted in terms of polling probabilities.  The natural measure of uniformity for any such distribution is the relative entropy between the indirect and direct weights of representation, also equal here to the deficit in Shannon entropy of the indirect distribution, compared to that of the direct distribution.  Optimal apportionments therefore minimize relative entropy, or equivalently maximize Shannon entropy, subject to suitable constraints on the total number of seats to be apportioned as well as lower and upper bounds on each state's allotment.  The resulting method of entropic apportionment enjoys not only an in-principle mathematical justification, but very reasonable in-practice performance, on both real-world census data and artificial test cases, similar to the recommendations of the widely-recommended  Huntington or Webster methods.  In fact, the figures-of-merit associated with both of these historic methods can be viewed rational-function approximations to the more fundamental relative entropy measure.   Where they disagree, entropic apportionment offers superior performance (by its own standards of course, but often also in regards to intuitive judgement).

These three methods share (albeit non-exclusively) various appealing features.  Each figure-of-merit can be written in terms of the weights of representation across individuals, which we regard as more fundamental than either the average districts sizes or the state-level quotas, but can also be re-expressed in terms of either of the latter set of variables if desired, or explicitly in terms of only the apportionments $\bv{a}$, the population counts $\bv{p}$, and the number of seats $R$, as required by the U.S.\ Constitution.  Like all global optimization methods, they not only can single out the optimal apportionment, but can be used to compare or rank any feasible apportionments.  By relying on optimization of an additive, convex objective function,  these methods can be implemented via a simple ``greedy'' algorithm which assigns seats sequentially, based on an easy-to-assess local optimization criterion.  However, if desired, any of these optimization methods can also be reformulated as a divisor or ranking method.  Because of additional structural properties enjoyed by these objective functions, the greedy allocation automatically builds optimal allocations for all House sizes up through the final value $R$, and is also self-consistent under sub-division, in the sense that if the apportionment procedure were repeated on any subset of the states, allocating to that subset the same sub-total of seats those states received in the full apportionment, each state would end up with the same number of seats as it did before.

However, unlike the Huntington-Hill or Webster-Willcox methods, relative entropy provides an unambiguous choice for the normalization of the figure-of-merit with respect to $R$, so if desired, also allows us to assess or choose an optimal House size along with the optimal apportionment of these seats to the various states.  Subsequent to apportionment itself and any proposed choice of intra-state district boundaries, the same relative entropy can also objectively measure the degree of inequity in the size of these Congressional districts.

Relative entropies can also be decomposed in a natural way into between-group and within-group contributions.  So if of interest, we can also separately quantify inequities due successively to: (i) the constraints that each state must receive a whole number of representatives within prescribed bounds; (ii) the possibility that some states may receive sub-optimal numbers of seats; (iii) the requirement that Congressional districts partition the state population into contiguous, non-overlapping districts each containing a whole number of represented individuals; and (iv) the existence of variation in the sizes of proposed districts beyond the unavoidable minimum.  Numerical values for entropies can be interpreted in terms of: the information cost due to the mismatch inherent in presuming (\ie, ``coding'' for) democratic equality but actually employing representative government; or equivalently, as the deficit in surprisal experienced or uncertainty resolved relative to a state of maximal surprise (upon choosing an individual at random) versus choosing via Congressional delegations; or, the difficulty of statistically discriminating samples drawn by polling from the indirect versus direct probability distributions.

\newpage
In reflecting on the Congressional apportionment problem, legal philosopher, First Amendment scholar, and civil rights advocate Zechariah Chafee, Jr.\ once noted \cite{chafee:1929} that ``the preservation of a respect for the law will in the long run be best obtained by the adoption of the plan which is least likely to produce a sense of unfairness in those who are forced to obey legislation.''  While public perception is hard to predict, and perhaps harder to direct or correct, we can say that, in a precise information-theoretic sense, entropic apportionments are least likely to be detectable as unfair by those who are asked to distinguish.

Because quantitative conceptualizations of thermodynamic entropy were not introduced until the work of Clausius, Boltzmann, and Gibbs in the middle to late decades of the 19th Century, and the Theory of Information was not developed by Claude Shannon and others until the middle of the 20th Century, the method of entropic apportionment would not have occurred to our 18th-Century Founders, nor would it have been computationally practical even had it been known, but we like to think it might have appealed to their Enlightenment sensibilities.  Indeed, Shannon's theory did not emerge until after the 1941 statute which has entrenched the Huntington-Hill method into our decadal political dynamics and discourse.  In the light of a more modern and more precise understanding of proportionality that information theory provides, perhaps it is past time to re-think our now antiquated rules for apportionment.

%
%
%


\newpage
\section*{\uppercase{References}}

\bibliography{casl_references}

%
%

\appendix

\renewcommand{\thesubsection}{\thesection.\arabic{subsection}}
\renewcommand{\thesubsubsection}{\thesubsection.\arabic{subsubsection}}

\makeatletter
\renewcommand{\p@subsection}{}
\renewcommand{\p@subsubsection}{}
\makeatother

\newpage
\section{\uppercase{Background and Legal Context}}\label{background}

\subsection{The House of Representatives, the Census, Apportionment, and Districting}

In the framing of the U.S.\ Constitution \cite{constitution:1789}, the ``Great Compromise'' provided the United States with a bicameral legislature, in which all states receive exactly equal representation in the Senate regardless of relative population, but representation ``according to their respective numbers'' of inhabitants in the House of Representatives, usually taken to mean that representatives are to be apportioned amongst the states at least approximately in direct proportion to their represented populations, which in turn are to be determined by ``actual enumeration'' in a decennial census.\footnote{Representatives themselves must be inhabitants of the state represented at the time of election, must be U.S.\ Citizens, and must be at least twenty-five years of age at the start of their service.}  Representatives are to be elected every two years ``by the People'' of the state in question.\footnote{Currently, \textit{at minimum} all residents of the corresponding U.S.\ state who are adult (i.e., eighteen years of age or over) citizens and free of felony convictions are entitled to vote for congressional representatives.  The rights to vote of adult citizens with felony convictions vary by state, and also depend on whether the individual is incarcerated, on probation, on parole, or unconditionally released.}  Given our federal system of government, implicit in Article I is the underlying assumption that each member of Congress represent people within one and only one state, and therefore that each state be granted a whole number of representatives exclusive to that state.  In addition, the Constitution requires that every state is to be given at least one representative, but otherwise no more than one representative for every thirty thousand represented residents.  For states with more than one representative, current statute further requires that each representative actually represent exclusively a subset of the state's population residing within a contiguous geographic district, but this has not always been the case.

In regards to both federal governance and mathematical analysis, this decennial Constitutional obligation raises a number of interesting questions, including how to ``best'' achieve the following:
\begin{enumerate}
\setcounter{enumi}{-1}
\item  obtain official and hopefully accurate enumerations (census) of the populations of the various states relevant for the purposes of apportionment;
\item \textbf{decide on the total number of Congressional Representatives to serve in the House of Representatives} or an acceptable range of sizes, until the next apportionment;
\item \textbf{apportion these Representatives amongst the several states}, within allowed bounds, according to the ``respective numbers'' of inhabitants;
\item for single-member district-based representation within states, determine the allowed sizes (or acceptable range of sizes) of each congressional district;
\item and choose constituencies within each state for each representative, typically in the form of contiguous but non-overlapping and exhaustive geographic residential districts;
\end{enumerate}
where each decision must remain compatible with requirements or constraints imposed by the U.S.\ Constitution, and where deciding on the precise meaning of ``best'' in each of these interrelated, if ultimately distinct, tasks has engendered an extensive and sometimes contentious debate.

Every decade, a census---now conducted by the Census Bureau within the Commerce Department---precedes  the reapportionment of representatives, and has of late created its own controversies, especially regarding the legal and practical meanings of the phrase ``actual enumeration,'' and the possible role, if any, for so-called \textit{imputation}, meaning the use of sampling and/or other statistical methods to help arrive at more reliable counts for the represented populations\footnote{Here what we will call ``represented populations'' refers to what the Census Bureau officially  terms ``populations for apportionment,'' namely the number of people that, under Constitutional requirements and current state and federal laws and regulations, should count toward Congressional apportionment---specifically, the total number of people who are deemed to usually or customarily reside in the state in question around the time the census is completed and are subject to U.S.\ jurisdiction, plus federal employees working oversees whose employing agencies declare that state as the employees' permanent residence, plus any spouses and dependents of such employees.} of each state which are to be used for the apportionment of Congressional seats.\footnote{Since, Constitutionally speaking, only U.S.\ states are entitled to voting representatives in Congress, the represented population does not include residents of the District of Columbia or of U.S.\ overseas territories, which now typically send to Congress \protect\textit{non-voting} delegates who lack full voting rights on the House floor but are often granted certain privileges at the committee level under House rules.  Under current law and executive branch policy, consistent with current Constitutional interpretation as to legislative and executive discretion, the populations for apportionment also exclude U.S.\ citizens living oversees who are not federal employees or their immediate family members, as well as temporary tourists to the U.S., and individuals living within the U.S.\ but who are not subject to its jurisdiction, namely those foreign nationals with diplomatic status.  The federal government grants limited Native American tribal sovereignty under the notion of ``domestic dependent nations'' but now also recognizes native Americans as U.S.\ citizens.  In particular, although Article I, Section 2, Clause 3 of the U.S.\ Constitution states that ``Representatives and direct Taxes shall be apportioned among the several States ... excluding Indians not taxed,'' since the time of the \textit{Revenue and Indian Citizenship Acts of 1924}, all Native Americans born within the territorial U.S.\ are citizens.  (At the time, almost two-thirds of Native Americans were already citizens by right as the result of previous laws and treaties, although of course efforts at effective disenfranchisement were not uncommon).  Thus, Native Americans living within the fifty U.S.\ States are now subject to federal taxation and also counted for apportionment.}

For the purposes of discussing apportionment itself as a distinct optimization problem, we sidestep the census/sampling debates,\footnote{However, in passing we cannot resist venturing a few opinions in this ongoing imputation controversy: namely, that we find it somewhat incomprehensible to imagine that the likes of Thomas Jefferson, Alexander Hamilton, John Adams, or James Madison, if they were alive today, would seriously object to the use of careful statistical techniques, if these methods could be expected to improve the accuracy and/or precision of population estimates.  In our reading, ``actual Enumeration'' does not require literal, one-by-one counting, and would not preclude sensible statistical corrections if these could improve the estimates, but rather disallows assigning counts by mere affirmation or  attribution without a good-faith effort to determine the numbers as accurately as possible.} and simply assume that essentially exact, or at least officially accepted, counts of represented populations within states are known from the most recent census. That is, we assume ``step 0'' above has been achieved successfully.  The actual reapportionment of representatives following the census is decided by the sitting Congress, which ostensibly has wide latitude in principle and legal precedent in its choice of assignment methodology, apart form the above Constitutional constraints as interpreted by the Courts.  Compared to other western democracies, the U.S.\ is somewhat unusual both in the level of involvement of the legislative branch itself in deciding issues concerning the subsequent apportionment of its own representatives, as well as in the frequent willingness of the judicial branch to intervene directly in these issues, and it is in this context that we suggest a new principle with some mathematical claim to greater fairness.

Here we primarily address the technical aspects of this \textit{inter-state apportionment problem}, namely, how to assign a whole number of representatives to each state in a manner that in some sense may be regarded as ``best.''  By this we mean a definite algorithm that is consistent with common sense and Constitutional requirements, is objective and transparent in its application; and whose output is arguably maximally ``fair'' or ``representative'' according to some sensible definition.  By deliberate choice, we mostly---but not completely---ignore the essential but overtly political issue, as to whether any particular method would likely be adopted by the existing Congress,\footnote{Particularly our current Legislative branch....} especially if considered \textit{after} a census is completed, when the consequences for all states of any proposed procedure could be predicted.
 
Apart from the requirements mentioned above and a few other hints mentioned in passing, the Constitution otherwise remains essentially silent on the details of how voters are to choose their representatives, or exactly whom individual representatives are to represent within a state, or how they are to be represented when the state is accorded more than one representative.  But Congressional statue has circumscribed state discretion in many of these areas.  In principle, at least, one might envision any number of complex schemes, including forms of collective, multi-member, or at-large representation, where each voter within a state might contribute to the selection of several or even all the state's representatives, who would then represent these constituents to some degree.

However, though not explicitly required by the Constitution, all states with multiple representatives are currently required by statute to use a partitioned representation scheme, whereby each state's population is divided into mutually exclusive and exhaustive, geographically contiguous but non-overlapping single-member districts, equal in number to the states apportionment of congressional seats, and each representative is selected by, and then represents, his or her individual district within the state.  The thinking seems to be that, with such a system, representatives may be more effective and more accountable in some sense if each answers to a smaller and clearly-defined constituency with some shared interests, that voters in turn need only inform themselves about a smaller number of candidates, and/or perhaps that such a partitioned system may be easier to administer.  It is not entirely clear whether the choice of geographically-contiguous boundaries is done to simplify districting, campaigning, and/or voting, to reduce opportunities for overt gerrymandering,\footnote{Clearly, the current system fails spectacularly at this.} or else to reflect a sincere (if perhaps debatable) expectation that shared voter concerns and experiences are correlated more strongly with geographic proximity than any other obvious demographic factors, such as party affiliation, type of employment, or socioeconomic status.
\newpage
In any case, the  reliance on (geographic) districting, and therefore periodic redistricting, creates an additional \textit{intra-state apportionment problem}, as to exactly how to partition the states population into districts.  Here, we also largely ignore the politically-charged question of specifically \textit{which} people to include in which district, or equivalently \textit{where} to draw the district boundaries,\footnote{In particular, debate continues as to the extent to which certain districts may be chosen deliberately with ``majorities of minorities,'' to better ensure that certain minority interests are represented somewhere and to comply with the Voting Rights Act, or whether districts should be chosen to the extent possible with a balance of interests and party affiliations to encourage a vigorous competition of ideas.  In practice, we currently see evidence in the U.S.\ of very significant gerrymandering at the level of Congressional districting, specifically the consequences of overt ``cracking'' (dispersing a group of voters into several districts to prevent them from reaching an actual majority), ``packing'' (combining as many like-minded voters into one district as possible to allow them to win small numbers of seats with large majorities but prevent them from affecting elections in other districts, which can be won by smaller majorities),  and ``stacking'' (grouping an apparent majority of lower-income, lower-education, younger, or otherwise lower-turnout voters into districts with more-educated, wealthier, whiter, or older voters who tend to turn out in greater numbers).  Discussion continues as to how to better detect, quantify, and limit gerrymandering at the Congressional level by parties controlling the state legislature or other bodies responsible for the re-districting.} but we do find that the prior question of \textit{how many} people should be assigned to each district within each state should not and cannot be completely separated from the question of how many seats each state should receive.

In addition, fair apportionment of representatives may depend on how representatives actually cast votes in the House, or how their votes are counted.  Historically, each full representative has been given one equal, binary vote in a majoritarian (or sometimes super-majoritarian) legislative system, but if ,as some have suggested, multiple voting or weighted voting were to be used, such a scheme might compensate, in a certain sense, for discrepancies in apportionment itself.  However, the Constitutionality of any such weighted voting schemes would be highly questionable, and in any event would not seem to stand an appreciable chance of adoption in practice, as many representatives would have to willingly surrender influence.  More is said about this below, but we remain extremely dubious of these suggestions.

Finally, although we focus here on apportionment in the context of partitioned representation and in particular single-member districts, note that the general method we advocate could easily accommodate multi-member districts, collective representation, or hybrid representational models.

\subsection{Represented Populations Versus Voting Populations}

There has been some debate, both historically and recently, about exactly whom Representatives are to represent. Partly this may be due to inaccurate language.  In the jurisprudence concerning the Equal Protection clause, one often hears of the ``one person, one vote'' principle.  But in the context of Congressional apportionment, we have argued that this might be more accurately described as the ``One Person, One \textit{Voice}'' principle.

The U.S.\ Constitution seems unambiguous in requiring that Representatives are to be allocated based on the represented populations of the states,\footnote{Again, the  population of a state is now understood to be the nominally permanent population of that state under U.S.\ jurisdiction excluding certain felons, as determined by the most recent census, or more specifically what are called populations for apportionment.} not their respective numbers of actual voters, or eligible voters, or registered voters, or adults, or U.S.\ citizens, or citizens plus documented immigrants, etc.  This was reaffirmed in the 14th Amendment, because in the debate preceding its ratification, Congress explicitly considered and \textit{rejected} proposals to instead allocate seats to states on the basis of voter populations.  As written and ratified, the 14th Amendment does distinguish citizens from residents, saying ``All persons born or naturalized in the United States, and subject to the jurisdiction thereof, are citizens of the United States and of the State wherein they reside,'' but also explicitly declares, ``nor shall any State deprive any person of life, liberty, or property, without due process of law; nor deny to any person within its jurisdiction the equal protection of the laws,'' and then asserts, ``Representatives shall be apportioned among the several States according to their respective numbers, counting the whole number of persons in each State....''

\newpage
Although the actual case focused on on districting for state-level legislatures, this principle was re-affirmed by the Supreme Court in the recent \textit{Evenwel v.~Abbott} decision \cite{ginsburg:2016}, where Justice Ginsburg,\footnote{Now known across the internet as ``The Notorious R.B.G.'' thanks to Shana Knizhnik.} writing for the majority, asserted that ``The Framers recognized that use of a total-population baseline served the principle of representational equality....  this Court recognized in \textit{Wesberry} ... the Constitution's plain objective of making equal representation for equal numbers of people the fundamental goal for the House of Representatives.''  She goes on to say:
\begin{quote}
Settled practice confirms what constitutional history and prior decisions strongly suggest. Adopting voter-eligible apportionment as constitutional command would upset a well-functioning approach to districting that all 50 States and countless local jurisdictions have long followed. As the Framers of the Constitution and the Fourteenth Amendment comprehended, representatives serve all residents, not just those eligible to vote. Nonvoters have an important stake in many policy debates and in receiving constituent services. By ensuring that each representative is subject to requests and suggestions from the same number of constituents, total-population apportionment promotes equitable and effective representation
\end{quote}
In the \textit{Federalist Papers} \cite{hamilton:1788}, James Madison declared that ``it is a fundamental principle of the proposed constitution that ... the aggregate number of representatives allotted to the several states, is to be ... founded on the aggregate number of inhabitants,'' while ``the right of choosing this allotted number in each state, is to be exercised by such part of the inhabitants, as the state itself may designate.''  That is, the total number of inhabitants were to form the basis for allotting apportioning representatives, even though only some of those inhabitants might be eligible to participate in the choice of said representatives.\footnote{Again, at the time of ratification and for almost  a century thereafter, certain inhabitants, namely slaves, were counted only as three-fifths of a person.   And the 13th, 14th, 15th, and 19th Amendments have now further restricted the ability of states to limit those who are entitled to vote for the representatives.}  Alexander Hamilton declared: ``There can be no truer principle than this---that every individual of the community at large has an equal right to the protection of government.''

In Congressional debate surrounding the 14th Amendment, Representative James Blain argued \cite{ginsburg:2016}, ``no one will deny that population is the true basis of representation; for ... non-voting classes may have as vital an interest in the legislation of the country as those who actually deposit the ballot.''  (Of course, others have attempted to deny this, hence the extensive debate, but the side favoring  a total-population basis won out over those favoring voter-eligible populations).  Introducing the final version of the Amendment on the Senate floor, Senator Jacob Howard declared,
\begin{quote}
the `basis of representation is numbers...; that is, the whole population except untaxed Indians [sic] and per-sons excluded by the state laws for rebellion or other crime... The committee adopted numbers as the most just and satisfactory basis, and this is the principle upon which the Constitution itself was originally framed, that the basis of representation should depend upon numbers; and such, I think, after all, is the safest and most secure principle upon which the Government can rest.  Numbers, not voters; numbers, not property; this is the theory of the Constitution.
\end{quote}
And in its famous \textit{Westbury} \cite{westbury:1964} decision, the Court ruled, ``The debates at the Convention make at least one fact abundantly clear: that when the delegates agreed that the House should represent people, they intended that in allocating Congressmen [sic] the number assigned to each state should be determined solely by the number of inhabitants ... While it may not be possible to draw congressional districts with mathematical precision, that is no excuse for ignoring our Constitution's plain objective of making equal representation for equal numbers of people the fundamental goal for the House of Representatives.''

\newpage
So despite some very weak arguments\footnote{Short of advocating for a Constitutional amendment, it would seem that counterarguments must turn on strained or absurd readings of phrases like  ``they reside,'' or ``any person within its jurisdiction.''  Some try to argue that non-citizens, or undocumented immigrants, are somehow not subject to the jurisdiction of the federal government or of the government of the state in which they clearly reside.  Really the only sorts of persons who are actually excluded are temporary visitors whose regular place of residence is elsewhere, or foreign nationals with diplomatic immunity who are therefore not subject to the full weight of U.S. law.} to the contrary,\footnote{See, for example, reference \cite{krabill_fielding:2012}.} it should be clear that, Constitutionally speaking, the Representatives are expected to represent all inhabitants of their respective districts within their respective states, not just those who voted for them, or those who voted at all, or those who could have voted, or those enjoying full citizenship, or those who were born in or legally immigrated to the U.S.\footnote{Currently, the most  obvious difference between those individuals who who are counted in deciding how many representatives each state receives, and those individuals who actually get to vote, include legal minors and resident aliens, as well as certain felons.  Historically, the most hypocritical and unjust of such democratic discrepancies, articulated in the infamous ``Three-Fifths'' compromise,'' was finally overturned by the 13th, 14th, and 15th Amendments, while another notably egregious instance, that of women's suffrage, was finally rectified in the 19th Amendment. Indigenous peoples are now in effect regarded as dual citizens of the U.S.A.\ and of their respective Tribal Nations.}  Likewise, all inhabitants are entitled to representation, and indeed to equal representation to whatever extent possible, whether they can or do vote.  This is what we shall mean by the \textit{One Person, One Voice} principle, and it very much informs our choice of apportionment method.

\newpage
\section{\uppercase{Some Mathematical Notations and Definitions}}\label{parameters}

Here we define and explicate in more detail various quantities and symbolic notation used throughout this text.
  
\subsection{Populations for Apportionment}  

Following standard if a bit sloppy convention, we use ``population'' to refer either to some identified collection of persons (i.e., the populace of a country, state, or district, regarded as a set or an ordered sequence), or to the \textit{number} of such individuals therein (i.e., the population count).  Hopefully the intended meaning will be clear from context.

Let $p_s$ denote the \textit{represented population} (\textit{count}),\footnote{These are the individuals residing in or otherwise legally associated with the various states \textit{at the time of the most recent census, as outlined above.}.  In between censuses, of course people are born and die, move between states or districts, change their legal residency or citizenship, etc.  So we do \textit{not} mean to imply that, going forward, representatives \textit{only} represent the individuals contributing to the count $p_s$, but rather, that these were the ones upon which the amount of representation was decided.} also known officially as the \textit{population for apportionment}, as determined by the most recent U.S.\ Census, of the $s$th state out of a total of $S$ states, for $s = 1, \dotsc, S$, where of course $S = 50$ currently.\footnote{Note once again that under the U.S.\ Constitution, U.S.\ territories and the District of Columbia have no voting members in the Senate or House.  They may be granted non-voting observers, and via the 23rd Amendment, Washington, D.C. has been granted (currently three) votes in the Electoral College.}  The \textit{total} represented population across all states is then given by the sum
\be
P \equiv \sum\limits_{s = 1}^{S} p_s,
\ee
which, again, actually excludes some people residing within the territorial U.S., but also includes some federal workers and their dependents living abroad.  The average state population is then just $\bar{p} = P/S$.
 
 A non-trivial apportionment problem arises when representatives are to be divided across multiple states, and presumably every such state houses a non-vanishing fraction of the total population.\footnote{Although not technically forbidden by the U.S.\ Constitution, a completely unpopulated territory, if there were still, or ever, such a region under U.S.\ jurisdiction, would presumably never be admitted into the Union as a separate state---there would be nobody to represent, and nobody to represent them.}  So in actual practice the represented populations of relevant states may be assumed to satisfy the inequalities
\be
0 < p_{\stext{min}} \equiv \min\limits_{s} \bigl[p_1, \dotsc, p_S \bigr] \le p_s \le p_{\stext{max}} \equiv  \max\limits_{s} [p_1, \dotsc, p_S ] < P,
\ee
for all $s = 1, \dotsc, S$.

For notational efficiency, we also collect the states' represented populations into an ordered $S$-tuple: $\bv{p} = (p_1, \dotsc, p_S)$,  Within the $s$th state, all $p_s$ represented persons may be indexed (without regard to district membership) by $n = 1, \dotsc, p_s$, using some lexicographic or other ordering convention.

Constitutionally speaking, one further adjustment to the populations for apportionment could in principle arise. Section 2 of the 14th Amendment declares:\footnote{This clause was included out of a concerns that, with the voiding of the Three-Fifths Compromise, southern states would earn more Congressional representation, but would continue to disenfranchise their (adult male) African-American populations.  In fact, even though southern States continued to deploy various pretexts to suppress the African-American vote, this clause was never invoked, and now the 1965 Voting Rights Act technically makes such denials illegal.  Evidently, the 19th Amendment included no similar clause for women because it more explicitly denied states the ability to abridge the right to vote based on sex.}
\begin{quote}
But when the right to vote at any election for the choice of electors for President and Vice President of the United States, Representatives in Congress, the executive and judicial officers of a state, or the members of the legislature thereof, is denied to any of the male inhabitants of such state, being twenty-one years of age, and citizens of the United States, or in any way abridged, except for participation in rebellion, or other crime, the basis of representation therein shall be reduced in the proportion which the number of such male citizens shall bear to the whole number of male citizens twenty-one years of age in such state.
\end{quote}
This seems to mean more than just disallowing states from counting disenfranchised inhabitants towards apportionment.  If $\sigma_s$ is the \textit{ratio} of adult male non-felon citizens residing within the $s$th state who are denied the right to vote in federal elections, 
divided by the total number of adult male\footnote{Here, ``adult male'' meant a male 21 years of age or older, but the 26th Amendment denied states the ability to violate voting rights of citizens in a federal election based on age, for anyone 18 years or older, but did not explicitly update this clause, and again the 19th Amendment extended voting rights to women also without replacing this clause.  So technically, it appears that if a state were to deny the right to vote on grounds other than age or sex to certain non-felon citizens who happened to be in the $18$--$20$ age range, then the population for apportionment would not be penalized.  Of course any such overt disenfranchisement could run afoul of statutory constraints such as the Voting Rights Act.}  residents of that state, then the total population for apportionment is to be reduced from $p_s$ to $p_s' = (1-\sigma_s) p_s$. 

\subsection{Apportioned Congressional Representatives}\label{apportioned_representatives}

The \textit{apportionment} $a_s$ of the $s$th state is the whole number congressional of representatives or ``seats'' in the U.S.\ House of Representatives legally assigned or allotted to that state, for a total House size of
\be
R \equiv \sum\limits_{s = 1}^{S} a_s
\ee
voting representatives in all.

Additionally, the U.S.\ Constitution (Article I, Section 2)  would appear to require that
\be
1 \le a_s \le \max\bigl[ 1, \lfloor p_s /  \Dmin \rfloor  \bigr] \hspace{28pt} \mbox{ for } s = 1, \dotsc, S,
\ee
where $\Dmin =  30\, \hspace{-0.44pt}000$ is in effect a mandated minimum representation ratio, and a \textit{minimum district size} for any states with multiple representatives.  This in turn implies that the total number of voting representatives $R$ to be apportioned must lie within some range
 \be
 R_{\stext{lb}} \le R \le  R_{\stext{up}},
\ee
where Constitutionally-implied lower and upper bounds on the House size $R$ are given respectively by the number of states
\be
R_{\stext{lb}} = S,
\ee
at the lower end, and by
\be
R_{\stext{ub}}  =  \sum_{s = 1}^{S}   \max\bigl[ 1, \lfloor p_s /\Dmin \rfloor  \bigr],
\ee
at the upper end, regardless of how the representatives are otherwise apportioned.  Currently it is the case that $p_s \gg \Dmin$ for all $s = 1, \dotsc, S$, so that $S < R_{\stext{ub}} \le   \lfloor P/\mathscr{D}_{\stext{min}} \rfloor$.  Of course, for parliamentary reasons (e.g., expectation of more efficient conduction of House business), or out of political calculation, or simply institutional inertia or tradition, the actual number of voting representatives $R$ may be partially or completely constrained prior to their actual apportionment, to some smaller interval:
 \be
 S \le  R_{\stext{min}} \le  R \le R_{\stext{max}} \le R_{\stext{ub}}.
 \ee
Since 1911,\footnote{Apart from a brief interlude when Hawaii and Alaska first became states.} Congress has in fact decreed that $R_{\stext{min}} \! = R = R_{\stext{max}} = 435$, although nothing but convention or precedent requires that this exact if arbitrary total be enshrined in law.\footnote{In fact, a very strong argument can (and has) been made \cite{baker:2009,thirtythousand:2010,kromkowski:1991,conley:2011,bartlett:2014,shugart:2014} that freezing the House size at $435$, over a period of history during which the population of the U.S.\ has essentially tripled, leads to significant distortions from the ideal of proportional representation, and growing malapportionment.  This number has been fixed by the \textit{Permanent Apportionment Act of 1929}, which itself was a belated attempt to resolve a crisis following the 1920 Census, when rural and nativist representatives in the U.S.\ House and Senate, concerned over support for Prohibition and the influence of ``urban'' and ``foreign'' elements in America's rapidly growing cities (sound familiar?), simply ignored Constitutional mandates and blocked reapportionment based on the latest population counts that revealed increased urbanization. As of the 2010 Census, the U.S.\ House of Representatives had about one representative for every $710\hspace{1.1pt}000$ people, one of the highest ratios amongst any popularly-elected national assemblies or parliaments, and over $23$ times higher than the Constitutionally mandated minimum. And because of the mandated minimum of one representative per state, large differential changes in state populations coupled with a fixed House size has led to a situation where the relative variation in district sizes between large and small states has become considerably greater than any judicially tolerated variation of district sizes within states.  However, it tends to be the lower-density, rural, Republican-leaning stares which benefit from this malapportionment, so we should not expect improvements any time soon.  Choice of House size is also discussed in Section~\ref{house_size} .}  Because, Constitutionally speaking, each state is also entitled to at least one representative, it follows that for any feasible choice of $R$, the $s$th state's allotment will necessarily lie in the range
\be
1  = \lambda_s \le a_s \le  u_s =  \max\bigl[ 1, \lfloor p_s /  \Dmin\rfloor  \bigr]
\ee
for all $s = 1, \dotsc, S$, where we have introduced the lower bound $\lambda_s = 1$ and upper bound $u_s$ on the $s$th state's possible apportionments.\footnote{Actually, for Congressional apportionment, we can impose possibly tighter upper bounds, namely $u_s = \min\Bigl[ \max\bigl[ 1, \lfloor p_s / \Dmin\rfloor  \bigr], R - S + 1 \Bigr]$, but this is unnecessary if we are already imposing the lower bounds.  For party-list apportionment problems, we could instead relax these bounds, setting $\lambda_s = 0$ and $\lambda_u = R$, which are then always satisfied and hence ignorable.}  In current practice, with the House size fixed at $R = 435$ and state populations what they are, the upper bounds are not terribly relevant, but the lower bounds very much are, since the populations of the smallest states fall substantially below the average district size overall.

This overall average delegation size, or average number of representatives per state, is of course just $\bar{a} = \frac{R}{S}$.  For later convenience, we can also define an ordered $S$-tuple of apportionments $\bv{a} = (a_1, \dotsc, a_S)$, referred to as the apportionment vector.  Similarly, we can define $\bv{\lambda} = (\lambda_1, \dotsc, \lambda_S)$ and $\bv{u} = (u_1, \dotsc, u_S)$, as the $S$-tuples of allowed lower and upper bounds.

\subsection{Congressional Constituencies} 
 
\subsubsection{District-Based Representation}    
  
With standard \textit{partitioned representation}, or equivalently \textit{single-member district-based representation}, the $k$th congressional representative in the $s$th state is to represent exclusively the inhabitants of a \textit{district} (entirely within the state) in which reside a whole number $d_{sk}$ of represented individuals.  The districts are assumed to contain exclusive and exhaustive sub-populations, each sending one representative to the House,\footnote{Represented individuals living abroad are associated with their permanent or previous congressional district.} so that the population of any state is decomposed as:
\be
p_s = \sum\limits_{k = 1}^{a_s} d_{s k}.
\ee
Mathematical and Constitutional constraints will further imply that
\be
\min[ p_s, \Dmin] \le d_{sk} \le p_{s} -  (a_s - 1)\Dmin.
\ee
Currently, all U.S.\ Congressional districts are much larger (by a factor of $20$ or more) than the allowed minimum, such that $d_{sk} \gg \Dmin$.

Nationwide, the overall \textit{average congressional district size} may be defined as
\be
\Dbar =  \frac{1}{R} \sum\limits_{s = 1}^{S} \sum\limits_{k = 1}^{a_s} d_{sk} =  \frac{1}{R} \sum\limits_{s = 1}^{S}  p_{s} =  \frac{P}{R},
\ee
while the \textit{intra-state average district sizes} are 
\be
\bar{d}_s  =   \frac{1}{a_s} \sum\limits_{k = 1}^{a_s} d_{sk}  =  \frac{p_s}{a_s},
\ee
and will satisfy
\be
 \min\bigl[p_s, \Dmin  \bigr]  \le \bar{d}_{s} \le p_s \hspace{28pt} \mbox{ for all } s = 1, \dotsc, S.
\ee
Note that neither the national average $\Dbar$ nor the  intra-state averages $\bar{d}_s$ are whole numbers in general, although it must be the case that $\sum\limits_{s = 1}^{S} \sum\limits_{k = 1}^{a_s} \bar{d}_s =  \sum\limits_{s = 1}^{S} a_s\,\bar{d}_s  =  R \, \Dbar = P$.  The average number of representatives per state can also be written as $\bar{a} = \tfrac{ \bar{p} }{ \Dbar }$.

We will typically use $k = 1, \dotsc, a_s$ to index the Congressional districts within a state.  Represented individuals within the $k$th district of the $s$th state will be indexed by $i = 1, \dotsc, d_{sk}$.

In the context of party-based representation, what we have called the average district size $\Dbar$ is instead known as the \textit{Hare quota} or \textit{simple quota} (not to be confused with the exact state quotas for representatives, to be defined below), named after British political scientist Thomas Hare.\footnote{Modifications of this sort of quota are also encountered in the literature and arise in various proportional representation and apportionment schemes.  For example, in our notation, the \textit{Hagenbach-Bischoff quota} is $\tfrac{P}{R+1}$, and the \textit{Droop quota} is given by $\tfrac{P}{R+1} + 1$, but we will not make use of them here.}  The Hare quota measures the ideal number of persons (or valid votes, in party-list representation) required to ``deserve'' one representative, and populations (or vote counts) are sometimes expressed in units of Hare quotas.\footnote{In the apportionment context, populations measured in units of the Hare quota will be numerically equal to the exact state quotas defined below....} 

For later convenience, we may also define for each state a \textit{district residual}
\be
\eta_{s} = p_s \!\! \! \mod a_s = a_s\bigl(  \tfrac{p_s}{a_s} - \lfloor  \tfrac{p_s}{a_s}  \rfloor \bigr) = a_s\bigl( \bar{d}_s - \lfloor  \bar{d}_s  \rfloor \bigr)
\ee
as the minimal number of individuals left over if we try to divide the $s$th state's represented population into exactly $a_s$ equally-populated districts.  Obviously each $\eta_s$ must be a nonnegative integer satisfying $0 \le \eta_s < a_s$, while
\be
p_s = a_s \bar{d}_s = a_s \lfloor \bar{d}_s \rfloor + \eta_s
\ee
for each $s = 1 , \dotsc, S$.

\subsubsection{District Sizes versus Congressional Constituencies}

Like with ``population,''  a reference to a ``district'' can mean either the size of the district (number of represented persons), or the actual identity of the district---the set of represented persons residing within or otherwise legally assigned to the district at the time of the census.  Suppose that all represented persons (at the time of the census).  We have assumed that represented persons within the $s$th state are indexed uniquely by $n$ for $n = 1, \dotsc, p_s$.\footnote{That is, we presume some invertible mapping or lookup table between the ordered pair of indices $(s,n)$  and some collection of distinguishing legal information unambiguously singling out an individual, such as full name, date and place of birth, residential addresses at time of census, Social Security Number if available, biometric data, etc.}  Once  district boundaries are drawn, we also assume that districts within the $s$th state may be unambiguously indexed by $k$ for $k = 1, \dotsc, k$, and that using the census data and the adopted district boundaries, each censused individual can be unambiguously assigned\footnote{For example, as of the 2010 Census, the author resided in the CA-$13$.} to a district via some mapping  $k = \EuScript{K}(n,s)$, such that the censused residents can be partitioned into exclusive and exhaustive Congressional constituencies $\C_{s_k} = \{ (s,n) \colon \EuScript{K}(n,s) = k \}$, where the cardinality of the set $C_{s_k}$ is just the corresponding district size $d_{sk}$: $\lvert \C_{s_k} \rvert = d_{sk}$. Intra-district indices $i = 1, \dotsc, d_{sk}$ can then be uniquely assigned to represented individuals within any district $(s,k)$, by some specified one-to-one mapping.\footnote{For example, a simple convention would assign $i = 1$ to the smallest $n$ in the district, $i = 2$ to the next smallest, etc.}

\newpage
\subsubsection{At-Large Representation and Other Representation Schemes}

Historically, a few states instead sometimes chose to elect representatives from the state \textit{at large}, leading to 
\textit{non-partitioned} or \textit{state-wide} representation.\footnote{The Constitution is largely silent on how representation is handled within states, except for demanding that Representatives are to be elected by the people of the state, and equal protection requirements that have been interpreted to demand that districts be closely matched in size when partitioned representation is used. By statute, at-large representation has been explicitly allowed or disallowed at various points in U.S.\ history, but mostly the latter.  Most recently, as of 2006, states with two or more apportioned representatives must be partitioned into geographically contiguous, exclusive and exhaustive districts within the state, each of whose voters elects a single representative.  So currently, the only ``at-large'' members are those from states with only a single representative.  Equivalently, we can interpret the latter scenario in terms district-based representation, but where the state in question has only a single district covering the state as a whole.}  In order to also accommodate this possibility within a uniform notation, we may define \textit{effective district sizes} $\tilde{d}_{sk}$, where
\be
\tilde{d}_{sk} = \begin{cases}
d_{sk} & \text{ if representation is district-based}\\
\bar{d}_s = \tfrac{p_s}{a_s} & \text{ if representation is at-large}
\end{cases},
\ee
in effect adopting the \textit{convention}, in the case of at-large representation, of  formally ``dividing'' the population $p_s$ into $a_s$ ``virtual'' districts each of effective size $\tilde{d}_{sk} = \bar{d}_s$, even though in actual practice these would not be actual geographic or administrative units and would need have no demographic reality nor any direct political significance, nor indeed necessarily even be associated with an integral number of persons.

Of course, hybrid representation schemes can be envisioned in principle---for example, some  districts could have multiple representatives, or districts could be  grouped into meta-districts with their representatives, or the state could have some other combination of district-based and at-large representation.\footnote{Political obstacles aside, it might be worth re-considering some of these schemes.  For example, having fewer districts but with multiple representatives could, under certain voting rules (such as instant-runoff), greatly reduce effects of gerrymandering.}.  Because these sorts of hybrid representational systems would seem to be somewhat exotic politically, and to our knowledge have not been used historically in the U.S., and are currently disallowed by statute, we will not pursue specific ideas further here, but if circumstances arose, our methodology could easily be generalized to allow for more elaborate or esoteric representation scenarios.

We also collect the average district sizes into $\bar{\bv{d}} = (\bar{d}_1, \dotsc, \bar{d}_S)$, and the state-level districts sizes into a nested sequence of $S$ $a_s$-dimensional ordered tuples $\bv{d}_{s} = (d_{s\, 1}, \dotsc, d_{s\,a_s} )$, and finally embed the latter into an array $\D$ with dimensions $S \times \max\limits_{s} \bigl[a_s \bigr]$, zero-padded as necessary, such that $[ \D ]_{sk} = d_{sk}$ for $S$ and $k$ in the allowed ranges, and zero otherwise.

\subsection{State Quotas}

For the $s$th state, the corresponding state-level \textit{quota} $q_s$, also called the \textit{exact quota} or the \textit{standard quota} of representatives, is the ideal number of representatives to which the $s$th state is entitled assuming a total of $R$ representatives and a goal of exactly proportionality in representation in the absence of other constraints.  That is, the state-level quota $q_s$ is determined by the condition
\be
\frac{q_s}{R} = \frac{p_s}{P},
\ee
which says that the proportion of representatives ``deserved'' by a state is equal to the proportion of that state's population relative to the whole population.  The $s$th state's quota can be expressed in any of the forms
\be
q_s \equiv \frac{\,p_s}{P} R =  \frac{R}{P} p_s =  \frac{\, p_s }{\Dbar},
\ee
and will satisfy
\be
0 \le q_s \le R  \hspace{28pt} \mbox{ for each } s = 1, \dotsc, S,
\ee
where the inequalities are strict in the usual case when the state's population satisfies $0 < p_s < P$ strictly.  As with other state-level quantities, we can collect the quotas into the $S$-tuple $\bv{q} = (q_1, \dotsc, q_S)$.

In this context, $\Dbar$ is also known as the \textit{standard divisor}, because $\Dbar$ provides the ideal ratio of represented persons per representative and is the starting point for so-called \textit{divisor} methods for apportionment, discussed in Appendix~\ref{typologies}.

In general, the quotas $q_{s}$ will not be exact integers---this is the principle reason why we face an apportionment \textit{problem}---although the quotas of various states are necessarily in rational ratio to each other, $\frac{q_{s}}{q_{s'}} = \frac{p_{s}}{p_{s'}}$, and it must be the case that $\sum\limits_{s=1}^{S} q_{s} = \sum\limits_{s=1}^{S} a_{s} = R$ exactly.

Given an apportionment, the ratios $\tfrac{a_s}{q_s}$, or their reciprocals $\tfrac{q_s}{a_s}$, measure the relative deviation between the actual and ideal levels of representation for each state.  For example, the average within-state district sizes are related to the  national average district size by
\be
\bar{d}_s = \frac{p_s}{a_s} = \frac{q_s}{a_s} \Dbar.
\ee

Following standard terminology, we also define the \textit{lower quota} $ \lfloor q _s \rfloor $ as the floor\footnote{The \textit{floor} of a nonnegative real number is the largest integer less than or equal to that number.} of the quota $q_s$, meaning the lower quota is defined as the exact quota rounded down to the nearest integer, and similarly we define the upper quota $\lceil q _s \rceil$ as  the ceiling\footnote{The \textit{ceiling} of a nonnegative real number is the smallest integer greater than or equal to that number.} of the quota, which is to say the exact quota rounded up to the nearest integer.  The lower and upper quota are integers bracketing (non-strictly) the exact quota, 
\be
\lfloor q _s \rfloor \le q_s \le \lceil q _s \rceil,
\ee
and will differ by exactly one representative, except in the unlikely event that the quota itself is an exact integer, in which case the lower, upper, and exact quotas would all coincide.

One declares an apportionment to \textit{satisfy lower quota} if and only if $a_{s} \ge \lfloor q_s \rfloor$ for all $s = 1, \dotsc, S$, and to \textit{satisfy upper quota} if and only if $a_{s} \le \lceil q_s \rceil$ for all $s = 1, \dotsc, S$. The apportionment is said to \textit{satisfy quota} or be \textit{on quota} if and only if $\lfloor q_s \rfloor  \le a_s \le  \lceil q_s \rceil$ for all $s = 1, \dotsc, S$, so that every state receives a whole number of representatives equal to either its lower or upper quota, and therefore a number of representatives within less than one representative of its exact quota.\footnote{Since the number of representatives is necessarily an integer while the exact quota is generically not, a better terminology might have been ``near quota,'' but we conform to conventional usage, and the phrase ``near quota'' has taken on a different meaning.  Also note that the ``on quota'' condition is equivalent to the constraints that  $| a_{s} - q_{s} | < 1$ for all $s = 1, \dotsc, S$.  Notice that this definition does not permit any wiggle room in the unlikely case that $q_s$ does happen to be an exact integer, since then the upper and lower quotas would coincide, rather than differ by one representative.  A more sensible, or at least more forgiving, definition might have been instead: $| r_{s} - q_{s} | \le 1$ for all $s$, but again we shall stick with the more restrictive, but conventional notion.}  Otherwise, the apportionment is said to be \textit{off quota} or to \textit{violate quota}, for one or more states.

Despite Balinski and Young having once referred to the state-level quota as a ``fundamental measure of fairness,'' \cite{balinski_young:1974}, and elsewhere declaring that ``any apportionment should satisfy quota'' \cite{balinski_young:1975}, we contend that violation of quota is not problematic \textit{per se}, because states do not have entitlements to House members; rather, the people of the various states do\footnote{Following the 17th Amendment enacted during America's Progressive era, Senators are now also elected by the people of the respective states, but in the Senate there is of course no possibility nor expectation of equality of representation amongst inhabitants, but rather equality between states.}.  Satisfying quota is simply not a necessary or even appropriate criterion.  Of course, the quotas themselves are just state populations expressed in units of the average district size, so quotas will obviously play some direct or indirect role in any apportionment method.  We will want apportionments to be close to quotas in some sense, but we need not obsess over whether the state-level apportionments are all within one unit or less of their respective quotas.  Whether an off-quota apportionment will be better than an on-quota choice must depend on the absolute number of represented persons affected by the difference in seats.

\subsection{``Fair'' Shares}

The notion of a state's so-called \textit{fair share} of representatives is closely related to the quota, but explicitly takes into account the Constitutionally-mandated lower bound, and, when actually relevant, the upper bound.  That is, since the Constitution entitles the people of each state to at least one representative, if the quota for a small state happens to fall below unity, then, so the argument goes, the quota itself can no longer be considered the state's ``fair share,'' if by fair we were to mean what the Constitution demands or suggests.  Rather, the fair share of representatives would be exactly unity for this state. Likewise, for any states whose quotas lie above their allowed upper bounds, fair shares should coincide with the mandated upper bounds.  And fair shares for all other states are to be determined self-consistently by dividing up the \textit{remaining seats} in an ostensibly self-consistent manner, based on the ratios of the remaining state populations.

We are altogether skeptical of the utility or even sensibility of this notion, because the concept seems to deliberately conflate the representation to which a state is legally entitled and what would be morally or democratically ``fair'' from the perspective of equal representation.  Nonetheless, we will briefly describe the determination of the fair shares as articulated by Balinski and Young.

The fair shares $\f_s$, $s = 1, \dotsc, S$ may be defined and calculated recursively.  First, exact quotas $q_1, \dotsc, q_S$ are calculated for all states, using the fractional populations $p_1/P, \dotsc, p_S/P$ multiplied by the full congressional house size $R$.  If any state's quota falls below that state's allowed lower bound $\lambda_s$, then it is assigned the lower bound as its fair share, or else if its quota lies above the allowed upper bound $u_s$, its fair share is taken to be $u_s$.  Then \textit{only} for the remaining states, \textit{adjusted} quotas $q_s' = p_s R'/P'$ are calculated using the fractional populations $p_s/P'$ relative to the aggregate population $P'$ of the only the remaining states whose fair shares have not yet been determined, multiplied by the remaining available seats $R'$ which were not allotted as fair shares previously. Then, if any of these adjusted quotas are at or out of mandated bounds, those states are assigned integral fair shares equal to the appropriate bound, and the process is repeated, obtaining further modified quotas $q''_s$, etc, until all states have either been assigned a lower or upper bound as a fair share, or else some non-integral fair share lying strictly within the allowed bounds.

Fair shares can also be calculated iteratively rather than recursively.  The fair share of a state may be expressed simply as
\be
\f_s = \text{median}[\lambda_s,\, \alpha \,q_s,\, u_s],
\ee 
where the common scale factor $\alpha > 0$ is to be chosen self-consistently (via iterative trial and error), so that $\sum\limits_{s=1}^{S} \f_s = R$ exactly.

Analogous to the terminology introduced for state quotas, we can compare the apportionments $a_s$ to $\lfloor \f_s \rfloor$ and $\lceil \f_s \rceil$ and speak of apportionments either satisfying or violating fair share, or just lower fair share, or just upper fair share, for one or more states.

Balinski and Young (in \cite{young:1985}) identify the fair shares with the ``states' exact entitlements'' and assert that ``if the [fair] shares are all integers, then they must constitute the unique acceptable apportionment.''  Many authors, including this one, would disagree.  To our thinking, the fair shares greatly over-emphasize the role payed by the lower/upper bounds constraints, and can deviate far too much from actual proportionality to be called ``fair'' in any democratic sense.  If, because of a mandated lower bound, some state must receive more than its quota of seats, then some other state must receive less than its quota, but this does not mean that, morally or democratically speaking, the people of the under-represented state were any less entitled to their due proportion of representatives.

\subsection{Weight of Representation and Related Notions}

\subsubsection{Voting Strength} 

Effective voting strengths, voting weights, voting power indices,  etc., can and have been defined in a number of distinct ways in various approaches to analyzing elections, governance, and collective choice.

Several commonly-used measures of \textit{voting strength} \cite{balinski_laraki:2010,grillidicortona:1987,saari:1995,taylor_pacelli:2010} attempt to quantify the probability that any one individual could change the outcome in an election.  This is the basic idea behind the \textit{Penrose-Banzhaf-Coleman} power index, for example, and also underlies recent analyses of elections by statistician Andrew Gelman.  A related idea, used to define the \textit{Shapley-Shubik} power index, is to assess the fraction of possible voting patterns in which the individual could be said to cast the deciding vote.

However, these estimates are largely based on extremely over-simplified probabilistic models of voter preferences, and for typical national elections would lead to probabilities that are so small that we may have very little intuition as to how to meaningfully assess differences or asymmetries in these numbers.

More importantly, these quantities are intended to measure \textit{voting} power, and are not really relevant in the present context, because Congressional apportionment is to be made in proportion to numbers of represented inhabitants, not just voting constituents, and representatives are to represent voters and non-voters alike.

\subsubsection{ Representational Surplus and Deficiency} 

In the literature on Congressional apportionment, a notion of representational surplus \cite{balinski_young:2001} (of the state with the smallest average district size as compared to the state with the largest average district size) has been defined as follows:
\be
\surplus(\bv{a}, \bv{p}) =   a_{\stext{mo}} - a_{\stext{mu}}  \tfrac{ p_{\stext{mo}}  }{ p_{\stext{mu}} } = a_{\stext{mo}} - a_{\stext{mu}}  \tfrac{ q_{\stext{mo}}  }{ q_{\stext{mu}} },
\ee
where $p_{\stext{mo}}$ is the population of the \textit{most over-represented} state (with the smallest average district size) and $a_{\stext{mo}}$ is its (actual or proposed) number of representatives, while $p_{\stext{mu}}$ is the population of the \textit{most under-represented} state (associated with the largest average district size), with $a_{\stext{mu}}$ its allotted number of seats.

Similarly, the representational deficiency has been defined as 
\be
\deficiency(\bv{a}, \bv{p})  =  a_{\stext{mu}} - a_{\stext{mo}}  \tfrac{ p_{\stext{mu}}  }{ p_{\stext{mo}} } = a_{\stext{mu}} - a_{\stext{mo}}  \tfrac{ q_{\stext{mu}}  }{q_{\stext{mo}} }.
\ee
From our perspective, these are both also very state-centric rather than inhabitant-focused notions, and do not appear entirely natural from the perspective of the democratic right to equal representation  of an individual inhabitant.

\subsubsection{Shares of a Representative}

The reciprocals of the district sizes give the individual \textit{shares of a representative}, meaning the average (typically fractional) number of representatives per represented inhabitant of that district.  If the districting partition were known, the intra-state values would be given by:
\be
\tilde{\share}_{sk} = \tfrac{1}{\tilde{d}_{sk}} = \begin{cases}
\share_{sk} = \frac{1}{d_{sk}} &\text{ if representation is single-member-district-based}\\
\,\frac{1}{\bar{d}_s} \, = \tfrac{a_s}{p_s}  &\text{ if representation is at-large}
\end{cases}
\ee
for all represented individuals within the $k$th (real or virtual) district of the $s$th state.
Before the districting partition is established, we make use of the statewide average share of a representative, which is just  
\be
\bar{\share}_s = \frac{1}{\bar{d}_s} = \frac{a_s}{p_s} =  \frac{a_s}{q_s} \frac{R}{P} = \frac{a_s}{q_s} \frac{1}{\Dbar},
\ee
per person, while 
\be
\bar{\share} = \frac{ \sum\limits_{s=1}^{S} p_s  \tfrac{a_s}{p_s} }{\sum\limits_{s=1}^{S} p_s}  = \frac{R}{P} = \frac{1}{\Dbar}
\ee
is the overall national average share of a representative.

When summed over all individuals within all districts, such shares will of course sum to the total number of representatives:
\be
\sum\limits_{s=1}^{S} \sum\limits_{k=1}^{a_s} \sum\limits_{i=1}^{d_{sk}} \tilde{\share}_{sk} = \sum\limits_{s=1}^{S} \sum\limits_{k=1}^{a_s}  1 = \sum\limits_{s=1}^{S} a_s = R.
\ee
For any given apportionment, associated individuals could be considered under-represented to the extent that $\tilde{\share}_{sk} < \bar{\share}$, and over-represented if $\tilde{\share}_{sk} > \bar{\share}$.

Notice that more uniform shares, or equivalently, probabilities, can arise not because of positive information regarding near equality of district sizes, but ignorance over possible assignments of represented persons to districts, in the face of which the distributions must be taken to be permutation symmetric.  We discuss the probabilistic interpretation next.

\subsubsection{Normalized Representational Shares, Weights of Representation, and Polling Probabilities}

These shares of a representative rightly focus on the legitimate interests of represented individuals rather than supposed entitlements of states.  But it will be useful to first renormalize, then re-interpret these shares.

\paragraph{Weights of Representation:}  Simply by dividing by the total House size $R$, we can renormalize so as to obtain \textit{(normalized) shares of representation}, or \textit{representational shares} for short, given by
\be
\tilde{w}_{sk} =  \tfrac{\tilde{\share}_{sk}}{R} = \begin{cases}
w_{ski} = \frac{1}{R d_{sk}}                   &\text{ if representation is single-member-district-based}\\
\;\, \bar{w}_s \,\,= \tfrac{1}{R \bar{d}_s}    &\text{ if representation is at-large}
\end{cases}
\ee
for every represented individual within the $k$th (real or virtual) district within the $s$th state.  They  are nonnegative rational numbers summing to unity:
\be
\sum\limits_{s=1}^{S} \sum\limits_{k=1}^{a_s} \sum\limits_{i=1}^{p_s} \tilde{w}_{sk} = 1,
\ee
and directly reflect the (relative) \textit{weight of representation} across individuals within the various states.  In order to avoid confusion with the unnormalized shares of a representative, we will mostly employ this latter terminology, or the variant \textit{representational weights}.

Note that evaluating the representational weights $\tilde{w}_{sk}$ presumes knowledge of any district partitioning, while the quantity 
\be
\bar{w}_s = \frac{1}{R \bar{d}_s} =  \frac{1}{R} \frac{a_s}{p_s} = \frac{a_s}{R} \frac{1}{p_s} = \frac{a_s}{q_s} \frac{1}{P}
\ee 
may be interpreted as the state-level average weight of representation (averaged across all represented inhabitants of the $s$th state), and
\be
\bar{w} =  \frac{1}{R \bar{D}} = \frac{1}{P}
\ee
is the overall national average weight of representation, equal to the average fraction of the total represented population associated with a single individual.

Throughout our approach to optimal apportionment, the over-arching goal will entail equalizing the individual weights of representation as much as possible, or equivalently minimizing deviations from the democratic ideal value of $\bar{w}$, consistent with Constitutional constraints, and in a quantifiably precise and defensible sense.

\paragraph{Polling Probabilities:}  in order to motivate what we regard as the correct measure of representational equality, we shall highlight a more fundamental way to understand these {weights of representation}, explicitly as probabilities.  In order to emphasize the probabilistic interpretation, we will introduce a parallel notation.

Because our main concern is with assessing and balancing democratic influence, we can naturally think about effective sampling or \textit{polling probabilities}, as if we were to elicit the opinion of the public on some matter before the legislature, hopefully in some maximally unbiased way.  In an ideal direct democracy, every represented individual would have an equal probability
\be
\bar{\pi} = \tfrac{1}{P} =  \bar{w}
\ee
of being the individual surveyed in any given instance.  We will call these the democratically ideal polling probabilities or the \textit{direct polling probabilities}, as if we were participating in a direct democracy,  This uniform probability distribution may be  associated with uniform random sampling across all $P$ represented individuals, irrespective of state or district of residency.

If instead this ``voice of the people'' is to be filtered through Congress, the analogous weight may be taken to be the probability of polling an individual in a two-stage process, whereby first a Congressional Representative is randomly selected, then a represented individual is randomly chosen within the sub-population associated with that Representative.  That is to say, in effect we first sample Representatives from states, then sub-sample individuals represented by the chosen delegate. 

Uniform sampling of Congress means that each representative has an equal probability $\tfrac{1}{R}$ being selected, corresponding to a probability $\sum\limits_{k=1}^{a_s} \tfrac{1}{R} = \tfrac{a_s}{R}$ for the $s$th state's delegation overall.  But to infer the appropriate intra-state polling probabilities, and thereby the overall sampling probabilities for individuals, we must take care to specify what is known (or assumed) or not, regarding the apportionment and districting.\footnote{In the Bayesian framework adopted here \cite{cox:1946,cox:1961,jaynes:1983,jaynes:2003}, all probabilities are understood as conditional probabilities, interpreted as consistent degrees of belief, based on some definite if partial background information.  Probabilities are interpreted not as states of an objectively stochastic Nature, but as states of mind, or at least as prescriptions for a rational mind in the face of uncertainty.}
 
First consider the simplest case, that of at-large representation, where the entire state population is collectively represented by the entire delegation of that state.  If the apportionments (and populations on which they are based) are specified, then the overall indirect polling probabilities are the same for all represented inhabitants of that state, namely  
\be
{\pi}_{sn} = \frac{a_s}{R} \frac{1}{p_s}  = \frac{1}{R} \frac{1}{\bar{d}_s} =  \frac{a_s}{q_s} \frac{1}{P} = \bar{w}_s.
\ee
Note that ${\pi}_{sn} \ge 0$, and $\sum\limits_{s = 1}^{S} \sum\limits_{n = 1}^{p_S} {\pi}_{sn} = \sum\limits_{s = 1}^{S}  \tfrac{a_s}{R} = 1$, so these constitute a well-defined (nonnegative and normalized) probability distribution over the possible polling outcomes of individual (censused) persons.

But what if instead we knew the actual population \textit{counts} $\bv{p}$, and the actual apportionments $\bv{a}$, but not the identities of who\footnote{That is, we explicitly lack knowledge as to which specific, named individuals live in which states.} actually resides in which state?  Then for an arbitrarily chosen individual, the polling probability under this state of knowledge must account for all possible assignments of this person to the various states:
\be
\sum\limits_{s=1}^{S} \frac{p_s}{P} \sum\limits_{n=1}^{p_s} {\pi}_{sn} = \sum\limits_s  \frac{p_s}{P} \frac{a_s}{R} \frac{1}{p_s} =  \frac{1}{P} \frac{1}{R}  \sum\limits_s a_s = \frac{1}{P} = \bar{w},
\ee 
where the probability that any particular represented individual chosen at random inhabits the $s$th state (knowing nothing except the state populations counts) would be ${p_s}/{P}$.  So under these conditions the polling probabilities would all revert  to the uniform direct-democracy values.

Now consider the case of (single-member) district-based representation.  If not only the apportionments $\bv{a}$, but the assignments of particular individuals to particular districts within particular states were somehow known, then the polling probabilities would be
\be
\pi_{ski} =  \tfrac{1}{R} \tfrac{1}{d_{sk} }  = w_{sk}
\ee
for the $i$th individual within the $k$th district of the $s$th state.  These are of course also nonnegative and sum to unity: $\sum\limits_{s = 1}^{S} \sum\limits_{k = 1}^{a_s} \sum\limits_{i = 1}^{d_{sk}} \pi_{ski} = 1$.

But during the apportionment procedure itself, that is, at the stage when seats are assigned to states, we cannot really know who will end up in what district, or even exactly how many people will be in each district, so we should sum over all relevant possibilities, accounting for the fact that, in the absence of further information, the probability of an arbitrary resident of a state ending up in a given district should be taken to be proportional to the size of that district.  The resulting probability becomes
\be
\bar{\pi}_{sk} = \sum\limits_{ \{\C_{s_k} \} }  \P(\C_{1_{1}},\dotsc, \C_{1_{a_1}}, \dotsc, \C_{S_{1}}, \dotsc, \C_{S_{a_s}} \given \bv{a}, \bv{p} ,\binfo)\, \sum\limits_{k = 1}^{a_s} \frac{d_{sk} }{p_s} \frac{1}{R} 
\frac{1}{d_{sk}}  = \frac{a_s}{p_s} \frac{1}{R} = \frac{a_s}{q_s} \frac{1}{P} = \bar{w}_{s},
\ee
whatever the adopted probability distribution over possible choices of Congressional constituencies $\C_{s_k}$, as long as the latter distribution is nonnegative and normalized.

If we gain (or assume) additional information about the exact sizes of the districts, and retain knowledge of who resides in which state but not who resides in which district, then the distribution $\P(\C_{1_{1}},\dotsc, \C_{1_{a_1}}, \dotsc, \C_{S_{1}}, \dotsc, \C_{S_{a_s}} \given \bv{a}, \bv{p} ,\binfo)$ concentrates probability mass on partitions consistent with these sub-totals, but nothing else changes, so we still arrive at the probabilities $\bar{\pi}_{sk}$.  If instead we know the population counts but do not even know who resides in which state, then once again the probabilities would revert back to $\bar{\pi}$ even if we could predict the district sizes.


These various polling distributions can also be motivated and derived using the Principle of Maximum Entropy, which is what we use to select the apportionment $\bv{a}$ itself---refer to Appendix~\ref{entropy} below for some background, and the main text (Section~\ref{entropic_apportionment}) for details on this approach.

It is of course a truism of American politics that in actual practice, different represented individuals within each state or district may effectively wield very different degrees and kinds of influence over the political process, or may enjoy very different levels of access to or responsiveness from their representative.  Ought they be assigned different weights of representation?  In the real world, representatives tend to make more time for those constituents who write bigger checks, or those who might mobilize more voters, or those whose ideologies resonate with the politician or party.  However, building this realization into our rules for apportionment would be antithetical to the ``One Person, One Voice'' principle enshrined in 14th Amendment jurisprudence and political discourse, would otherwise be unpalatable to almost everyone, and would be effectively impossible to quantify in any precise way.  For the purposes of inter-state apportionment, we see no justifiable and Constitutional alternative but to presume a best case in which all members of a district or sub-population are in effect equally represented by their representative.  Indeed, for the purposes of Congressional apportionment, democratic principles in general and the 14th Amendment in particular would seem to demand this presumption (or perhaps just pretense) of even-handedness.\footnote{On the other hand, it has even been suggested (see for example \cite{toplak:2009}) that we could account for residual differences in district sizes by weighting the votes of the Congressional Representatives themselves.  With freedom to choose these weighs arbitrarily, the argument goes that we could compensate for any inequities in apportionment of representatives to states or in differences in actual or effective sizes of congressional  districts within or across states.  Polling probabilities would then also be chosen proportional to the same weights.  However, we cannot take this idea seriously. Unequal voting rights or voting weights in Congress would appear dubious Constitutionally, and certainly problematic politically, and it seems highly unlikely that any such scheme would ever be considered under House rules.  Various phrases like ``concurrence of two-thirds'' (Article I, Section 5), ``the Yeas and Nays of the Members of either House'' (\textit{ibid}.), and ``two thirds of that House'' (Article I, Section 7) would all seem to implicitly presume that Congressional Representatives are to enjoy equally-weighted, binary votes on legislative matters.  And regardless of whether allowed by the Constitution, it seems extremely unlikely that Congress would actually adopt such an unwieldy procedure for the sake of any sort of argued theoretical fairness.  In terms of day-to-day operations, would variable weighting hold only for votes on legislation on the House floor, or also in committees?  Since half of all Congressional districts must fall below average in size, which of the corresponding Representatives would voluntarily vote to change House rules so as to assume less influence than their colleagues in larger districts?  And regardless of influence over floor votes, Congressional representatives with greater than average district sizes do not have any more time or resources to provide constituent services, which are often at least as important to their constituents as their voting patterns. This idea must be dismissed as impractical, or even nonsensical.}


\newpage
\section{\uppercase{Problems and ``Paradoxes'' of Apportionment}}\label{paradoxes}

Historically, realization of various ``paradoxes'' of apportionment \cite{balinski_young:1977r,balinski_young:1979,balinski_young:1982,cipra:2010}  has played a large role in the debate over methodology.  Of course, as the long as a proposed assignment $\bv{a} = (a_1, \dotsc, a_S)$ is feasible, consisting of a non-negative integral number of seats allotted to each state, and satisfying all needed constraints, there can be nothing truly paradoxical in a strict logical sense.  What is meant is that the outcome may be ostensibly counterintuitive or undesirable based on some intuition or other criteria of fairness not built into the apportionment method itself.

To better situate these so-called paradoxes, we begin with some natural-sounding criteria which are not difficult for an apportionment strategy to satisfy.  Indeed, essentially any of the (non-lottery-based) schemes that have been proposed will satisfy the following desirable consistency properties:
\begin{enumerate}[a.]
\item \textit{Permutation Symmetry}:  if the population counts (or equivalently, state labels) are permuted, then the corresponding permutation of the seat assignments must be an acceptable apportionment under the method;
\item  \textit{Population Consistency}: no state can receive fewer representatives than another state with a strictly smaller population (unless their minimum acceptable apportionments favor the smaller state and force this behavior);
\item \textit{Ceteris Paribus Monotonicity}:  if the populations of all states but one are fixed, then the number of representatives apportioned to that state will be a non-decreasing function of the state's population;
\item  \textit{Homogeneity}:  if the populations of all states are increased in fixed proportion, then the apportionments for a fixed house size can remain unchanged---that is, the apportionments $\bv{a}$ for the populations $\bv{p}$ and $\alpha \bv{p}$ should be (or in the presence of ties, can be) the same, for any nonnegative constant $\alpha$;
\item \textit{Perfect Proportionality}: if it happens that there exists a feasible apportionment for which $\bv{a} \propto \bv{p}$ exactly, then the method should assign such an apportionment.
\end{enumerate}  

But here are some properties that might sound nearly as plausible or as desirable as those above, but which apportionment schemes may or may not always satisfy:
\begin{enumerate}[i.]
\item \textit{House Monotonicity}:  If the total house size is increased (keeping all state populations fixed), no state's apportionment should decrease;
\item \textit{Population-Pair Monotonicity}:  under a change in populations, if a state $A$'s \textit{relative} population increase (percentage change since the previous census) exceeds that of state $B$, then state $A$ should not lose seats while state $B$ holds steady or gains seats;\footnote{Balinski and Young \cite{balinski_young:1974,balinski_young:1975} offer a slightly different definition of what they refer to as \textit{population monotonicity}: supposing populations $\bv{p}$ for $S$ states lead to an apportionment $\bv{a}$ and populations $\tilde{\bv{p}}$  for $\tilde{S}$ states lead to an apportionment $\tilde{\bv{a}}$, if for any pair of states $s$ and $s'$ it is the case that $\tilde{p}_s \ge p_{s}$ but $\tilde{p}_{s'} \le p_{s'}$ and at least one of the inequalities is strict, then  (i) $\tilde{a}_s \ge a_{s}$ and/or (ii) $\tilde{a}_{s'} \le a_{s'}$.  But if the apportionment method is also homogeneous, then this basically says that no state that grows relative to a second state gives up seats to the second state.}
\item \textit{New-State Consistency}:  If a new state is added with (non-negative) population $p_{S+1}$, bringing the total population from $P$ to $P' = P + p_{S+1}$,  then there exists some nonnegative increase $\Delta R$ to the previous house size $R$, close to $\tfrac{p_{S+1}}{P} R$, such that the apportionment rule applied to $R' = R + \Delta R$ seats delivers $\Delta R$ representatives to the new state while leaving the apportionments of all other states unchanged;
\item \textit{Quota Non-Violation}:  each state receives a number of representatives equal either to its lower quota or upper quota (unless otherwise required by lower or upper bound constraints);
\item \textit{Fair Share Non-Violation}:  the apportionment satisfies
$\lfloor \f_s \rfloor \le a_s \le \lceil \f_s \rceil$, or equivalently $\abs{ a_s - \f_s } < 1$, for all states $s = 1, \dotsc, S$;
\item \textit{Near Fair Shareness}: no transfer of a seat between any pair of states can bring both states nearer their fair shares in absolute value; that is, there is no pair of states $s$ and $s'$ such that $\f_s - (a_s - 1) < a_s - \f_s $ and $a_{s'} + 1 - \f_{s'} <  \f_{s'} - a_s$;
\item \textit{Avoidance of Strategic Splitting}: there is no incentive for a state (or party, etc.) to split into two or more sub-groups in the expectation of increasing the total number of allotted seats.
\end{enumerate}  

How seriously should we take violations of any of these latter properties?  Many methods can violate quota or fair share in principle, though to our knowledge, no U.S.\ Congressional apportionment has done so practice.  Certainly opinions differ as to the importance of satisfying either quota or fair share.  Many, including this author, would remain entirely nonplussed by such outcomes.

Most apportionment algorithms that enjoy anything resembling decent proportional accuracy and avoid significant large-state bias may be subject to some extent to risks of strategic splitting, but this is more of a potential problem with party-list systems than Congressional apportionment.\footnote{Only one existing U.S.\ state has ever split, when West Virginia separated from Virginia, and that was in the context of secession and the Civil War, not over apportionment concerns.  According to the Constitution, splitting of states would require approval of both the state involved and the U.S.\ Congress as a whole.}  The idea is that a large party might divide into smaller parties, in such a way that each party would have just enough votes to get a seat, by whichever method is in use.  But taking advantage of such in-principle opportunities would typically require very accurate prediction of expected vote totals.

Historically, it has been violation of properties \textit{i}., \textit{ii}., or \textit{iii}. that are regarded as ``paradoxes,'' although no such violations actually lead to inconsistencies in any single-case apportionment; rather, one must look at proposed apportionments under different House sizes, trends from past apportionments, or different (counterfactual) population distributions to observe the surprising behavior associated with violating any of these three properties.  From our point of view, these so-called ``paradoxes'' are all rooted in presumptions about the entitlements of states, rather than rights of representation of individuals, so we are not as bothered as are some commentators---though admittedly, violation of House Monotonicity in particular (known as the ``Alabama Paradox,'' as discussed next) would trouble the intuition.  In any case, the entropic apportionment method advocated here will remain automatically immune to all three.
 
\subsection{The Alabama Paradox}

A violation of House Monotonicity is known as the \textit{Alabama Paradox}, and first surfaced after the 1880 Census, regarding the number of seats to be received by Alabama via Hamilton's method, under different house size scenarios.  It was noticed by C.W.~Seaton, chief clerk of  the U.S.~Census Office, that Alabama would receive $a_{\stext{AL}} = 8$ seats with a House size of $R = 299$, but only $a'_{\stext{AL}} = 7$ seats with a larger house size of $R' = 300$.  More generally,  the paradox arises whenever, given fixed populations, an increase in the total number of proposed representatives would lead to a decrease in the number of representatives apportioned to some particular state.

This is in our view the most serious of the so-called paradoxes of apportionment, but also the easiest to avoid once quota methods are eschewed.  
While Huntington \cite{huntington:1928} asserted that ``No method can be regarded as satisfactory which is subject to the Alabama Paradox,'' some other prominent scholars are nonplussed even by this ``paradox''---Birkhoff for example \cite{birkhoff:1976} contends that ``there is no real reason for requiring apportionment to be house-monotone. The objective should be to minimize inequity.''  Though we are certainly sympathetic to the idea of minimizing inequity above all else, our proposed method for doing so will also and automatically maintain house monotonicity, so we can have our distributional cake and eat it too on this particular issue.

\subsection{The Population Paradox}

The \textit{Population Paradox} refers to a violation of Population-Pair Monotonicity, and first arose around 1900, when it was discovered that Virginia had grown faster in relative terms than Maine since the previous census, in the sense that the ratio of the population of Virginia compared to Maryland increased,\footnote{If $p_v$ and $p_m$ denote the previous (nonnegative) populations of Virginia and Maryland, and $p'_v$ and $p'_m$ their more recent populations, note that $\tfrac{p'_m - p_m}{p_m} > \tfrac{p'_v - p_v}{p_v}$ if and only if $\tfrac{p'_m}{p_m} > \tfrac{p'_v}{p_v}$, and if and only if $\tfrac{p'_m}{p'_v} > \tfrac{p_m}{p_v}$.} but nonetheless Virginia lost a seat and Maryland gained a seat compared to the previous apportionment, using the very same House size.

\subsection{The New State Paradox}

The New State Paradox was discovered in 1907, when Oklahoma became a state.  The paradox arose when the addition of this new state, together with its expected share of additional congressional seats (expected, that is, by almost any of the measures considered, given its population and the previous apportionments of existing states), with the intent to leave all apportionments to other states unchanged, nonetheless affected the number of seats apportioned to other states, with Maine gaining a seat and New York losing a seat.

\subsection{Balinski-Young Impossibility Theorem}
Kenneth Arrow famously proved, and others have elaborated \cite{balinski_laraki:2010,taylor_pacelli:2010,szpiro:2010,saari:1995}, how no voting scheme can be perfectly democratic, in the sense of possessing all of the plausible properties we might want in a fair procedure for social choice.  Since voting and apportionment are closely related, it is not surprising that any one apportionment scheme cannot exhibit all of the plausible properties that have been suggested or advocated for fair allocation.

In particular, in 1982, Balinski and Young \cite{balinski_young:1982} proved an impossibility result for apportionment.  Specifically, with three or more states, they showed that either an apportionment method can violate quota, or else it will be susceptible to the Alabama and/or Population paradoxes.  Indeed, the quota and population-pair-monotonicity criteria are mutually incompatible.

Again, we are unpersuaded by arguments for maintaining quota, so are not terribly bothered by this impossibility result.

\subsection{``Bias''}

In the apportionment literature, the concept of \textit{bias} refers to some identifiable property of certain states, or membership in some identifiable subset of states, being correlated (in some sense) with greater under-representation or over-representation (in some sense).  Since the only features of states which are supposed to enter the apportionment algorithm are the populations counts themselves, bias in relation to the size of states has been widely discussed, though in principle other properties of the populations could also be associated with bias.\footnote{For example, whether populations are even or odd, or prime or composite (or more generally, with many divisors or few), could effect tendencies for states to end up over-represented or under-represented.}   

There are several different senses in which methods have been judged for bias against large or small states.  Balinski and Young \cite{balinski_young:1975} first define bias towards small/large states in a comparative sense between apportionment schemes.  In their approach, an apportionment method $A$ is said to favor large states over another method $B$ if: apportionment $A$ gives as least as many seats at some House size $R'$ to the larger of any pair of states as does method $B$ under a possibly different House size $R$, whenever the total seats allotted by $A$ to this pair of states at House size $R'$ is equal to the total number allotted by $B$ at house size $R$.  A method favors large states in an absolute sense if the method favors large states over any rival method.
 
This notion of bias is actually both overly strong and overly narrow.  More generally, we can think in terms of the probabilities or tendencies of ending up over-represented.  For instance, a method can be said to have a bias towards large states (or against small states) when $p_{s} > p_{s'}$ implies that $\P(a_{s} > q_{s}) > \P(a_{s'} > q_{s'})$.  Bias towards small states (or against large states) is defined in an analogous but opposite manner.  Since results will depend on the entire distribution of populations across states, this sort of bias is typically investigated via Monte Carlo simulations, where bias towards large states manifests in a noticeable positive correlation between states' population size and the chance of ending up over-represented.

Other definitions of bias might focus instead on the relative degree of over-representation $\tfrac{a_s - q_s}{q_s}$, or the relative deviation in average district sizes, $\tfrac{\bar{d}_s - \Dbar}{\Dbar} = (\tfrac{q_s}{a_s} - 1)$, but the plethora of possibilities to define or measure bias perhaps hints at the very \textit{ad hocness} and mis-focus of the concept.  From our perspective, these sorts of notions of bias towards small/large states privilege the states, rather than the people.  At minimum, a better measure would directly involve deviations in the weights of representation of individuals, because we ought to be concerned with disproportionate representation of people, nor states.  More could be said about this, but there is a deeper sense in which the whole notion of \textit{post-facto} signatures of bias towards some identified group of states involves moving the motivational goalpost.  As Balinski and Young \cite{balinski_young:1977r} themselves point out,
\begin{quote}
Nevertheless given any method it is an almost irresistible temptation to analyze particular numerical solutions by adding and subtracting different combinations of the numbers to show that the method is in some peculiar sense unfair to certain groups of states. Thus one may question whether a particular solution gives more than a just share to the `larger' states versus the `smaller' states (or the `middle' states) or to the North versus the South, or to the states with large fractions versus those with small fractions, and so forth. These investigations may generally be called ones of `bias' and they purport to establish empirically that certain `new' principles are violated; principles which by the very nature of the case are different from those already agreed upon as defining the method. For the notion of bias to even make sense, a normative principle
must be postulated; one may then ask what methods (if any) satisfy this principle instead of other principles.
\end{quote}
That is to say, if one is going to be bothered if actual results of an apportionment end up possessing a purportedly undesirable feature, or lack a purportedly favorable one, then presumably one should have built into the apportionment method itself some reward or penalty enforcing adherence to (or avoidance of) this property, instead of relying on whatever principles were actually used to motivate adoption or construction of the method.  After all, some states are destined to receive less than their exact quota of seats, and others more, and the point of a principled apportionment \textit{method} is to obviate \textit{ad hoc} or self-serving arguments after the fact as to why various states or other sub-populations have been unfairly treated.


\newpage
\section{\uppercase{A Typology of Apportionment Methods}}\label{typologies}

Apportionment rules can be categorized in various ways, and often any one method can be interpreted or implemented in more than one way.  One useful high-level classification is between quota methods, divisor methods, ranking methods, and optimization methods, though there is significant overlap between these.

To better delineate and differentiate methodologies, we take a closer look at a seemingly elementary mathematical procedure which underlies many apportionment algorithms, as well as much of the political debate surrounding apportionment:  that of rounding real numbers to integers. 

\subsection{Rounding Functions and Rounding Thresholds}

In the end, the apportionment problem always comes down to a question of rounding, of mapping fractions to whole numbers in some way which may be considered most equitable, or at least satisfactorily fair.  Since the fundamental task of apportionment consists of projecting rational numbers to integers, a choice of \textit{rounding function} is central to many apportionment methods.

In fact, both the so-called quota and divisor apportionment methods may be defined by a choice of a \textit{divisor} $D$, or target district size, and a \textit{rounding rule} or function.  By \textit{rounding function} $\round(x)$ we shall mean here a function which projects nonnegative real numbers into nonnegative integers in some sensible fashion.  Intuitively, any well-defined rounding function $\round \colon \{x \in \realsymbol \,\vert\, x \ge 0\}  \to \naturalsymbol$ ought to possess the following elementary characteristics:
\begin{enumerate}[i.]
\item \textit{monotonicity}: $0 \le x \le y$ implies $0 \le \round(x) \le \round(y)$;
\item \textit{idempotency}: $\round\bigl( \round(x) \bigl) = \round(x)$;
\item \textit{proximateness}: $\round(x) \in \{ \lfloor x \rfloor , \lceil x \rceil \}$.
\end{enumerate}
Together these imply that along the nonnegative real axis, the value of a rounding function jumps by exactly one integer exactly once within each unit interval, but is otherwise flat.  Simple examples of rounding functions include:  the floor function (always rounding $x$ down to $\lfloor x \rfloor$), the ceiling function (always rounding $x$ up to $\lceil x \rceil$), and standard rounding to the nearest integer.\footnote{In the latter case, an additional convention must be adopted for what to do precisely at the midpoint between two integers---for instance one can round up at midpoints by using $\lfloor x + \tfrac{1}{2} \rfloor$, or round down using $\lceil x - \tfrac{1}{2} \rceil$, or use parity (e.g., round to the even number), or randomize.}  Equivalently, any rounding function may be uniquely defined by the discrete locations where it jumps.  That is, we can generate a rounding rule by specifying an associated \textit{rounding threshold}\footnote{What we call the rounding threshold $\theta(x)$ is also referred to as a rounding cutoff, breakpoint, rounding criterion, round-up point, or critical fraction.} denoted by $\theta(x)$, that should satisfy, for all $x \ge 0$, the following properties:
\begin{enumerate}[a.]
\item $\lfloor x \rfloor \le \theta(x) \le \lceil x \rceil$;
\item $\theta(x)$ is non-decreasing;
\item $\theta(x)$ is piecewise smooth,
\end{enumerate}
such that any nonnegative $x$ is rounded down to the floor $\lfloor x \rfloor$ or up to the ceiling $\lceil x \rceil$ depending on whether $x$ is less than or greater than the threshold $\theta(x)$, respectively (with some further deterministic or randomized convention in the case where $x  =  \theta(x)$ exactly but $\lfloor x\rfloor < \lceil x \rceil$ strictly):
\be
\xi(x) = \begin{cases}
\lfloor x \rfloor &\text{ if } x < \theta(x)\\
\lceil x \rceil &\text{ if } x > \theta(x)\\
\end{cases}.
\ee
Typically $\theta(x)$ is chosen as some simple function of $\lfloor x \rfloor$ and $\lceil x \rceil$, such as the minimum, maximum, arithmetic mean, geometric mean, quadratic mean,\footnote{So far as we know, an apportionment method based on root-mean-square rounding has never been advocated seriously.
} harmonic mean,\footnote{Given $x \ge 0$ and $y \ge 0$, the arithmetic mean of the pair is defined as $\A(x,y) = \tfrac{1}{2}(x + y)$,
the geometric mean is $\G(x,y) = \exp \bigl[ \tfrac{1}{2} (\ln x + \ln y) \bigr] = \sqrt{x y}$, the harmonic mean is $\H(x,y) = \bigl[ \tfrac{1}{2}( x^{-1} + y^{-1})  \bigr]^{-1} = \tfrac{2 x y}{x + y}$, and the quadratic mean (root-mean-squared) is $\Q(x,y) = \bigl[ \tfrac{1}{2}(x^2 + y^2) \bigr]^{1/2}$.  For nonnegative arguments these means satisfy the chain of inequalities $\min[x,y] \le \H(x,y) \le \G(x,y) \le \A(x,y) \le \Q(x,y) \le \max[x,y]$, with equality for strictly positive arguments if and only if $x = y$.  If either argument vanishes ($xy = 0$) then $\min[x,y] = \H(x,y) = \G(x,y) = 0$, while $\A(x,y) = \Q(x,y) = 0$ if and only if both arguments vanish (i.e., $x = y = 0$).} etc.

Of course, in certain apportionment methods, the choice of rounding threshold may have to depend on additional parameters besides the number $x$ to be rounded itself, particularly when we want to impose some sort of collective constraint on the rounded values, such as limit on their sum.

\subsection{Quota Rules}
  
As the name suggests, \textit{quota methods} \cite{balinski_young:1977r,balinski_young:1982} are engineered to satisfy quota, and deliver to each state either its lower quota or upper quota (unless the rounded quota itself lies outside prescribed bounds, in which case the assignment must be shifted to saturate the relevant bound).

In a pure quota method, each state is assigned either its lower quota or upper quota by rounding the exact quota $q_s$ up or down.  If the total house size is fixed in advance, the rounding cannot be achieved by a universal rounding function, but must depend on some sort of prioritization criterion across the states, in order that the total number of representatives achieves its predetermined target.  Note that each state is to appear at most once on any such actual or implicit rounding prioritization list, so can only receive at most one seat more than its lower quota.  If the total number of seats is not fixed precisely in advance, then a uniform rounding rule could be applied to all states, leading however to a House size that may differ from the original target used to calculate the quotas.

With additional lower or upper bounds, a pure quota method must be modified to accommodate these constraints: either a state is instead assigned the closest boundary value if the rounded quota lies outside the allowed range, or else one works with the fair shares rather than the original exact quotas, and the rounding algorithm is applied only to those states with fractional fair shares while assigning the remaining seats, in which case the method might be better described as a fair-share method rather than a quota method.

Many quota methods are prone to the trifecta of Alabama, Population, and New State paradoxes.  There is one quota method (introduced by Balinski and Young, and described below) which avoids the Alabama paradox, but all quota methods remain susceptible to the Population Paradox.  

As to the desirability of quota methods, we find ourselves in agreement with Huntington \cite{huntington:1928} on this point, who asserted  that ``it is a common misconception that in a good apportionment the actual assignment should not differ from the exact quota by more than one whole unit,''  and also concur with Edelman \cite{edelman:2006} (p. 338), who wrote
\begin{quote}
Should we really be concerned with how close a state is to quota?  If the Court's concern is adherence to ``one person, one vote,'' then the relevant unit of analysis is the representation of citizens, not the total representation of the state.  Under this inquiry, the state has no real claim at all....
\end{quote}
and
\begin{quote}
Further buttressing this argument is the federal statue requiring states to elect their representatives by single-member districts.  This not only removes from the states the decision of how to elect the representatives, but ties each representative to a sub-population of the state, rather than the state itself.  This makes it difficult to view representatives as being representatives of the state \textit{qua} state, rather than as a representative of those people within the district.
\end{quote}
From the perspective of proportionality alone, we see no particular reason to worry more about satisfying quotas for individual states rather than any other identifiable partitions of the population.  But it can be impossible to satisfy quotas at both the level of single states and groups of states.  See Table~\ref{table:aggregate_quota} in Appendix~\ref{tables} for a simple example.

Furthermore, in intuitively sensible apportionments, the optimal number awarded to any one state cannot always be determined by that state's quota alone.  Table~\ref{table:quota_not_enough_a} in Appendix~\ref{tables} presents a simple example where, for a fixed house size, the difference in assignments to State $B$ does not arise from any differences in $B$'s quota but instead depends on small shifts in population amongst other states. 

From our perspective, the problem with quota methods is not so much their propensity for ``paradoxes,'' but the misguided goal itself.  Directly or indirectly, all methods with any pretense to proportionality end up using the exact quotas as apportionment targets in some sense,\footnote{After all, the exact quotas are simply the state populations, measured in units of the average district size $\Dbar$.} but fastidious insistence on assigning seats by rounding up or down exact quotas seems rooted in the idea of states being entitled to seats, rather than the people of the various states deserving equal representation.

\subsection{Divisor Methods}

Quota methods are based on rounding the exact quotas $q_s = {p_s}/{\Dbar}$ calculated using the standard divisor $\Dbar$ but possibly adjusting the rounding procedure depending on some prioritization rule to ensure all constraints are satisfied.  In contrast, \textit{divisor methods} \cite{balinski_young:1982,lauwers:2005,agnew:2008} instead use a fixed rounding method for all states, but adjust the divisor as necessary to achieve the target House size.

Given some modified or ``sliding'' divisor $D$ which can differ somewhat from the standard divisor $\Dbar$, we first define $Q_s = {p_s}/{D}$ for $s = 1, \dotsc, S$, referred to as the modified quotas, pseudo-quotas, or else \textit{quotients}.\footnote{We will follow the latter convention, to avoid confusion with the exact state quotas.}  These quotients are then rounded up to $\cQ$ or down to $\fQ$ based on the adopted rounding rule (and replaced if necessary by a prescribed lower or upper bound if the results would otherwise lie outside the mandated range), then $D$ may be adjusted and the process repeated until the total number of seats thereby assigned is exactly equal to some predetermined target.  This clever technique of a sliding divisor was introduced by W.F.\ Willcox, so is sometimes called a ``Willcox divisor.''

In a traditional divisor method, the output of the rounding rule for each state depends only on the value of $Q_s = p_s/D$ for that state itself.  In a so-called generalized divisor method, the rounding function can also depend parametrically on the total number of states $S$ and/or the total House size $R$---but not directly on the populations of other states.  Either type will be referred to as a divisor rule here.

In practice, typically one can begin with the choice $D = \Dbar$, and adjust iteratively as needed by trial and error, increasing $D$ if too many seats were tentatively assigned, and decreasing $D$ if too few seats were assigned.  However, one can also implement a divisor algorithm starting with a very large $D \gg \Dbar$, leading to a minimal number of assignments consistent with any prescribed lower bounds, then decreasing $D$ gradually as the total number of seats thereby tentatively apportioned increases one by one (with some randomization or other convention adopted in the unlikely event of exact ties).  As long as $R$ is consistent with the lower and upper bound constraints, and one uses a properly-defined rounding function (with suitable allowance for exact ties), this sort of implementation guarantees that there will always be a range of values for $D$ that produce an apportionment with exactly $R$ seats, and that the resulting apportionments will always be identical (apart from instances involving exact ties) for any choice of the sliding divisor $D$ within this range.

It may seem counterintuitive to make use of a divisor $D$ other than the overall average district size $\Dbar$, and then have to adjust it iteratively.\footnote{The divisor $D$ essentially becomes a \textit{maximum} allowable intra-state average district size, in the absence of lower or upper bound constraints.}  However, when divisor methods were first suggested, the House size typically was \textit{not} fixed in advance.  Instead, a target value for $D$ would be chosen, resulting in apportionments $\bv{a}$ obtained by rounding the quotients, and the total House size $R = \sum\limits_s a_s$ would fall where it may.  The resulting national average $\Dbar = P/R$ would tend to be near but not in general exactly equal to the originally chosen target size $D$.

Any consistent divisor method can sometimes violate quota even in the absence of lower or upper bound constraints, but is guaranteed to be house monotone.  Additionally, pair-population monotonicity will always be satisfied by an apportionment rule if and only if the rule is equivalent to a traditional or generalized divisor method \cite{balinski_young:1982}.

\subsection{Ranking and Prioritization Methods}

In ranking or prioritization methods, each state receives some initial allotment of seats $a'_{s_0}$, then the remaining seats are doled out one at a time, based on the relative values of some \textit{ranking index}\footnote{Confusingly, some sources instead refer to the ranking indices as quotients as well, because in many methods they take the simple form $\rho = \tfrac{p_s}{\chi(a'_s)}$, for some function $\chi(a)$ for which $a\le \chi(a) \le a+1$.  But it is then the quantities $\chi(a)$, and not the prioritization indices $\rho$, that are most closely related to the quotients as used in a divisor method.  The ranking index $\rho$ itself is instead more closely related to the divisor $D$.} 
 $\rho(a'_s, p_s, P, R, R')$, which determine the priority for the next seat to be distributed, given the seat counts $a_s'$ assigned up to that point, the state's population $p_s$, and possibly the total population $P$, total number of seats to be apportioned $R$, and the number $R'$ apportioned so far.  That is, at each stage, the state with the largest value of the prioritization score or ranking index $\rho(a'_s, p_s ; P, R, R')$ receives the next seat, then its cumulative apportionment and incremental ranking are updated, and the process is continued until all seats have been distributed.  In the absence of symmetry-breaking lower/upper bound constraints, note that permutation invariance demands that the same functional \textit{form} of ranking function $\rho(a,p,P,R,R')$ should be used for all states.

Both divisor methods and quota methods can be interpreted or implemented as ranking methods.  For quota methods, we have already seen how which of the states' quotas are rounded up or down must depend on a prioritization list, if the total house size is to be fixed ahead of time.  As for divisor methods, by again imagining a procedure involving gradually decreasing the sliding divisor $D$ as seats are assigned one by one, we can see directly how to translate between a divisor rounding rule and an equivalent ranking index:  $\rho(a'_s, p_s, P, R, R')$ can be set equal to the largest value of the divisor $D$ at which the quotient $Q_s$ would be rounded up to $a'_s + 1$.  (Here, the ranking index must be expressible as a function of $a'_s$ and $p_s$ only).

Certain ranking methods automatically guarantee that each state receives at least one seat,\footnote{Authors disagree on whether this property should be regarded as a feature or a bug.  We consider it a definite strike against any apportionment method, which ought to be directly applicable, without \textit{ad hoc} modification, to cases where the lower bound constraints may differ from unity. For example, in party-list elections, the minimum is often intentionally left at zero to avoid assigning seats to parties with very small support.  Either one employs a method that does not automatically assign one seat, or else all parties below a certain threshold of support must be explicitly excluded before apportionment of seats to the remaining parties.  Also, explicit lower bounds or minimal apportionments may exceed one.  For example, before 1988 when France switched to single-member constituencies, their National Assembly required at least two deputies from each department.} but in general, representational lower/upper bound constraints may need to be enforced explicitly.  Any state that has achieved its upper bound can be subsequently re-assigned a low priority score ensuring it never receives further seats.  To enforce nontrivial lower bound constraints, some ranking methods assign to states their lower bounds initially.  In other approaches, states all start out without any seats, but if any states would end up below their allowed minimums, priorities are modified in the final stages so that the last assigned seats are shifted to the states that would otherwise fall below allowed lower bounds.  For ranking functions $\rho(a,p; P,R,R')$ which are monotonically decreasing functions of $a$, the final apportionments will not depend on which of these modifications (i.e., shifting the initial versus final assignments) is used to enforce the constraints (although the path taken to the final allocation might differ).  Essentially all ranking functions  that have been seriously suggested enjoy this rank-monotonicity property, because if the ranking function is intended to reflect which state most ``deserves'' the next seat, intuition suggests that a state should deserve its first seat more than its second seat, its second more than its third, etc.

Other variants of ranking-based apportionment are possible.  For example, ``fair shares'' can be determined, and the ranking-based allocation can be used to distribute remaining seats only to those states with non-integral fair shares.

\subsection{Optimization Methods}

Optimization-based apportionment methods include Pareto optimization, local hill-climbing, (constrained) global searches, etc.  Although in principle any apportionment method can be cast \textit{a posteriori} as some sort of  global optimization method, this framing may not always be natural in practice; and conversely, not just any optimization principle will lead to a computationally tractable integer programming problem that can be implemented with straightforward algorithms.  However, most of the traditional apportionment methods that have been proposed do admit a natural formulation in terms of the optimization of some simple \textit{objective} function that numerically encodes the explicit or implicit goal of apportionment.

\subsubsection{Pairwise Comparison}

The mathematician E.V.\ Huntington \cite{huntington:1921,huntington:1928,huntington:1941} investigated a class of methods based on a notion of \textit{stability} under \textit{pairwise} comparisons involving hypothetical exchanges of a \textit{single} seat, or what we might also call pairwise Pareto optimality.  The idea is to find an apportionment for which any further exchanges of a seat from one state to another state cannot improve fairness according to some adopted criterion, or so-called comparison  ``test'' (of approximate proportionality).  We can imagine the apportionment as proceeding from some initial allocation to the final, optimal allocation by a sequence of feasible pairwise exchanges, each judged to improve matters, until no further pairwise exchanges are possible that both satisfy the constraints and lead to any less unequal representation.

Specifically, the $s$th state can be said to be over-represented (in an absolute sense) if $a_s > q_s$, or equivalently if $\bar{d}_s < \Dbar$, and under-represented if $a_s < q_s$, or equivalently $\bar{d}_s > \Dbar$.  But for any pair of states, we may say that state $o$ is over-represented \textit{relative} to state $u$, or compared to state $u$, if and only if state $o$ has higher average share of representation than state $u$, or equivalently if and only if state $o$ has a smaller intra-state average district size than state $u$:
\be\nonumber
\text{state } o \text{ is over-represented relative to state } u \text{ if and only if } \tfrac{a_o}{p_o} > \tfrac{a_u}{p_u}, \text{ or equivalently, }  \bar{d}_o < \bar{d}_u,
\ee
and likewise, we can say state $u$ is under-represented relative to state $o$ whenever state $o$ is over-represented relative to state $u$.

Huntington introduced the notion of a so-called fairness test, or pairwise comparative test of inequality, associated with some \textit{comparison function} $T(a_o, p_o, a_u, p_u)$ which, for any pair of states $o$ and $u$: (i) is equal to zero if and only if $a_o/p_o = a_u/p_u$, and (ii) is positive if and only if $a_o/p_o > a_u/p_u$, and (iii) whose deviation from zero is taken to quantify the extent to which state $o$ is over-represented relative to state $u$. The larger the value of $T(a_o, p_o, a_u, p_u)$, the more unfair we are to regard the gap in the shares of representation of the $o$th and $u$th states, or the greater we are to regard the mismatch in the representation afforded to these two states.

An apportionment is said to be stable with respect to a given pairwise fairness test if and only if, for every pair of states with unequal shares of representation, switching one representative from the relatively over-represented state (provided it is not at its lower bound) to the relatively under-represented state (provided it is not already at its upper bound) would (a) reverse the order of per-capita representation but also (b) increase, or leave unchanged, the degree of unfairness.  That is, stability requires
\be
T(a_u + 1, p_u, a_o-1, p_o) \ge T(a_o, p_o, a_u, p_u) ,
\ee
for every pair of states for which both $a_o/p_o > a_u/p_u$ and the proposed swap would be compatible with the lower/upper bound constraints.\footnote{These definitions are written as conventionally employed in the later apportionment literature.  It is not actually clear whether Huntington himself would have considered properties (ii) and (a) to be strictly \textit{essential} to the notion of a pairwise comparison test, although all of the tests he mentions do in fact share them, as will most other reasonably choices for the figure of merit.}  In a pairwise comparison-based approach, an apportionment would be considered permissible if and only if it satisfies all constraints (i.e., lower/upper bounds and total House size) and is stable with respect to all feasible exchanges of one seat between any pair of states, a feasible swap being a swap that would not violate any of the said constraints.

However, it is not immediately clear that greater equalization in the relative representation between two states cannot worsen the relation between one of these states and a third state, and in fact not all reasonable-looking choices for a pairwise fairness test $T(a_o, p_o, a_u, p_u)$ necessarily lead to unique, stable apportionments, even in the absence of ties resulting from commensurate populations.  Some reasonable-looking choices of the test function can result in intransitive orderings where apportionment $\bv{a}_{\stext{A}}$ is judged more fair than $\bv{a}_{\stext{B}}$, $\bv{a}_{\stext{B}}$ is more fair than $\bv{a}_{\stext{C}}$, but $\bv{a}_{\stext{C}}$ is more fair than $\bv{a}_{\stext{A}}$.  Huntington called a comparison test \textit{workable} if it is immune to such inconsistencies.  And though pairwise comparisons are perhaps useful for characterizing or comparing apportionments, comparison tests do not lead directly to particularly efficient constructive algorithms (because each state must be repeatedly compared to every other state), but are instead implemented in practice as divisor or ranking methods.  In fact, Huntington showed how various simple choices for $T(a_o, p_o, a_u, p_u)$ involving rational functions of $a_o$ and $a_u$ lead to the five traditional divisor methods.  Further, Huntington suggested that these were the \textit{only} workable pairwise comparison methods, but this would seem to be incorrect if we allow for more complicated comparison functions.

Huntington contended that these sort of pairwise fairness tests reflect mathematically what inevitably occurs politically after any apportionment is proposed: namely, that Congressional delegates, Governors, or other state politicians calculate average intra-state district sizes, and if their district size is larger than that of some other state, raise the question of whether perhaps a seat should be transferred to their state.  ``The size of the House being fixed,'' Huntington contended \cite{huntington:1941}, ``the debate always comes down, in the last analysis, to this question: `should or should not such a transfer be made?'{}''  But from our perspective, this attitude (i) again risks overemphasizing the prerogatives of states over the rights of people, and (ii) blurs the fact that both mathematical and moral calculations should be distinct, to the maximum extent possible, from any inevitable but purely political maneuverings.  Just because this is how individual states will try to frame their criticism does not necessarily mean that pairwise comparisons offer the best way to assert or resolve questions of fairness.  And of course, (iii) neither Huntington's family of fairness measures, with which he characterized the five traditional divisor methods, nor the particular choice of fairness test he favored, involving the relative difference in district sizes (or equivalently, relative difference in representational shares), and leading to the Method of Equal Proportions, provides the only definition or even necessarily the most natural definition of even a local notion of optimality, 
notwithstanding the fact that his framework has been extremely influential and has tended to dominate much of the subsequent discussion of apportionment methods in Congress and within the Census Bureau.

In particular, global optima must also be local optima, so any of the global optimization methods of the sort mentioned below could \textit{a fortiori} also be used to assess fairness of possible exchanges of seats between states, even if the comparative assessments induced by the global figure-of-merit may not always lead naturally to a formulation of a comparison test satisfying all of the criteria outlined above.
In particular, although true for all of the traditional divisor methods (and, as it turns out,  the entropic method),  it does not actually seem inevitable that any viable notion of global optimality need result in a pairwise comparison function whose sign reflects which state is relatively over-represented, nor necessarily lead to optima in which any feasible swap of any one seat must always result in a reversal of relative over-representation as defined above.  But the figure-of-merit itself will still offer a natural measure of deviations from global optimality, which would seem to be even better than mere pairwise Pareto optimality.

\subsubsection{Pairwise Consistency}

Balinski and Young \cite{balinski_young:1977s, balinski_young:1977r}  situate Huntington's pairwise comparison methods  within a certain sub-class of ranking methods\footnote{They call such methods ``{Huntington methods},'' not to be confused with \textit{the} Huntington method of equal proportions, which is one member of this class.} defined iteratively as follows.  Letting $\rho(a,p)$ be a ranking index that assumes the same functional form for all states and that depends only on that state's  population and apportionment so far, and letting $a'_s(p_s,R')$ denote the current allocation of seats to the $s$th state out of a total of $R'$ seats allocated so far based on this ranking index, the apportionment rule is:
\begin{enumerate}
\item $a'_s\bigl(p_s, \sum\limits_s \lambda_s\big) = \lambda_s, \text{ for } s = 1, \dotsc, S$ 
\item if $s^*(R')$ is some one state for which $\rho\bigl(a_{s*(R')}(p_{s^*(R')},R'),p_{s^*(R')} \bigr) \ge \rho\bigl(a_s(p_s,R'),p_s \bigr)$ for every $s \in\{ 1, \dotsc, S\}$, then $a'_{s^*(R')}\bigl(p_{s^*(R')}, R'+1\big) = a'_{s^*(R')}\bigl(p_{s^*(R')}, R'\big) +1$,
and $a'_{s'}\bigl(p_{s'}, R'+1\big) = a'_{s'}\bigl(p_{s'}, R'\big)$ for all $s' \neq s^*(R')$.
\end{enumerate}
Such methods are clearly House monotone, and are also \textit{pairwise-consistent} in the sense that: (i) as the House size is incremented by one seat, the decision as to which state within any pair of states deserves the seat more will depend only on the respective populations of these two states and the number of seats already allocated to each of these two states (and possibly on the total $P$ and $R$), but not on the individual populations of or current apportionments to the other states; and (ii) any ties based on the comparisons lead to equally acceptable apportionments.  In fact, they proved the converse, namely that the only methods that are both House monotone and pairwise-consistent in this sense can be interpreted as ranking methods of the sort just described.  Balinski and Young consider this property to be a fundamental feature of any reasonable apportionment procedure, dictated by common sense.  The entropic apportionment method will always be pairwise consistent in this sense.

\subsubsection{Global Optimization Methods}

It seems what we really want is not an apportionment which is somehow minimally unfair with respect to feasible swaps of one seat between any pair of states, but one which is globally optimal under any feasible exchanges of any seats amongst any number of states.  Some authors have referred to this as minimizing a \textit{total error} or \textit{total inequity} of apportionment.

As the terminology suggests, global optimization methods seek a (constrained) global optimum of some so-called \textit{figure-of-merit}, \textit{variational potential}, or \textit{objective function},\footnote{\textit{Figure-of-merit} is a general term often used when the optimum corresponds to a maximum.  \textit{Variational potential} is often encountered in physics when the optimum corresponds to a minimum. \textit{Objective function} is a more general terminology common in operations research, where the goal or objective is to optimize this function.  In economics and decision theory, one usually speaks of maximizing a \textit{utility function}, \textit{reward} function, \textit{profit function}, or \textit{fitness function}, or of minimizing a \textit{loss function} or \textit{regret function} or \textit{cost function}.  The terminology of a \textit{scoring rule} is also used in economics, statistics, decision theory, psychology, and other fields when eliciting probabilistic predictions.  In mathematics, one often refers simply to a \textit{maximand}, \textit{minimand}, or \textit{optimand}.} either maximizing a global \textit{fairness} function $\Phi(\bv{a}; \bv{p}, R, S)$ which measures the overall degree of proportionality, uniformity, or equality of the proposed apportionment $\bv{a}$ given the populations $\bv{p}$, or equivalently, minimizing some \textit{unfairness}, inequity, or anti-democratic loss function $\U(\bv{a}; \bv{p}, R, S)$ which provides a total measure of overall disproportionality, inequality, or deviation between the proposed apportionment $\bv{a}$ and the ideal or perfectly fair allocation.

Mathematically, the choice of maximizing fairness or minimizing unfairness is purely a matter of convenience or convention, since the measures can be related by 
\be
\U(\bv{a}; \bv{p}, R, S) = -\alpha(R,P,S)\, \Phi(\bv{a}; \bv{p}, R,S) +  b(R, P, S),
\ee
for some nonnegative scale factor $\alpha(R,P,S)$, and offset $b(R, P, S)$, that are independent of how the total population or total seats are distributed amongst the states.  Typically, the chosen objective function will not incorporate directly any penalties for violating the constraints, so optimization is to be performed subject to additional explicit constraints on the total number of representatives and allowed minimum and maximum allocations for each state.\footnote{While we are denoting by $\U(\bv{a}; \bv{p}, R, S)$ the ``unfairness'' function to be minimized, this should not be confused with a utility function, which we would normally want to maximize rather than minimize.  In fact, given the goal of equal representation, we can regard the unfairness function as a disutility or \textit{loss} function $\mathcal{L}(\bv{a}; \bv{p}, R, S)$, which is in turn related to but not to be confused with the \textit{Lagrangian} function introduced below.}

While we often speak of \textit{the} global constrained optimum, in some cases we ought to refer more accurately to \textit{a} global constrained optimum, since some population distributions for which the $(p_1, \dotsc, p_S)$ are not relatively prime can admit non-unique optima for certain choices of $R$, between which we must choose by some arbitrary convention---typically the seat is awarded to the larger state, or else assigned at random.  However, instances of exact ties will be rare for non-artificial examples, particularly if we have some freedom in adjusting $R$ sensibly.

In principle, any deterministic (but for ties) apportionment method must be equivalent to the optimization of \textit{some} objective function defined over possible assignments of seats.  Given a finite house size $R$ and a finite number of states $S$ with finite populations $\bv{p}$, there are a (perhaps combinatorially large but) finite number of possible apportionments, and each possibility can be assigned some numerical score which picks out as optimal the apportionment(s) generated by the desired rule or procedure. But it may not be practical to construct such a function explicitly and in advance, justify it naturally, or optimize it efficiently without already knowing the answer.\footnote{Indeed, a good optimization method reverses this logic, and will not just pick out one optimal apportionment, but provide a meaningful way to quantify deviations between a proposed apportionment and the ideal, or to compare or rank different proposed apportionments.}

Nonetheless, undoubtedly to many physicists, engineers, and statisticians, just this sort of \textit{variational} approach will definitely seem the most obvious and natural way to frame and solve an apportionment problem.  Any actual apportionment thereby obtained will by construction be guaranteed to be the ``best'' possible given the populations and constraints, and attention can turn to motivating, deriving, or justifying the best definition of ``best.''  The debate thereby shifts from what sort of apportionments might be generated in various individual cases, which can be hard to assess directly or compare in marginal or boundary cases, or from which properties exactly are to be maintained or which ``paradoxes'' are to be avoided (given that impossibility theorems dictate that no method can possess all properties that have been considered desirable), to the fundamental question of what sort of overarching principle of balance, equity, equality, uniformity, fairness, or proportionality in apportionments we ought to maintain. The subsequent output of the algorithm is then to be accepted as optimal, wherever it happens to fall.  Thus it is somewhat surprising that relatively little of the historical or contemporary literature on apportionment has actually focused on such a global variational framework, given its advantages and its generality.

Indeed, global optimization approaches have been criticized by Balinski and Young \cite{young:1985} and others on the grounds that ``choice of objective function remains \textit{ad hoc}.''  But any of the apportionment approaches described above also rely on particular choices for rounding methods, ranking indices, or pairwise comparison tests.  At least with a global optimization method, it is clear exactly what one is choosing---namely, a universal notion of \textit{overall} fairness, proportionality, uniformity, or equality of representation of the proposed apportionment.  Of course, any number of such measures of equity of representation might be invoked, so some further axioms, desiderata, or characterizations need to be put forward to justify a particular choice of objective function.  Our intent throughout this work is of course to advocate for one particular functional form, based on information theoretic entropy, which we contend is the most natural and defensible way to measure equity or uniformity of apportionment.

``Intuitively, `minimizing inequity' is what the apportionment problem is all about,'' Balinski and Young \cite{balinski_young:1977r} elsewhere admit. ``The real problem is to determine what `inequity' means.''  We agree, but contend that there is after all a natural and unique way to define and quantify the overall level of inequity in this context.  They go on to assert:
\begin{quote}
the axiomatic approach to apportionment proceeds by making a choice concerning the principles which any fair apportionment should satisfy, and then identifying that method (or methods) that satisfy the principles. The advantage of beginning with agreed-upon fairness principles is that subsequent squabbles over particular numbers resulting from these principles are avoided.
\end{quote}
We concur, with one modification, or at least an important change of emphasis.  As mentioned above, we need not and should not impose the sorts of principles Balinski and Young have in mind directly at the level of the apportionment procedure itself or its output---many such principles have been suggested and debated, not all can be satisfied simultaneously, and it is difficult to know how to separate out the compelling from the merely plausible or the largely \textit{ad hoc} principles.  Instead, we contend that the optimal apportionment should indeed literally minimize (subject to relevant constraints) a global measure of inequity in the weights of individual representation, and then impose suitable properties on this \textit{measure} that pick out a unique functional form.  The apportionment procedure follows automatically from the inequity measure, not the other way around.  Here we depart partly but sharply from the contentions of Balinski and Young \cite{balinski_young:1977r}, who claim:
\begin{quote}
The lessons of history clearly point to the necessity of arriving at a fundamental understanding of the properties of methods.  Put in other terms, political apportionment must be based on principles of fair division rather than on \textit{ad hoc} choices of measures of inequity. Thus axiomatics finds a political role!
\end{quote}
On the contrary, its precisely the role of (hopefully compelling) ``principles of fair division'' to uniquely characterize a non \textit{ad hoc} quantification of fairness or inequity of distribution, and then optimal apportionments will by definition optimize this measure (subject to required constraints).  In our view, some pretty fundamental principles do single out an essentially unique measure of inequity, and the apportionment method associated with minimizing this measure naturally \textit{inherits} various ``good'' properties from this figure-of-merit.  
 
Many of the historic apportionment methods turn out to be equivalent to minimizing (subject to constraints on lower/upper bounds and on total seats) an objective function of \textit{additive} form
\be
\U(\bv{a}; \bv{p}, R, S) = \sum\limits_{s=1}^{S} \psi(a_s; q_s, R, P, S),
\ee
in which the global measure of apportionment inequity simply accumulates separate contributions from each state's proposed assignment.  Any such  optimization method is automatically pairwise consistent---whether a swap of a seat between any pair of states will be judged advantageous does not depend on the details of how the remaining population or seats are distributed amongst other states not involved in the proposed swap.\footnote{To be clear, this does \textit{not} mean that one state's quota by itself determines its final optimal apportionment---one state's optimal apportionment can still depend on shifts in population between other states, because of the nature of the constraints.  See Table~\ref{table:quota_not_enough_a} in Appendix~\ref{tables} for an example.}  Permutation invariance\footnote{Which is arguably demanded by the Constitution, and certainly by basic intuitions of democratic fairness....} dictates that the summand $\psi(a; q; R,P,S)$ should assume the same functional form for all states.  And If the variational principle is to do its job, the value of $\U(\bv{a}; \bv{p}, R)$ should increase as $\bv{a}$ deviates in feasible directions from $\bv{q}$ in some prescribed sense.

 But beyond these basic features, any further mathematical properties of the objective function should be derived or defended.\footnote{In particular, note that we do not actually require that $\psi(a; q,R,P,S)$ has an absolute minimum at $a = q$ in the absence of constraints.  The budget on $\sum_s a_s$ may play an essential role here in shifting the location of the optimum, as is the case when using relative entropy.}  The exact choice for $\psi(a; q, R,P,S)$ must of course be justified, and certainly not all reasonable-looking measures will lead to efficient computational algorithms.  But as we will show, if $\psi(a; q, R, P,S)$ additionally satisfies a certain natural discrete \textit{convexity} condition, then local optimality under all feasible pairwise swaps entails (constrained) global optimization, so the latter may be  straightforwardly achieved by a simple greedy algorithm, leading in turn to an equivalent ranking or divisor method that additionally is guaranteed to be House monotone and free of the other historic ``paradoxes.''

Additionally, for several of the prominent variational methods, including the relative entropy, the functions $\psi(a; q, R,P,S)$ can be chosen to be of a form which, perhaps apart from a pre-factor and offset \textit{common to all states}, depends only on the quantities $a$ and $p = \tfrac{q}{R} P$.  Under such circumstances, the greedy optimization has two further features, namely that: (i) it is recursively consistent when the apportionment problem is sub-divided, and (ii) it automatically constructs optimal apportionments for all House sizes up to the specified $R$.  See  See Section~\ref{recursive_HWE} and Appendix~\ref{optimization} for more details.

Huntington \cite{huntington:1928} specifically criticized the use of such additive objective functions not just for alleged arbitrariness but on the grounds that ``a total or average error may be reasonably small, while at the same time the error affecting some particular state may be shockingly large; and a gross injustice done to a particular state could hardly be successfully defended on the grounds that `on the average' the others states  are fairly treated.''  This criticism strikes us as particularly surprising given that all of the ``workable'' comparison methods Huntington himself studied, including his own favored method of Equal Proportions, can be derived from exactly these sorts of variational principles with additive objective functions.  

\subsubsection{Minimax Methods}

But as an alternative, Huntington and others have also mentioned \textit{minimax} criteria, resulting in apportionment rules of the type
\be
\bv{a}= \argmin\limits_{\bv{a}} \max\limits_{s} \zeta(a_s, q_s; R, P, S),
\ee
or 
\be
\bv{a}= \argmin\limits_{\bv{a}} \max\limits_{s,s'} \gamma(a_s, q_s; a_{s'}, q_{s'}; R, P, S),
\ee
where we are to minimize the maximum value of some discrepancy measure either across all states or all pairs of states, subject as usual to constraints on the value of $R$ and the lower and upper bounds on each $a_s$.

We tend to be a bit skeptical of minimax approaches.  If the goal is to apportion representatives across the entire partitioned population, it is most natural to do this so as to achieve a global (constrained) optimum in some accepted measure of proportionality or equity that is sensitive to the populations of, and seat assignments granted to, all states, not just the most under-represented or over-represented states.  \textit{Contra} Huntington, the whole point is that we must balance inevitable tradeoffs in assigning seats to one state versus to any of the others.  And with a sensible choice of a convex, additive, global equity function satisfying permutation symmetry, and conforming to prescribed constraints on the total house size $R$ and lower bounds $\lambda_s$, it will simply be the case that the resulting constrained optima automatically avoid the sort of  ``gross injustice'' feared by Huntington.

Furthermore, minimax criteria seem especially unappealing in the apportionment context, because such measures are essentially insensitive to how disproportionality of representation is spread amongst those states which happen not to be the most misrepresented.\footnote{Minimax rules are commonly encountered in decision theory, where pessimists might want to act so as to minimize the cost of the worst possible outcome.  But in decision theory contexts, possible choices or actions are typically mutually exclusive and possible outcomes are mutually exclusive, so only one scenario unfolds, and in some situations we might be justified in focusing on containing the risks of the worst-case scenario---although Bayesian decision avoids use of minimax procedures directly, in lieu of expected utility maximization.  In any case, in the apportionment problem, all states receive some seat assignment, so worrying about none but the most under-represented and/or over-represented state does not seem particularly sensible.}  Nor will such minimax methods necessarily lead to unique solutions even in the case of completely non-commensurate (i.e., relatively prime) population counts, so in the end a minimax rule must often be supplemented with secondary criteria to break what would otherwise be many ties.\footnote{That being said, certain choices of \textit{convex} minimax criteria, combined with a constraint on the total number of representatives, can lead to unique apportionments.  Hamilton's method is such a case.}

\subsubsection{Geometric Interpretations}

Apportionment methods, particularly when derived via optimization or variational principles, can often be understood geometrically.  In particular, the exact quotas $\bv{q} \in \rationalsymbol^{S}$ can be though of as an $S$-dimensional vector constrained to lie on the hyperplane defined by $R - \sum\limits_{s} q_s = 0$ (or on some feasible subset of this hyperplane, for which all lower and upper bound constraints are also satisfied).  In any allowed apportionment, the exact quota vector $\bv{q}$ is to be replaced with some lattice point, i.e.,  an $S$-dimensional vector $\bv{a} \in \integersymbol^{S}$ with all nonnegative integer components but which also lies on the same feasible hyperplane as does the exact quota vector $\bv{q}$.

Many of the well-known methods to be described in Appendix~\ref{historical_methods} below have reasonably simple geometric interpretations,  For example, the Hamilton-Vinton Method chooses  the lattice point closest to $\bv{q}$ on the feasible part of the hyperplane (which also assigns at least one representative to each state), where ``closest'' is measured in the usual Euclidean $\mathcal{L}^2$ metric (or in fact for any $\mathcal{L}^{\mu}$ metric with $\mu \ge 1$).  The Webster Method chooses the lattice point on the hyperplane nearest to the line that passes through both the origin and $\bv{q}$ (and which also assigns at least one representative to each state), again measured in the Euclidean metric.  And it turns out that the Huntington-Hill method, currently mandated by federal statute, is equivalent to finding the closest feasible point to the exact quota in the Hilbert projective metric:
\be
h(\bv{a}, \bv{q} ) = \log \max\limits_{s,s'} \frac{a_{s} q_{s'} }{ a_{s'} q_{s} },
\ee
assuming all components $a_s$ and $q_s$ are positive.

While this geometric language may help us visualize the operation of some apportionment rule, it does not necessarily help us adjudicate between competing methods, because \textit{a posteriori} any number of metrics or geometries can be imposed arbitrarily on the space of apportionment vectors, and it is not at all obvious \textit{a priori} which is the most natural geometric structure.  As Balinski and Young \cite{balinski_young:1977r} note, ``to say it is desirable to `minimize the length of the inequity vector in [some] Euclidean $S$-space' begs the question.''

Indeed, our contention is that we should really look, not directly in the space of quotas and apportionments, but rather in the space of polling probability distributions, or normalized weights of representation across all individuals.  While the relative entropy provides the most natural measure of discrepancy or divergence in such a space of distributions, it is well known that it does not describe a true metric, because in general it is not symmetric and does not satisfy the triangle inequality.  But in comparing a proposed distribution to the ideal distribution, symmetry is already broken, so a metric structure is not required, only some appropriate notion of divergence between distributions, or perhaps of \textit{projection} of an ideal distribution onto a feasible sub-space of realizable distributions consistent with constraints.\footnote{Nevertheless, in the \textit{information geometry} approach developed by Amari and others, an actual metric tensor can be associated with a manifold of probability distributions.  Known as the Fisher metric, this turns out to be proportional to the Hessian matrix (i.e., matrix of second derivatives) of the relative entropy.  However, given the Fisher metric, the induced ``$\alpha$-divergences,'' and in particular the Kullback-Leibler numbers, assume a more important role than does, say, the geodesic distance.}


\subsection{Lottery Methods}

All methods discussed so far are deterministic in the absence of some randomized tie-braking.  In contrast, various lottery-based or randomized apportionment methods have been suggested \cite{grimmett:2004} because they can ensure proportionality ``on average'' or ``in the long run.''  Of course many different randomized methods can achieve \textit{expected} proportionality in the statistical sense, so further criteria, such as minimizing variance, must be invoked to single out a particular strategy.  For example, one could imagine constructing (or simulating on a computer) a roulette wheel having $S$ slots with widths proportional to the state populations.  The wheel could be spun $R$ times, and seats assigned by where the ball lands.  A better (or at least lower-variance) alternative would be to first give each state the integer part of its quota (or perhaps fair share) and then only use the roulette wheel to allocate the additional seats, with the slot widths proportional to the states' fractional remainders of the quotas (or of the fair shares).  Or, one could imagine placing colored marbles into an urn (specifically, $p_s$ marbles of the $s$th color for the $s$th state), and drawing $R$ marbles with or without replacement.

But the job of any apportionment scheme is to make the best \textit{singe-case} assignment of seats for the actual set of populations obtained, not in some imagined long-run repetition of random draws.  We only get to make one apportionment every decade, so what may or may not average out in the long run is irrelevant.  And randomized methods can violate fundamental desiderata of fairness, with smaller states sometimes receiving more representatives than strictly larger states.  Our own intuition rebels at the use of randomization where not strictly necessary.

Of course there can arise situations where some sort of randomization may be unavoidable, in order to break certain ties.  In various methods relying on ranking or prioritization or optimization, there is always a possibility of an exact tie, for particular population ratios.  Sometimes (but not always) this can be resolved by a further deterministic rule, for example dictating that in the case of a tie in a quota or ranking method, the seat always goes to the larger state.  But any proposed scheme should be able to work unambiguously in a hypothetical case where two or more states happen to have exactly the same populations, but must receive an unequal number of representatives.  If for instance we ever must apportion an odd number of seats amongst an even number of states of exactly equal population, there is really no other fair option but randomization, unless it is permissible to go back and change the House size to avoid the incommensurability altogether.  Using some other characterization of the states, such as their lexicographic ordering, not only strikes us as arbitrary and unsatisfying, but should probably be regarded as un-Constitutional, violating the ``according to their respective numbers'' clause which would seem to implicitly preclude use of any information about the states apart from their population counts, in making decisions about apportionment.

In short, our sense is that in apportionment, randomization should be avoided altogether, except where it becomes essential.

\newpage
\section{\uppercase{Some Specific Apportionment Methods}\label{historical_methods}}

For comparison and context, here we provide descriptions and discussions of several methods that have been actually used, or seriously suggested, for apportionment in the U.S.\ House of Representatives.  While this catalog is not meant to be entirely exhaustive, it does cover most methods that have been employed or debated historically or discussed more recently:

\subsection{Hamilton Method}
(Introduced by Alexander Hamilton, 1792; also known as the Method of Largest Fractions, the Method of Greatest Remainders, or the Method of Largest Fractional Remainders).

Initially, each state is assigned its lower quota  $\lfloor q_s \rfloor$, then the states are sorted in order of decreasing values of the $(q_s -  \lfloor q_s \rfloor )$, and, going down the list of states, an additional representative is assigned to each state in order of these fractional remainders, until a total of $R$ representatives have been assigned.  (Exact ties at the cutoff may have to be broken by some other criterion.  For example, under a tie the extra representative can be given to the larger state.  In the off-chance of states with exactly the same populations, such ties should be broken with a coin flip or other randomized choice).

In the absence of ties, the Hamilton Method can also be thought of as rounding to the nearest integer \textit{after} shifting the exact quotas $q_s$ by a common offset:
\be
a_s = \lfloor q_s + \Delta_{\stext{H}} + \tfrac{1}{2}  \rfloor,
\ee
where $\Delta_{\stext{H}}$ is a state-independent constant chosen self-consistently so that $\sum\limits_{s=1}^{S} a_s = R$.

In terms of a global variational principle, Hamilton's method minimizes
\be\label{hamvar1}
\U_{\stext{H}}(\bv{a}; \bv{p}, R,S; \ell) = \sum\limits_{s=1}^{S} \bigl\lvert a_s - q_s\bigr\rvert^{\ell}
\ee
for any fixed constant $\ell \ge 1$ defining a choice of norm.

Birkhoff \cite{birkhoff:1976}  introduced the notion of what he called \textit{binary fairness}, which insists  that it should not be possible to transfer a representative from a state $s$ to a state $s'$ and reduce $\lvert a_s - q_s \rvert + \lvert a_{s'}- q_{s'} \rvert$.  But looking at the figure-of-merit \eqref{hamvar1} for the case $\ell = 1$,  we can see that Hamilton's method will be the only apportionment method satisfying this binary fairness principle.  The principle itself strikes us as \textit{ad hoc} and state-centric, so we are not bothered by violations of it.

A proposed apportionment based on Hamilton's Method led to the very first U.S.\ presidential veto, by George Washington in 1792.  However, Hamilton's method was used subsequently starting in the 1850 census, but by the early 1880's its many serious flaws became too apparent to ignore.\footnote{Sorry, Hamilton fans, or \textit{Hamilton} fans.}  Hamilton's method can be subject to any and all of the Alabama Paradox, the Population Paradox, and the New States Paradox.  In its original formulation, this method can also assign zero seats to a small state, but can be modified in an obvious way (see below) to incorporate constraints involving lower and/or upper bounds.

More critically from our perspective, like all quota methods, Hamilton's method overemphasizes the idea that it is states that are entitled to representatives, rather than people from the various states.  And at least at first glance, rounding based only on the fractional parts of the quotas might seem to ignore the absolute numbers of persons who thereby become under-represented or over-represented in the population.  On the other hand, choosing $\ell = 1$, we notice that the Hamilton apportionment also follows from minimizing the equivalent expression:
\be
\U_{\stext{H}}(\bv{a}; \bv{p}, R,S;1) = \tfrac{1}{P} \sum\limits_{s=1}^{S} p_s \bigl\lvert  \tfrac{1}{\bar{d}_s} - \tfrac{1}{\Dbar} \bigr\rvert,
\ee
which is the average over all represented individuals of the absolute deviation of the (intra-state average) shares of a representative from the democratic ideal. So whether or not Hamilton's method accounts for the number of individuals affected by the rounding might seem to depend on the form in which the variational principle is written.  One moral might be that neither the mere plausibility of a proposed inequity function, nor perhaps the appearance of implausibility in one particular mathematical form, will be sufficient to make a principled choice: we shall require deeper desiderata than mere reasonable appearance if we are to pick out a unique objective function to optimize.


It is not hard to see that the Hamilton apportionment also follows from the minimax solution to $\min\limits_{\bv{a}} \max\limits_s\,  \lvert a_s - q_s \rvert$,  subject to constraints on the non-negativity of all allotments $a_s$, as well as on the total house size $\sum\limits_{s=1}^{S} a_s = R$.

\subsection{Hamilton-Vinton Method}
(Also known simply as Vinton's Method, the Hare-Niemeyer method or else, somewhat confusingly, occasionally by any one of the other names associated with Hamilton's original method, such as the Method of Greatest Remainders).

The original Hamilton Method was rediscovered by Vinton but adjusted to accommodate the requirement that each state receive at least one representative but no more than a specified maximum.  Any states with zero lower quotas are placed at the top of the priority list before the remaining representatives are apportioned.  (And any states which are already at their upper bounds can be moved to the bottom of the list).

More generally, we can start each state out with  $a'_s = \max\bigl[\lambda_s, \lfloor q_s \rfloor \bigr]$ representatives, then round up the states with largest values of $(q_s - a'_s)$.  These are automatically negative for those states whose quotas are below the minimal apportionments, ostensibly placing these states at the bottom of the priority list.  Likewise, states whose upper quotas would lie above the prescribed upper bound can be given low priority.

However, apart from incorporating lower/upper bound constraints, Vinton's Method otherwise suffers from the same shortcomings as the pure Hamilton Method, including susceptibility to the Alabama paradox.

\subsection{Lowndes Method}
(Suggested by William Lowndes, a Representative from South Carolina, in 1822.  Also known as the Method of Largest Relative Fractions).

This is similar to the Hamilton-Vinton Method, except that after an initial assignment  given by $a'_s = \max\bigl[\lambda_s, \lfloor q_s \rfloor \bigr]$, states are sorted in order of decreasing values of the relative fractional parts $(q_s -  a'_s )/\lfloor q_s \rfloor$, and the remaining seats are awarded in this order.  This too can be easily modified to incorporate upper bound constraints.

Lowndes argued that, at the margin, an extra seat was much more valuable to a small state (with fewer representatives) than a larger state (which presumably already has many representatives).  While undoubtedly a larger delegation is more valuable politically in terms of the state's overall influence in Congress and the Electoral College, clearly the point should be to assign seats based on fairness to the people, not political value to a state.

Originally devised to try to maintain quota but reduce perceived bias against small states, the Lowndes Method is still prone to the Alabama paradox.
  
 \subsection{Modified Vinton Method}
 
After the Alabama paradox came to light, various (unsuccessful) attempts were made to modify the Hamilton-Vinton method to produce an apportionment rule that both satisfied quota and was house monotone.  For example, what became known as the Modified Vinton Method based its prioritization in rounding on $(q_s - \lfloor q_s \rfloor)/p_s$ rather than on $(q_s - \lfloor q_s \rfloor)$ as in the original Vinton method.  However, this failed to immunize against the Alabama paradox.
  
\subsection{Hill's Quota Method}
(Discussed by J.A.~Hill in 1910, according to \cite{balinski_young:1977s}.  Also known as the Method of Alternating Ratios.  Similar but not equivalent to the Huntington-Hill Method, described below).
 
Each state is initially assigned $a'_s = \max\bigl[\lambda_s, \lfloor q_s \rfloor \bigr]$ representatives.  Remaining seats are distributed one at a time based on a priority ranking of eligible states by $\tfrac{p_s}{ \sqrt{a'_s (a'_s + 1)} }$, until the specified house size is reached.  States that would receive more than their upper quota, or other upper bound, can instead be pushed to the bottom of the priority list.

The method satisfies quota unless lower or upper bound constraints override, but it is susceptible to the Alabama paradox.

Essentially, the method is a variant of the Huntington-Hill scheme discussed below, only with additional constraints imposed to maintain state-level quotas.

\subsection{Balinski-Young Quota Method}

In 1974, mathematicians M.L.~Balinski and H.P.~Young  \cite{balinski_young:1974,balinski_young:1975} proposed a method which (i) satisfies quota (if the quota lies within the allowed lower and upper bounds), (ii) is immune to the Alabama paradox, and (iii) is consistent under the accumulation of seats, in the sense that, as the house size is increased, relative claims for an extra seat between two eligible states depend only upon their respective populations and their current apportionments (and possibly on the total population and house size), but not the individual populations of other states.
 
Their method is most easily described and implemented as a ranking method, where all states start off with no seats, and seats are assigned one by one.   If $a_s'$ is the $s$th state's allotment of seats at a given stage at which a total of $R'$ seats have been assigned so far, where $0 < R' < R$, remaining seats are assigned one at a time based on a priority list ranking states in decreasing order of $\tfrac{p_s}{a'_s + 1}$, amongst all states which are deemed currently \textit{eligible} to receive the next seat, in the sense that $a_s' \le \tfrac{p_s}{P} (R'+1)$, which is to say, the state has not yet exceeded what would be its upper quota for a house size of $R'+1$.  Though not obvious, it turns out this rule will also guarantee each state at least its lower quota of seats.
 
Instead of using quotas, Balinski and Young actually came to advocate using a variant of this method based on fair shares, applying the ranking method described above to any and all states whose fair shares turned out to be non-integral.\footnote{Later still, they appear to have recanted and seemed to end up endorsing \cite{balinski_young:1980} Webster's method.}

\subsection{Randomized Quota Method}

As mentioned above, all methods may need to be augmented with some sort of randomization in the case of certain ties that cannot be broken by any objective standard, for example where an odd number of representatives must be apportioned to an even number of states with exactly equal  populations, for there is then no non-arbitrary deterministic criteria to decide between them.   But some methods relay on randomization more generally.

To generate a quota method which is unbiased between large and small states, a randomized rounding schemes has been suggested, where each state's quota $q_s$ is independently and at random rounded up with probability $q_s - \lfloor q_s \rfloor$, and down otherwise, and the overall random assignment across all states is simply repeated until the total number of apportioned representatives first agrees with the target house size $R$.

However, in any one realization, this procedure could give more representatives to some state than to a strictly more populous state, which is hard to justify based on any sort of assurances of fair behavior in the long run.  Again, wherever and whenever possible, an apportionment method really ought to be deterministic.

\subsection{Jefferson Method}
(Suggested by Thomas Jefferson, in 1792; also known as the Method of Greatest Divisors, the Modified Lower Quota Method, the Method of d'Hondt, or the Hagenbach-Bischoff Method.  Used for the United State Congress until 1840).

Historically, Jefferson's Method was the first of many divisor methods.  Jefferson's method tentatively assigns each state $\lfloor Q_s \rfloor$ representatives, where $Q_s = p_s/D$, and then adjusts the divisor $D$ as necessary, always rounding down the quotients calculated using the modified divisor, until the total number of representatives assigned is exactly $R$.

Jefferson publicly argued that the fractional remainders of the $Q_s$ were simply ``unprovided for'' by the Constitution, so should be dropped.\footnote{It is unclear whether this argument was entirely sincere.  Perhaps more likely, he privately surmised that the method would tend to favor large states like Virginia.}

This pure Jeffersonian strategy does not ensure that each state receive at least one representative, so the last representatives assigned as $D$ is decreased from a large initial value would have to be shifted to the states without representatives. (More generally, enough seats can be shifted to ensure that each state receives any consistent prescribed lower bound $\lambda_s$, and no state exceeds its upper bound $u_s$).

Interpreted instead as a ranking method, the Jefferson Method (modified to ensure each state's representation falls within allowed bounds) is equivalent to the following iterative process:  begin by giving each state its minimum allowed number of representatives (for U.S.~Congressional seats, exactly one for each state) so as to take care of the lower bounds from the start.  Maintain a list of states in decreasing order of $\tfrac{p_s}{a_s' + 1}$, where $a_s'$ is the state's allotment of seats at the current iteration; remove any states which have reached their upper bound; and assign additional representatives from this list in decreasing order, incrementing the selected $a_s'$ and re-sorting after each assignment.  Note that the ranking score $\tfrac{p_s}{a_s'+1}$ would be the average district size within the $s$th state if the next representative were to be awarded to that state.

As a pairwise comparison method, the Jefferson method is equivalent to seeking apportionments which are stable with respect to the fairness test
\be
T(a_o, p_o, a_u, p_u) = a_o \tfrac{p_u}{p_o} - a_u = a_o \tfrac{q_u}{q_o} - a_u,
\ee 
closely related to the notion of so-called representational deficiency introduced above, but which, to our taste, lacks symmetry, and certainly any obvious rationale, compared to any number of other possible comparison tests.

The Jefferson method can violate upper quota, but never lower quota in the absence of active lower or upper bound constraints, and in fact turns out to be the only divisor method guaranteed to satisfy lower fair-share.  It is also the only divisor method which can be said to be fully immune to the temptations of strategic splitting.  But as widely acknowledged, it is heavily biased in favor of larger states, in the sense that a large state with a quotient of, say, $Q_s = 50.1$ would be rounded down to $a_s = 50$, involving a small relative change, but a small state with a quotient of $Q_{s'} = 1.9$ would be rounded down to $a_{s'} = 1$, a very large relative drop.

\subsection{Adams Method}
(Suggested by John Quincy Adams in or around 1832, but never adopted for Congressional apportionments.  Also known as the Method of Smallest Divisors, or the Modified Upper Quota Method).

Given a divisor $D$,  Adams' method works much like the mirror image of Jefferson's method, by instead assigning the ceiling $\lceil Q_s \rceil$ of the quotients as the allotted representatives, and adjusting $D$ as necessary, recalculating tentative seat assignments by always rounding up all quotients using the modified divisor, until the total number of representatives assigned is $R$. 

This automatically guarantees each state receives at least one seat, but for apportionment problems where the imposed lower bounds differ from unity, the algorithm can be modified by shifting the last assigned seats as $D$ is lowered.

As a ranking method, the Adams method turns out to be equivalent to ranking states by $\tfrac{p_s}{a_s'}$, the average intra-state district sizes given the representatives apportioned so far, so the state with the largest average district size based on the previous apportionments earns the next representative.

As a pairwise comparison method, the Adams Method is equivalent to seeking apportionments which are stable with respect to the fairness test
\be
T(a_o, p_o, a_u, p_u) = a_o - a_u  \tfrac{p_o}{p_u} =  a_o - a_u  \tfrac{q_o}{q_u},
\ee 
which is closely related to the notion of so-called representational surplus mentioned above, and is in a sense dual to, but also just as \textit{ad hoc} as, the comparison test associated with the Jefferson method.

Late in his career, Walter F.~Willcox, formerly a professor at Cornell University, president of the American Statistical Association, and regarded as one of the first professional statisticians in the United States, apparently switched his professional loyalties from Webster's Method, which he had long championed, to Adams' Method.  He contended that Adams' Method of Smallest Divisors ``secures the smallest average population per district and the narrowest range between the largest and smallest average size of district.''  By ``average population per district,'' Willcox apparently does not mean the natural average district size, $\tfrac{1}{R} \sum\limits_{s=1}^{S} \sum\limits_{k=1}^{a_s} d_{sk} = \tfrac{P}{R} = \Dbar$, which must be the same for all apportionments with the same values of $R$ and $P$, but rather the \textit{unweighted} average (over states) of their respective intra-state average district sizes, $\tfrac{1}{S} \sum\limits_{s=1}^{S} \bar{d}_s = \tfrac{1}{S} \sum\limits_{s=1}^{S} \tfrac{p_s}{a_s}$, which does not reflect the multiplicity of districts within each state.  The ``range'' in average district sizes to which he refers is just $\max\limits_s \bar{d}_s - \min\limits_s \bar{d}_s = \max\limits_s \tfrac{p_s}{a_s} - \min\limits_s \tfrac{p_{s}}{a_{s}} =  \max\limits_{s,s'} \bigl[ \tfrac{p_s}{a_s} - \tfrac{p_{s'}}{a_{s'}} \bigr]$, which is insensitive to the uniformity of any districts intermediate in average size between the very largest and smallest.  So these criteria do not seem particularly natural.  And anyway, Adams' method does \textit{not} actually guarantee an apportionment minimizing either of these quantities---it is straightforward to find counter-examples.  Adams' method may tend to do somewhat better at minimizing these than does either Webster's or Huntington's methods (to be discussed below), but that is hardly a compelling argument.  If one's primary objectives are to make Congressional districts as small and as nearly equally-sized as possible, then presumably one should explicitly optimize on those measures.

However, the Adams' apportionment does minimize the maximum average intra-state district size, so the simple optimization criteria $\min\limits_{\bv{a}} \max\limits_{s} \tfrac{p_s}{a_s}$, subject to $\lambda_ s \le a_s \le u_s$ for $s = 1, \dotsc, S$ and $\sum\limits_{s=1}^{S} a_s = R$, will generate  the Adams apportionment, essentially uniquely accept for exact ties.  This follows from the ranking-based formulation of the method:  Let $\bv{a}$ be the Adams apportionment, for which state $m$ possesses the maximal average district size: $p_m/a_m \ge p_s/a_s$. Consider any distinct, feasible apportionment $\bv{a}'$, for which there must be at least one state $z$ which has strictly fewer seats than in the Adams' apportionment, i.e., $1 \le a'_z < a_z$.  In the Adams' apportionment,  state $z$'s last seat would have been awarded at some specific stage of the ranking algorithm, at which point state $m$ would have had $a^*_m \le a_m$ seats.  Since state $z$ won the seat at that stage, it follows that $p_z/(a_z-1) \ge p_m/a^*_m$, and since $a'_z  \le a_z - 1 < a_z$, we may infer
\be
\frac{p_z}{a'_z}  \ge \frac{p_z}{a_z-1} \ge  \frac{p_m}{a^*_m} \ge  \frac{p_m}{a_m},
\ee
so the maximum average district size must go up, or at least stay the same, under any other feasible apportionment.

Adams' Method can violate lower quota, but never upper quota in the absence of upper or lower bound constraints, and in fact turns out to ne the only divisor method guaranteed to satisfy upper fair-share.  It favors smaller states in the same sense that the Jefferson method favors larger ones---smaller states are the ones that tend to end up more over-represented in a relative sense.

\subsection{Dean Method}
(Communicated by James Dean, a former professor of Astronomy at Dartmouth College, to Daniel Webster.  Also known as the Method of Harmonic Means).

Given a divisor (target district size) $D$, Dean's Method \cite{bell:2009} may be thought of as choosing the apportionments $a_s$ so that the average intra-state districts sizes $\bar{d}_s = \tfrac{p_s}{a_s}$ are all as close to $D$ as possible in absolute value. However, this does not lead to a simple global optimization method in the sense defined above, because in minimizing $\sum\limits_{s=1}^{S} \lvert \tfrac{p_s}{a_s} - D \rvert$ we must also adjust $D$ self-consistently to ensure $\sum\limits_{s=1}^{S} a_s = R$.

However, Dean's method does turn out to be equivalent to rounding each quotient $Q_s$ based on a threshold $\theta(Q_s)$ equal to the harmonic mean of $\lfloor Q_s \rfloor$ and $\lceil Q_s \rceil$.  As with other divisor methods, $D$ can then be adjusted iteratively until the number of representatives apportioned, $\sum\limits_{s=1}^{S} a_s$, exactly equals $R$, the predetermined House size.

As a ranking rule, Dean's Method prioritizes seats based on decreasing values of  $p_s \tfrac{2a'_s + 1}{2 a'_s (a'_s+1)}$, which looks complicated, but is just the population $p_s$ divided by the harmonic mean of the seat count $a'_s$ so far and what would be the next seat total, $a_s' + 1$.

As a pairwise comparison method, the Dean Method seeks the apportionment which is stable with respect to the fairness test
\be
T(a_o, p_o, a_u, p_u) = \tfrac{p_u}{a_u} - \tfrac{p_o}{a_o} = \tfrac{P}{R} \bigl( \tfrac{q_u}{a_u} - \tfrac{q_o}{a_o} \bigr),
\ee
which reveals that Dean's Method favors smaller absolute differences in average district size under any feasible proposed pairwise swap of seats.  At the very least, this fairness test appears somewhat more symmetric and more natural than the fairness tests for either Jefferson's or Adams' methods.

In an influential 1929 report on the Census and apportionment \cite{bliss:1929}, the U.S.\ National Academy of Sciences apparently suggested, erroneously, that Dean's Method minimizes the difference between the largest and smallest intra-state average district sizes.  It is not hard to find counter-examples to this claim.  It seems that the committee may have confused Huntington's notion of stability under pairwise swaps of seats with the requirements for a true global optimum.  Likewise, the Dean apportionment does \textit{not} necessary minimize the quantity $\sum\limits_{s=1}^{S}\sum\limits_{s'=1}^{S} \abs{ \bar{d}_{s} - \bar{d}_{s'} }$, nor $\sum\limits_{s=1}^{S} \abs{ \bar{d}_s - \Dbar }$, nor $\sum\limits_{s=1}^{S} \bigl( \bar{d} - \Dbar \bigr)^2$.  Using the ranking-based characterization, it is possible to construct a minimax criteria generating the Dean apportionment, but the result does not have any particularly natural interpretation, and in fact we have been unable to find any compelling and simple global optimization principle leading to Dean's Method.

Dean's Method tends to have a slight bias in favor of smaller states, but not as pronounced as Adams' Method.

\subsection{Webster Method}
(Suggested by Daniel Webster, circa 1832, and used in the apportionments after the 1840, 1880, 1900, and 1910 censuses.  Explored and re-formulated by Walter F.\ Willcox around 1910, after which it was also known as the Webster-Willcox Method, or the Method of Major Fractions, or the Method of Arithmetic Means.  Known in Europe as the Sainte-Lagu\"e Method, as it was popularized for the purposes of party-list proportional representation in the latter's influential 1910 article).

Another divisor method, the Webster Method \cite{balinski_young:1980} instead chooses the $a_s$ so that state representational shares $\tfrac{1}{\bar{d}_s} = \tfrac{a_s}{p_s}$ are all as close to $\tfrac{1}{D}$ as possible in absolute value.  It is not hard to show that this is equivalent to using as rounding threshold the arithmetic mean of $\lfloor Q_s \rfloor$ and $\lceil Q_s \rceil$, which entails that the quotients $Q_s$ are simply rounded to the nearest whole number:
\be
a_s = \lfloor Q_s + \tfrac{1}{2} \rfloor,
\ee
then $D$ can be adjusted as necessary until the total apportionment reaches the predetermined target $R$.
 
As a ranking method, the Webster method amounts to ranking states and assigning additional seats according to the ratios $\tfrac{p}{a_s' + \tfrac{1}{2}}$. Here $a'_s + \tfrac{1}{2}$ is of course the arithmetic mean of the current seat total $a_s'$ and what would be the next seat total $a_s' + 1$.

As a pairwise comparison method, the Webster method is associated with the fairness test
\be
T(a_o, p_o, a_u, p_u) = \tfrac{a_o}{p_o} - \tfrac{a_u}{p_u} = \tfrac{R}{P}\bigl( \tfrac{a_o}{q_o} - \tfrac{a_u}{q_u}  \bigr),
\ee
meaning it favors equalizing the absolute differences between the shares of representation under possible pairwise swaps.

As a global optimization method, Webster apportionments can be generated by minimizing (subject to constraints) the objective function
\be
\U_{\stext{W}}( \bv{a}; \bv{p}, R,S) = \tfrac{1}{P} \sum\limits_{s=1}^{S} p_s \bigl(  \tfrac{1}{\bar{d}_s} - \tfrac{1}{\Dbar} \bigr)^2 = \tfrac{R}{P^2} \sum\limits_{s=1}^{S}  \tfrac{ (a_s - q_s  )^2}{q_s},
\ee
which is the mean-squared deviation in best-case shares of representatives (over all individuals), or equivalently can be interpreted as a quantity proportional to the \textit{chi-squared} statistic $\chi^2$ measuring the discrepancy between the ``observed'' distribution $\bv{a}$ of seats and the ``expected'' proportional distribution $\bv{q}$, across all states.

Birkhoff \cite{birkhoff:1976} also introduced a notion of what he called \textit{binary consistency}, and what Balinski and Young \cite{balinski_young:1982} later labeled \textit{relative well-roundedness}. They define a state's apportionment to be \textit{over-rounded} if $a_s > q_s + \tfrac{1}{2}$, and \textit{under-rounded} if $a_s < q_s - \tfrac{1}{2}$.    An apportionment $\bv{a}$ is said to be relatively well-rounded if there is no pair of states such that one state is under-rounded and another over-rounded.  It turns out that the Webster method is the only method which is house monotone, pairwise consistent, and always relatively well-rounded.  Since we are unmoved in general by arguments about nearness to quotas \textit{per se} and in particular by an argument that presumes without justification an \textit{ad hoc} definition of over or under-rounding that ignores the absolute sizes of the states, we are not bothered that the entropic method can violate this property.

The Webster Method can also violate quota in either direction, but in actual practice rarely does so for non-contrived examples.  It is apparently the only divisor method guaranteed to be near fair-share, which underlies its eventual advocacy by Balinski and Young.\footnote{But only after some earlier enchantment with their own quota/fair share  method....}  Computer simulations have suggested that the Webster method very slightly favors large states, but it appears to have the smallest bias of any of the historic divisor methods, by various measures.

 \subsection{Huntington Method}
(Explored and advocated in detail starting around 1921 by Harvard mathematician and logician Edward V.\ Huntington, the third President of the Mathematical Association of America, modifying an earlier approach of Joseph A.\ Hill, a Census Bureau statistician.  Also known as the Huntington-Hill Method, and the Method of Equal Proportions).

This method has been used for Congressional apportionment starting in 1929, and since 1941, has in fact been mandated by ongoing Congressional statute.  
 
Huntington's goal was to keep the {ratio} of one state's persons-per-representative to that of another state as close to unity as possible under possible pairwise swaps. In this approach, seats are assigned to the states so that no transfer of a seat can reduce the \textit{relative} difference in per-capita representation between those states, or equivalently the \textit{relative} difference in average district size, where relative discrepancies are to be determined by dividing the absolute value of the difference by the smaller of the two values. 

That is to say, as a pairwise comparison method, the Huntington-Hill approach demands stability with respect to the fairness test
\be
T(a_o, p_o, a_u, p_u) = \tfrac{a_o p_u}{a_u p_o} - 1 = \tfrac{a_o q_u}{a_u q_o} - 1,
\ee
meaning it favors reducing the \textit{relative} differences between the district sizes or equivalently, relative differences between the shares of representation.  The fact that this method is ``self-dual'' in this sense and does not require a choice between using district sizes or shares of representation (reciprocal district sizes) to assess inequality of representation in the comparison test was a major selling point for Huntington.

As a divisor method, the algorithm chooses $a_s$ by rounding the quotient $Q_s$ based on a threshold given by the geometric mean of $\lfloor Q_s \rfloor$ and $\lfloor Q_s \rfloor$, then adjusts $D$ as necessary and recalculates the $a_s$ until the total number of  representatives assigned is $R$.  And just like the other divisor rules, this approach may be modified to incorporate lower/upper bounds on the apportionments.  By default, it must assign at least one seat to each state, even in the absence of any explicit lower bound constraints, which in our view speaks strongly against its universality. 

As a ranking rule, additional seats are apportioned according to the magnitude of $\tfrac{p_s }{\sqrt{a_s' (a_s'+1)}}$, where the denominator $\sqrt{a'_s (a'_s+1)}$ is  the geometric mean of the current seat count $a'_s$ and the next seat count $a'_s+1$.

In terms of global optimization, the Huntington method can be shown to minimize (subject to constraints) the function
\be
\U_{\stext{H}}( \bv{a}; \bv{p}, R,S) = \tfrac{1}{R}\sum\limits_{s=1}^{S}  a_s \bigl( \bar{d}_s - \Dbar \bigr)^2 = \tfrac{P^2}{R^3} \sum\limits_{s=1}^{S}  \tfrac{ (a_s - q_s  )^2}{a_s},
\ee
which is the mean-squared deviation of (best-case) district sizes across all states, or equivalently can be seen as a sort of dual to the chi-squared function generating the Webster apportionment, in that the roles of the $\bv{a}$ and $\bv{q}$ are reversed \textit{vis-\`a-vis} the Webster objective function.

The Huntington method can violate quota, but to date has not done so in practice for any U.S.~Congressional apportionments.  Theory and simulations suggest that it tends to favors small states slightly.  

\subsection{Condorcet Method}
(Proposed by Nicolas de Caritat, marquis de Condorcet).

Condorcet's apportionment method is similar to (but predates) Webster's method, except that the rounding threshold occurs $40\%$ into the interval between $\lfloor Q_s \rfloor$ and $\lceil Q_s \rceil$, rather than $50\%$, apparently based on intuition for what might better reduce bias between large and small parties.\footnote{In party-list systems, others have advocated shifting in the other direction, and using a threshold of $70\%$, so as to build in a higher hurdle for very small parties to receive a seat.}

\subsection{Smith Method}
(Suggested in online discussions by Warren Smith \cite{smith:2007a,smith:2007b}, circa 2007).  

Several divisor methods may be seen as special cases of the rule
\be
a_s = \lceil Q_s - \Delta \rceil,
\ee
where $Q_s = p_s/D$ as usual, and $\Delta$ is a constant offset.  For example,  the methods of Webster, Adams, and Jefferson correspond to the choices $\Delta = \tfrac{1}{2}$, $\Delta = 0^{+}$, and $\Delta = 1^{-}$, respectively.

Smith has recommended a rule of this form but with a different intermediate value for $\Delta$.  Based on a simple (if questionable) probabilistic model of state populations,\footnote{The populations for apportionment are assumed to be IID exponentially distributed random variables.  This is, to be sure, the maximum entropy distribution given an overall mean state population, but clearly, in the U.S., we have strong prior information suggesting that state populations are not identically distributed, and migration problematizes the assumption that they are independently distributed.} minimizing the expectation value of an additive measure of bias from rounding suggests an optimal value for $\Delta$ (in the absence of lower/upper bound constraints):
\be
\Delta = \bar{a}\, \ln \bigl[ \bar{a} (1 - e^{-\frac{1}{\bar{a}}}) \bigr]^{-1},
\ee
where $\bar{a} = \tfrac{R}{S}$ is the average number of seats per state, which for current U.S.\ parameters ($S = 50$ and $R = 435$) leads to $\Delta \approx 0.495211\ldots$, close to the Webster limit. 

However, as stated this approach will not guarantee a one-seat minimum per state.  Instead, $\Delta$ can be set to zero for the first round of seat assignments, then subsequently set to
\be
\Delta =  \bar{a} \bigl\{   \ln \tfrac{\bar{a}}{\bar{a} - 1} - \ln \bigl[ \bar{a} (e^{\frac{1}{\bar{a}}} -1) \bigr]   \bigr\},
\ee
provided $0 < S < R$ strictly.  For $\bar{a} = 435/50 = 87/10$, this yields a larger value $\Delta = 0.557505\ldots$.  Different lower bounds would lead to still different values of $\Delta$, which strikes us an odd sort of dependency to build into one's rounding criteria.

Smith also discusses a number of other optimality criteria, leading to different choices for the offset $\Delta$. 

\subsection{Burt-Harris-Edelman Method}
(Popularized in 2006--2008 by Paul Edelman \cite{edelman:2006}, a mathematics professor at Vanderbilt.  The same method had been proposed earlier by O.~Burt and C.~Harris, Jr.\ in 1963 \cite{burt:1963}, and critiqued by E.J.~Gilbert and J.A.~Shatz by 1964 \cite{gilbert:1964}.  Also known as the Minimum Total Variation Method).  
 
The National Academy of Science erroneously attributed to Dean's Method the property of minimizing the range of average district sizes across states.  Willcox erroneously attributed this property to Adams' Method.  But this criterion can be used to define a (non-unique) apportionment strategy.
 
Edelman tried to argue that Supreme Court jurisprudence in intra-state Congressional districting cases singles out a specific numerical measure of disparity in district size as the key criterion in inter-state apportionment as well.  This claim is highly debatable, as the Supreme Court has never invalidated any of the traditional methods that have been used for Congressional apportionment, nor would a unique mathematical measure of deviation in district size obviously follow from any of their decisions to date.  Yes, the Supreme Court has ruled that Congressional districts within states should be chosen ``as nearly as practicable'' to be equal in size within states, and has permitted only ``limited variances which are unavoidable despite a good-faith effort to achieve absolute equality, or for which justification is shown,'' but these standards were never put forward as explicit criteria for apportionment of representatives amongst U.S.\ states as opposed to districting within states, and of course any number different criteria might be used to measure inequality in district size within or across states.
  
Nevertheless, based on a small body of case law regarding apportionment, and somewhat more case law regarding intra-state Congressional districting, Edelman argued for (constrained) minimization of one particular inequity function, what he calls the total variation in district size, but what we would call the scaled range of intra-state average district sizes.  Specifically, for single-member district-based representation, his optimal apportionment minimizes with respect to $\bv{a}$ (subject to constraints on House size and lower and upper bounds) the difference 
\be
\U_{\stext{E}}(\bv{a}; \bv{p}, R) =  \frac{\max\limits_{s} \bar{d}_s - \min\limits_{s} \bar{d}_s }{\Dbar} = \tfrac{R}{P} \bigl[  \max\limits_{s} \tfrac{p_s}{a_s} - \min\limits_{s} \tfrac{p_s}{a_s}  \bigr] =  \max\limits_{s} \tfrac{q_s}{a_s} - \min\limits_{s} \tfrac{q_s}{a_s} ,
\ee 
where in the first expression, the overall normalization by $\Dbar$ can be ignored if the House size $R$ is fixed \textit{a priori}.  Again, this optimality condition is essentially the same as one of the primary criteria eventually emphasized by Willcox, if mis-attributed by him to the Adams method.

However, one can certainly raise several objections to this measure of equitable apportionment.  Notably, it is largely insensitive to the absolute number of individuals whose share of representation might be affected by a shift of seats, and appears completely insensitive to variations in district sizes that do not affect the largest and smallest average district sizes across all states. 

In particular, the latter insensitivity means that this criterion can easily fail to single out a unique apportionment.  Edelman recognized this shortcoming and suggested three different supplementary principles for breaking ties amongst apportionments that minimize his total variation measure: minimizing the standard deviation in district sizes, minimizing the mean absolute deviation in district sizes, or recursively minimizing as needed the second largest deviation, third largest deviation, etc., between under-represented and over-represented districts.  But even Edelman could muster no Constitutional argument as to which of these secondary criterion should be preferred in the event of ties.

The Burt-Harris-Edelman Method appears strongly biased in favor of smaller states.  For example, applied to the 2000 U.S.\ Census data, it would have reproduced the same apportionment as Adams' Method (well known to favor small states) and violated lower quota for the three largest states (California, New York, Texas), while neither the Huntington nor Webster methods would have violated quota.
  
\subsection{Ossipoff-Agnew Method}
(Proposed by Michael Ossipoff in online discussions beginning in 2006 \cite{ossipoff:2013}.  Essentially the same method was derived and analyzed independently by Robert Agnew in 2008 \cite{agnew:2008}.  Following Agnew, this apportionment rule is also referred to as the Method of Identric Means, and is equivalent to what we call Entropic Apportionment).

\subsubsection{ Divisor Rule}
This is another divisor method also intended, according to Ossipoff, to minimize bias with respect to large or small states.  Quotients are to be rounded up at or above the threshold given by:
\be
\theta(Q) = \tfrac{1}{e} \tfrac{\cQ^{\cQ} }{\fQ^{\fQ}},
\ee
with the convention that $0^0 = 1^0 = 1$.  That is, the rounding threshold is the so-called identric mean\footnote{Given any $x > 0$ and $y > 0$, the identric mean of this pair of positive numbers may be defined as $\mathcal{I}(x,y) = \exp \tfrac{\int_{x}^{y} \!dz\, \ln z}{ \int_{x}^{y} \!dz\, } =  \exp\bigl[\tfrac{y\ln y - x \ln x + x - y}{y - x} \bigr] = \tfrac{1}{e} \bigl(\tfrac{y^y}{x^x} \bigr)^{1/(y-x)}$.  This is reminiscent of the more familiar geometric mean, except the latter is defined using sums rather than integrals:  $\G(x,y) = \exp\bigl[  \tfrac{\ln x + \ln y }{1+1}\bigr] = \exp\bigl[  \tfrac{\ln x y }{2} \bigr] = \sqrt{xy}$.  As such, the notion of a geometric mean can be naturally extended to any finite set of nonnegative numbers, but the identric mean cannot, since the numbers to be averaged appear as integration limits of a one-dimensional integral over an interval.} of the floor and ceiling.  As with many other divisor methods, the pure strategy does not ensure that every state receives its minimal number of representative, but the last (or for that matter, the first) representatives assigned as $D$ is decreased could be shifted to states without enough representatives.

\subsubsection{Global Optimization Rule}

With good marketing in mind, Ossipoff called this the ``Bias-Free'' (BF) Method.  He was, however, not entirely clear regarding what was to be intended by this assertion, but it appears that he focused on a measure of bias assuming a uniform distribution of state populations, which seems quite questionable.  But whatever his original motivation, by our analysis he does seem to have stumbled upon what, based on more fundamental grounds, we argue is the optimal method for apportionment, since this method is actually equivalent to the constrained global optimization of the relative entropy, as discussed in the main text.  As a global optimization method, it therefore minimizes
\be
\K(\bv{a}, \bv{q}; R,S) = \sum\limits_{s=1}^{S} \tfrac{a_s}{R} \log \bigl( \tfrac{a_s}{q_s} \bigr),
\ee
subject to lower bound, upper bound, and/or total House size constraints.

\subsubsection{Ranking Method}

This turns out to be equivalent to a ranking procedure relying on the prioritization indices
\be
\rho(a'_s, p_s) = p_s \tfrac{ e \,  {a'_s}^{a'_s} }{(a'_s + 1)^{a'_s+1} }.
\ee
although the common factor of $e$ could be dropped, or a common factor of $1/P$ introduced, or we could use any monotonically increasing function of these indices, such as the logarithm---we have defined these indices to be linear in the populations $p_s$ so as to conform to the same conventions introduced for the historical methods.

\subsubsection{Comparison Test}

We can also show that a global optimum is stable with a certain form of Huntington comparison test.  First, suppose that, at an interior\footnote{By this, we mean that state $o$ is not at its lower bound, and state $u$ is not at is upper bound, so that a shift in one seat from $o$ to $u$ can be entertained.} optimum,
$\tfrac{a_{o}}{q_{o}} > \tfrac{a_u}{q_{u}}$.  If it is also the case that $\tfrac{a_{o}-1}{q_{0}} > \tfrac{a_u + 1}{q_{u}}$, then $\tfrac{a_{o}-x}{q_{o}} > \tfrac{a_{u} + x}{q_{u}}$ for all real-valued $x$ in the range $0 \le x \le 1$.  But then it would follow that
\be
\tfrac{d}{dx} \bigl[ (a_{o}-x) \log\bigl( \tfrac{a_{o}-x}{q_{o}}\bigr) + (a_{u}+x) \log\bigl(\tfrac{a_{u}+x}{q_{u}} \bigr) = 
\log \bigl(  \tfrac{a_{u}+x}{q_{u}}  \tfrac{q_{o}}{a_{o}-x}   \bigr) < 0,
\ee
for all $x \in [0,1]$, so upon integrating, we could conclude that
\be
\bigl[ (a_{o}-1) \log\bigl( \tfrac{a_{o}-1}{q_{o}} \bigr)+ (a_{u}+1) \log\bigl(\tfrac{a_{u}+1}{q_{u}} \bigr)  \bigr] - \bigl[a_{o} \log \bigl(\tfrac{a_{o}}{q_{o}}\bigr) + a_{u} \log\bigl( \tfrac{a_{u}}{q_{u}} \bigr) \bigr] < 0,
\ee
contradicting the assumption that the original apportionment was optimal with respect to the relative entropy.

So starting from an optimum, a shift of one seat from will reverse the sense of relative over-representation in the affected pair of states, but must not decrease the relative entropy, meaning
\be
 a_{o} \log \bigl(\tfrac{a_{o}}{q_{o}}\bigr) + a_{u} \log\bigl( \tfrac{a_{o}}{q_{o}} \bigr)   \le
(a_{o}-1) \log\bigl( \tfrac{a_{o}-1}{q_{o}} \bigr)+ (a_{u}+1) \log\bigl(\tfrac{a_{o}+1}{q_{o}} \bigr)  .
\ee
This is beginning to look like a Huntington comparison test, but we need to ensure that the associated comparison function will vanish if and only if $\tfrac{a_o}{q_o} = \tfrac{a_u}{q_u}$ and that its sign correctly indicates the direction of relative over-representation.  Using the fundamental logarithm inequality, we may verify that
\be\label{TK}
{T}_{\K}(a_0, q_o; a_u; q_u) =  \sgn(  q_u a_o  - q_o a_u ) \, \bigl[  a_o \ln( \tfrac{a_o}{q_o} ) + a_u \ln( \tfrac{a_u}{q_u}) + (q_o - a_o) + (q_u - a_u)   \bigr]
\ee
defined in terms of natural logarithms, will do the job.  The term in square brackets is always nonnegative, vanishing if and only if $\tfrac{a_o}{q_o} = 1 = \tfrac{a_u}{q_u}$, implying both states have received their exact quota.  The \textit{signum} function vanishes for other instances of identical relative representation, where of $\tfrac{a_o}{q_o} = \tfrac{a_u}{q_u}$, and otherwise encodes the direction of relative over-representation.  If desired, we can convert to the conventional form using the substitutions $q_o = \tfrac{R}{P} p_o$ and $q_u = \tfrac{R}{P} p_u$ .

\subsection{Logarithmic Mean Method}
(Discussed by Robert Agnew in 2008 \cite{agnew:2008}.  This method is equivalent to what we call Dual Entropic Apportionment).
 
This is yet another divisor method, discussed by Robert Agnew, based on taking as the rounding threshold the so-called logarithmic mean\footnote{Given any $x > 0$ and $y > 0$, the logarithmic mean of this pair of numbers may be defined as $\L(x,y) = \Bigl[ \tfrac{\int_{x}^{y} \!dz\, z\inv}{ \int_{x}^{y} \! dz\, } \Bigr]^{-1} = \tfrac{y - x}{\ln \tfrac{y}{x}}$.  This is reminiscent of the harmonic mean, which, however, is defined using sums rather than integrals:  $\H(x,y) = \bigl[  \tfrac{ x\inv + y\inv }{1+1}\bigr]^{-1} = \tfrac{2 x y }{x + y}$.  Thus the harmonic mean can be generalized to any number of arguments, but, much like the identric mean, the logarithmic mean cannot, because $x$ and $y$ appear as integration limits of a one-dimensional integral over an interval.} of $\fQ$ and $\cQ$,
\be
\theta(Q) =  \tfrac{\cQ - \fQ}{ \ln (\cQ/\fQ) }.
\ee
Equivalently, as a ranking method, the method of logarithmic means relies on the indices 
\be
\rho(a'_s, p_s) = p_s \log \bigl( \tfrac{a'_s + 1}{a'_s} \bigr),
\ee
and as a global optimization method, it minimizes the ``dual'' Kullback-Leibler divergence, that is, the divergence between $\bv{q}/R$ and $\bv{a}/R$, given by 
\be
\tilde{\K}(\bv{a}, \bv{q}; R,S) = \sum\limits_{s=1}^{S} \tfrac{q_s}{R} \log \bigl(  \tfrac{q_s}{a_s} \bigr).
\ee
subject to any constraints on upper or lower bounds and total House size.

Because the relative entropy is not symmetric, in various contexts in physics, statistics, information theory, or machine learning, there can be not infrequently some confusion as to which relative entropy (primal or dual) should be used in a given situation.  For apportionment, we believe the answer is clearly $\K = \sum\limits_{s} \tfrac{a_s}{R} \log \bigl(  \tfrac{a_s}{q_s} \bigr)$ rather than the dual $\tilde{\K} = \sum\limits_{s} \tfrac{q_s}{R} \log \bigl(  \tfrac{q_s}{a_s} \bigr)$.  First, when the goal is equality of representation across inhabitants (or across votes in party-list systems), minimizing the relative entropy $\K$ is equivalent to maximizing the Shannon entropy $\S$ for the indirect polling distribution, which is a direct measure of the uniformity of the weights of representation.  Second, notice that $\tilde{\K}$ introduces an infinite penalty when $a_j = 0$ but $q_j \neq 0$.  So $\tilde{\K}$ cannot be used for party-list apportionment, or in other situations where zero seats may be awarded, and hence fails as a universal method.  On the other hand, the dual entropy does \textit{not} directly penalize apportionments for which $q_s = 0$ but $a_s \neq 0$, so it would allow (in the absence of additional constraints) apportioning representatives where none are deserved.
 
 The dual entropic method also leads to a Huntington comparison test.  Using an argument analogous to the previous derivation for the case entropic apportionment, we are led to a comparison function of the form
\be\label{TdK}
{T}_{\tilde{\K}}(a_0, q_o; a_u; q_u) = \sgn(  a_o q_u -  a_u q_o )  \, \bigl[  q_o \ln( \tfrac{q_o}{a_o} ) + q_u \ln( \tfrac{q_u}{a_u}) + (a_o - q_o) + (a_u - q_u)  \bigr],
\ee 
 where the logarithms are here taken to be natural (i.e., base-$e$) logarithms.  The term in square brackets is always nonnegative, vanishing if and only if $\tfrac{q_o}{a_o} = 1 = \tfrac{q_u}{a_u}$.  The \textit{signum} function vanishes for other occurrences of $\tfrac{q_o}{a_o} = \tfrac{q_u}{a_u}$, and otherwise indicates the sense of relative over-representation.
 

\subsection{Summary}
  
Table~\ref{summary} summarizes features of the most important apportionment methods. 
  
\begin{table}[t!]
\begin{small}
\begin{tabular}{lll|cccc}
\hline
 Apportionment		  &  			&			& Rounding  		& Ranking 		& Pairwise  			& Global  	\\
Method 		  &  	(also known as)		&			& Threshold  		&  Index 						&  Comparison 			&  Minimum 	\\
\hline\hline
 Hamilton-Vinton		  & Largest Fractions  & (LF)		&  	$\lfloor q_s + \Delta_{\stext{H}} + \tfrac{1}{2}  \rfloor$, \text{ where}
		&  	  & 	  	&				$\sum\limits_s \lvert a_s - q_s\rvert$	 \\
		
		& Greatest Remainders & (GR) & $\Delta_H$ ensures $\sum\limits_s a_s = R$ \\
\hline
 Adams		  & Smallest Divisors  & (SD)		& $\cQ$ 				& $\tfrac{p_s}{a'_s}$ & 	$a_o - \tfrac{p_o}{p_u} a_u $  	&				$\max\limits_s \tfrac{p_s}{a_s}$		 \\
\hline
 Dean 		  &  Harmonic Means & (HM) 	& $\tfrac{2 \fQ \cQ}{\fQ + \cQ}$ & $p_s \tfrac{2a'_s + 1}{2 a'_s (a'_s+1)}$ & $\tfrac{p_u}{a_u} - \tfrac{p_o}{a_o}$   			&		$\max\limits_s \tfrac{p_s}{\sqrt{a_s(a_s+1)}}$					 \\
\hline
Huntington-Hill   & Geometric Means & (GM) \\  
   	  		 & Equal Proportions & (EP) 	& $(\fQ\cQ)^{\frac{1}{2}}$ & 	$\tfrac{p_s}{\sqrt{a'_s (a'_s + 1)}}$	&	$\tfrac{p_u}{p_o} \tfrac{a_o}{a_u} - 1$   &	$\tfrac{P^2}{R^3} \sum\limits_s \tfrac{(a_s - q_s)^2}{a_s}$						 \\
 \hline
 Dual-Entropic 	  & Logarithmic Means & (LM) 	& $\tfrac{\cQ - \fQ}{ \ln ( \cQ/\fQ)}$ &	$p_s \log \bigl( \tfrac{a'_s + 1}{a'_s} \bigr)$ 	&    $T_{\tilde{\K}}$   & $\sum\limits_s \tfrac{q_s}{R} \log \bigl(  \tfrac{q_s}{a_s} \bigr)$	\\
\hline
\textbf{Entropic} & Identric Means & (IM)  		& $\tfrac{1}{e} \bigl(  \tfrac{ {\cQ}^{\cQ} }{ {\fQ}^{\fQ} } \bigr)^{\frac{1}{ \cQ-\fQ}}$  & $p_s \tfrac{ e \,  {a'_s}^{a'_s} }{(a'_s + 1)^{a'_s+1} }$	&  $T_{\K}$    & $\sum\limits_s \tfrac{a_s}{R} \log \bigl(  \tfrac{a_s}{q_s} \bigr)$			\\	
\hline
Webster-Willcox 		  &  Arithmetic Means & (AM) \\
 			 & Major Fractions & (MF)  & $\tfrac{\fQ + \cQ}{2}$  &  $\tfrac{p_s}{a'_s + \frac{1}{2}}$ &  $\tfrac{a_o}{p_o} - \tfrac{a_u}{p_u}$ 	&	$\tfrac{R}{P^2} \sum\limits_s \tfrac{(a_s - q_s)^2}{q_s}	$				\\
\hline
Jefferson 		  & Greatest Divisors & (GD) 	& $\fQ$ & $\tfrac{p_s}{a'_s + 1}$ & $\tfrac{p_u}{p_o} a_o - a_u$								&		$\max\limits_s \tfrac{p_s}{a_s+1}$						\\
\hline\hline
\end{tabular}
\end{small}
\caption{Comparison of apportionment methods. As in the main text, $p_s$ denotes the population of the $s$th state, $Q_s = p_s/D$ denotes the quotient relative to a chosen divisor $D$, $a'_s$ represents the apportionment at a given stage of seat distribution via ranking, $a_o$ and $p_o$ refer the proposed apportionment and population of the relatively over-represented state, while $a_u$ and $p_u$ refer to the relatively under-represented state, and $q_s = p_s R/P$ represents the exact quota, where $R$ is the total house size and $P$ is the total population. The Huntington comparison functions for the primal and dual entropic methods are given by equations \eqref{TK} and \eqref{TdK}, respectively.  The Hamilton-Vinton approach is a quota method; all others are divisor methods.  The SD, HM, EP, and LM methods automatically give at least one seat to each state as long as $R \ge S$; the other methods can allot zero seats to a sufficiently small state unless lower bound constraints are imposed explicitly.  In all cases, global minimization refers to a constrained minimum of the indicated inequity function subject to constraints $0 \le \lambda_ s \le a_s \le u_s$ for $s = 1, \dotsc, S$, and $\sum\limits_{s} a_s = R$.}\label{table:divisor_methods_summary}\label{summary}
\end{table}
 
In addition to the five traditional divisor methods discussed by Huntington, we have included the two others mentioned by Agnew, one of which (the Method of Identric Means) is equivalent to our recommended entropic apportionment method at the level of awarding seats.

Notice that the SD and GD methods, and also the HM and MF methods, and the primal entropic (IM) and dual entropic (LM) methods, are dual in the sense of the roles played by the $a_s$ and $q_s$ (proportional to the $p_s$) in their respective pairwise comparison tests.  The EP method is self-dual in this sense, which was a principal selling point for Huntington.

The MF and EP methods, and also the IM and LM methods, are dual in a different sense, that of the roles played by the $a_s$ and $q_s$ in the global optimization criteria, while the GR method is self-dual in this sense.

However, in our view, neither of these notions of duality is particularly important, because the apportionments $\bv{a}$ and quotas $\bv{q}$ play asymmetric roles in the representational weights, and in the relative entropy itself.

 
 %
 %
 %
 
\newpage
\section{\uppercase{Shannon Entropy and Relative Entropy---Motivations, Interpretations, and Characterizations}}\label{entropy}

Entropy is the central unifying concept in both statistical mechanics \cite{jaynes:1983,tribus:1961,wehrl:1991,buck:1991,jaynes:2003} and information  theory \cite{shannon:1948a,shannon:1948b,shannon:1949,khinchin:1957,fano:1961,pierce:1980,cover_thomas:1991,mackay:2003}.  Many different arguments have all led to the adoption of the \textit{Shannon entropy} as the essentially unique quantification of the \textit{amount of uncertainty} or \textit{degree of uniformity} in one probability distribution, and the relative entropy, or Kullback-Leibler divergence \cite{kullback:1959}, as a measure of discrepancy between or divergence from one distribution and another.  Derivations of or justifications for what is now called the Shannon entropy pre-date Shannon, going back to the combinatoric arguments of Boltzmann and Wallis.  Axiomatic characterizations and uniqueness theorems were pioneered by Shannon and Weaver, continued by Fadeev, Kolmogorov, Khinchin, and many others, and have been streamlined or modernized by a number of authors, leading to a large body of results now summarized in a number of books and review articles on information measures.  See for example \cite{aczel:1975,aczel:1984,aczel:1987,ebanks_sahoo:1998,arndt:2001,karmeshu:2003,csiszar:2008,sharma:2010,wibral:2014}.

By a \textit{characterization} of entropy we mean a set of (hopefully compelling, or at least plausible) desiderata, axioms, or properties that we demand of our measure of uncertainty, and that lead uniquely to the Shannon entropy functional.  Similarly, a characterization of relative entropy is list of properties to be possessed by a measure of divergence,  leading to the Kullback-Leibler number.

\subsection{Entropy as Expected Surprisal}

How much information we would gain, or uncertainty we would resolve, upon learning that some event occurs may be naturally identified with how surprising, unexpected, or improbable the event was considered before observing it.
 
Our degree of surprise may be identified with the amount of information we actually receive when learning that the outcome in fact happened.  If there are $N$ equally likely, mutually exclusive and exhaustive possibilities, then clearly one would gain $\log_2 N$ bits of information upon learning which outcome actually obtains, equivalent to learning the answers to $\log_2 N$ independent yes/no questions.  

More generally, consider some event to which we assign a probability $\p$ of occurrence, where $0 \le \p \le 1$.  We seek an expression $H(\p)$ for the \textit{surprisal}, or quantified amount of surprise that would be reasonable to experience upon learning the outcome does indeed happen, or equivalently, the amount of information acquired when learning that it indeed occurs.  It is natural to demand that this surprisal satisfies the following conditions:
\begin{enumerate}
\item \textit{non-negativity}:
\be
H(\p) \ge 0 \text{ for any } \p \text{ satisfying } 0 < \p \le 1,
\ee
because learning that some event actually happened should not make us any more ignorant about that event than we were before learning that it occurred;\footnote{A slightly stronger condition would assert that events which are not certain \textit{a priori} should surprise us when they happen at least a little, so that $H(\p) > 0$ for any $\p$ such that $0 < \p < 1$, and $H(0) = +\infty$ since we would be infinitely surprised upon observing an event known to be impossible, but we can make do with the weaker criterion.  Alternatively, we could also replace this property with the condition that the surprisal should be non-decreasing for small positive values of $\p$.}


\item \textit{additivity}: if $\p = \mu \theta$ for some  $\mu$, $\theta$ satisfying $0 < \mu \le 1$ and $0 < \theta \le 1$, then
\be
H(\mu \theta) = H(\mu) + H(\theta),
\ee
because surprisal toward compound events involving \textit{independent} sub-events should be \textit{additive}.  That is, if an overall outcome, of probability $\p = \mu \theta$,  is equivalent to the joint occurrence of two independent contributing outcomes, the first with probability $\mu$, and the second with probability $\theta$ independently of the first, then 
we can either learn of the occurrence directly or infer it from the joint occurrence of the two contributing events. The overall information we acquire in learning of the joint event should therefore be equal to the information acquired upon learning the first contributing event occurs plus the information acquired upon learning the second contributing event occurs;
 
\item \textit{standard normalization}: optionally, the surprisal may be normalized so that
\be
H(\tfrac{1}{2}) = 1 \text{ bit},
\ee
because if we assigned exactly equal odds to an event occurring or not, we will gain exactly one bit of information upon learning the outcome.  This normalization is strictly optional, but it should always be possible by a standard change of units involving multiplication by a pre-factor.
\end{enumerate}

Then it turns out that the only class of functions $H(\p)$ satisfying the first {two} requirements are multiples of the logarithm:
\be
H(\p) = \kappa \log \tfrac{1}{\p} = - \kappa \log \p,
\ee
for some positive constant $\kappa$ which basically determines the units in which surprisal will be measured---if we do adopt use of bits, we can set $\kappa = 1$ and take base-$2$ logarithms, such that $H(\p) = -\log_2(\p)$.

This measure of surprisal automatically enjoys several other features which did not need to be assumed, including: (i) continuity, implying small changes in the probability lead to small changes in the surprisal, and in fact (ii) smoothness, for $0 < \p \le 1$, (iii) 
strict monotonicity, such that $H(\p) < H(\mu)$ whenever $0 \le \mu < \p \le 1$, such that events of strictly greater probability correspond to strictly less surprisal, and (iv) sensible boundary behavior, where $H(0) = +\infty$, meaning we should be infinitely surprised upon seeing an event known to be impossible, but $H(1) = 0$, meaning we would experience no surprisal upon observing an event certain to occur.

Now consider $N$ possible outcomes, labeled by $j = 1,\dotsc, N$.  We wish to quantify our amount of uncertainty when our knowledge tells us that the events are mutually exclusive and exhaustive, \ie, that one and only one of them will occur, but not which.  The best we can do is assign probabilities $\p_1, \dotsc, \p_N$, where $\p_j$ represents the rational degree of belief that the $j$th outcome will obtain, conditional on our limited knowledge.  The Shannon entropy measures this uncertainty by taking the \textit{expectation value of the surprisal} over these mutually exclusive and exhaustive possibilities:
\be
\S(\p_1, \dotsc, \p_N) = \kappa \sum\limits_{j=1}^{N} \p_j \log \frac{1}{\p_j} = - \kappa \sum\limits_{j=1}^{N} \p_j \log \p_j.
\ee
Equivalently, when measured in bits, the Shannon entropy may be interpreted as the \textit{average} number of yes/no questions that we would need to have (truthfully) answered in order to learn which outcome obtained.

The weakness of this approach is a lack of clear motivation as to why we rely on the \textit{average} surprisal, and not the median, mode, or some other measure of central tendency or typicality across the possibilities.
 
\subsection{Acz\'el-Ng Characterization}\label{SubAppendix:Aczel}

Many characterizations of information-theoretic entropy rely on principles such as \textit{strong additivity} or \textit{recursivity} (see below) which also build in a preference for the arithmetic average.  But it would be preferable to derive rather than assume such structural properties.  It turns out that the usual average is singled out as the only operation consistent with a deeper principle that we already invoked in the definition of surprisal itself, that of additivity over independent information.

Perhaps the most elegant and compelling characterization theorem for Shannon entropy is that due to Acz\'el and Ng \cite{aczel_ng:1974,aczel:1975}.  Again assume $N$ possible outcomes $j =1, \dotsc, N$, known to be mutually exclusive and exhaustive.  We seek to quantify the uncertainty by some some real-valued function $\H_N(\p_1, \dotsc, \p_n)$ of the probabilities, presumed to satisfy the following properties:
\begin{enumerate}
\item \textit{symmetry}:
\be
\H_N(\dotsc, \p_j, \dotsc, \p_k, \dotsc) = \H_N(\dotsc, \p_k, \dotsc, \p_j, \dotsc)
\ee
for every pairwise transposition of the outcome probabilities, and by extension, for every permutation of the probabilities, or their labels.

The measure of uncertainty should depend only on the probabilities of the exclusive and exhaustive possibilities, not how we happen to label them.  Hence, the measure of uncertainty should be invariant under all permutations of the probabilities, and for a fixed number of possibilities $n$, the measure $\H_N$ is a function of the \textit{multiset} of probabilities $\{ \p_j \}$.

\item \textit{expansibility}:
\be
\H_{N+1}(\p_1, \dotsc, \p_N, 0) = \H_N(\p_1, \dotsc, \p_N).
\ee
Appending to the list of possibilities an outcome known to be impossible should not change the uncertainty.
\item \textit{near-certainty}:
\be
 \lim\limits_{\p \to 0^+} \H_2(1-\p,\p) = 0.
\ee
In the limit of only one viable possibility, the amount of uncertainty should vanish.

\item \textit{additivity}:
\be
\H_{NM}\bigl( \{\p_j w_k \} \bigr) = \H_N\bigl(  \{ \p_j \} \bigr) + \H_M\bigl( \{ w_k \} \bigr).
\ee
If we perform two independent experiments, or ask two independent questions, the total information gained should be the sum of the information gained from each separately.  If perhaps somewhat less self-evident than the other desiderata, this should seem quite natural to all who have grown up in the digital age, and are accustomed to the idea of information accumulating additively in disk drives, Dropbox accounts,  data charges in cellphone plans, etc.

\item \textit{sub-additivity}:
\be
\H_{NM}\bigl( \{\p_{jk} \} \bigr) \le \H_N\bigl( \bigl\{ \sum\limits_{k=1}^{M} \p_{jk} \bigr\} \bigr) + \H_M\bigl( \bigl\{ \sum\limits_{j=1}^{N} \p_{jk} \bigr\} \bigr),
\ee
where $\sum\limits_{k=1}^{M} \p_{jk}$ and $\sum\limits_{j=1}^{N} \p_{jk}$ are the marginal probabilities associated with the joint probabilities $\p_{jk}$ defined over the outcomes $(j,k)$ for $j  = 1, \dotsc, N$ and $k = 1, \dotsc, M$.  This means that we cannot gain more information from asking dependent questions than from asking independent questions.

\item \textit{actuality of uncertainty}:
\be
\H_{N}(\tfrac{1}{N}, \dotsc, \tfrac{1}{N}) \ge \H_N(1,0, \dotsc, 0), \text{ with equality if and only if } N = 1.
\ee
In the state of maximal uncertainty about the outcome, we have more uncertainty than in a state of maximal knowledge about the outcome.  This requirement was not mentioned explicitly by Acz\'el, but appears necessary simply to rule out the trivial case where the measure of uncertainty is identically zero.
\end{enumerate}

The only type of function satisfying these desiderata is the Shannon entropy:
\be
\H_{N}(\p_1, \dotsc, \p_N) = \kappa\, \S(\p_1, \dotsc, \p_N) = - \kappa \sum\limits_{j=1}^{N} \p_j \log \p_j,
\ee
for some choice of a positive constant $\kappa$.  Once again, this overall scaling reflects the remaining freedom to choose the base of the logarithms\footnote{If we set $\kappa = 1$ and use base-$2$ logarithms, the entropy (missing information) is thereby measured in binary digits (``bits'');  if we take $\kappa = 1$ and use natural (base-$e$) logarithms, the entropy is instead said to be measured in natural units, or ``nats;'' and if we set $\kappa = 1$ and use base-$10$ logarithms, the entropy may be measured in (decimal) digits.  Retaining the choice of base-$10$ logarithms but instead setting $\kappa = 10$ leads to information measured in ``decibans,'' a unit coined by Alan Turing \cite{good:1983} in analogy to the more familiar``decibels,'' but named after the town of Banburry (near to Bletchley Park, the center of British code-breaking activity during World War II) from which his group ordered special stationery used for computations.  In thermodynamics, conventionally one uses natural logarithms but sets $\kappa$ equal to Boltzmann's constant $\kappa_{\stext{B}}$, and  in \textit{SI} units measures entropy in joules per kelvin.} and/or the units in which the uncertainty/missing information is measured.  In the main text, unless otherwise noted we will measure entropies in bits, by taking $\kappa = 1$ and $\log x = \log_2 x$.

\subsection{Tikochinsky-Tishby-Levine Characterizations}

Some characterizations of Shannon entropy more directly emphasize its \textit{operational} role in assigning or inducing probability distributions from certain limited information, rather than starting with its interpretation as quantifying missing information.

For instance, Tikochinsky, Tishby, and Levine \cite{tikochinsky:1984a,tikochinsky:1984b}  justify probability assignment via entropy maximization using three different arguments, but in each case demanding certain reasonable behavior from the assignment algorithm when using data consisting of average-value data from presumably \textit{reproducible} experiments.

Specifically, they consider the task of assigning probabilities over mutually exclusive and exhaustive alternatives, given prescribed average values of some variables defined on those alternatives.\footnote{This is not the most general type of inference task, but any general method should work in this sub-class of problems, and this is sufficient to single out the entropy.}  The experiments are assumed to be reproducible, in that they could be independently repeated any finite number $N$ of times.  The assignment procedure is postulated to be \textit{universal}, in that data of a given kind should be handled in the same way, and different experiments are not arbitrarily subjected to different procedures  or rules for data processing or probability assignment.  Provided the data are non-contradictory, the resulting probability assignment should be unique, or else the procedure would be regarded as incomplete.

\subsubsection{Uniform Consistent Inference}

Given that an experiment can be reproduced, if $N$ repetitions of the experiment are performed, there are two ways to induce probabilities over the extended hypothesis space regarding the set of all possible outcomes.  Either we assign probabilities to the single-trial outcomes based on prescribed expectation values, and then extend to the distribution over the $N$-fold repetitions using a multinomial distribution, or we can apply the \textit{same} assignment algorithm directly to the extended hypothesis space over the $N$ independent repetitions, prescribing the sample averages.  

If we demand that the resulting probability assignments should be the same under either strategy, then the single-trial probability assignment must be that which maximizes the Shannon entropy using the prescribed average values as constraints.

\subsubsection{Most Stable Inference}

Given any procedure to map expectation-value data to a probability distributions, small statistical errors in the inputs (the specified average values) should lead to small changes in the probability assignment.  Among all assignments agreeing with the original average-value constraints, the maximum entropy assignment is also the \textit{least sensitive} to small errors in this input data, in the specific sense of saturating the Cramer-Rao inequality.  This make intuitive sense, since the maximum entropy distribution  is always the most  ``uniform'' distribution consistent with the constraints, so small changes in the values of average values should lead to less appreciable changes in the probabilities, compared to any other distribution agreeing with the constraints.

\subsubsection{Sufficient Inference}

In statistical inference, \textit{sufficient statistics} are functions of the data which convey all of the information in the data relevant for estimating some parameter or parameters of interest---that is, if the posterior probability distribution for the parameter(s) depends on the data only through the value of the (jointly) sufficient statistics, for any choice of the prior distribution.  

There is no guarantee that any such ``compression'' of the information in the data relevant to parameter estimation will be possible.  But if we are to assign probabilities based only on average values of certain variables, then it is natural to demand that the sample averages for these variables should end up as sufficient statistics for the parameters consisting of the corresponding expectation values over the assigned distribution.  The Pitman-Koopman-Darmois theorem tells us that the only probability distributions which admit sufficient statistics belong to what is known as the exponential family, and it follows that the assigned distributions must be maximum entropy distributions if the sample averages are to become sufficient statistics for the expectation values. 

\subsubsection{Some Points of Connection}

Given the goal of inducing probabilities from averages, these authors present three complementary arguments all leading to the same, unique procedure, namely entropy maximization.  Each of the three approaches invokes the notion of reproducibility of experiments, but this requirement is translated in different ways into relations between measured sample averages and calculated expectation values under the probability distribution assigned.  In the consistency approach, it is demanded that averaging (if only in the sense of a \textit{gedanken} experiment) of the sample average over possible samples should yield the expectation value over the assigned distribution.  In the second, stability-based approach,  the expectation value of the distribution is equated to one particular measured value of the sample average, but it is recognized that in so doing there may be a statistical error (due to the finite size of the sample), whose effects should be minimized.   The third, sufficiency, approach stems from the assumption that the sample average of some variable(s) may be all the information that can be extracted from the observations regarding the corresponding expectation value(s).

\subsection{Gull-Skilling Characterization}

Another characterization that focuses on the role of Shannon entropy as a variational \textit{maximand}, to be used in assigning or approximating probability distributions, was suggested by Gull and Skilling \cite{gull:1984}. Originally developed in the context of astronomical image processing, their axiomatic approach does not even presume that the distribution of interest is necessarily interpreted in terms of probability.  

Suppose that a non-negative, additive, discrete distribution (what they call a ``reconstruction'' or ``representation'') $\bv{x} = (x_1, \dotsc, x_N)$  is to be sought by maximizing some function over``trial'' reconstructions $\bv{x}$, subject to certain constraints (including normalization if relevant) based on measurements or prior information.  Their derivation is based on the following assumptions:
\begin{enumerate}[i.]
\item \textit{variational universality}:  the same variational maximand is to be used on all problems of a given size $N$;
\item \textit{output-extensibility}:  the maximum for smaller $N$ coincides with the maximum for larger $N' > N$ when the extra cells are constrained to be zero;
\item \textit{scale invariance}:  the choice of units should not affect the shape of the reconstruction.  This implies that the maximand can be taken to be a function of the relative \textit{proportions} $\rho_j = \frac{x_j}{\sum\limits_i x_i}$, for $j = 1, \dotsc, N$.
\item \textit{differentiability}:  since we will be seeking constrained maxima, for technical reasons the maximand is assumed to be twice continuously differentiable in its arguments (for all positive values of all of the proportions);
\item \textit{subset independence}:  knowledge of the relative proportions within some cells should not affect the relative proportions in the remaining cells (apart from overall normalization), except insofar as there is prior knowledge of correlations.  That is, if, say, $\rho_j + \rho_k$ is fixed, the optimal value of the remaining proportions $\rho_{\ell}$ for $\ell \neq j$, $\ell \neq k$  should not depend on the ratio $\frac{\rho_j}{\rho_k}$, unless explicitly constrained to do so.

Together, these assumptions require that the variational maximand be of \textit{additive} form
\be
R_N(\bv{\rho}) = \sum\limits_{j=1}^{N} \phi_j(\rho_j)
\ee
for some suitably well-behaved functions $\phi_j(\rho_j)$;
\item \textit{permutation-invariance}   In the absence of prior information favoring some cells over others, cells should be treated on an equal footing, implying that the function should be invariant with respect to permutations of the cell labels.  This means the maximand simplifies to
\be
R_N(\bv{\rho}) = \sum\limits_{j=1}^{N} \phi(\rho_j).
\ee
for some one scalar function $\phi(\rho)$;
\item \textit{marginal independence}:  if the cells ``factorize'' in some natural way (e.g., as a two-dimensional image), the information constraining the structure within one sub-class should not impose any structure in any other sub-class, unless one has additional prior knowledge regarding such correlations.  That is, the reconstructed proportions themselves should factorize into the product of marginal distributions in the absence of explicit constraints introducing correlations.

This turns out to require that
\be
\phi(\rho_j) = A \rho_j \log \rho_j + B \rho_j + C,
\ee
so that
\be
R_N(\bv{\rho}) = A \sum\limits_{j=1}^{N} \rho_j \log \rho_j + B \sum\limits_{j=1}^{N} \rho_j + C N = A \sum\limits_{j=1}^{N} \rho_j \log \rho_j + (B + C N),
\ee
for some constants $A$, $B$, and $C$.  $A$ must be positive if $R_N(\bv{\rho})$ is to have a maximum rather than a minimum, but the values of $B$ and $C$ do not effect the location of the maximum, and can be set to zero.
\end{enumerate}
Garrett \cite{garrett:1999} offers a similar derivation, only instead of marginal independence, demands that the variational principle always outputs nonnegative proportions, even when not explicitly constrained to be so.

\subsection{Baez-Fritz-Leinster Characterization}

Shannon entropy over a discrete outcome space can also be characterized by considering not the amount of information itself, but the \textit{change} in informativeness associated with some  measure-preserving ``data-processing'' transformation.

Specifically, suppose $\X = \{x_1, x_2, \dotsc, x_N \}$ is the set of original, ``latent'' mutually exclusive and exhaustive states,
with probabilities $\bv{\p} = (\p_1, \p_2, \dotsc, \p_N )$, and $\Y = \{y_1, y_2, \dotsc, y_M \}$ is a space of outcomes to be actually measured, also assumed mutually exclusive and exhaustive, and where it is understood  here that we only include outcomes with non-zero chance of observation.  In a \textit{measure-preserving} transformation $f\colon \X \to \Y$, the number $M$ of actually realizable outcomes in $\Y$ must be less than or equal to the number $N$ of states in $\X$, while the corresponding probabilities $\bv{\m} = (\m_1, \m_2, \dotsc, \m_M)$ for the $\Y$ outcomes are simply inherited from the latent $\X$ states, by the law of total probability:
\be
\m_k \;= \!\!\!\! \sum\limits_{j \colon \! f(x_j) = y_k}\!\!\!\!\!\!\! \p_j 
\ee
for each $k  = 1, \dotsc, M$.

Intuitively, such a transformation might represent an ideal type of measurement that can involve deterministic course-graining, but cannot add further ``randomness'' or uncertainty, and hence cannot increase the \textit{informativeness} of the subsequent measurement as to the latent state, as compared to measuring this state directly.

Suppose $\Delta h$ is intended to quantify this loss in potential informativeness of the measurement regarding the unknown latent state, which we refer to as ``information loss'' for short\footnote{Note however, that the uncertainty as to the measurement outcome cannot increase under such a transformation.}.  Baez, Fritz, and Leinster \cite{baez:2011} suggest that a natural set of constraints on $\Delta h$ is:
\begin{enumerate}[i.]
\item \textit{additivity under composition}: if the measure-preserving transformation is effected in two (or more) stages, then the total information lost in the whole transformation is the sum of the information loss in each stage; 
\item \textit{continuity}: if the original probabilities over $\X$ are changed slightly, then the information-loss changes only slightly;
\item \textit{convex linearity}: if a (possibly biased) coin is used to decide whether to effect one measure-preserving transformation or another, then without knowing the result of the coin flip, the information loss will be the expected loss, averaged with respect to the outcomes of the coin flip.
\end{enumerate}
They show that the only measure satisfying these conditions is
\be
\Delta h = c\bigl[\S(\bv{\p}) - \S(\bv{\m})\bigr],
\ee
where $\S(\bv{\p})$ is the Shannon entropy over the latent $\X$ states, $\S(\bv{\m})$ is the Shannon entropy over the measured $\Y$ outcomes, and $c > 0$ is a positive constant (which just serves to change units).

\subsection{Further Properties and Characterizations of Shannon Entropy}

The {Shannon entropy}  $\H_N(\p_1, \dotsc, \p_N) = - \kappa\S = -\kappa \sum\limits_{j = 1}^{N} \p_j \log \p_j$ also enjoys several additional mathematical properties, including: 
\begin{enumerate}
\item \textit{continuity}: $\H_N(\p_1, \dotsc, \p_N)$ is a continuous function of all $\p_j$  for $0 \le \p_j \le 1$.  
\item \textit{smoothness}:  In fact, $\H_N(\p_1, \dotsc, \p_N)$ is a smooth function at all points away from boundary points where one or more of the $\p_j$ vanish;
\item \textit{measurability}: $\H_N(\p_1, \dotsc, \p_N)$ is a Lebesgue measurable function of all $\p_j$  for $0 \le \p_j \le 1$. 
This property is weaker than, but a consequence of, the continuity property;
\item \textit{lower boundedness}:\footnote{In working with entropies, it is convenient and natural to adopt the convention that $0 \log 0 = \lim\limits_{x \to 0^+} x \log x = 0$.} $\H_N(\p_1, \dotsc, \p_N) \ge \H_N(1, 0, \dotsc, 0) = 0$, with equality if and only if $\p_j = \delta_{jj'}$ for some fixed $j' \in \{1, \dotsc, N\}$;
\item \textit{upper boundedness}:  $\H_N(\p_1, \dotsc, \p_N) \le \H_N(\tfrac{1}{N}, \dotsc, \tfrac{1}{N}) = \kappa\, \log N$, with equality if and only if  $\p_1 = \dotsc = \p_N = \tfrac{1}{N}$;
\item \textit{monotonicity}: $\H_N( \tfrac{1}{N}, \dotsc, \tfrac{1}{N}) > \H_M( \tfrac{1}{M}, \dotsc, \tfrac{1}{M})$ for $N > M$; 
\item \textit{strict additivity}: any joint entropy over the distribution $w_{jk}$ for $j = 1, \dotsc, M$, and $k = 1, \dotsc, N$, satisfies:
\be
\H_{MN}\bigl( \{ w_{jk} \} \bigr) \le \H_M\bigl( \bigl\{ \sum\limits_k w_{jk} \bigr\} \bigr) + \H_N\bigl( \bigl\{ \sum\limits_j w_{jk} \bigr\} \bigr),
\ee
with equality if and only if $w_{jk} = \bigl( \sum\limits_{k' = 1}^{N} w_{jk'} \bigr) \bigl( \sum\limits_{j' = 1}^{M} w_{j'k} \bigr)$.  That is, the entropy of a joint distribution is additive only for independent  events, and strictly sub-additive otherwise;
\item \textit{recursivity}: the entropy over $\mathcal{M}$ possibilities can decomposed as:
\be
\H_{\mathcal{M}}\bigl( \{ \p_{j_k} \} \bigr) =  \H_{N}\bigl(   \{ \p_{j} \} \bigr)  + \sum\limits_{j = 1}^{N} \p_j \H_{{M}_j}\bigl(  \bigl\{  \tfrac{\p_{j_k}}{\p_j}  \colon k = 1 , \dotsc, M_j \bigr\}  \bigr),
\ee
where $\mathcal{M} = \sum\limits_{j}^{N} M_j$, $M_j$ are the number of outcomes aggregated into the $j$th group (where the groups are assumed non-empty and non-overlapping), and $\p_{j} = \sum\limits_{k = 1}^{M_j} \p_{j_k}$ is the overall probability for the $j$th group of possibilities.  If the fine-grained events are aggregated into course-grained events,  then the total (fine-grained) entropy can be decomposed into the entropy of course-grained possibilities plus the average conditional entropy of the fine-grained sub-possibilities constituting each course-grained possibility;
\item \textit{strong additivity}: if $\p_{jk}= w_{j} \, \m_{k|j}$ is a decomposition of  joint probabilities using the probabilistic product rule, in which $w_{j} = \sum\limits_k \p_{jk}$ for $j = 1, \dotsc, N$ are marginal probabilities  and hence $\m_{k|j} = \tfrac{\p_{jk}}{w_{j}}$, $k = 1, \dotsc, M$ are the associated conditional probabilities for each $j$, then
\be
\H_{NM}\bigl( \{ \p_{jk} \} \bigr)  = \H_{N}\bigl( \{ w_{j} \} \bigr) + \sum\limits_{j=1}^{N} w_j \H_{M}\bigl( \{ \m_{k|j} \colon k = 1, \dotsc, M \} \bigr),
\ee
which is to the say, the entropy of a joint distribution can be decomposed into the sum of the entropy of one marginal distribution and the average (with respect to the marginal distribution) over the entropy of the associated conditional distribution;\footnote{Note that recursivity and strong additivity are closely related.}
\item \textit{concavity under mixing}: if $(\mu_1, \dotsc, \mu_M)$ is a normalized mixing distribution (satisfying $\mu_k \ge 0$ and $\sum\limits_{k=1}^{M} \mu_k = 1$), and $\{ \p_{j_k} \}$ are a finite sequence (for $k = 1, \dotsc, M$) of probability distributions over the outcomes $j = 1, \dots, N$, then:
\be
\H_{N}\bigl( \bigl\{  \sum\limits_{k=1}^{M} \mu_k \p_{j_k} \colon j = 1, \dotsc N \bigr\} \bigr) \ge \sum\limits_{k = 1}^{M} \mu_k \, \H_{N}\bigl( \{  \p_{j_k} \colon j = 1, \dotsc N \} \bigr),
\ee
 which just says that the entropy is a concave function of its arguments, so that the entropy of a mixture df distributions cannot be less that the weighted average of the entropies of each contributing distribution.  This is a corollary of Jensen's inequality;
\item \textit{concavity under smoothing}: if $\Theta_{jk}$ $j,\,k \in\{1, \dotsc, N\}$ is a \textit{doubly stochastic} matrix (satisfying $\Theta_{jk}\ge 0$, $\sum\limits_j \Theta_{jk} = 1$, \textit{and} $\sum\limits_k \Theta_{jk} = 1$), then   for any probability distribution $\p_1, \dotsc, \p_N$, 
\be
\H_{N}\bigl( \{  \sum\limits_k  \Theta_{jk}  \p_k  \colon j = 1, \dotsc N \} \bigr) \ge \H_N  \bigl( \{ \p_j \colon j = 1, \dots, N \}\bigr),
\ee
which captures the notion that entropy  is non-decreasing under any \textit{smoothing} or \textit{convolution} of the probability distribution.  This also represents a certain sort of  convexity property, but complementary to the mixing convexity---mixing involves a weighted average of several distributions, while smoothing involves a sort of weighted moving average of one distribution.
\end{enumerate}

Various subsets of these or related properties can also be used to uniquely characterize the Shannon entropy.  For example, a well-known theorem of Fadeev shows that the only function that jointly satisfies \textit{symmetry}, \textit{continuity}, and a binary case of \textit{recursivity}, namely
\be
\H_{N}(\p_1, \p_2, \dots, \p_N) = \H_{N-1}(\p_1+\p_2, \p_3, \dotsc, \p_N) +  (\p_1 + \p_2) \, \H_{2}\bigl( \tfrac{\p_1}{\p_1+ \p_2}, \tfrac{\p_2}{\p_1+\p_2}\bigr),
\ee
must in fact be a multiple of the Shannon entropy.  Khinchin proved that \textit{expansibility}, \textit{upper boundedness}, \textit{continuity}, and \textit{strong additivity} uniquely characterize the Shannon entropy.  Shannon himself used \textit{monotonicity}, \textit{continuity}, and \textit{strong additivity} to characterize entropy.

\subsection{Relative Entropy and its Characterizations}

We turn to consideration of  the \textit{relative entropy}\footnote{\textit{Nota bene}: some authors define the relative entropy with an extra minus sign, i.e., as the additive inverse of what we have called the relative entropy, in order to make it look more like the Shannon entropy.  However, our sign convention is more prevalent, and ensures that the relative entropy, like the Shannon entropy, is nonnegative when defined over a countable space of mutually exclusive and exhaustive possibilities, and that $\K$ measures an average information \textit{gain}.} $\K = \K(\bv{\p}; \bv{\m}) =  \sum\limits_j \p_j \log \tfrac{\p_j}{\m_j}$, which has been rediscovered many times, and therefore is known by many names:  it was called the \textit{directed divergence} when introduced into statistics by Kullback and Leibler in 1951, the \textit{discrimination information} by Kullback in his influential book on statistics and information theory, the \textit{Kullback-Leibler number} or \textit{Kullback-Leibler divergence} by many who learned about it from that book but somehow ignored his advice on nomenclature, the \textit{information gain} by R\'enyi, the \textit{error} by Kerridge, the \textit{decibannage} by Turing during early use in World War II for cryptanalysis, and then the \textit{expected weight of evidence} by his assistant, the statistician I.J.~Good, who also mentioned \cite{good:1983} that another accurate if awkward terminology might be \textit{binegentropy}.\footnote{Some other authors like Shore and Johnson have also referred to the relative entropy as the \textit{cross entropy}, but that terminology is now usually reserved instead for a different quantity,  $\EuScript{C} = -\sum_j \p_j \log \m_j$, what Kerridge called the \textit{inaccuracy}.}

Relative entropy has been used both as as a scalar measure of discrepancy, difference, departure, or \textit{divergence} of one probability distribution from another, and as a measure of \textit{information gain} associated with updating from one probability distribution to another.  From an information-theoretic perspective, these are really the same thing, because if we are to quantify by a single real number how different are two probability distributions (over the same space of mutually exclusive and exhaustive possibilities), it is natural to use the amount of information gained about the outcome in updating from one distribution to another.  So we will use the terms divergence or information gain more or less interchangeably.

\subsubsection{Relative Entropy as Expected Information Gain}

If Shannon entropy measures average surprisal, then relative entropy measures the average change in surprisal.  Specifically, suppose we acquire some new information leading us to update our state of knowledge from a prior probability distribution, $\bv{\m} = (\m_1, \dots, \m_N)$, to a posterior probability distribution, $\bv{\p} = (\p_1, \dotsc, \p_N)$, regarding some set of mutually exclusive and exhaustive possibilities.  How much information \textit{about} the potential outcome was acquired?

For the $j$th possibility, the change in surprisal in updating from the prior probability $\m_j$ to  the posterior probability $\p_j$ is 
\be
\Delta H_j =  \bigl[ \log \tfrac{1}{\p_j} - \log \tfrac{1}{\m_j} \bigr] =  \bigl[ \log \m_j - \log \p_j \bigr] =  \log \tfrac{\m_j}{\p_j},
\ee
which, depending on the actual nature of information obtained, can of course be positive, zero, or negative.  But because of the Gibbs inequality, the expectation value (with respect to the posterior distribution) of the change in surprisal over all possibilities cannot be positive---that is, on average, the new information must reduce our surprise:
\be
\bigl \langle \Delta H \bigr\rangle_{\rho} = \sum\limits_j \p_j \log \tfrac{\m_j}{\p_j} \le 0,
\ee
with equality if and only if we do not learn anything relevant, such that $\p_1 = \m_1, \dotsc, \p_N = \m_N$.

We then identify the expected amount of information gain with the expected loss of surprisal upon acquiring the information:
\be
\K(\bv{\p}; \bv{\m}) = \K(\p_1, \dotsc, \p_N; \m_1, \dots \m_N) = - \bigl\langle \Delta H\big\rangle = \sum\limits_j \p_j \log \tfrac{\p_j}{\m_j} \ge 0,
\ee
with equality if and only if the distributions are equal, i.e., $\p_1 = \m_1, \dotsc, \p_N = \m_N$.

Here, it is crucially important to distinguish the change in expected surprisal $\Delta \S$---that is, the difference in Shannon entropies, which can be of either sign\footnote{At first, this may seem counter-intuitive. After all, any change in our probabilities, and corresponding entropies, presumably is to be based on new knowledge, and requires some non-zero number of bits of relevant conditioning information to be learned.  But remember that Shannon entropy measures uncertainty or missing information \textit{as to the outcome of the event}, not the total amount of information received or processed in reaching our probabilistic judgement as to the possible outcomes.  Obtaining extra information about something can definitely make one less certain about something related.  For example, here in Berkeley, California in July, our prior degree of belief that it will rain today, based on general climate patterns and past meteorological experience, is well below $50\%$, and our entropy regarding the rain/non-rain dichotomy is therefore well below one bit. But if we receive (many bits of) information detailing dropping barometric pressures, satellite photos indicating incoming thunderclouds, etc., the conditional probability of rain may creep up towards, yet remain below, 50\%, thereby increasing our entropy regarding the possibility or not of rain.  Of course, at the same time this information obviously lowers our uncertainty as to the actual atmospheric pressure, and the existence of incoming clouds.  However, the \textit{expected} change (under various possible evidentiary findings) in expected surprisal is negative, and can be expressed in terms of another quantity from information theory, the \textit{mutual information}, which is a special case of relative entropy.}---from the expected loss in surprisal, $\K$---which is always nonnegative.

\subsubsection{Relative Entropy as Discrimination Information or Expected Weight of Evidence}

The idea of relative entropy as information gain is reinforced by its interpretation in statistical testing as an expected \textit{weight of evidence} in favor of one hypothesis over another , or the amount of information provided by the evidence relevant to discriminating one hypothesis from another.  If hypothesis $A$ assigns probabilities $\bv{\p}_A$ to certain possible observable outcomes, and hypothesis $B$ assigns probabilities $\bv{\p}_B$ to the same set of outcomes, then if the $j$th outcome were to be observed, our posterior odds in favor of hypothesis $A$ relative to $B$ would be updated by multiplying by the likelihood ratio $\tfrac{\p_{A_j}}{\p_{B_j}}$.  The logarithm of this factor is the relative \textit{weight of evidence} $\log \tfrac{\p_{A_j}}{\p_{B_j}}$ in favor of hypothesis $A$ relative to hypothesis $B$.  It is particularly convenient to work with logarithms because (i) the weight of evidence becomes additive for independent observations, and (ii) human perception of uncertainty (very much like our perception of the loudness of sounds, the brightness of light, pressure on our skin, or other psychophysical responses) seems to work on something resembling a logarithmic scale, giving our brain more ``dynamic range'' when faced with probabilities that are either very small or very close to unity.

\textit{Before} seeing the data, we can ask how informative or discriminating an experiment or observation is \textit{expected} to be, in the sense of how much weight of evidence we expect it to provide in favor of the correct hypothesis. The \textit{expected weight of evidence} in favor of hypothesis $A$ against hypothesis $B$, if hypothesis $A$ is in fact true, is then the average (with respect to the probabilities assigned by hypothesis $A$) of the weights of evidence, which are the logarithms of the Bayes factors:
\be
\K(\bv{\p}_A; \bv{\p}_B) = \sum\limits_j \p_{A_j} \log \tfrac{\p_{A_j}}{\p_{B_j}},
\ee
which is just the relative entropy once again.

\subsubsection{Acz\'el-Ng Characterization of Relative Entropy}

One of the more compelling standard characterizations for relative entropy is provided by theorems of Acz\'el and Ng \cite{ng:1974,aczel:1975}.

Suppose we demand that the measure of information gain, or divergence, $\D_N(\p_1, \dotsc, \p_N; \m_1, \dots \m_N)$ between two probability distributions over the same set of mutually exclusive and exhaustive possibilities satisfy the following properties:
\begin{enumerate}
\item \textit{labeling symmetry}:
\be
\hspace{-10pt} \D_{N}( \dotsc, \p_j, \dotsc, \p_k, \dotsc ;  \dotsc, \m_j, \dotsc, \m_k, \dotsc ) =    \D_{N}( \dotsc, \p_k, \dotsc, \p_j, \dotsc ;  \dotsc, \m_k, \dotsc, \m_j, \dotsc ) 
\ee
for all pairwise transpositions, and by extension all permutations, as long as the same permutation is applied to both the prior and posterior probabilities.  This says that the measure should not depend on arbitrary choices we made in labeling the possibilities.

\item \textit{extensibility}:
\be
\D_{N+1}( \p_1, \dotsc, \p_N, 0 ; \m_1, \dotsc, \m_N,0) =  \D_{N}( \p_1, \dotsc, \p_N ; \m_1, \dotsc, \m_N).
\ee
Including outcomes already known to be impossible \textit{a priori} should not change the amount of information gain.

\item \textit{nilpotence}:
\be
\D_{N}( \p_1, \dotsc, \p_N ; \p_1, \dotsc, \p_N) = 0.
\ee
Unless our probabilities over the outcomes change, we have gained no information about the outcome.
\item \textit{ordering}:
\be
\D_{N}(1,0,\dotsc, 0;  \tfrac{1}{N}, \dotsc, \tfrac{1}{N}) > 0 \text{ for } N \ge 2.
\ee
If we go from a state of maximum uncertainty to maximum knowledge regarding the outcome, we must have gained information.

\item \textit{continuity}:
\be
\D_N(\p_1, \dotsc, \p_N; \m_1, \dotsc, \m_N)  \text{ is a continuous function }
\ee
of each $\m_j$ for $0 < \m_j  \le 1$ and for every $\p_k$ in $0 \le \p_k \le 1$. 
Small changes in any of the probabilities should lead to small changes in the measure of information gain.\footnote{Actually, only Lebesgue measurability is needed in the proofs, but continuity seems a more natural, if mathematically stronger, requirement.}

\item \textit{additivity}:
\be
\begin{split}
\D_{NM}(\p_1 w_1, &\dotsc, \p_1 w_M, \dotsc, \p_N w_M ; \m_1 \mu_1, \dotsc, \m_1 \mu_M, \dotsc, \m_N \mu_M) = \\
&\phantom{+}\D_N(\p_1, \dotsc, \p_N; \m_1, \dotsc, \m_N) + \D_M(w_1, \dotsc, w_M; \mu_1, \dotsc, \mu_M).
\end{split}
\ee
Information gain should be additive when performing fully independent experiments or asking independent questions.

\item \textit{branching}:
\be
\begin{split}
\D_{N}( \p_1, \dotsc, \p_N ; \m_1, \dotsc, \m_N) &= \D_{N-1}(\p_1+ \p_2, \p_3, \dotsc, \p_N; \m_1 + \m_2, \m_3, \dotsc, \m_N) \\
&+  \J_N(\p_1, \p_2; \m_1, \m_2) 
\end{split}
\ee
for some sequence of functions $\J_N(\p_1, \p_2; \m_1, \m_2)$.  When we aggregate possibilities, total information gain should be expressible in terms of that between the aggregated categories plus that within the aggregated categories.

\end{enumerate}

Then the only family of functionals satisfying all these properties is:
\be
\D_{N}( \p_1, \dotsc, \p_N ; \m_1, \dotsc, \m_N) =  \K( \p_1, \dotsc, \p_N ; \m_1, \dotsc, \m_N)= \kappa \sum\limits_{j = 1}^{N} \p_j \log \tfrac{\p_j}{\m_j},
\ee
for some choice of a positive scaling constant $\kappa > 0$ reflecting the choice of units and/or base of logarithm.

\subsubsection{Some Further Properties and Characterizations of Relative Entropy}

In addition to those features invoked previously, the \textit{relative entropy} enjoys the following mathematical properties\footnote{Here and elsewhere, it is convenient to adopt the following natural conventions: $x \log \tfrac{x}{0} = +\infty$ for any $x > 0$; $0 \log \tfrac{0}{x} = 0$ for any $x > 0$; and $0 \log \tfrac{0}{0} = 0$.}  (all of which can be derived from the definition, and not all of which are independent):
\begin{enumerate}
\item \textit{smoothness}: $\K(\p_1, \dotsc, \p_N; \m_1, \dotsc, \m_N)$  is a smooth function of all of its arguments, apart from boundary points where one or more probabilities vanish;
\item \textit{measurability}: $\K(\p_1, \dotsc, \p_N; \m_1, \dotsc, \m_N)$ is (Lebesgue) measurable function of all arguments. This is a weaker condition than continuity, but implied by the latter;
\item \textit{positive definiteness}: For any probability distributions over the same space of possibilities, 
\be
\K(\p_1, \dotsc, \p_N; \m_1, \dotsc, \m_N) \ge 0,
\ee
with equality if and only if the two distributions are identical, i.e.,  $\p_1 = \m_1, \dotsc, \p_N = \m_N$.  This follows from the so-called Gibbs inequality,\footnote{The Gibbs inequality (also known as the Shannon inequality or the Gibbs-Shannon inequality) says that for any pair of probability distributions $\p_1, \dotsc,\p_N$ and $\m_1, \dotsc, \m_N$ over $N$ possibilities, $\sum\limits_{j = 1}^{N} \p_j \log \p_j \ge \sum\limits_{j = 1}^{N} \p_j \log \m_j $, with equality if and only if the distributions are the same, i.e., $\p_1 = \m_1 \dotsc \p_N = \m_N$.  The Gibbs inequality follows from the Jensen inequality, or more simply from the fundamental logarithm inequality: $\ln x  \le x-1$ for all $x \ge 0$, with equality if and only if $x = 1$.  This in turn can proven by considering integrals of $1/x$.} and provides the underpinnings of much of the structure of statistical mechanics;
\item \textit{sum property}:
\be
\K(\p_1, \dotsc, \p_N; \m_1, \dotsc, \m_N) = \sum\limits_{j} \mathcal{G}(\p_j, \m_j),
\ee
for some function $\mathcal{G}(p,m)$ of two nonnegative variables.  This says that the total discrimination information is the sum of separate contributions from each outcome.
\item \textit{restriction-monotonicity}:
$\K_{M}(1/N, \dotsc, 1/N, 0, \dotsc, 0;  \tfrac{1}{M}, \dotsc, \tfrac{1}{M})$ is an increasing function of $M$ and a decreasing function of $N$, for any integers satisfying $1 \le N \le M$.  This says that information is gained when the number of equally likely possibilities is reduced from $M$ to $N$, and the bigger the reduction the more the gain;
\item \textit{convexity}: $\K(\p_1, \dotsc, \p_N; \m_1, \dotsc, \m_N)$  is concave-up in all arguments or combinations of arguments.
\item \textit{joint-convexity}:  an important special case of the above is:
\be
0 \le \K\bigl(   \sum\limits_i  \mu_i \,\bv{\p}_i ;  \sum\limits_i  \mu_i\,  \bv{\m}_i  \bigr) \le \sum\limits_i \mu_i \, \K\bigl( \bv{\p}_i ; \bv{\m}_i\bigr),
\ee
for any nonnegative normalized mixing weights $\mu_i$ satisfying $\mu_i \ge 0$, $\sum\limits_i \mu_i = 1$, and any finite sequence of probability distributions
$\bv{\p}_i = (\p_{1_i}, \dotsc, \p_{N_i})$ and $\bv{\m}_i = (\m_{1_i}, \dotsc, \m_{N_i})$ over the same space of $N$ mutually exclusive and exhaustive possibilities.
\item \textit{smoothing-monotonicity}: If $\theta_{j|k}$ is a stochastic matrix\footnote{Notice that $\theta_{j|k}$ must be stochastic but is \textit{not} required to be \textit{doubly} stochastic here.} satisfying $\theta_{j|k} \ge 0$ and $\sum\limits_j \theta_{j|k} = 1$, and $\tilde{\p}_j = \sum\limits_k \theta_{j|k} \p_k$ and $\tilde{m}_j = \sum\limits_j \theta_{j|k} \m_k$, are ``smoothed'' distributions, then
\be
0 \le \K_N(\tilde{\p}_1, \dotsc, \tilde{\p}_N; \tilde{m}_1, \dotsc, \tilde{m}_N) \le \K_N(\p_1, \dotsc, \p_N; \m_1, \dotsc, \m_N),
\ee
so that divergence between distributions decreases under (parallel) smoothing of both distributions;
\item \textit{smoothing chain rule}:  with the same notation as above,
\be
\begin{split}
\K_N(\p_1, \dotsc, \p_N; \m_1, \dotsc, \m_N) &= \K_N(\tilde{\p}_1, \dotsc, \tilde{\p}_N; \tilde{m}_1, \dotsc, \tilde{m}_N) \\
&+\sum\limits_j  \tilde{p}_j \, \K_N\bigl( \tfrac{ \theta_{j1} \p_1  }{\tilde{\p}_j} , \dotsc,   \tfrac{\theta_{jN} \p_N }{\tilde{\p}_j} ; \tfrac{ \theta_{j1} \m_1  }{\tilde{m}_j}, \dotsc, \tfrac{\theta_{jN} \m_N  }{\tilde{m}_j}  \bigr),
\end{split}
\ee
which is somewhat reminiscent of the strong additivity property of Shannon entropy.
\item \textit{conditional chain rule}:  denoting a joint probability distribution by
\be
\bv{\p} \otimes \bv{\theta} =
(\p_1 \theta_{1|1}, \dotsc, \p_1 \theta_{M|1}, \dotsc, \p_N \theta_{1|N}, \dotsc \p_N \theta_{M|N} ),
\ee  
in terms of a marginal distribution $\bv{\p}$ over $N$ ``primary'' outcomes and a conditional distribution $\bv{\theta}$ over $M$ ``secondary'' possibilities given any one of the primary outcomes, then 
\be
\K_{NM}(\bv{\p} \otimes \bv{\theta} ; \bv{\m} \otimes \bv{w}) = 
\K_{N}(\bv{\p} ; \bv{\m}) + \sum\limits_j \p_j \, \K_{M}( {\theta}_{1|j}, \dotsc, \theta_{M|j} ; {w}_{1|j}, \dotsc, w_{M|j}).
\ee
This expresses overall information gain (regarding both the primary and secondary  outcomes) in terms of the information gain over the primary outcomes plus the average of the conditional information gain over the secondary possibilities given each primary outcome;
\item \textit{composition}: if the probabilities are grouped so that $w_1 = \p_1 + \dotsb + \p_r$, $w_2 = \p_{r+1} +\dotsb + \p_N$, $\mu_1 = \m_1 + \dotsb + \m_r$, and $\mu_2 = \m_{r+1} + \dotsb + \p_{N}$, then
\be
\begin{split}
\K_{N}( \p_1, &\dotsc, \p_r, \p_{r+1}, \dotsc, \p_N ; \m_{1}, \dots, \m_{r}, \m_{r+1}, \dotsc, \m_{N})
= 
\K_2(w_1, w_2; \mu_1, \mu_2) \\
&+ w_1 \K_{r}\bigl(  \tfrac{\p_{1}}{w_1}, \dotsc, \tfrac{\p_{r}}{w_1}   ;   \tfrac{\m_{1}}{\mu_1}, \dotsc, \tfrac{\m_{r}}{\mu_1}  \bigr)
+ w_2 \K_{N-r}\bigl(  \tfrac{\p_{r+1}}{w_2}, \dotsc, \tfrac{\p_{N}}{w_2}   ;   \tfrac{\m_{r+1}}{\mu_2}, \dotsc, \tfrac{\m_{N}}{\mu_2}  \bigr)
\end{split}
\ee
for any integer $r$ satisfying $1 \le r < N$.  This just says that information gain is equal to the information gain at the group-level plus the expected information gain within each group, and is closely related to the conditional chain rule property;
\item \textit{recursivity}:
\be
\begin{split}
\K_{N}( \p_1, \p_2, \dotsc, \p_N ; \m_1, \m_2,  \dotsc, \m_N) =
&\phantom{+} \K_{N-1}( \p_1+ \p_2, \dotsc, \p_N ; \m_1+ \m_2,  \dotsc,  \m_N) \\
&+ (\p_1+\p_2) \,\K_2\bigl( \tfrac{\p_1}{\p_1+\p_2}, \tfrac{\p_2}{\p_1+\p_2} ; \tfrac{\m_1}{\m_1+\m_2}, \tfrac{\m_2}{\m_1+\m_2} \bigr),
\end{split}
\ee
which is closely related to the previous composition and  chain rules properties---a generalization includes all three properties as special cases is:
\be
\begin{split}
\K_{\mathcal{M}}\bigl(  \bigl\{\{ \p_{j_k}  \}_{k = 1}^{M_j} \bigr\}_{j=1}^{N} ;  \bigl\{\{ m_{j_k}  \}_{k = 1}^{M_j} \bigr\}_{j=1}^{N}  \bigr) &=  
\K_{N} \bigl(  \{ \p_j  \}_{j = 1}^{N} ;  \{ m_j  \}_{j = 1}^{N}   \bigr) \\
&+\sum\limits_{j=1}^{N} \p_j \, \K_{M_j} \bigl(  \bigl\{   \tfrac{\p_{j_k}}{\p_j} \bigr\}_{k=1}^{M_j}   ;   \bigl\{ \tfrac{\m_{j_k}}{m_j} \bigr\}_{k=1}^{M_j} 
\bigr),
\end{split}
\ee
where $\mathcal{M} = \sum\limits_{j}^{N} M_j$ is the total number of possibilities, aggregated into non-empty, non-overlapping groups of $M_j$ outcomes each, for $j = 1, \dotsc, N$, and $\p_{j} = \sum\limits_{k = 1}^{M_j} \p_{j_k}$ and $m_{j} = \sum\limits_{k = 1}^{M_j} \p_{j_k}$
are, respectively, the overall posterior and prior probabilities for the $j$th group of possibilities, for every $j = 1, \dotsc, N$.  If the fine-grained events are aggregated into course-grained events,  then the total (fine-grained) relative entropy can be decomposed into the relative entropy of course-grained possibilities plus an average conditional relative entropy of the fine-grained sub-possibilities constituting each course-grained possibility.
\end{enumerate}

Various subsets of these and other properties have also been used to uniquely characterize the relative entropy.  For example, starting with the \textit{continuity}, \textit{symmetry}, \textit{nilpotence}, \textit{extension-monotonicity}, and \textit{composition} properties, Hobson \cite{hobson:1969} proved that the only functions which satisfies all of these conditions are positive multiples of the relative entropy.  Kannappan and Ng \cite{kannappan:1972,kannappan_rathie:1973,kannappan:1974,kannappan:1988} instead uniquely characterized relative entropy entirely in terms of \textit{recursivity}, \textit{symmetry} only for the $N = 3$ case, \textit{measurability}, and \textit{nilpotence} only for the $N = 2$ case.  Other combinations of axioms are possible---see for example \cite{taneja:1974,haaland:1979} for further discussion.

\subsubsection{Johnson Characterization}

Another compelling characterization is due to R.\ Johnson \cite{johnson:1979}, based on some earlier efforts by Kashyap. First, we mention a few more criteria, slightly different than those introduced above, but wherein $\bv{\p}$ and $\bv{m}$ continue to represent nonnegative, normalized distributions over the same set of mutually exclusive and exhaustive possibilities:
\begin{enumerate}
\item \textit{finiteness}:  $\D(\bv{\p}; \bv{p}) < \infty$;
\item \textit{positivity}: $\D(\bv{\p}; \bv{m}) \ge 0$, where the inequality is strict if $\bv{p} \neq \bv{m}$;
\item \textit{semiboundedness}: $\D(\bv{\p}; \bv{m}) \le  \D(\bv{\p}; \bv{\p})$, where the inequality is strict if $\bv{p} \neq \bv{m}$.
\end{enumerate}
Then Johnson has shown that \textit{measurability}, \textit{additivity}, \textit{finiteness}, and either \textit{positivity} or \textit{semiboundedness} together entail that
\be
\D_N(\bv{\p}; \bv{m}) = B\, \K(\bv{\p}; \bv{m}) + C\, \K(\bv{m}; \bv{\p})
\ee
for some constants $B \ge 0$ and $C \ge 0$, that require further boundary conditions to pin down.  For a measure of information gain in going from the prior distribution $\bv{m}$ to the posterior $\bv{\p}$, it makes sense to also demand that $\D$ become infinite if there is some outcome for which $\p_j > 0$ but $m_j = 0$, but not the other way around.  This requires $C = 0$.  To obtain a non-trivial measure, we then require  $B > 0$.

\subsubsection{Operational Desiderata: Shore-Johnson, Caticha, and Other Variational Characterizations}\label{SubAppendix:ShoreJohnson}

As with the Shannon entropy itself, some characterizations of relative entropy have emphasized its operational purpose in assigning distributions rather than just its information-theoretic meaning in quantifying information and information gain.

Assigning probability distributions via entropy maximization, or relative entropy minimization, originated in the development of statistical thermodynamics by Boltzmann and Gibbs, and generalizes Laplace's Principle of Insufficient Reason, which itself formalizes widespread intuitions regarding probability and symmetry that trace back at least to the birth of probabilistic thinking in games of chance.  In the second half of the 20th century,  E.T.~Jaynes \cite{jaynes:1983,jaynes:2003} clarified and extended the role of maximum entropy reasoning in statistical mechanics, generalized and championed the concept as a basic ingredient of rational inference, and inspired the application of the so-called Principle of Maximum Entropy (MAXENT) in a number of new areas in physics, astronomy, geology, chemistry, biology, economics, and other fields.

Since entropy measures uncertainty or missing information, Jaynes argued that assigning a prior probability distribution over some hypothesis space by maximizing entropy, subject to whatever constraints are known, reflects the most honest representation of our partial knowledge, since by definition any other distribution of smaller entropy would embody less uncertainty, and thereby pretend to information we do not in fact possess.

Subsequently, various author endeavored to translate this insight into more precise mathematical terms, by formulating compelling axioms which lead to the Principle of Maximum Entropy or Minimum Relative Entropy as a general rule for assigning prior probability distributions.  That is, if we seek to assign probability distributions via some \textit{variational principle}, based on optimization of some functional of the distribution, what properties should be required, and to what extent do these pick out a unique variational ``potential'' to minimize?

\paragraph{Shore-Johnson Axioms:}  Shore and Johnson \cite{shore_johnson:1980,shore_johnson:1981} make a case for the following consistency axioms (in which possible events or outcomes are cast in the language of states of a system or systems), which reflect the idea that if a problem can be solved in more than one equivalent way, the answers should be the same.  Stated informally, their axioms are:
\begin{enumerate}
\setcounter{enumi}{-1}
\item \textit{universality}:  the same general optimization framework should apply to all cases, only using different constraints and priors to reflect the specifics of the problem;
\item \textit{uniqueness}:  given a well-posed prior distribution and consistent constraints leading to a convex set of feasible distributions, the posterior distribution should be uniquely determined via constrained optimization;
\item \textit{coordinate invariance}:  our choice of coordinates systems or labels for the possibilities over which the probability distribution is defined should not matter;
\item \textit{system independence}:  it should not matter whether independent information about independent systems is incorporated separately (in marginal distributions) or jointly (in a joint distribution);
\item \textit{subset independence}:  it should not matter whether one treats information pertaining only to certain subsets of a system in terms of a conditional probability distributions for the subsets, or in terms of the full distribution for the full system.
\end{enumerate}
Shore and Johnson then argue how these requirements necessarily\footnote{Strictly speaking, their argument confines its attention to problems where the constraints limit the feasible distributions to some closed, convex subset.  The apportionment problem does not actually involve minimization over a closed, convex set of distributions, since the allowed distributions constitute a discrete number of possibilities dotting the space of probability distributions, as they involve the apportionment of whole numbers of seats.  However, if a variational principle is  favored in the case of convex constraints, it would seem that we should also use it in the case of non-convex constraints, even if the optimal solution is then not guaranteed to be unique.} lead to a distribution $\p_1, \dotsc, \p_N$ minimizing the relative entropy $\K(\bv{\p}) = \sum\limits_j \p_j \log \tfrac{\p_j}{\m_j}$ subject to certain constraints embodying the available information,  in terms of some prior distribution $\m_1, \dotsc, \m_N$ to which the probability distribution would relax in the absence of any further information.

\paragraph{Skilling Axioms:}  Skilling \cite{skilling:1988} has characterized relative entropy using arguments along the same line as employed to derive the Shannon entropy, contending that essentially the same axioms ought to apply in essentially any situation where we wish to generate or compare approximations to any non-negative, additive distribution $\bv{f} = (f_1, \dotsc, f_n)$, even if not naturally interpreted as a probability distribution \textit{per se}.\footnote{Focusing at the time in image processing applications, Skilling called such a distribution a ``scene,'' and any one estimate of it an ``image.''}  By additive, we mean here that the ``weight'' attached to some set is always the sum of the weights of the members of the set.

Based on several different arguments, Skilling showed\footnote{See also his contribution in \cite{buck:1991} for a more streamlined argument.} that the nonnegative distribution $\bv{f}$ should be chosen so as to maximize a functional of the form 
\be
\EuScript{H}(\bv{f}; \bv{\m})  = \sum\limits_j \bigl[ f_j - \m_j - f_j \log \tfrac{f_j}{\m_j}  \bigr] ,
\ee
subject to whatever further constraints on $\bv{f}$ are known,
and given some ``prior model'' $\bv{\m}$ to which $\bv{f}$ should relax in the absence of constraints.  This just reduces to $-\K(\bv{f}; \bv{\m})$ when the distributions are both normalized so that $\sum\limits_j f_j = \sum\limits_j \m_j  = 1$.

\paragraph{Caticha Axioms:}  Caticha \cite{caticha:2003,caticha:2003b,caticha:2007,caticha:2008,caticha:2009,caticha:2010,caticha:2012,caticha:2014} has developed a very similar axiomatization to that of Shore and Johnson, in terms of properties to be satisfied by the variational principe used to assign probability distributions given certain information as constraints.  Guided by what he refers to as the \textit{Principle of Minimal Updating} (PMU), namely that beliefs are to be updated only to the minimal extent required by any new information, he invokes the following desiderata:
\begin{enumerate}
\item \textit{locality}:  local information has local effects, in the sense that if new information refers only to some sub-domain, then the distribution conditional on being outside that sub-domain should not change.  An important special case is one of \textit{idempotence}:  when there is no new information there is no reason to change one's mind;
\item \textit{coordinate invariance}:  the content of the assignment should not depend intrinsically on our choice of labels or coordinates used to specify the possibilities;
\item \textit{independence}: when systems are known to be independent, it should not matter whether they are treated jointly or separately,
\end{enumerate}
which together lead again to relative entropy minimization.  Here universality of the variational principle is left implicit, while his coordinate invariance axiom is the same as that of Shore and Johnson, his independence axiom is essentially the same as their system independence axiom, and his locality axiom is closely related to their subset independence axiom.\footnote{Some form of the independence assumption plays a central role in the Acz\'el, Shore-Johnson, Caticha, and many other axiomatizations of entropy and relative entropy. Its connection to apportionment may seem obscure, but really the connections is so obvious that is goes unspoken---apportioning seats in Switzerland should have no effect on apportioning seats in the U.S.A.  Such separability is imposed by hand with most figures-of-merit, but arises as it were automatically for entropy.  If we simultaneously optimize the joint relative entropy (constraining each county's legislative seats to be granted to districts form that country) then we naturally arrive at the same answer as if we optimize separately.}

\subsubsection{Baez-Fritz Relative Entropy Characterization}

As with the entropy, relative entropy can also be characterized in terms of properties expected under \textit{measure-preserving transformations} \cite{baez:2014}.  Consider the same setup as in the earlier Baez-Fritz-Leinster entropy characterization, only now we add another ingredient: some sort of  background hypothesis leading to (i) measurement probabilities $\bv{q}$ over the outcomes in $Y$, and (ii) ``transition'' probabilities $\theta_{j|k}$ about  the possible latent state of the system given the result of the measurement.  The latter are assumed to be consistent with the measure-preserving transformation $f\colon \X \to \Y$, in the sense that $\theta_{j|k} = 0$ unless $f(x_j) = y_k$.

Together, these lead to a ``prior'' distribution over the latent states via
\be
\m_j = \sum\limits_j  \m_k \theta_{j|k} 
\ee
But new knowledge will lead in general to a different distribution $\bv{\p}$ over the latent states.  It is demanded that the quantification of information gain $G$ associated with this new knowledge satisfy three criteria.

First, the information gain $G$ should be \textit{additive} under the composition of such measurements.  Second, the information gain is \textit{lower-semicontinuous} with respect to the underlying probability distributions, meaning if $\bv{\p}_i$ and $\bv{\m}_i$ are sequences of normalized distributions converging to $\bv{\p}$ and $\bv{\m}$, respectively, in the limit as $i \to \infty$, then $G(\bv{\p}; \bv{\m}) \le \lim\limits_{i \to \infty} \inf\limits_{ i' \ge i} G(\bv{\p}_{i'}; \bv{\m}_{i'})$.  (This means that if the problem is changed very slightly, the gain is either close to or below its original value).  Third, the information gain is \textit{convex linear}.  Intuitively, this means that if we are to flip a (biased) coin to decide whether to perform one measurement process or another, the  information gain is the expectation of the gain, averaged  with respect to the unknown result of the coin flip.  Fourth, the information gain vanishes when the hypothesis is ``ideal,'' so that $\bv{\p} = \bv{\m}$.  Then it can be shown that $G(\bv{\p}; \bv{\m}) = \K(\bv{\p}; \bv{\m})$ for some choice of the positive constant $\kappa>0$.

\subsubsection{Information Geometry and Amari's Characterization}\label{SubAppendix:Amari}

Beginning perhaps with the pioneering work of statisticians R.A.~Fisher and C.R.~Rao, many authors have explored connections between probability theory and differential geometry, culminating in the work of Amari \cite{amari:2000,amari:2010,amari:2016} and his co-workers on what is now known as \textit{Information Geometry}, which makes statistical inference look something like general relativity.\footnote{Or possibly vice versa?  The connection between gravitation and information has become an area of active contemporary research in theoretical physics.}  So as not to take us too far afield, this brief summary simply presumes some basic familiarity with differential geometry, but may be safely skipped or skimmed.

The key idea is to consider a \textit{parameterized} family of probability distributions,\footnote{For probabilities over a finite number of possibilities, as in the applications studied here, the probabilities $\bv{\p} = (\p_1, \dotsc, \p_N)$ themselves can be thought of as the parameters.} then interpret the parameters as coordinates on a Riemannian manifold of distributions, in order to invoke the usual notions of metrics, connections, etc.\ developed in the context of differential geometry.  Because each ``point'' in the manifold corresponds to one well-defined probability distribution, any geometric structure must be compatible with this additional probabilistic structure, which greatly narrows the consistent possibilities.  In fact, it is natural to demand that the metric and connections are invariant under transformations associated with so-called \textit{sufficient statistics},\footnote{Recall that a sufficient statistic is a mapping of the raw data which still contains all of the information relevant to estimating the parameters of the likelihood distribution.  If $\bv{x}$ represents the full data and $\bv{\theta}$ represents the parameters determining the likelihood function $\P(\bv{x} \given \bv{\theta})$ , then $\bv{y} = \bv{F}(\bv{x})$ is a sufficient statistic if the posterior predictive probabilities for the parameters are identical whether conditioned on $\bv{y}$ or $\bv{x}$:  $\P(\bv{\theta} \given \bv{y}) =  P(\bv{\theta} \given \bv{x})$.} and this requirement alone singles out the \textit{Fisher information matrix} as the only natural metric tensor,\footnote{In the case of a discrete probability distribution $p^1, p^2, \dotsc$, where the probabilities themselves can be identified as natural coordinates on the manifold, the Fisher metric tensor is $g_{\mu \nu} =  \tfrac{ \delta_{\mu \nu} }{ \sqrt{p^{\mu} p^{\nu}}}$.} and also singles out one family of torsion-free connections parameterized by a single real number, known as the $\alpha$-connections, where the connections corresponding to $\pm \alpha$ are dual with respect to the Fisher metric, while the $\alpha = 0$ case corresponds to the Levi-Civita connection for this metric.

Furthermore, it is known that there is a direct correspondence between these connections and so-called \textit{directed divergences}, defined in information geometry as functionals (of pairs of probability distributions) satisfying a positive-definiteness property, namely, that for any pair $\bv{\p}$, $\bv{\m}$ of  ``points'' (i.e., probability distributions) in the manifold,
\be
\text{\textit{positive-definiteness}: } \D_{\alpha}(\bv{\p}; \bv{\m}) \ge 0, \text{ with equality if and only if } \bv{\p} = \bv{\m} \text{ (almost everywhere)}.
\ee 
The $\alpha$-divergences corresponding to the $\alpha$-connections are not in general symmetric with respect to interchanging the roles of distributions $\bv{\p}$ and $\bv{\m}$, and do not in general satisfy a triangle inequality, so cannot represent true distances on the manifold of probability distributions,\footnote{The fact that the most natural ways to quantify how probability distributions differ are \textit{not} interpretable in general in terms of distances in the manifold, geodesic or otherwise, has led some authors, particularly John Skilling \cite{skilling:2014,skilling:2015}, to remain skeptical of imposing a metric structure on probability spaces, or to critique the usefulness of the geometric approach altogether.  In recent years, Skilling has argued that when it comes to probability distributions, there is a useful notion of ``from/to''---namely, the relative entropy---but not ``between.''} with the exception of the self-dual $\alpha = 0$ case, which is known as the Hellinger distance.\footnote{In the case of discrete probability distributions, the Hellinger distance can be written as $\mathscr{H}(\bv{\p},\bv{\m}) = 2 \sum\limits_k ( \sqrt{\p_k} - \sqrt{\m_k})^2$.  Interestingly, this is reminiscent of the Hilbert-space distance that arises in quantum mechanics.}  The $\alpha = -1$ case corresponds to the relative entropy or Kullback-Leibler divergence $\K(\bv{\p}; \bv{\m})$, and the $\alpha = +1$ case corresponds to its dual $\tilde{\K}(\bv{\p}; \bv{\m}) = \K(\bv{\m}; \bv{\p})$.  These two $\alpha$-divergences are in fact the only choices which are \textit{additive} for independent events, and once again the $\alpha = -1$ case is usually to be preferred if $\bv{\m}$ is regarded as the prior, default, reference, or source distribution, and $\bv{\p}$ is the posterior, destination, or variational trial distribution, because this ensures that impossible (zero probability) events remain impossible.\footnote{And in applications to legislative apportionment, the minimization of $\K$ as variational potential is applicable to party-list representation without non-zero lower bounds, while the dual measure $\tilde{\K}$ is not.}

This overtly {geometric} approach to entropy has been further explored in the context of Bayesian inference by Carlos Rodr\'iguez (see for instance \cite{rodriguez:1991,rodriguez:1998}), and in the context of thermodynamics  by a number of authors, including Weihold, Ruppeiner, Salamon, Gilmore, Levine, and Crooks.  See for example \cite{levine:1986,andresen:1988,crooks:2007}, and references therein.  A somewhat similar statistical framework has also been explored by Ole Barndorff-Nielsen \cite{barndorff-nielsen:1978}.

\subsubsection{Information Elicitation and a Scoring-Rule Characterization}

Another illuminating approach to motivating relative entropy, explored by a number of scholars in probability, statistics, economics, psychology, and artificial intelligence, involves elicitation of honest information about uncertain outcomes.

By the early 1950s, statistician I.J.\ Good and many others had begun to explore notions of epistemic utility, or the \textit{value} of information (as opposed to just the amount of information or information change).  Around the same time,  Glenn Brier observed that then-standard sorts of evaluations for tasks like weather forecasting actually introduced incentives for rational forecasters to mis-report or misrepresent their actual probabilistic predictions (e.g., the ``chance of heavy, light, or no rain tomorrow morning  in Philadelphia.'').  If above all, we desire rational information-handling and utility-maximizing agents to honestly report predictions,\footnote{This task is similar to problems in collective social choice, where we want to encourage honest voting rather than strategic voting.} presumably in support of future decisions by other parties, what sort of utility function should we try to impose on the forecaster?  

That is, suppose that, based on relevant data available to her, a rational forecaster adopts a probability distribution $\bv{\p}= (\p_1, \dotsc, \p_n)$ over $n$ mutually exclusive and exhaustive possible outcomes, but chooses to report publicly a distribution $\bv{\m}= (\m_1, \dots, \m_n)$.  With nothing at stake, there is not necessarily any incentive for honest reporting, or with arbitrary or exogenous stakes, there may arise incentives for distorted reporting.  But if the payoffs can be shaped, honesty can be encouraged.

Obviously a forecast should come before observation of the outcome, but before either, imagine that it is credibly made known to the forecaster that the outcome will subsequently be observed and compared to her probabilistic predictions $\bv{m}$, and if the $k$th outcome actually obtains, she will be rewarded with a utility $u_k(\bv{\m})$ that depends on the actual outcome and her reported probabilistic predictions.  In this context, the collection of these utility payoffs are referred to as a \textit{prediction scoring rule}.  The idea is to induce incentives so that a rational forecaster tells us what she really thinks, not what she thinks we want to hear.

\newpage
With this goal in mind, we might think that a good prediction scoring rule\footnote{``Psychological'' or personalistic Bayesians like de Finetti and Savage have considered similar scoring rules designed to elicit \textit{coherent} probabilities. However, here we are implicitly assuming that the forecaster is rational, and will always report probabilities that are nonnegative and sum to unity over the mutually exclusive and exhaustive hypothesis space.} should satisfy the following properties:
\begin{enumerate}
\item \textit{smoothness}:  each $u_k(\m_1 \dotsc, \m_n)$ varies smoothly as a function of each $\m_j$ when $0 < \m_j < 1$;
\item \textit{propriety}:\footnote{This is also known as a \textit{reproducing} property in the literature.} given any fixed ``internal'' probability distribution $\bv{\p}$, the function $\bar{U}(\bv{\m}; \bv{\p}) = \sum\limits_{k=1}^{n} \p_k\, u_k(\m_1, \dotsc, \m_n)$ is maximal (as a function of possible reported distributions $\bv{\m}$) when $\bv{\m} = \bv{\p}$. This says that in order to \textit{maximize expected utility}, the forecaster can report her actual probabilities.
\item \textit{strict propriety}:  For a fixed distribution $\bv{\p}$, the function $\bar{U}(\bv{\m}; \bv{\p}) = \sum\limits_{k=1}^{n} \p_k\, u_k(\m_1, \dotsc, \m_n)$ has a \textit{unique} absolute maximum at $\bv{\m} = \bv{\p}$ amongst all nonnegative, normalized distributions.  This says that a rational forecaster \textit{must} report her actual probabilities in order to maximize expected utility.
\item \textit{prediction-monotonicity}: if in the probability distributions $\bv{\m}' = (\m_1' , \dotsc, \m_n')$ and $\bv{\m} = (\m_1, \dotsc, \m_n)$, it is the case that $\m_k' > \m_k$, then $u_k(\m_1', \dotsc, \m_n') > u_k(\m_1, \dotsc, \m_n)$.  This would mean that greater reward is offered for a more accurate prediction that placed higher probability on the outcome actually realized, however such increased probability comes at the expense of probabilities over the unrealized possibilities;\footnote{However appealing this sort of property, keep in mind that the \textit{primary} goal here is to elicit honest prediction, not accurate prediction.}
\item \textit{symmetry across unrealized alternatives}: if $\pi(\,)$ is a permutation of the indices $(1, \dotsc, n)$, then
$u_{\pi\inv(k)}(\m_{\pi(1)}, \dotsc, \m_{\pi(n)}) = u_k(\m_1, \dotsc, \m_n)$.  This is saying much more than just (i) rewards should not depend on arbitrary re-labeling of the outcomes (assuming which label corresponds to which actual outcome is known to forecaster and evaluator), but moreover that (ii) the score is affected by how much of the probability mass was  assigned to the unrealized outcomes,\footnote{Note that the remainder of the unit probability mass must have been assigned to the correct alternative.} or possibly by the set of magnitudes of probabilities assigned to the unrealized outcomes, but not on which of these probabilities is assigned to which amongst the unrealized outcomes;\footnote{An equivalent way to express this symmetry constraint is that, for all $k$, $u_k(\m_1, \dotsc, \m_n) = f(\m_k, \m_1, \dots, \m_{k-1}, \m_{k+1}, \dotsc, \m_n)$ for some one function $f(\m_1 ; \m_2, \dotsc, \m_n)$ that is symmetric in all but the first argument.  We might easily imagine other situations where this sort of symmetry would not be desirable, for example if accurate prediction of some possible outcomes is already known to be more important than others.  But again, here we are supposing that the goal is honest elicitation of predictions across all possible outcomes, without regard to the nature of the outcome itself, the idea being that these probabilities of an informed expert could subsequently serve as input for many different possible decisions problems where different end users might have very different utilities over the outcomes.}
\item \textit{irrelevance under non-occurrence}: $u_k(\m_1, \dotsc, \m_n) = \phi_k(\m_k)$ is a function only of $\m_k$, for each $k = 1, \dotsc, n$.  This says that the reward depends only on which outcome obtained and the reported probability assigned to that outcome, but not on how the remaining probability mass was distributed across the unrealized outcomes.\footnote{Once again, one can imagine situations where this would not be desirable--for example, if some wrong answers are deemed to be better than others.  But in the current context it seems to be a compelling feature of fairness to the forecaster that once it is known that outcome $k$ obtains, the appraisal should be based on how likely that one outcome was judged.  Note that without the reward-symmetry property, \textit{irrelevance} does not by itself demand that the reward for accurate responses be the same across all all realized outcomes.}  This property is also referred to as \textit{locality}.
\end{enumerate}
Then it can be shown that the only prediction scoring rules satisfying \textit{smoothness}, \textit{strict propriety}, and \textit{irrelevance} are of the form
\be
u_k(\m_k) = \kappa \log(\m_k) + b_k,
\ee
for some positive coefficient $\kappa > 0$ (the same for all $k$), and real offsets $b_k$.  Monotonicity need not be assumed, but is enjoyed by such scoring rules.

If we additionally demand \textit{symmetry}, then $b_k = b$ for all $k = 1, \dotsc, n$.  So the expected utility becomes
\be
\bar{U}(\bv{\m}; \bv{\p}) =  \sum\limits_{k=1}^{n} \p_k\, [ \kappa\, \log \m_k + b] =  b + \kappa \sum\limits_{k=1}^{n} \p_k \, \log \m_k, 
\ee
which, apart from an overall additive shift $b$, is just proportional to what is known as the \textit{cross entropy} in information theory.  Because additive constants do not affect the location of the optimum, the constant $b$ can be chosen arbitrarily in principle, (as long as it remains independent of $\bv{\m}$),
although in practice,  $b$ may need to be chosen sufficiently large (via cash rewards for example), so that the forecaster agrees to participate at all.  Ignoring such concerns here, it will be convenient, for any fixed $\bv{\p}$, to choose $b$ to be proportional to the Shannon entropy of the forecaster's actual distribution, 
\be
b = \S(\bv{\p}) = -\kappa \sum\limits_{j = 1}^{n} \p_j \, \log \p_k,
\ee 
so that the expected utility becomes equal to the additive inverse of the (dual) relative entropy:
\be
\bar{U}(\bv{\m}; \bv{\p}) = -\kappa \sum\limits_{k=1}^{n} \p_k \log \bigl[ \tfrac{\p_k}{\m_k} \bigr] = - \K(\bv{\p}, \bv{\m}) = -\tilde{\K}(\bv{\m}, \bv{\p}) \le 0,
\ee
with equality if and only if $\bv{\m} = \bv{\p}$.  Notice however that the ``trial'' distribution $\bv{\m}$ appears here as the reference measure (in the denominator of the logarithm), so the utility plummets infinitely if an outcome predicted to be impossible actually does turn out to occur.\footnote{This forces a rational forecaster to declare non-zero probability to all outcomes under consideration unless an outcome is logically impossible.  Statistician D.V.\ Lindley called this ``Cromwell's Rule.''}

We may conclude that the log-score, leading to the dual relative entropy, is essentially the only smooth, symmetric, strictly proper prediction scoring rule that depends only on the quality of prediction accorded to what turns out to be the true hypothesis.  This sort of log scoring rule has been explored by a number of authors, including Good, Savage, McCarthy, de Finetti, and Shufford, Albert, and Massengill.\footnote{More generally, for smooth, proper scores, the difference in expected proper score between honest and dishonest reporting is a so-called \textit{Bregman divergence}, discussed below.}

\subsubsection{Lattice Valuations and the Knuth-Skilling Characterization}\label{SubAppendix:KnuthSkilling}

A recent axiomatization of entropies developed by Knuth and Skilling \cite{knuth_skilling:2012} just might provide the strongest justification for entropic apportionment, in the sense of requiring the \textit{fewest} overtly probabilistic assumptions.  Their axioms and arguments do however involve some significant mathematical technicalities involving lattice theory, so are only summarized briefly here. 

Knuth and Skilling build on the pioneering work of the physicist Richard Cox \cite{cox:1946,cox:1961}, who motivated the usual rules of the probability calculus from more basic desiderata of logical consistency, and established that probability theory is essentially the unique quantification of rational degree of belief and the unique extension of deductive logic to cases of uncertainty, where information may be incomplete, and the truth value of propositions may not be known.  Knuth and Skilling begin with what appears to be a somewhat more general situation, seeking natural constraints\footnote{The authors speak of respecting certain lattice \textit{symmetries}, but it might be more accurate to say that the demand is that valuations embody structure-preserving homomorphisms between the lattice and the numerical assignments on the lattice.} to be satisfied by any \textit{valuations} on the elements of a \textit{distributive lattice}.\footnote{Bayesian probabilities live on a so-called \textit{Boolean lattice} of propositions, which is closed and consistent under all logical disjunctions (ORs), conjunctions (ANDs), and negations (NOTs), and their combinations.  Knuth and Skilling employ the broader construct of a \textit{distributive lattice} equipped with notions of meet, join, and partial order (which in many cases can be identified with set-theoretic intersection, union, and inclusion, respectively).  Boolean lattices are special cases of distributive lattice, but by focusing on the latter, the authors manage to motivate the rules of probability theory without direct recourse to the use of logical negation.  In fact, they do not even make use of all of the defining properties of a distributive lattice, so it may turn out that their conclusions will hold for a still broader class of lattices or partially ordered sets.}  They are able to infer that compatibility with the underlying structure\footnote{They make much use of ordering and associativity properties, but need not assume commutativity of the measure from the start.  Additivity, and thereby commutativity, for the valuations are derived rather than postulated.  They do make much use of an independence postulate, which again says that independent problems can be analyzed separately or together, and conclusions should be the same.} of the lattice of possibilities requires that the valuations be equivalent (up to an invertible mapping) to the usual notion of nonnegative, additive measures.

\newpage
Each such measure is uniquely determined by  the values assumed on the ``atomic'' elements just above the ``bottom'' of the lattice, and if these atomic valuations are normalized so as to sum to unity, the valuations will satisfy all of the mathematical rules of probability, whether or not they are naturally interpreted as probabilities in the sense of degrees of rational belief or as degrees of partial implication.\footnote{In a Bayesian framework like that of Knuth and Skilling, all probabilities are \textit{conditional} degrees of belief, so technically probabilities are constructed as ratios of what the authors call measures, rather than measures themselves.} 

Apportionment naturally possesses just this sort of distributive lattice structure.  In the case of Congressional apportionment, the atomic elements of the lattice may be identified with individual inhabitants, and can be joined into districts, the districts into states, and if desired, the states into regions, etc.  Or in the case of party-list representation, the votes can be grouped by parties, the parties into coalitions, etc.  Shares of representation are naturally represented by nonnegative numbers which can be assigned to any subset, and at each level of aggregation, the share of representation for any group is just the \textit{arithmetic sum} of the shares of its distinct members.

This suggests that in such applications we can, and indeed should, make use of the further mathematical machinery which emerges from the Knuth and Skilling lattice axioms. Having inferred the sort of measures that can be naturally defined on a distributive lattice, they then look at valuations on these valuations---real-valued functions of the measures that can be used to characterize a measure, to compare or rank these measures, or to assign these measures through an associated variational principle.

First, one seeks a functional which quantifies the divergence of some ``destination''  measure $(\p_1, \dotsc, \p_N)$  from a ``source'' measure $(\m_1, \dotsc, \m_N)$ that existed before the constraints that lead to the $\p_j$ were imposed.   Given a measure $(\p_1, \dotsc, \p_N)$ over the atoms of the lattice, compatibility with the underlying distributive lattice structure turns out to limit consideration to variational potentials of the form:
\be
\Gamma(\p_1, \dotsc, \p_N) = \sum\limits_{j = 1}^{N}  \bigl[A_j +  B_j \p_j  + C_j (\p_j \log \p_j - \p_j)\bigr] 
\ee
for some choice of real-valued parameters $A_j$, $B_j$, and $C_j$.

Given that we seek a quantified notion of divergence or discrepancy between measures, we demand that the function $\Gamma$ is to achieve its minimum when (and only when) $\p_j = \m_j$ for all $j = 1, \dotsc, N$.
This requires that $B_j = -C_j \log \m_j$ and $C_j > 0$.  As a matter of convenience, we might as well choose this minimum value to be zero,\footnote{This just serves to fix an overall additive offset.} which further entails that $A_j = - C_j$, and hence
\be
\Gamma(\p_1, \dotsc, \p_N ; \m_,\dotsc, \m_N) = \sum\limits_{j = 1}^{N}  C_j \bigl[ \m_j -  \p_j  + \p_j \log \tfrac{\p_j}{\m_j}  \bigr]. 
\ee
The coefficients $C_j$ now clearly represent some sort of intrinsic weighting of the possibilities.  In most applications, the different outcomes $j = 1, \dotsc, N$ are to be treated equivalently \textit{a priori}, requiring that the $C_j$ all assume a common value $C_j = C$.  In other words, we demand symmetry under \textit{parallel} permutations of the $\p_j$ and also the $\m_j$. This is particularly salient in cases of apportionment, where $j$ will index different inhabitants (in the case of Congressional apportionment) or different votes (in the case of party-list representation), and so the equality of the $C_j$ follows from our basic democratic motivations.  The possible functions are thereby further reduced to
\be
\Gamma(\p_1, \dotsc, \p_N ; \m_,\dotsc, \m_N) = C \sum\limits_{j = 1}^{N} \bigl[ \m_j -  \p_j  + \p_j \log \tfrac{\p_j}{\m_j}  \bigr],
\ee
and finally, if both measures are normalized so that $\sum\limits_j \p_j = \sum\limits_j \m_j = 1$, we are led to
\be
\Gamma(\p_1, \dotsc, \p_N ; \m_,\dotsc, \m_N) = C\, \sum\limits_{j = 1}^{N}  \p_j \log \tfrac{\p_j}{\m_j},
\ee
which is recognized as the relative entropy, scaled by some factor $C > 0$.

A related, if distinct task involves quantifying the \textit{uniformity} of a single measure over the atomic possibilities.  Assume the measure $(\p_1, \dotsc, \p_N)$ has been normalized\footnote{The lattice is assumed to be generated from a finite number of ``atoms,'' so this is always possible by a simple re-scaling if necessary.} such that $\sum\limits_j \p_j  = 1$.   Here we require the function $\Gamma(\p_1, \dotsc, \p_N)$ to be permutation invariant, to achieve its minimum when the measure is maximally non-uniform, namely when any one value is unity and the rest of the values are zero, and to achieve its maximum when the measure is completely uniform, namely when all of the $\p_j$ are equal. Again as a matter of convenience, we can choose the minimum to be zero.  Together, these constraints require $A_j = 0$ and  $B_j = C_j = C = -\abs{C}$ for some constant $C < 0$, resulting in
\be
\Gamma(\p_1, \dotsc, \p_N) = -\abs{C} \,\sum\limits_j \p_j \log \p_j,
\ee
which is recognized as being proportional to the Shannon entropy.

\subsection{Interpretation of Entropy and Relative Entropy in Communication Theory}

In Shannon's original application to communication theory, entropy represents the (receiver's) uncertainty as to which message will be sent, and therefore the amount of uncertainty that is expected to be removed if the message is received unambiguously.  Of course the total number of  bits used to express or transmit the message may tend to be greater, because of inefficient coding, unavoidable redundancy in an otherwise convenient or familiar language, or the need for deliberate error correction in noisy channels, but the entropy represents the minimum number of bits to which the message could be reliably compressed on average (in the absence of noise).

But in order to achieve good compression, one must be well informed about the source of possible messages.  While the Shannon entropy $\S(\bv{\p})$ represents best-case compressibility given message probabilities $\bv{\p}$, relative entropy $\K(\bv{\p} ; \bv{\m}) $ represents the cost in terms of expected extra bits required when the coding is optimized for message probabilities $\bv{\m}$ but messages are predicted to actually occur with probabilities $\bv{\p}$.  It is a measure of inefficiency in coding or transmission, or equivalently if more optimistically, of potential savings if information about the source is gained.

\subsection{Relationships Between Entropy and Relative Entropy}

Whether we consider Shannon entropy or relative entropy more fundamental is largely a matter of taste.  If we adopt either of these functionals, the other arises almost inevitably.

\subsubsection{From Entropy to Relative Entropy}

Consider $K$ ``course-grained'' possibilities, assumed mutually exclusive and exhaustive, and indexed by $k = 1, \dots, K$.  Imagine we learn that each can be refined into $n_k$ ``fine-grained'' exclusive sub-possibilities which we label by $j = 1, \dotsc, n_k$ for each $k$, for a total of $N = \sum\limits_k n_k$ fine-grained possibilities in all.  Knowing nothing else, the principle of maximum entropy assigns equal probabilities $\tfrac{1}{N}$ to all of the fine-grained possibilities, leading to probabilities
\be
g_k =  \sum\limits_{j = 1}^{n_k} \tfrac{1}{N} =  \tfrac{n_k}{N} = \tfrac{n_k}{\sum\limits_{k} n_k} 
\ee
for the course-grained possibilities.  These are rational numbers, but mathematically speaking we may imagine problems involving refinements of arbitrary granularity, so the $g_k$ can be arbitrarily close to any discrete probability distribution.  

Our uncertainty as to which fine-grained possibility will obtain is quantified by the (maximal) entropy
\be
\S_{N} = -\sum\limits_{k = 1}^{K} \sum\limits_{k = 1}^{n_k} \tfrac{1}{N} \log \tfrac{1}{N} = \log N,
\ee
while our uncertainty as to the course-grained outcome is quantified by the entropy
\be
\S_K = -\sum\limits_{j = 1}^{K}  g_k \log g_k = \S_N -  \sum\limits_{k = 1}^{K} g_k \log n_k.
\ee
But suppose we receive additional information implying (usually via Bayes' rule) that the probabilities of the course-grained possibilities become $\p_k$ for $k = 1, \dotsc, K$, where still $\p_k \ge 0$ and $\sum\limits_k \p_k = 1$, so that our uncertainty as to the course-grained outcome changes to $\S'_K = -\sum\limits_k \p_k \log \p_k$.
If we learn nothing further about which of the $n_k$ fine-grained possibilities may be more or less likely given that the course-grained outcome $k$ occurs, then each course-grained probability $\p_k$ must be spread uniformly over all $n_k$ fine-grained possibilities, which is to say that the conditional probabilities over the fine-grained outcomes, given any course-grained outcome $k$, remain uniform, and equal to $\P(j\given k) = \tfrac{1}{n_k}$.  Our overall uncertainty as to the fine-grained outcome becomes
\be
\S = -\sum\limits_{k = 1}^{K} \sum\limits_{j = 1}^{n_k} \p_k \tfrac{1}{n_k} \log \bigl[ \p_k \tfrac{1}{n_k} \bigr]
= -\sum\limits_{k = 1}^{K}  \p_k  \log \bigl[ \tfrac{\p_k}{N g_k} \bigr]
= \log N - \sum\limits_{k = 1}^{K}  \p_k  \log \bigl[ \tfrac{\p_k}{g_k} \bigr]
= \S_N - \K(\bv{\p}; \bv{g}),
\ee
which is simply the difference between the maximum possible entropy $\S_N$ and the relative entropy $\K(\bv{\p}; \bv{g})$ between the updated distribution $\bv{\p} = (\p_1, \dots, \p_K)$ and the prior distribution $\bv{g} = (g_1, \dotsc, g_K)$ over the course-grained possibilities.  So $\K$ can be interpreted as the amount by which the fine-grained Shannon entropy has been reduced by learning the information taking our course-grained probabilities from $\bv{g}$ to $\bv{\p}$. The distribution $\bv{g}$ is sort of a ``density of states'' or degeneracy factor that properly accounts for the fine-grained multiplicity of the course-grained outcomes in the evaluation of the overall entropy.

So from this point of view, probability assignment via the Principle of Minimum Relative Entropy can be viewed as just a special case of the Principle of Maximum Entropy, when the specified hypothesis space is a course-graining of some more primitive space of elementary possibilities. 

\subsubsection{From Relative Entropy to Entropy}

Conversely, suppose we start with the Kullback-Leibler divergence
\be
\K(\bv{\p} ; \bv{\m}) = \sum\limits_{j=1}^{N} \p_j \log \tfrac{\p_j}{\m_j},
\ee
accepted as a measure of relative information, of a probability distribution $\bv{\p} = (\p_1,\dotsc, \p_N)$ over some set of mutually exclusive and exhaustive possibilities, \textit{relative to} another distribution $\bv{\m} = (\m_1, \dotsc, \m_N)$ over the same space of possibilities, or equivalently, as the quantified \textit{information gain} in updating to probabilities $\bv{\p}$ starting from prior probabilities $\bv{\m}$.  

We then seek a measure of overall \textit{uncertainty}, or \textit{missing information}, associated with a single distribution $\bv{\p}$.
In this context, a prior state of minimal knowledge is one in which we know nothing except that there are $N$ mutually exclusive and exhaustive possibilities, leading to the probability assignment  $\m_j = \tfrac{1}{N}$ for all $j = 1, \dots, N$.  Under this state of ignorance, any other distribution would pretend to knowledge we do not have, namely that certain outcomes are more likely than others.  This is known as the Principle of Indifference, or Principle of Insufficient Reason, but is really just an elementary symmetry requirement in the absence of symmetry-breaking information.

On the other extreme, a state of \textit{maximal} knowledge would obviously correspond to learning which one of the mutually exclusive and exhaustive  outcomes actually obtains, taking us from the prior distribution $\bv{\m}$ to a probability distribution $\bv{\mu} = (\mu_1, \dotsc, \mu_J)$ concentrated on exactly one of the possible outcomes, say the $\ell$th:  
\be
\mu_j = \delta_{j \ell} = 
\begin{cases} 
	1 &\text{ if } j = \ell \\
	0 &\text{ if } j \neq \ell
\end{cases},
\ee
where $\delta_{j \ell}$ is the usual Kronecker delta symbol.

Thus, $\K(\bv{\mu}; \bv{\m})$ quantifies the maximal information obtainable starting in the prior state of maximal ignorance.  If our actual state of knowledge is instead described by the distribution $\bv{\p}$, then the missing information that would be required to reach a state of maximal knowledge should be the difference between the maximal information obtainable and the information already gained in going from $\bv{\m}$ to $\bv{\p}$:
\be
\begin{split}
\Delta \K &= \K(\bv{\mu}; \bv{\m}) - \K(\bv{\p}; \bv{\m}) = \sum\limits_j \mu_j \log \tfrac{\mu_j}{\m_j} - \sum\limits_j \p_j \log \tfrac{\p_j}{\m_j}
= \sum\limits_j \delta_{j \ell} \log \tfrac{\delta_{j\ell}}{1/N} - \sum\limits_j \p_j \log \tfrac{\p_j}{1/N} \\
&= \log N - \sum\limits_j \p_j  \log N -  \sum\limits_j \p_j \log \p_j  = - \sum\limits_j \p_j \log \p_j  = \S(\bv{\p}),
\end{split}
\ee
which is just the Shannon entropy of the final probability distribution $\bv{\p}$.

From this complementary point of view, probability assignment via the Principle of Maximum Entropy can be viewed as just a special case of the Principle of Minimum Relative Entropy when the background measure over the specified hypothesis space is assumed uniform. 

\subsubsection{Entropy and Bregman Divergence}

Another connection between Shannon entropy and relative entropy may be summarized by saying that the relative entropy is the \textit{Bregman divergence} of the entropy.  The notion of Bregman divergence frequently arises in contemporary machine learning applications, as well as other applications involving convex optimization.  Given a continuously-differentiable and strictly convex function defined on a convex set, the associated Bregman divergence may be defined as the difference between that function evaluated at one point in the set, and its first-order Taylor approximation evaluated at the point but centered at some reference point.
In the context of discrete probability distributions, if $F(\bv{\p})$ is the original (convex and differentiable) function defined on the probability simplex, then the associated Bregman divergence generated by $F(\bv{\p})$ is
\be
\mathcal{B}_F(\bv{\p}, \bv{m}) = F(\p_1, \dotsc, \p_N) - F(m_1, \dotsc, m_N) - \sum\limits_{k = 1}^{N} (p_k - m_k)
\, \tfrac{\del}{\del m_k} F(m_1, \dotsc, m_N).
\ee
This divergence is \textit{nonnegative definite} for nonnegative, normalized distributions, is always \textit{convex} with respect to the first argument $\bv{p}$, and enjoys a number of additional useful properties relevant in statistical estimation and convex optimization problems.

Since spaces of normalized discrete probability distributions are naturally closed and convex, and the \textit{negentropy} $F(\bv{\p}) = -\S(\bv{\p})$ is convex and differentiable at interior points, we can define a Bregman divergence of the negentropy with respect to some reference distribution $\bv{\m}$, and the result is just the relative entropy $\K(\bv{\p}; \bv{\m})$:
\be
\begin{split}
-\S(\p_1, &\dotsc, \p_N) + \S(m_1, \dotsc, m_N) + \sum\limits_{k = 1}^{N} (p_k - m_k)
\, \tfrac{\del}{\del m_k} \S(m_1, \dotsc, m_N) \\
&= \kappa \sum\limits_{k = 1}^{N} \bigl[ p_k \log p_k - m_k \log m_k - (p_k - m_k) \log m_k   - \tfrac{1}{\ln b}(p_k-m_k) \bigr]\\
&= \kappa \sum\limits_{k = 1}^{N} \bigl[ p_k \log p_k  - p_k \log m_k  \bigr] = 
\kappa \sum\limits_{k = 1}^{N} p_k \log \tfrac{p_k}{m_k},
\end{split}
\ee
where here $b$ represents the base of the logarithm used in defining $\S$, but drops out since it only appears explicitly in a term which vanishes because $\sum\limits_k p_k = \sum\limits_k m_k = 1$.

\subsection{Relative Entropy and Csisz\'ar $f$-Divergences}\label{on_fdivergences}

The relative entropies belong to a wider class of functionals known as Csisz\'ar $f$-divergences\footnote{Such measures of divergence are also known as \textit{Csisz\'ar-Morimoto} divergences, or  \textit{Ali-Silvey} divergences.} which are intended to measure the discrepancy or divergence between two probability distributions $\bv{\p} = (\p_1, \dotsc, \p_N)$ and $\bv{\m} = (\m_1, \dotsc, \m_N)$ defined over the same space of $N$ mutually exclusive and exhaustive outcomes.  The $f$-divergence represents an expectation value (with respect to the reference probability distribution $\bv{\m}$) of some real-valued function $f(\,)$ of the relative odds ratio $\tfrac{\p_k}{\m_k}$:
\be
\D_f(\bv{\p}; \bv{\m}) = \sum\limits_{k=1}^{N} \m_k \, f\bigl( \tfrac{\p_k}{\m_k} \bigr).
\ee
To generate a well-defined $f$-divergence, the function $f(x)$ must possess two properties,\footnote{Additionally, we adopt the conventions that $0\, f\bigl(\tfrac{0}{0} \bigr) = 0$, $0\, f\bigl(\tfrac{0}{x} \bigr) = x \lim\limits_{y \to 0^+} f(y)$, and $0 \,f\bigl(\tfrac{x}{0} \bigr) = x \lim\limits_{y \to \infty} \tfrac{f(y)}{y}$.} namely: (i) it must be convex (\ie, concave up) for all $x > 0$, and (ii) it must satisfy the boundary/normalization condition $f(1) = 0$.  Because valid inputs must be normalized probability distributions, notice that the generating function $f(x)$ for any particular divergence measure is not unique, as we can add to $f(x)$ any linear ``gauge'' function of the form $g(x) = \gamma\, (x -1)$, which always averages to zero but does not alter the convexity properties\footnote{Preservation of convexity follows because the second derivative of a linear function vanishes everywhere.} of $f(x)$, nor the value of $f(1)$.

All $f$-divergences satisfy the following properties:
\begin{enumerate}
\item \textit{positive definiteness}:  $\D(\bv{\p} ; \bv{\m}) \ge f(1) = 0$, with equality if and only if $\bv{\p} = \bv{\m}$;
\item \textit{Markov convexity}:
$\D_f\bigl(\tilde{\bv{\p}}; \tilde{\bv{\m}}\bigr) \le D_f\bigl(\bv{\p}; \bv{\m}\bigr)$,
where $\tilde{\bv{\p}}$ and $\tilde{\bv{\m}}$ are transformed from $\bv{\p}$ and $\bv{\m}$ respectively by the same stochastic transition matrix $\Theta^{k}_{\;j}$, and the equality holds if and only if the transition is associated with a sufficient statistic for $(\bv{\p}, \bv{\m});$\footnote{Meaning that, given the transition matrix $\Theta^{k}_{\;j}$, $\bv{\p}$ can be inferred unambiguously from $\tilde{\bv{\p}}$, and $\tilde{\bv{\m}}$ from $\bv{\m}$.  Note that re-labeling symmetry is an important special case of this property.}
\item \textit{joint convexity}: for $0 \le \lambda \le 1$, 
\be
\D_f\bigl( \lambda \bv{\p} + (1-\lambda) \bv{\p}'; \lambda \bv{\m} + (1-\lambda)\,\bv{\m}'\bigr) \le \lambda \,\D_f\bigl(\bv{\p} ; \bv{\m} \bigr) + (1-\lambda) \D_f\bigl( \bv{\p}'; \bv{\m}' \bigr)\
\ee
when $\bv{\p}$, $\bv{p'}$, $\bv{\m}$, $\bv{\m}'$ are all probability distributions over the same space.
\end{enumerate}

Some commonly-encountered examples are listed in Table~\ref{f_divergences}.  The $\chi^2$-divergence
\be
\D_{\chi^2}(\bv{p}; \bv{m}) = \sum\limits_{k = 1}^{N} m_k \bigl( \tfrac{p_k}{m_k} - 1\bigr)^2 =  \sum\limits_{k = 1}^{N} \frac{(p_k - m_k)^2}{m_k}
\ee
is commonly encountered in statistics applications, more out of historical precedent and continued convenience than anything else more fundamental. 
The \textit{total variation distance} represents a standard $\mathcal{L}^1$ distance on the space of normalized, nonnegative probability distributions  over the $N$ exclusive and exhaustive ``elementary'' outcomes,
\be
\D_{\stext{TV}}(\bv{p}; \bv{m}) = \tfrac{1}{2}\sum\limits_{k=1}^{N} \,\bigl\lvert p_k - m_k \bigr\rvert,
\ee
but can also be interpreted as the 
the maximum absolute difference in probability assigned by the two distributions to any (possibly compound) event.  The Hellinger distance
\be
\D_{H}(\bv{p}; \bv{m}) = \sum\limits_{k = 1}^{N} \bigl( \sqrt{p_k} - \sqrt{m_k} \bigr)^2
\ee
is a standard $\mathcal{L}^2$ Hilbert-space metric, but using the square roots of the probabilities rather than the probabilities themselves, so is somewhat reminiscent of quantum mechanics, as mentioned earlier.  As alluded to above, the Amari $\alpha$-divergences are the only $f$-divergences compatible with an arguably natural differential geometric structure on the manifold of probability distributions, and amongst these, the Kullback-Leibler divergences (corresponding to $\alpha = \pm 1$) enjoy additional information theoretic properties, while the $\alpha = 0$ case is the only true metric, and is proportional to the Hellinger distance.

\begin{table}[h!]
\begin{tabular}{|c|cl|}
\hline
Type of $f$-Divergence & Generating Function & $f(x)$  \\
\hline\hline
Kullback-Leibler divergence &  $\kappa \, x \, \log x$  & $(\kappa  > 0)$  \\
\hline
\;\; dual Kullback-Leibler divergence\;\; & $-\kappa \,\log x$ & $(\kappa  > 0)$  \\
\hline
Hellinger distance & $( \sqrt{x} - 1)^2$ & or \\
 \mbox{}  & $2(1 - \sqrt{x})$ & \\
\hline
total variation distance  $\vphantom{\Bigl[}$ & $\tfrac{1}{2} \lvert x - 1\rvert$ & \\
\hline
$\chi^2$ divergence  $\vphantom{\Bigl[}$ & $(x-1)^2$ & or \\
 \mbox{}  & $x^2 - 1$ & \\
\hline
dual $\chi^2$ divergence  $\vphantom{\Bigl[}$ & $x(1-x\inv)^2$ & or \\
 \mbox{}  & $x\inv - x$ & \\
\hline
\mbox{} & 
$-\ln x$ & if  $\alpha = -1$  \\
Amari $\alpha$-divergences & $\tfrac{4}{1 - \alpha^2} \bigl( 1 - x^{\frac{1 + \alpha}{2}} \bigr)$ & if $\lvert \alpha \rvert \neq 1$ \\
\mbox{}  & $x \ln x$ & if $\alpha = +1$ \\
\hline
\end{tabular}
\caption{Csisz\'ar $f$-divergences commonly encountered in probability theory.  Given two probability distributions $\bv{\p} = (\p_1, \dotsc, \p_N)$ and $\bv{\m} = (\m_1, \dotsc, \m_N)$ over the same space of $N$ mutually exclusive and exhaustive outcomes,  the $f$-divergence is given by $\D_f(\bv{\p} ; \bv{\m}) = \sum_{k = 1}^{N} \m_k\, f\bigl( \tfrac{\p_k}{\m_k}\bigr)$. The generating function $f(x)$ for any one type of divergence between normalized distributions is not unique, but these representatives are perhaps the simplest. The family of Amari $\alpha$-divergences are closely related to what are known as R\'enyi divergences, which in turn are based on the R\'enyi entropy, a generalization of the Shannon entropy.  The Amari divergences for $\pm \alpha$ are duals.  Notice that the Kullback-Leibler divergences are special limiting cases of the Amari $\alpha$-divergences corresponding to $\alpha \to \pm 1$ (provided we measure the entropies in nats rather than bits), while the Hellinger distance corresponds (apart from an overall factor of $2$) to the self-dual case $\alpha = 0$.}\label{f_divergences}
\end{table}
 
Although certain types of $f$-divergences (like the Hellinger distance and the total variation distance) are true metrics, in general an $f$-divergence need not be symmetric under interchange of $\bv{\p}$ and $\bv{\m}$, and need not satisfy a triangle inequality.  But given a generating function $f(x)$, notice that we can naturally define a \textit{dual} generating function by $\tilde{f}(x) = x f( x\inv )$, which satisfies $\tilde{f}(1) = f(1) = 0$ while sharing the same convexity properties as $f(x)$ itself.\footnote{This is easy to see if $f(x)$ is twice-differentiable, for then $\tilde{f}''(x) = x^{-3} f''(x\inv) \le 0$ whenever $x > 0$.}  The corresponding dual $f$-divergence just reverses the roles of the distributions $\bv{\p}$ and $\bv{\m}$:
\be\label{fdivergence2}
\D_{\tilde{f}}(\bv{\p}; \bv{\m}) = \D_f(\bv{\m}; \bv{\p}).
\ee
Since two successive interchanges of $\bv{p}$ and $\bv{m}$ should do nothing, this duality relation had better be idempotent, and indeed we find $\tilde{\tilde{f}}(x) = x \tilde{f}( x\inv) = x \bigl[x\inv f\bigl( (x\inv)\inv \bigr) \bigr]= f(x)$.  An $f$-divergence is symmetric (with respect to interchanging \textit{any} pairs of distributions $\bv{\p}$ and $\bv{\m}$) if  and only if it can be generated by an $f(x)$ that is self-dual in the sense that $\tilde{f}(x) = f(x) + \gamma\, (x-1)$ for some constant $\gamma$, and for all $x > 0$.  The Kullback-Leibler number $\K$ and dual Kullback-Leibler number $\tilde{\K}$ are dual but not self dual in just this sense, while the Hellinger distance is self-dual, as is the total variation distance.  In general, the Amari $\alpha$-divergence is dual to the $(-\alpha)$-divergence.

While we maintain that various information-theoretic considerations single out the relative entropy as particularly fundamental, (constrained) minimization of any $f$-divergence between the indirect and direct polling probabilities could be used as a workable Congressional apportionment method.  Indeed, several existing methods can be cast in terms of minimizing a suitable $f$-divergence between the direct (democratically ideal) distribution $\bv{m}$ and indirect (Congress-mediated) polling probability $\bv{p}$.  In particular, the identric mean method is associated with the Kullback-Leibler divergence, the logarithmic mean method with the dual Kullback-Leibler divergence, the Webster-Willcox method with the $\chi^2$ divergence,  the Huntington-Hill method with the dual $\chi^2$ divergence, 
and Hamilton's method with the total variation distance.\footnote{Instead of or In addition to relying on what we regard as dubious arguments based on satisfying quota, the Hamilton method could have been motivated by the property that it minimizes the absolute value of the difference between the net indirect and direct weights of representation for the \textit{worst-case} grouping of individuals, consistent with other constraints.}

Any apportionment method based on minimizing an $f$-divergence does enjoy the following appealing feature:
Our desideratum to focus on the weights of representation across represented individuals, rather than the entitlements of states, dictates that we try to minimize some measure of the difference between the direct polling probabilities $\m_{sn} = \tfrac{1}{P}$ and the indirect polling probabilities\footnote{Once again, we are assuming ideal sampling within states, since the district partitioning is not known as this stage.} $\bar{\pi}_{sn} = \tfrac{1}{R} \tfrac{1}{\bar{d}_s} = \tfrac{a_s}{R} \tfrac{1}{p_s}$, for all represented individuals in all states.

But one could also just try to minimize a discrepancy between a state-level sampling distribution $(\tfrac{a_1}{R}, \dots, \tfrac{a_S}{R})$ and the democratic ideal $(\tfrac{q_s}{R}, \dots, \tfrac{q_s}{R})$.  In general, using the same measure of discrepancy, we would arrive at different answers for the optimal choice of the $a_s$.  While in our view the former is clearly the correct approach in principle, it is reassuring that if we use an $f$-divergence, then we must obtain the same optimal apportionment $\bv{a}$ either way.  This is because for these polling distributions,
\be
\sum\limits_{s = 1}^{S} \sum\limits_{n=1}^{p_s} \tfrac{1}{P} \, f\bigl(\tfrac{\frac{a_s}{R}\frac{1}{p_s}}{\frac{1}{P}} \bigr) 
= \sum\limits_{s = 1}^{S} \tfrac{p_s}{P} \, f\bigl(\tfrac{a_s P}{p_s R}  \bigr)
= \sum\limits_{s = 1}^{S} \tfrac{1}{R} \tfrac{p_s R}{P} \, f\bigl( a_s \tfrac{P}{p_s R}  \bigr) 
= \sum\limits_{s = 1}^{S} \tfrac{q_s}{R} \, f\bigl(\tfrac{a_s}{q_s}  \bigr) 
= \sum\limits_{s = 1}^{S} \tfrac{q_s}{R} \, f\bigl(\tfrac{a_s/R}{q_s/R}  \bigr),
\ee
so the \textit{same} $f$-divergence can be written using the same generating function either at the level of individual weights of representation, or in terms of the state sampling probabilities, as long as we assume ideal (district-averaged) sampling probabilities within states.

However, several examples of the ``dual'' generating functions, including those associated with the dual entropic and the Huntington-Hill methods, suffer from a mathematical divergence $\lim\limits_{x \to 0^+} f(x) = +\infty$ and so would deliver an infinite penalty if some $a_s = 0$ but $q_s > 0$, and hence cannot generally be used for applications such as party-list voting, and thus are disfavored according to our universality desideratum.   In contrast, certain ``primal'' divergences, including those associated with the entropic and Webster apportionment methods, instead satisfy $\lim\limits_{x \to 0^+} f(x) < +\infty$, but $\lim\limits_{x \to +\infty} f(x) = +\infty$, and would assign infinite penalty only if some $q_s = 0$ but $a_s > 0$.  Our view is that this latter behavior is more reasonable in a variational principle generating apportionments.  After all, if it is considered undesirable or unacceptable for groups with positive quotas to receive nothing, it is straightforward to enforce lower bounds with explicit constraints during optimization.  But it would be not just undesirable, but non-sensical, to give seats or votes to groups with zero quotas---Congressional apportionment in the U.S.\ should never assign representatives to Canadian provinces, for example.

The Kullback-Leibler ($\K$-based) and Webster ($\chi^2$-based) approaches differ most noticeably in how these measures decompose under different levels of aggregation or subdivision.  Suppose for instance that the subsets $\mathscr{S}_r$, for $r = 1, 2,\dotsc$, partition the states $\{1, \dotsc, S \}$ into non-empty, exclusive and exhaustive groups,\footnote{Since the sets are exhaustive, there must be at least one set in the partition, and since we are requiring the subsets to be both non-empty and exclusive, there can be at most $S$ sets.} say, by regional affiliation, or historical electoral college voting patterns, or really any criterion.  The figure-of-merit for the Webster method can be thought of as a population variance of the individual weights of representation, and can be decomposed into between-group  and within-group contributions, as follows:
\be
\begin{split}
\tfrac{1}{P} \sum\limits_{s =1}^{S} \sum\limits_{n =1}^{p_s}  \bigl( \tfrac{a_s}{R} \tfrac{1}{p_s} - \tfrac{1}{P}  \bigr)^2 &=
\sum\limits_r  \tfrac{\nu_r }{P}  \bigl( \tfrac{\sigma_r}{R} \tfrac{1}{\nu_r}  - \tfrac{1}{P}  \bigr)^2
+
\sum\limits_r   \tfrac{\nu_r}{P} \sum\limits_{s \in \mathscr{S}_r} \tfrac{p_s}{\nu_r} \bigl( \tfrac{a_s}{R} \tfrac{1}{p_s} - \tfrac{\sigma_{r}}{R} \tfrac{1}{\nu_r}  \bigr)^2  \\
&=  \tfrac{1}{P}\sum\limits_r  \tfrac{1}{\nu_r }  \bigl( \tfrac{\sigma_r}{R}  - \tfrac{\nu_r}{P}  \bigr)^2
+
\tfrac{1}{P} \sum\limits_r    \tfrac{\sigma_r}{R} \sum\limits_{s \in \mathscr{S}_r} \tfrac{1}{p_s}  \tfrac{\sigma_r}{R}  \bigl( \tfrac{a_s}{\sigma_r}  -  \tfrac{p_s}{\nu_r}  \bigr)^2,
\end{split}
\ee
where
\bsub
\begin{align}
\sigma_r &= \sum\limits_{s \in \mathscr{S}_r} a_s,  \;\;\text{ and } \\
\nu_r &= \sum\limits_{s \in \mathscr{S}_r} p_s
\end{align}
\esub
are, respectively, the number of seats and the number of represented individuals associated with the $r$th group.   But the weighted averaging that emerges, as required to maintain the equality, is not truly recursive. And while all terms in the first and second formulations are easily interpretable, the weights appearing in the last expression are a bit hard to understand. 

In contrast, the relative entropy can be decomposed recursively as
\be
\begin{split}
\sum\limits_{s =1}^{S} \sum\limits_{n =1}^{p_s} \tfrac{a_s}{R}\tfrac{1}{p_s} \log\bigl[ \tfrac{a_s}{R} \tfrac{P}{p_s}  \bigr] &=
\sum\limits_{s =1}^{S}  \tfrac{a_s}{R} \log\bigl[ \tfrac{a_s/R}{q_s/R} \bigr] \\
&= \sum\limits_r  \tfrac{\sigma_r}{R} \log\Bigl[ \tfrac{ \frac{\sigma_r}{R}}{ \frac{\nu_r}{P}} \Bigr]
+ \sum\limits_{r}  \sum\limits_{s \in \mathscr{S}_r} \sum\limits_{n =1}^{p_s}  \tfrac{a_s}{R} \tfrac{1}{p_s} \log \Bigl[ \tfrac{ \frac{a_s}{R}\frac{1}{p_s}  }{ \frac{\sigma_r}{R}\frac{1}{\nu_r} } \Bigr] \\
&= \sum\limits_r  \tfrac{\sigma_r}{R} \log\bigl[ \tfrac{\sigma_r/R}{\nu_r/P} \bigr]
+ \sum\limits_{r}  \tfrac{\sigma_r}{R} \sum\limits_{s \in \mathscr{S}_r} \tfrac{a_s}{\sigma_r} \log \bigl[ \tfrac{a_s/\sigma_r}{p_s/\nu_r} \bigr],
\end{split}
\ee
which provides a more natural sum of between-group and averaged within-group contributions.  Furthermore, the same sort of recursivity holds if we refine rather than course-grain the partitions, so that relative entropy provides a fully consistent and universal measure of discrepancy between ideal and actual weights of representation within individual states, as well as between states or groups of states, or any other subdivisions of the population that might be of interest.

Another unique feature for the relative entropy is that it is the only measure on the space of normalized, nonnegative probability distributions that is both an $f$-divergence and a Bregman divergence.  Moreover, both of these divergences are generated from the same function, in that $\K(\bv{\p}; \bv{m})$ is the $f$-divergence given by $\sum\limits_{j=1}^{N} \p_j \, f\bigl( \tfrac{\p_j}{m_j}  \bigr)$, and the Bregman divergence is associated with a  Taylor expansion of $F(\bv{\p}) = \sum\limits_{j=1}^{N} f(\p_j)$, both divergences involving the same function $f(x) = \kappa \,x \log x$.


\newpage
\section{\uppercase{Lorenz Curves and Inequality Indices}}\label{inequality}

An extensive scholarship on measurement of \textit{distributional inequality} in income, wealth, or other resources provides yet another route to the entropic figure-of-merit \cite{dalton:1920,foster:1964,theil:1967,elteto:1968,atkinson:1970,rothschild:1973,allison:1978,bourguignon:1979,shorrocks:1980,young:1985,young:1994,tsui:1999,lugo_maasoumi:2008,cowell:2011}.  We start with a review of the Lorenz curve and its properties, and then discuss quantitative, summary measures of the inequality embodied in an asset distribution or its associated Lorenz curve, finally settling on the Theil, or entropic, index, as the most natural.

\subsection{The Lorenz Curve as a Visual Representation of Inequality}

What is now called a \textit{Lorenz Curve} was introduced in 1905 by American economist M.O.\ Lorenz to represent and visualize concentration or inequality of wealth or income distributions, but it can also be used for distributions of essentially any (additive, divisible, ratio-scale) asset or resource, including representational weight in a legislative government.\footnote{Other examples include distributions of financial debt or energy consumption across households, racial or political segregation, biomass or reproductive fecundity of organisms or biodiversity of species in evolutionary ecology, prevalence of word use in a language, or sizes of cities in a country, etc.}  

Suppose we rank order all individuals in a population by asset level, from lowest (asset-poorest) to highest (asset-richest). For each rank, we may determine the proportion of the population at that rank or below, and also the proportion of the total assets controlled by individuals at that rank or below.  A Lorenz curve itself plots the first fraction against the second fraction, so that $L(F)$ specifies what fraction of the total assets are possessed, controlled, or associated with units constituting a certain fraction $F$ of the population, measured up from the bottom of the distribution when the units are sorted in ascending order from asset-poorest to asset-richest.\footnote{The Lorenz curve is hence also closely related to the concepts of ``rank-size''  or ``rank-frequency'' distributions.}  Considering, for example, the wealth distribution across all households in a certain country at a certain time, $L(0.8) = 0.2$ would mean that the bottom $80\%$ of households together possess only $20\%$ of the wealth, the prototypical ``Pareto'' rule.

A Lorenz curve can be used to characterize either a continuous univariate probability distribution, a discrete probability distribution, or some finite population of discrete units.\footnote{Depending on the context, these observed individuals may or may not be considered a random sample from a larger super-population, in a statistical sense}  Though much of the theory has developed for the continuous case, we are interested in the latter case, which requires a bit of care in its construction if we are to usefully treat populations where many different individuals may have exactly the same asset level, without introducing excessive course-graining into the curve.  Here we will rely on a standard definition based on piecewise linear interpolation.\footnote{Other interpolating functions (such as exponentials or power laws) could be better justified, but would make little visible difference on the scales at which we would ever plot a Lorenz curve for the represented U.S.\ population, with hundreds of millions of individuals.}

\subsubsection{Empirical Lorenz Curve for Voting Weight Distributions}

Though ultimately interested in measuring inequality in political representation, in order to emphasize how this task fits squarely within a larger class of applications, we will talk more or less interchangeably in terms of distribution of  ``representational weight'' or more generally of any  ``resource'' or ``asset''  which is quantifiable on a nonnegative, ratio scale, but  also is additive across individuals, such that the asset level of any identifiable group is the sum of the asset levels associated with the members of that group, and the \textit{proportion} of assets associated with a group is the sum of their assets divided by the total amount of assets within the observed population as a whole.

Consider some nonnegative resource levels $\bv{x} = (x_1, \dotsc, x_P)$ distributed amongst a population of $P$ individuals or units.  The corresponding \textit{fractional asset shares}, or \textit{allotment proportions}, are
\be
\rho_j = \frac{x_i}{P \bar{x}} \text{ for } j = 1, \dots, P,
\ee
where
\be
\bar{x} = \tfrac{1}{P}\sum\limits_{j=1}^{P} x_j
\ee
is the overall average asset level within the population.  The fractional shares therefore satisfy
\bsub
\begin{align}
\rho_j &\ge 0,  \\
\sum\limits_{j=1}^{P} \rho_j &= 1,
\end{align}
\esub
and, if convenient, can be interpreted as probabilities, associated with random sampling of elementary units of the assets, and asking to whom they belong.  However, we emphasize that a probabilistic interpretation is not required for these proportions.

In the context of representational equality, the population consists of $P = \sum\limits_{s=1}^{S} p_s$ represented individuals across all states, and the ``assets'' may be taken to be the corresponding weights of representation:
\be
\bv{x} = \bv{w} = (w_{1{1}}, \dotsc, w_{1{p_1}}, \dotsc, w_{S{1}}, \dotsc, w_{S{p_S}}),
\ee
with population average
\be
\bar{x} = \bar{w} = \tfrac{1}{P}.
\ee
At the pre-districting stage, all represented individuals within a given state will have the same (averaged) weight of representation, but here we can consider an arbitrary distribution.  These weights have already been normalized by construction, but we would obtain the same curve if we started instead with the (unnormalized) shares of a representative, and then took care to consider the allotment proportions---that is to say, the Lorenz curve will be invariant under mere changes in the units in which we measure the assets.

Individual asset levels (counted according to their multiplicity) can be re-sorted into non-decreasing order,
\be
\min\limits_{j} \{ x_{j} \} = x_{(1)} \le x_{(2)} \le \dotsb  \le x_{(P-1)} \le x_{(P)} = \max\limits_{j} \{ x_j \},
\ee
which are known as the empirical \textit{order statistics}.  In our case, $x_{(\ell)} $ denotes the representational weight of the $\ell$th least-represented (or more generally, least-resourced) individual.  In the event of multiple individuals with the same weight, we may add some tiny random jitter or adopt some convention for mapping between these sorted weights and the actual individuals so represented, although the actual Lorenz curve will be independent of such choices.

Next, we consider quantiles of the ranked population, and also fractions of the aggregated assets associated with these sub-populations.  Specifically , for non-negative integer $\ell$, let 
\be
\Phi_{\ell} =
\begin{cases}
0 & \text{ if } \ell \le 0 \\
\frac{\ell}{P}  &\text{ if }  1 \le \ell \le P \\
1 & \text{ if } \ell \ge P
\end{cases}
\ee
denote the \textit{cumulative} fraction of the population made up of the $\ell$ least-resourced individuals, and define
\be
\Lambda_{\ell} = \begin{cases}
0 & \text{ if } \ell \le 0 \\
\sum\limits_{j=1}^{\ell} \rho_{(j)} = \tfrac{ \sum\limits_{j=1}^{\ell} x_{(j)} }{ \sum\limits_{j=1}^{P} x_{(j)} } = \sum\limits_{j=1}^{\ell} x_{(j)} &\text{ if }  1 \le \ell \le P \\
1 & \text{ if } \ell \ge P
\end{cases}
\ee
as the \textit{cumulative} proportion of weight\footnote{Representational weight $w_{sn}$ was already normalized to unity, so the cumulative fraction is just equal to the sum.}  carried collectively by the $\ell$ least-resourced individuals.

Using these population data, we can define the empirical Lorenz curve $L(F)$ over the interval $0 \le F \le 1$ as the continuous, piecewise-linear interpolant passing through the points:
\be
(0,0), \,(\Phi_1, \Lambda_1),\, \dotsc, \, (\Phi_P, \Lambda_P),\, (1,1).
\ee
That is, $L(F)$ is constrained to satisfy
\be
L(\Phi_{\ell}) = \Lambda_{\ell},
\ee
and has constant slope in between these points.

\subsubsection{Basic Properties of the Empirical Lorenz Curve}

Over the interval $0 \le F \le 1$,  an empirical Lorenz curve $L(F)$ possesses the following properties:
\begin{enumerate}
\item $L(F)$ is continuous;
\item $L(F)$ is piecewise linear;
\item L(F) is non-negative;
\item $L(F)$ is non-decreasing;
\item $L(F)$ is (non-strictly) convex;
\item $0 = L(0) \le \Theta(F-1) \le L(F) \le  F \le L(1)  = 1$,
\end{enumerate}
where 
\be
\Theta(x) = \begin{cases}
1 & \text{ if } x \ge 0 \\
0 & \text{ if } x < 0
\end{cases}
\ee
is the right-continuous\footnote{That is, here we are adopting the convention $\Theta(0) = \!\! \lim\limits_{x\to 0^+} \! \Theta(x) = 1$.} Heaviside step function.

In the continuous case, $L(F)$ can be derived directly from a continuous cumulative distribution function (CDF) $\Phi(x)$, and, like $\Phi(x)$, can itself be interpreted as a CDF.  Either we can randomly sample individuals and ask whether their weights of representation fall within a certain range, as described by $\Phi(x)$, or we can randomly choose units of representation, and consider the probability that they will be associated with some subset of individuals,\footnote{In the discrete empirical case, the possibility of exact ties in the weight of representation complicates this picture somewhat.  We can view the empirical Lorenz curve as a sample estimate for the CDF of some underlying continuous probabilistic model of the possible realizations of populations, and/or we can imagine adding tiny amounts of random ``jitter'' to the individual weights to make it almost certain that no exact ties will actually occur.} using $L(F)$.

In any case, the Lorenz curve $L(F)$ encodes information about the representational inequality across the population.  We see that  $L(F)$ is always bounded (non-strictly) from below by the step-function $L(F) = \Theta(F-1)$, which is referred to as the curve of \textit{perfect inequality}, and would reflect a limiting case
where a negligible fraction of the population controls all of the weight.\footnote{In the case of a finite discrete population, the most unequal distribution is not quite a step function, but corresponds to state of affairs where one dictatorial individual has all of the representational weight.  The empirical Lorenz curve $L(F)$ would vanish for $0 \le F \le 1 -\tfrac{1}{P}$, then rise linearly form $0$ to $1$ over the interval $1 - \tfrac{1}{P} \le F \le 1$, which is quite steep if $P \gg 1$.}
 
$L(F)$ is bounded (non-strictly) from above by the straight line $L(F) = F$, known as the line of \textit{perfect equality}, wherein all individuals would possess exactly the same weight, such that the weight possessed by any group is exactly proportional to the size of the group.

The Lorenz curve cannot rise above the line of perfect equality, nor sink below the curve of perfect inequality.   Where it falls in between reflects the degree of inequality in the distribution over the corresponding range of population quantiles.

The Lorenz curve may displayed graphically, and/or characterized in terms of some reduced sets of descriptive parameters.

\subsubsection{Conventional Summary Statistics of the Lorenz Curve}

Traditionally, the two parameters in most common use to characterize the shape of the Lorenz curve are the \textit{Gini coefficient} and the \textit{Lorenz asymmetry coefficient}.  

In the case of continuous distributions, the \textit{Gini} coefficient is defined s the ratio of two areas, namely the area between the line of perfect equality and Lorenz curve, and the area between the curves of perfect equality and perfect inequality:
\be
\Gini = \tfrac{\int\limits_{0}^{1}\! dF\, \bigl[ F - L(F) \bigr]}{\int\limits_{0}^{1}\! dF\, F }
= 2 \int\limits_{0}^{1}\! dF\, \bigl[ F - L(F) \bigr] = 1 - 2\int\limits_{0}^{1}\! dF\,  L(F).
\ee
Introduced by Italian sociologist and statistician\footnote{He also had degrees in mathematics, biology, and law.} Corrado Gini in 1912 (along with several other variants), the Gini coefficient serves as a measure of departure from the line of perfect equality and hence emerges as an overall reflection, or summary statistic, of inequality within the population.  The Gini coefficient necessarily falls in the interval $0 \le \Gini \le 1$, and is thus a normalized measure of distributional inequality, where larger values correspond to greater inequality, and the upper and lower bounds are achievable only in the case of perfect inequality or perfect equality, respectively. 

By integrating the interpolated curve, or equivalently by approximating an integral using a discrete numerical quadrature rule, the Gini coefficient for a finite, discrete population can be written directly in terms of the empirical order statistics:
\be
\Gini = \frac{1}{2} \frac{ \sum\limits_{j=1}^{P}  \sum\limits_{\ell=1}^{P} \bigl\lvert x_{j} - x_{\ell} \bigr\rvert }{ \sum\limits_{j=1}^{P}  \sum\limits_{\ell=1}^{P} x_{j}}
 =  \frac{1}{2P} \sum\limits_{j=1}^{P} \sum\limits_{\ell=1}^{P} \bigl\lvert \rho_{j} - \rho_{\ell} \bigr\rvert
=  \frac{1}{P} \sum\limits_{j=1}^{P} \sum\limits_{\ell < j} \bigl[ \rho_{(j)} - \rho_{(\ell)} \bigr],
\ee
which is one-half the \textit{relative mean difference} in  weight across all pairs of individuals, which is a standard measure of statistical dispersion.\footnote{Sometimes the Gini coefficient $\Gini$ for a discrete population is defined with an additional pre-factor of $\tfrac{P}{P-1}$, so that it assumes the maximum value of unity for the most unequal achievable distribution.}

Many different Lorenz curves can share the same Gini coefficient.  A second, complementary parameter, the \textit{Lorenz asymmetry coefficient} (LAC), is often reported, and may be defined in the continuous case by
\be
\text{LAC} = \Phi(\bar{x}) + \Lambda(\bar{x}),
\ee
where $\Phi(\bar{x})$ and $\Lambda(x)$ represent, respectively, the fraction of the population with  assets at or below average, and the fraction of the total assets associated with this part of the population.  This parameter measures asymmetry with respect to a prospective \textit{axis of symmetry} defined\footnote{Note that the axis of symmetry is not itself a valid Lorenz curve, but every Lorenz curve intersects this line exactly once.} by $(1-F)$.  Some Lorenz curves are symmetric under reflection about this axis, but others are not.  In the case of Lorenz curves derived from continuous distributions, it can be shown that the Lorenz curve is parallel to the line of perfect equality (i.e., has slope $\tfrac{d}{dF}{L(F)} = 1$) at the population quantile corresponding to the average asset level, $x = \bar{x}$.  If $\LAC > 1$, then the point on the Lorenz curve which is parallel to the line of perfect equality lies above the axis of symmetry.  This reflects a situation where the inequality in the population is due primarily to a small number of very over-represented individuals.   If $\LAC < 1$, then this  point of parallelism lies below the axis of symmetry, indicating that the inequality is due primarily to an overabundance of very under-represented individuals. If $\LAC = 1$, then this point lies on the axis of symmetry, and in fact, it can be shown that the entire Lorenz curve is symmetric under reflections about this axis of symmetry.  Lorenz curves corresponding to the line of perfect equality, and the curve of perfect inequality are both symmetric in this sense, as are certain other distributions, such as the log-normal.

When the Lorenz curve derives from a finite population, though these exact theorems no longer hold, an empirical estimate for  $\LAC$ can still provides a useful measure of asymmetry.  We may define
\be
\Phi(x) = \sum\limits_{j} \Theta(x - x_{(j)})
\ee
as the empirical CDF giving the fraction of the population with weights of representation at or below $x$, then estimate the population quantile corresponding to the mean level of representation as
\be
F_{\bar{x}} = \lim\limits_{\epsilon \to 0} \tfrac{1}{2} [ \Phi(\bar{x} - \epsilon) + \Phi(\bar{x} + \epsilon) ], 
\ee
and finally use the approximation
\be
\LAC \approx  F_{\bar{x}}  + L(F_{\bar{x}})
\ee
as the empirical LAC.

\subsection{Inequality Metrics and their Characterizations}

The Gini coefficient is perhaps the most familiar amongst a wide class of measures of distributional inequality known as (income) \textit{inequality metrics}\footnote{Income inequality metrics are not typically true metrics in the mathematical sense, and may measure distributional inequality of assets other than monetary income.} or \textit{inequality measures}, or \textit{inequality indices}.  These may be axiomatically characterized in a manner similar to measures of uncertainty as explored earlier---and as we will see, the most reasonable measure again just corresponds to the relative entropy.

\subsubsection{Defining Properties}

Once more considering a population of size $P$ with associated nonnegative asset levels $\bv{x} = (x_1, \dotsc, x_P)$, an {inequality metric} $\I_P(\bv{x})$ is a mapping from possible asset distributions into the real numbers, which is intended to quantitatively summarize or characterize the level of inequality in the distribution, where larger values indicate more distributional inequality, in some prescribed sense. 

Such an inequality measure should satisfy certain plausible axioms if it is to serve its intended purpose.  In the literature (mainly in economics, but also in ecology and other fields), four basic properties are generally required of any inequality metric:
\begin{enumerate}
\item \textit{anonymity} or \textit{symmetry}: the inequality metric does not depend on the labeling of individuals within the population, but only on the \textit{multiset} of asset levels.

Exactly who gets which slice of cake should not affect our objective judgement of inequality in the cake-slicing.  That is, $\I_P(x_1, \dotsc, x_P) = \I_P(x_{\pi(1)}, \dotsc, x_{\pi(P)})$ for any permutation $\pi(\,)$ of the integers $\{1, \dotsc, P\}$;
\item \textit{asset homogeneity}, \textit{mean independence}, or \textit{scale-invariance}:  the inequality metric is independent of the \textit{units} in which the asset $x$ is measured.\footnote{Of course, the \textit{same} choice of unit must be adopted across all individuals, if comparisons are to be meaningful.}  Inequality of the slices of the cake should not depend on the absolute size of the cake, nor on whether we measure sizes of slices in cubic centimeters or cubic inches.

That is, if asset levels are all multiplied by a common positive factor, the inequality metric should be unchanged: $\I_P(\tau \bv{x}) = \I_P(\bv{x})$ for any $\tau > 0$;
\item \textit{population independence} or \textit{population replication-invariance}: the inequality metric should not depend intrinsically on whether the population is large or small in number---the overall degree of fairness of cake-cutting should be quantifiable independently of the number of cake-eaters.

In particular, if we imagine replacing every individual in the original population with the \textit{same} number $n$ of ``children,'' dividing up \textit{equally} the individual asset level of their ``parent,''  then the overall measure of inequality should be unchanged: $\I_{nP}( \tfrac{1}{n}\bv{x}, \tfrac{1}{n}\bv{x}, \dotsc, \tfrac{1}{n}\bv{x}) = \I_{P}(\bv{x})$;
\item \textit{(weak Pigou-Dalton) transfer principle}: If some assets are transferred from an initially asset-richer person to an initially asset-poorer person, conserving assets overall but without the initially richer person ending up poorer than where the initially poorer person began,\footnote{Some authors adopt a seemingly weaker principle, considering only  hypothetical exchanges that do not reverse the relative ordering of the two individuals exchanging assets.  But since we have already demanded permutation symmetry, these conditions are actually equivalent.} (and no other exchanges are made), then the inequality metric should not increase.  Taking a bit of cake from someone with a large slice and giving it to someone with a thin slice should improve matters.

That is, if two individuals start out with assets $x_j$ and $x_k$, and exchange assets so as to end up with $x'_j$ and $x_k'$, respectively, where $x'_j + x'_k = x_j + x_k$ and $\lvert x'_j - x'_k \rvert \le \lvert x_j - x_k\rvert$, and no other exchanges are made, then $\I_P(\bv{x}') \le \I_P(\bv{x})$.  A stronger version of this principle would make these inequalities strict, and say that an exchange bringing two individual strictly closer in asset level  should strictly decrease the level of inequality.\footnote{Note however that the \textit{sensitivity} to transfers (by how \textit{much} some pairwise transfer would tend to improve equality) can vary amongst different choices for the metric, and at different asset levels for a given metric.}
\end{enumerate}

All such inequality measures have the property that if the Lorenz curve $L_A(F)$ for one distribution $\bv{x}_A$ falls strictly below the Lorenz curve $L_B(F)$ for another distribution $\bv{x}_B$, except at the endpoints where they must agree, then the first distribution ($\bv{x}_A$) will be judged as more unequal than the second distribution ($\bv{x}_B$), in the sense of having a larger value for the inequality metric.  However, for more general cases, where Lorenz curves might cross in the interval $0 < F < 1$, even the \textit{ordinal} rankings of the associated resource distributions can depend on the choice of the inequality metric.  Some further criteria will be required to single out an inequality index.

\subsection{Additional Properties}

Other quite reasonable properties, sometimes but not always imposed on inequality metrics, include:
\begin{enumerate}[i.]
\item \textit{non-negativity} or \textit{zero lower bound}: the inequality metric is always greater than or equal to zero;  
\item \textit{egalitarianism of zero}: the inequality metric vanishes when all the assets are distributed exactly equally.

Together these two properties just define a natural reference point in the limit of no inequality;
\item \textit{nonnegative definiteness}: the inequality metric is nonnegative, and zero if and only if the distribution of assets is perfectly uniform;
\item \textit{upper boundedness}:  the inequality metric achieves a finite upper bound for a distribution of maximum inequality, where one individual controls all assets\footnote{Inequality metrics are sometimes, but definitely not always, normalized such that this upper bound is either unity, or some number which approaches unity in the limit as the population size approaches infinity.}.  A stronger version of the principle holds that the upper bound is realized \textit{only} for the maximally unequal distribution where all but one of the individual asset levels $x_j$ vanish;
\item \textit{continuity}:  small changes in any (non-zero) asset levels lead to small changes in the inequality metric $\I_P(\bv{x})$;
\item \textit{smoothness}: the inequality metric $\I_P(\bv{x})$ is continuously differentiable with respect to all non-zero asset levels.  A stronger version may demand more orders of continuous differentiability;
\item \textit{aggregativity}: when the population is partitioned into exclusive and exhaustive sub-groups, the overall inequality can be expressed self-consistently as a function of the inequality within the various subgroups and of aggregated characteristics (\ie, asset levels and numerical size) across the groups.

Suppose we cut a cake, then further sub-divide these slices.  The overall inequality should depend on the inequality of the original slices, and, recursively, on the inequality within each slice after the second round of cuts.

Whenever we are interested in distributional inequality, almost inevitably we may be curious 
about the interplay of within-group and between group differences.\footnote{For instance, when assessing income inequality, demographic factors of interest might include country or region of residence, age, gender, race, education, employment status, political affiliation, etc.}  Aggregativity  embodies this intuition that there should be a coherent functional relationship between the level of inequality in the whole population and inequality in its constituent parts;  

\item \textit{subgroup decomposability} or \textit{additive decomposability}:  a narrower version of the aggregativity principle demands that the within-group and between-group contributions to the overall inequality should be \textit{additive}---because otherwise, it would be prohibitively difficult to interpret these different components of the inequality, or even to call them components, and to ask or answer \textit{how much} of the inequality can be attributed to variation within groups versus between groups.

That is, if the population is arbitrarily partitioned into $g$ disjoint, non-empty, non-overlapping sub-groups, of respective positive sizes $\bv{n}= (n_1, \dots, n_g)$, where $n_1 + \dotsb + n_g = P$, and with respective intra-group asset distributions  $\bv{x}_k$ such that $\bv{x} = \bv{x}_1 \oplus \dotsb \oplus \bv{x}_g$, then the inequality metric can be decomposed as a sum of a between-group contribution and within-group contributions:
\be
\I_P(\bv{x}) = \I_{P}(\overbrace{\bar{x}_1, \dotsc, \bar{x}_1}^{n_1}, \dotsc, \overbrace{\bar{x}_g, \dotsc, \bar{x}_g}^{n_g}) + \sum\limits_{k=1}^{g} \Omega_{k}(\bv{x}; \bv{n}) \, \I_{n_k}(\bv{x}_k)
\ee
for some set of nonnegative weights $\Omega_k(\bv{x}, \bv{n})$, the $\bar{x}_k$ representing the intra-group averages
\be
\bar{x}_k = \tfrac{1}{n_g} \sum\limits_{j \in \EuScript{S}_k} x_{j}, 
\ee
where $\EuScript{S}_k \subset \{1, \dotsc, P \}$ , $k =1, \dotsc, g$ indicate the set of individuals belonging to the various sub-groups, such that $n_1 \bar{x}_1 + \dotsb + n_g \bar{x}_g = P \bar{x}$.

Additive decomposability means that the  total inequality of a population may be expressed as the weighted sum of the inequality within subgroups of the population plus the (course-grained) inequality existing between these groups, where each component of inequality is quantified using the same sort of metric.  But the nature and meaning of the weighting coefficients is as yet left unspecified.
\end{enumerate}

We will see that demanding all of these further properties, in addition to the basic four axioms, and making some further sensible restrictions on the nature of the weights in a decomposition, will uniquely characterize an  entropic measure of inequality.

\subsection{Utility-Based Approaches}

Some economists and philosophers argue that the choice of an inequality metric must involve at a \textit{normative judgement} as to whether one distribution is to be preferred to another, and not just an empirical observation of  differences in asset levels, and that therefore the choice of $\I(\bv{x})$ should depend on the nature of the population's overall utility function, known as a social welfare function.  We tend to disagree with this assertion, thinking that the amount of inequality in a distribution is actually a distinct concept from how psychologically or sociopolitically undesirable such inequality is or ought to be collectively viewed.  However, these approaches have led to some interesting mathematical results.

The basic idea is to adopt a utility or social welfare function which builds in a preference for equality of asset distribution, then take the inequality measure to decrease monotonically with increases in this utility function.

One simple, if extreme, measure is the so-called \textit{Rawls's maximin criterion},
\be
\EuScript{R}(\bv{x}) = 1 - \min\limits_{j} \{ \tfrac{x_j}{\!\bar{x}} \colon j = 1 \dotsc, P  \}
= 1 - P\, \min\limits_{j} \{\rho_j  \colon j = 1 \dotsc, P  \},
\ee
which for a fixed population and a fixed total level of assets is sensitive only to the status of the worst-off individual in the population.

More generally, Suppose $U(\bv{x})$ is a collective utility, or social welfare, function over individuals' assets, which is: (i) \textit{continuous} in each argument, (ii)  \textit{monotonically increasing} in each argument, (iii) \textit{symmetric} under any permutations of the asset levels, and (iv) \textit{locally equality-preferring}, which means that $U(x_1, \dotsc, x_P)$ can be written as 
\be
U(x_1, \dotsc, x_p) = \Xi( \psi(x_1), \dotsc, \psi(x_P) ),
\ee
where $\Xi(y_1, \dots, y_P)$ is a symmetric (\ie, permutation-invariant), jointly concave function of $P$ arguments, and $\psi(x)$ is a concave function of one argument.

As we see it, the problem from the point of view of actual psychological or economic utility is that there is no reason whatsoever to assume that the utility function should be symmetric, in the sense of all individuals sharing the same quantitative preferences for the asset in question.  Nevertheless, we can strip away this interpretation of the utility, and just regard it directly as a measure of equality.
 
Then it can be shown that if an asset distribution $\bv{x}_A$ leads to an a Lorenz curve $L_A(F)$ that never falls below and somewhere lies above the Lorenz curve $L_B(F)$ for another distribution $\bv{x}_B$ (over the same populations), then $U(\bv{x}_A) > U(\bv{x}_B)$.  In terms of pairwise transfers, if a distribution $\bv{x}_A$ can be obtained from distribution $\bv{x}_B$ by a sequence of pairwise transfers from relatively asset-richer individuals to relatively asset-poorer individuals, then again $U(\bv{x}_A) > U(\bv{x}_B)$.

In many discussions, the utility function is for simplicity taken to be additive, 
\be
U(x_1, \dotsc, x_p) = \psi(x_1) +  \dotsc + \psi(x_P)
\ee
where $\psi(x)$ is continuous, strictly monotonic, and strictly concave.\footnote{This is, of course, only a proper subset of the class of utility functions described above.}  Strict monotonicity and concavity assumptions imply that as an individual's nonnegative asset level $x$ increases, $\psi(x)$ increases but at an ever decreasing rate, so that $\psi'(x)$ decreases while remaining positive.

A further simplifying assumption is that of \textit{constant inequality aversion}, which says that the relative decrease in $\psi'(x)$ for a very small relative increase in asset level $x$ is independent of the absolute asset level $x$.  This is measured by an inequality aversion parameter $\epsilon \ge 0$, characterizing the strength of society's ``yearning for equality''  \textit{vis-\`a-vis} uniformly higher total income.  Therefore $\psi(x)$ would satisfy the differential equation
\be
\frac{x \,\psi''(x)}{\psi'(x)} = \epsilon,
\ee
or 
\be
\psi_{\epsilon}(x) \propto \psi_1 + \psi'_1 \,\frac{x^{1- \epsilon} - 1}{1 - \epsilon}
\ee
for some integration constants $\psi_1$ and $\psi'_1 > 0$.

In his pioneering work, Dalton \cite{dalton:1920} suggested that  inequality can be indexed by how how far the actual \textit{average} social utility per individual falls short of potential average social utility if all income were distributed equally:
\be
\I_D = 1 - \tfrac{\bar{\psi} }{ \psi(\bar{x})  },
\ee
where $\bar{\psi} = \tfrac{1}{P} \sum\limits_{j=1}^{P} \psi(x_j)$.

For the particular choice $\psi_1 = 0$, this leads to the \textit{Dalton} family of inequality measures:\footnote{These are closely related to the so-called \textit{Tsallis} generalized entropies.}
\be
D_{\epsilon}(\bv{x}) = 1 - \frac{ \tfrac{1}{P}\sum\limits_{j=1}^{P} x_j^{1-\epsilon} -1 }{\bar{x}^{1-\epsilon} -1}
\ee
 for any $\epsilon \ge 0$.
 
But the Dalton inequality measures are not invariant with respect to affine transformations of the utility function,\footnote{If utility functions are to be revealed by preference orderings amongst possible allocations (including randomized mixtures thereof), then they can only be determined modulo affine transformations $U(\bv{x}) \to a \, U(\bv{x}) + b$ (for any positive $a$), which would change the unknowable origin and overall scale.} and/or changes of units used to measure assets.  While such changes would not effect the ordering properties of the inequality measure applied to distributions with the same mean asset level (\ie, for a given $\epsilon$, all such choices would lead to the same ordinal ranking of different distribution across the same population and with the same mean), this behavior is unacceptable in a fundamental measure of inequality.

Atkinson \cite{atkinson:1970} suggested instead using
\be
\I_A = 1 - \frac{ \psi\inv(\bar{\psi}) }{\bar{x}}
\ee
as the inequality measure, and demanded full scale invariance, which restricts the possibilities to
\be
A_{\epsilon}(\bv{x}) = \begin{cases}
1 - \tfrac{1}{\bar{x}} \Bigl( \tfrac{1}{P}\sum\limits_{j=1}^{P} x_j^{1-\epsilon}  \Bigr)^{\frac{1}{1-\epsilon}} &\text{ if } 0 \le \epsilon < 1, \text{ or } \epsilon > 1 \\
1 - \tfrac{1}{\bar{x}} \Bigl( \prod\limits_{j=1}^{P} x_j  \Bigr)^{\frac{1}{P}} &\text{ if } \epsilon = 1
\end{cases},
\ee
involving relative differences between the arithmetic mean and so-called H\"older generalized means of order $(1-\epsilon)$.  As such, for a fixed asset distribution $\bv{x}$, $A_{\epsilon}(\bv{x})$ increases monotonically with $\epsilon$.

Again, the $\epsilon$ parameter can be interpreted as the ``inequality aversiveness'' of the measure, quantifying the amount of utility presumed to be gained from redistribution.  At $\epsilon = 0$, the index $A_0(\bv{x}) = 0$  is indifferent to any differences in asset levels across individuals. In the limit as $\epsilon \to \infty$, infinite utility can be gained by redistribution from any state of inequality to the state of perfect equality.  In between, as $\epsilon$ rises, $A_{\epsilon}$ tends to become more sensitive to changes at the lower end of the asset distribution, and as $\epsilon$ falls, $A_{\epsilon}$ tends to become more sensitive to changes at the upper end of the asset distribution.
 At $\epsilon = 1$,  a small relative increase in the asset level leads to the same  relative decrease in the underlying marginal utility, regardless of the asset level.  Here the Atkinson measure just reduces to the relative difference between the arithmetic and geometric means of the asset levels across the population, which sounds reminiscent of the entropic figure-of-merit---in fact $A_1$ is just a monotonic function of the relative entropy.
 
 
\subsection{Some Other Inequality Measures}

Besides the Gini index, Rawls index, and Atkinson indices, many other inequality measures (all satisfying the basic four axioms, at least weakly, and some satisfying some of the other axioms) have been proposed in economics, ecology, and other contexts, including:

\subsubsection{$20$-$20$ Ratio}

As the name suggests, the $20$-$20$ \textit{ratio} simply measures how much wealthier (by ratio) the richest $20\%$ of the population is compared to the bottom $20\%$.  This is not bounded from above,  and does not satisfy the strong transfer principle, and is otherwise rather \textit{ad hoc}, but may be useful in a descriptive sense.

Obviously, variants of this idea can be defined for other quantiles---for example, the \textit{decile dispersion ratio} compares the assets of the top $10 \%$ and bottom $10 \%$ of the population. Empirically, in a discrete population, interpolation\footnote{Following standard practice, we have used linear interpolation to define the empirical Lorenz curve, but a better choice might involve Pareto interpolation, or maximum entropy interpolation.} may be needed to estimate these indices if individuals do not happen to fall exactly at the indicated percentages and percentiles.
 
 Any of these quantile-based indices are easily interpretable, but fundamentally arbitrary, and insensitive to changes in the middle of the distribution.

\subsubsection{Palma Ratio}

The \textit{Palma ratio} is the ratio of asset share of the upper $10 \%$ of the population divided by the share of the poorest $40 \%$.  This is bounded in neither direction, and only satisfies a weak transfer principle.  It was motivated by an empirical observation that middle class incomes typically tend to represent about half of gross national income, while the other half is split between the richest $10 \%$ and poorest $40 \%$ in  manner reflective of political forces and the direction of sociopolitical alignment of the middle class.  From our perspective this is perhaps useful as a summary description, but highly \textit{ad hoc} as a fundamental measure.

\subsubsection{\'Eltet\"o-Frigyes indices}

When characterizing inequality using a ratio of assets controlled by different portions of the population, a less arbitrary dividing line might be the mean asset level itself.

Letting $\bar{x}$ continue to represent the overall mean asset level across the entire population, define $\bar{x}_{+}$ as the mean asset level of all individuals richer than the overall mean, and  
similarly define $\bar{x}_{-}$ as the mean asset level of all individuals poorer than the overall mean.
 
Then the unnormalized \textit{\'Eltet\"o-Frigyes} indices \cite{elteto:1968} are the ratios of these means, namely
\bsub
\begin{align}
u_{E} &= \tfrac{\!\!\!\bar{x}}{\bar{x}_{-}} , \\
w_{E} &=  \tfrac{\bar{x}_{+}}{\!\!\!\bar{x}},   \\
v_{E}  &=  \tfrac{\bar{x}_{+}}{\bar{x}_{-}} = u_E \, w_E,
\end{align}
\esub
which range in value from $1$ to $+\infty$.  Note that only two of these three quantities are independent.  It is convenient to instead use standardized indices,
\bsub
\begin{align}
u'_{E} &= 1 - \tfrac{1}{u_E}  =    \tfrac{\bar{x} - \bar{x}_{-}}{\bar{x}},  \\
w'_{E} &= 1 - \tfrac{1}{w_E}  =  \tfrac{\bar{x}_{+} - \bar{x}}{\bar{x}_{+}},   \\
v'_{E}  &=  1 - \tfrac{1}{v_E} \,= \tfrac{\bar{x}_{+} - \bar{x}_{-}}{\bar{x}_{+}} = 1 - (1 - u'_E)(1 - w'_E),
\end{align}
\esub
which lie between $0$ to $1$, inclusive.  However, these indices still only satisfy a weak version of the transfer principle, and are not decomposable.

\subsubsection{Coefficient of Variation}

The \textit{coefficient of variation} is defined as the square root of the variance of the possessed asset levels, divided by the mean asset level:
\be
C(\bv{x}) = \frac{ \Bigl[ \frac{1}{P} \sum\limits_{j=1}^{P} (x_j - \bar{x} )^2 \Bigr]^{\frac{1}{2}}}{\frac{1}{P} \sum\limits_{k=1}^{P} x_k  } = \Bigl[ \tfrac{1}{P} \sum\limits_{j=1}^{P} \bigl( \tfrac{x_j}{\bar{x}} - 1)^2 \Bigr]^{\frac{1}{2}} = \Bigl[ \tfrac{1}{P} \sum\limits_{j=1}^{P} \bigl( \tfrac{x_j}{\bar{x}})^2     \; - \,1\Bigr]^{\frac{1}{2}}.
\ee
The coefficient of variation  is not bounded from above, but that is not terribly troubling.  Its square is subgroup-decomposable (though not in a particularly natural way, in our view). 

\subsubsection{Hoover index}

The \textit{Hoover index} is the proportion of all assets which would have to be redistributed so as to achieve a state of perfect equality.  This is also known as the \textit{Robin Hood} index or the \textit{Shutz} index.  Graphically, in the case of continuous distributions, it can be expressed as the largest vertical separation between the Lorenz curve $L(F)$ and the line of perfect equality $F$.  It is automatically bounded between $0$ and $1$.

The Hoover index for a discrete population can be written in the simple form:
\be
H =  \frac{1}{2} \frac{ \sum\limits_{j=1}^{P}  \lvert x_{j} - \bar{x} \rvert }{ \sum\limits_{j=1}^{P} x_{j}}  =   \frac{1}{2P} \sum\limits_{j=1}^{P}  \frac{\lvert x_{j} - \bar{x} \rvert}{\bar{x}} = \tfrac{1}{2P} \sum\limits_{j=1}^{P} \lvert  \tfrac{x_{j}}{\bar{x}} - 1 \rvert ,
\ee
so is proportional to the \textit{relative mean deviation}.  Though simple to interpret, the Hoover index only satisfies a weak version of the transfer principle, since any exchanges between a pair of individuals both below the mean or both above the mean do not produce any change in $H$.  Also it is non-aggregative.

\subsubsection{Logarithmic variance}

Another commonly encountered inequality measure is the \textit{logarithmic variance}, 
\be
v_{L} = \tfrac{1}{P} \sum\limits_{j=1}^{P} \bigl( \ln \tfrac{w_j}{\!\bar{w}} \bigr)^2,
\ee
which is nonnegative definite, but is not bounded from above, and is not aggregative.\footnote{In contrast, the \textit{variance of the logarithms} of asset levels has also been used as an inequality measure, but does not satisfy even a weak transfer principle for exchanges between individuals at sufficiently high asset-levels, so is unacceptable.}

\subsubsection{Herfindahl-Hirschman index}

The \textit{Herfindahl-Hirschman} index (HII) is just given by the sum of the squares of the fractional asset shares:
\be
\text{HHI} = \sum\limits_{j=1}^{P}  \rho_j^2 = \sum\limits_{j=1}^{P} \bigl( \tfrac{x_j}{P \bar{x}} \bigr)^2 =  \tfrac{1}{P^2} \sum\limits_{j=1}^{P} \bigl( \tfrac{x_j}{\!\bar{x}} \bigr)^2,
\ee
and in, say, a wealth distribution, just represents the probability that two dollars sampled at random belong to the same individual.  Note that amongst distributions with the same mean asset level, this induces the same ordering as the coefficient of variation.

In ecology, this is known instead as the \textit{Simpson} index,\footnote{The complementary quantity $(1 - \text{HHI})$ is known as the Gini-Simpson diversity index, not to be confused with the Gini \textit{coefficient} defined above.} and is used as a measure of species diversity.  In microbiology, it is known as the Hunter-Gaston index.  In quantum theory, it is used as the conventional measure of \textit{purity} of the quantum state, and it is also related (by exponentiation) to the order-$2$ \textit{R\'enyi generalized entropy}, also known as the \textit{collision entropy}.  Indeed, these R\'enyi entropies can be used to define an entire family of inequality measures, as discussed next. 

\subsubsection{R\'enyi Divergences}

The R\'enyi entropies of nonnegative order $\alpha$,
\be
H_{\alpha}(\bv{\rho}) = \tfrac{1}{1-\alpha} \log \sum\limits_{j=1}^{P} \rho_j^{\alpha},
\ee
were introduced as generalizations of the Shannon entropy, which corresponds to the $\alpha \to 1^+$ limiting case of R\'enyi entropy.\cite{vanerven:2014}  

And just as the R\'enyi entropies generalize the Shannon entropy, the \textit{R\'enyi divergences} generalize the standard relative entropy (\ie, Kullback-Leibler divergence).\footnote{As in the standard Shannon-Kullback case, the R\'enyi divergence between a given distribution and the uniform distribution is equal to the difference in their R\'enyi entropies.  However, we continue to maintain that the standard expressions for entropy and relative entropy are better motivated and behaved  for most applications, including fair apportionment.}  We can use any of the R\'enyi divergences between the realized (or proposed) and ideal asset distributions as an inequality metric.  The divergence of order $\alpha$ becomes
\be
R_{\alpha}(\bv{x}; P) = \tfrac{1}{\alpha -1} \log \sum\limits_{j=1}^{P} \rho_j^{\alpha} (\tfrac{1}{P})^{1-\alpha}
=  \tfrac{1}{\alpha-1}  \log \Big[ \tfrac{1}{P} \sum\limits_{j=1}^{P} \bigl( \tfrac{x_j}{\!\bar{x}}\bigr)^{\alpha} \Bigl].
\ee
Although the R\'enyi entropies are only (meaningfully) defined for $\alpha \ge 0$, it turns out that the divergences can be extended\footnote{Again the $\alpha = 1$ case must be arrived at by a limiting procedure, and corresponds to the usual Kullback-Leibler divergence.} to all real-valued $\alpha$.  These divergences are nonnegative definite, scale-invariant, replication-invariant, and also all aggregative, but not in what we view as a natural way, and they lack desirable convexity properties for $\alpha > 1$. 

Note that these divergences are closely related to the Atkinson indices defined above, and to the generalized entropy indices, to be defined below, as well as to the Amari divergences discussed previously.

\subsubsection{Generalized Entropy Indices}

For a suitable choice of a \textit{kernel function} $h(x)$, a natural inequality index can be constructed as the difference of the average value of $h(x)$ in the perfectly uniform distribution, $\rho_j = \tfrac{1}{P}$ for $j = 1 \dotsc, P$, and the average of the same function under the actual distribution, $\rho_j = \tfrac{x_j}{P \bar{x}}$, $j = 1, \dotsc, P$. That is, we can define
\be
\I_h(\bv{x}; P) = \sum\limits_{j=1}^{P}  \tfrac{1}{P} \,  h\bigl( \tfrac{1}{P} \bigr) - \sum\limits_{j=1}^{P}  \tfrac{x_j}{P\bar{x}} \,  h\bigl( \tfrac{x_j}{P\bar{x}} \bigr) 
= \sum\limits_{j=1}^{P}  \tfrac{x_j}{P\bar{x}}  \Bigl[  h\bigl( \tfrac{1}{P} \bigr) -  h\bigl( \tfrac{x_j}{P\bar{x}} \bigr) \bigr]
\ee
Such averages are automatically scale invariant, but additional structure must be imposed on $h(x)$ if this is to function as an inequality index.  In particular, if $h\bigl( \tfrac{1}{P} \bigr) -  h\bigl( \tfrac{x_j}{P\bar{x}}\bigr)  = f\big(  \tfrac{x_j}{\bar{x}} \bigr)$ for some convex function $f(y)$ satisfying $f(0) = 1$, then
\be
\I_h(\bv{x}; P) = \tfrac{1}{P} \sum\limits_{j=1}^{P} \tfrac{x_j}{\bar{x}} f\big(  \tfrac{x_j}{\bar{x}} \bigr)
\ee
can be interpreted as nonnegative-definite Csisz\'ar $f$-divergence, and is then also automatically replication invariant.

The parameterized family of kernels
\be
h_{\alpha}(\rho) = \begin{cases}
-\ln \rho &\text{ if } \alpha = 0 \\
\phantom{+} \frac{1- \rho^{-\alpha}}{\alpha (\alpha-1)} &\text{ if } \alpha \neq 0,1 \\
\phantom{+} \rho \ln \rho &\text{ if } \alpha = 1
\end{cases}
\ee
leads to the \textit{generalized entropy} indices, or \textit{Shorrocks} \cite{shorrocks:1980} indices, 
\be
E_{\alpha}(\bv{x}) = \begin{cases}
\tfrac{1}{P} \sum\limits_{j=1}^{P} \ln \tfrac{\bar{\! x}}{x_j}  &\text{ if } \alpha = 0 \\
\tfrac{1}{\alpha(\alpha - 1)}  \tfrac{1}{P} \sum\limits_{j=1}^{P} \Bigl[ \bigl(\tfrac{x_j}{\bar{x}} \bigr)^{\alpha}  -1 \Bigr]  &\text{ if } \alpha \neq 0,1\\
\tfrac{1}{P} \sum\limits_{j=1}^{P}  \tfrac{x_j}{\! \bar{x}} \ln \tfrac{x_j}{\! \bar{x}} &\text{ if } \alpha = 1 
\end{cases} \;\;\;\;,
\ee
defined for any real value of the parameter $\alpha$, which reflects a weighting given to disparities between asset levels in different parts of the distribution.  In general, lower values of $\alpha$ correspond to more sensitivity to inequality in the lower tail of the asset distribution, and for higher values to greater sensitivity in the upper tail.\footnote{Most authors have advocated values corresponding to $\alpha \le 2$, since larger value show little ``concern'' with inequality except amongst the richest echelon of the population.  In contrast, the $\alpha \to -\infty$ limit focuses entirely on the very bottom of the distribution, and induces the same rankings as Rawls's maximin criterion.}

It can be shown that these generalized entropy indices are the \textit{only} family of additively decomposable inequality metrics.  

This family is monotonically related to the R\'enyi divergences, and includes monotonic functions of the Atkinson indices as special cases, as well as other familiar indices.  For example, $E_2(\bv{w})$ is related to the coefficient of variation and the HHI index, and $E_0(\bv{w})$ is the mean log deviation, or so-called Theil $L$-index.  $E_1(\bv{w})$ is our favored choice, known in this context as the Theil $T$-index, discussed next.
 
\subsubsection{Theil Index}

Aggregativity, and especially decomposability, are very natural but also very stringent requirements.  Indices, such as the Gini coefficient, lacking any aggregativity property can exhibit some quite counterintuitive behavior, where under some redistributions of assets, inequality within \textit{every} subgroup goes up, the inequality across groups remains the same, yet the overall measure of inequality goes down.\footnote{This is vaguely reminiscent of what in statistics is called ``Simpson's paradox.''}

In our view, it is sensible, even essential, to have an inequality metric that is in fact {additively} decomposable into within-group and between-group contributions, so that we can meaningfully address questions of how much of the total inequality can be attributed to 
within-group or between-group differences, and even speak of the proportions of different contributions to the overall inequality.  Hence the parameterized family of generalized entropy indices enjoys a privileged status, not shared by the Gini coefficient or other non-decomposable indices.

But demanding \textit{additive} decomposability does not by itself buy much unless the coefficients $\Omega_k$ weighing the within-group contributions can be interpreted in a sensible way.   So for example, while additive decomposition of the familiar variance of a distribution underlies much of the logic of standard linear regression and analysis of variance (ANOVA), the meaning of the weighting coefficients is not entirely transparent.

Indeed, in our view, about the only reasonable choice compatible with permutation, scale, and  replication invariance should make the within-group contribution a \textit{weighted average} of the inequality of the sub-groups.  As we have seen, in information theory, this property is called \textit{recursivity}.  Demanding additive decomposability (together with the four primary axioms) restricts options to the family of generalized entropy indices, while it turns out that insisting on the stronger recursivity requirement singles out a pair of special cases, the Theil indices.

Straightforward calculation reveals that the sub-group weighting coefficients appearing in the decomposition of the generalized entropy index $E_{\alpha}(\bv{x})$ of order $\alpha$ are given by
\be
\Omega_{\alpha_k} = \tfrac{n_k}{P} \bigl( \tfrac{\,\bar{x}_k}{\bar{x}} \bigr)^{\alpha},  \;\; k = 1, \dotsc, g.
\ee
These weights are always nonnegative, but can only be interpreted as probabilities when they are also suitably normalized,
\be
\sum\limits_k \Omega_{\alpha_k} = 1,
\ee
and that only happens when $\alpha = 0$ or $\alpha = 1$

Furthermore, it can be shown that the normalization deficit $(1 - \sum\limits_k \Omega_{\alpha_k})$ is directly proportional to the between-group component itself,  so these two indices are the \textit{only} (permutation, scale, and replication invariant) inequality metrics additively decomposable in a manner where the within-group inequality components can be said to be truly independent of the between-group contribution.

The $\alpha = 0$ case corresponds to the \textit{Theil} $L$-Index \cite{theil:1967},
\be
\mathcal{L}(\bv{x}) = E_0(\bv{x}) = \tfrac{1}{P} \sum\limits_{j=1}^{P} \ln (\tfrac{\bar{\! x}}{x_j} )
\ee
also known as the logarithmic mean deviation.  In terms of (normalized) representational weight, this can be written as
\be
\mathcal{L}(\bv{w}) = \sum\limits_{j=1}^{P}  \tfrac{1}{P}  \ln \bigl(\tfrac{1/P}{w_j}\bigr)
=  \tfrac{1}{P} \sum\limits_{j=1}^{P} \ln (\tfrac{1}{w_j} ) - \ln(P) =  \tfrac{1}{P} \sum\limits_{j=1}^{P} \ln (\tfrac{1}{w_j})  = 
\sum\limits_{j=1}^{P}  \tfrac{1}{P}  \ln\bigl(\tfrac{\frac{1}{P}}{w_j}\bigr) = \tilde{\K}(\bv{w}),
\ee
which is just the \textit{dual} Kullback-Leibler divergence, measured in nats, between the uniform and actual distribution.

The $\alpha = 1$ case corresponds to the \textit{Theil} $T$-index
\be
T(\bv{x}) = E_{1}(\bv{x}) = \tfrac{1}{P} \sum\limits_{j=1}^{P}  \tfrac{x_j}{\! \bar{x}} \ln (\tfrac{x_j}{\! \bar{x}} ),
\ee
and is often just referred to as \textit{the} Theil index.  As a measure of inequality of representational weight, the Theil $T$-index reduces to
\be
T(\bv{w}) = \sum\limits_{j=1}^{P}  w_j  \ln ( P w_j ) = \ln P - \sum\limits_{j=1}^{P}  w_j  \ln(\tfrac{1}{w_j} )= 
\sum\limits_{j=1}^{P}  w_j  \ln\bigl(\tfrac{w_j}{\frac{1}{P}}\bigr) = {\K}(\bv{w}).
\ee
The Theil $T$-index is none other than the Kullback-Leibler divergence, or relative entropy (measured in nats, rather than bits) of the asset distribution relative to the perfectly equal (uniform) distribution, or equivalently, can be interpreted as the Shannon redundancy.   It has been used as a measure of resource inequality, non-uniformity, compressibility, segregation or stratification, and ecological diversity.   Apart from a change of units, it is precisely the entropic figure-of-merit we have advocated for apportionment.

As a measure of income or wealth inequality, the Theil $T$-index can be interpreted as the uncertainty associated with randomly sampling dollars of income and asking to whom they belong.  In terms of communications theory, it can be thought of as indicating under-utilized information capacity which reduces the effectiveness of price signals, or as measuring the ``redundancy'' of assets in some individuals, implying scarcity in others.  A high value for the Theil $T$-index indicates that total income is not distributed evenly among individuals, in the same way that an uncompressed text file does make equal use of words or characters.

Proposed by Dutch econometrician Henri Theil in his 1967 book on economics and information theory \cite{theil:1967}, both of these eponymous indices satisfy strong versions of all of the properties of inequality metrics suggested above.  The weights in the decomposition of the Theil $L$-index (dual relative entropy) are the \textit{population} shares $\tfrac{n_k}{P}$ of the respective sub-groups, while the weights in the Theil $T$-index are the \textit{asset} shares $\tfrac{n_k \bar{x_k}}{P\bar{x}}$ of the sub-groups.

As general measures of distributional inequality, one of these two should be the more natural choice, but which?  As we are interested in the distribution of assets across individuals, and not individuals across assets, an asset-weighted average across groups seems preferable, and this suggests the $T$-index.\footnote{Shorrocks \cite{shorrocks:1980} instead argues for the $L$-index over the $T$-index.  He notes that a question like ``how much of the observed income inequality is due to age?'' can have (at least) two interpretations.  It might be seeking: how much less income inequality (as a proportion of the total) would be observed if age variation were the only source of income variation?  This suggests that we group the population into age brackets, and calculate the ratio of the between-group inequality level (which is the inequality with within-bracket differences ignored) to the total inequality level.  But it might instead be asking: by how much (as a proportion of the total) would inequality decrease if age-income correlations were eliminated?  This suggests using in the numerator the difference between the total inequality level in the original distribution and the inequality in a distribution where inequality within each age bracket is left unchanged, but the mean income levels across age brackets are equalized.  The latter distribution will then have no between-group inequality, but is numerically equal to the average within-group inequality in the original distribution for the Theil $L$-index, but not the Theil $T$-index, since adjusting the within-group mean income levels will change the asset shares but not the population shares of the groups.  Thus, the two ratios are always the same under either interpretation for the $L$-index, but not the $T$-index.  However, we do not see this is as a compelling argument---obviously different questions may have different answers.  It will be better to carefully specify what one is seeking than to choose an index on the grounds that it happens to given the same answer to two distinct questions and so saves us from thinking harder about which question we are actually asking.}  Additionally, the $L$-index can exhibit some odd behavior.  For example, suppose in income distribution $\bv{x}_A$, one person makes a million dollars per year, a million people each earn one dollar per year, and one person receives one cent per year.  In income distribution $\bv{x}_B$, one person makes a million dollars and one cent per year, a million people each earn one dollar per year, and one person gets nothing.  The Theil $L$-index judges distribution $B$ to be \textit{infinitely} more unequal than distribution $A$.

It may be helpful to see how decomposition works out in the case of the $T$-index.  Specifically, consider a partition of all individuals into $g$ exhaustive, non-empty, non-overlapping sub-groups of respective sizes $n_1, \dotsc, n_g$, and respective distributions $\bv{x}_1, \dotsc, \bv{x}_g$, as defined above.  The intra-group Theil index of the $k$th subgroup is
\be
T(\bv{x}_j) = \sum\limits_{j \in \EuScript{S}_k} \tfrac{x_j}{n_k \bar{w}_k} \ln (\tfrac{x_j}{\bar{w}_k}),
\ee
where the total amount of assets controlled by the $k$th group is $n_k \bar{x}_k$, so $\tfrac{x_j}{n_k \bar{x}_k}$ represents the fractional share of the intra-group assets controlled by individual $j$ within group $k$, and 
\be
\frac{ \frac{x_j}{n_k \bar{x}_k}  }{ \frac{\bar{x}_k}{n_k \bar{x}_k} }
= \frac{ \frac{x_j}{n_k \bar{x}_k}  }{ \frac{1}{n_k} } = \tfrac{x_j}{\bar{x}_k}
\ee
represents the ratio of this individual intra-group fractional share to the ideal fractional share under equal distribution of intra-group assets, which would of course just be
\be
\frac{ \tfrac{1}{n_k} (n_k \bar{x}_k)}{n_k \bar{x}_k} = \frac{1}{n_k}.
\ee 
\newpage
The overall Theil index can then be written as
\be
\begin{split}
T(\bv{x}) &=  \sum\limits_{k=1}^{n_g}  \tfrac{n_k \bar{x}_k}{P \bar{w}} \,  \ln \tfrac{\bar{x}_k}{\! \bar{x}}
+ \sum\limits_{k=1}^{g} \tfrac{n_k \bar{x}_k}{P \bar{x}}  \, T(\bv{x}_g) \\
&= T(\overbrace{\bar{x}_1, \dotsc, \bar{x}_1}^{n_1}, \dotsc, \overbrace{\bar{x}_g, \dotsc, \bar{x}_g}^{n_g})   +    \sum\limits_{k=1}^{g} \tfrac{n_k \bar{x}_k}{P \bar{x}}  \, T(\bv{x}_g),
\end{split}
\ee
wherein $\tfrac{n_k \bar{x}_k}{P \bar{x}}$ represents the fractional share of total assets controlled by the $k$th group, while
\be
\frac{ \frac{n_k \bar{x}_k}{P \bar{x}} }{ \frac{ n_k}{P} }  = \frac{ \frac{n_k \bar{x}_k}{P \bar{x}} }{ \frac{ n_k}{P} } = \frac{\bar{x}_k}{\! \bar{x}}
\ee
represents the ratio of this achieved fractional share to the ideal share assuming perfect proportionality (taking into account the sizes of the groups).

For fixed population size $P$, the Theil $T$-index satisfies
\be
0 \le T(\bv{x})  \le \ln(P),
\ee
with equality at the lower bound if and only if all individuals have exactly the same asset level, and  equality at the upper bound if and only if one individual controls all of the assets.\footnote{Each of its components separately satisfies analogous inequalities.}  Some have criticized the Theil $T$-index for not having an upper bound of unity, but this is actually a point in its favor---surely, if one individual is to control all of the assets, then the total amount of inequality should grow monotonically with the number $(P-1)$ of individuals who are thereby left with nothing.\footnote{And if we were to renormalize $T(\bv{x})$ by dividing by $\ln(P)$, it would no longer satisfy the population replication property, nor a natural decomposability property.}


%
%
%
\newpage
\section{\uppercase{Some Optimization Lemmas}}\label{optimization}

Arbitrary combinatorial optimization problems can become arbitrarily difficult,\footnote{Some, in fact, are known to be NP hard, and even NP complete.} but in certain special cases, we can exploit analogies to the more familiar mathematical task of optimizing smooth functions.  Here we verify a few basic results which are relevant to optimal allocation in the apportionment problem, where many of the most commonly suggested apportionment methods may be derived from a variational principle minimizing (subject to constraints) some concave, additive objective function representing a total error of apportionment.\footnote{Or equivalently, maximizing a figure-of-merit reflecting the degree of fairness or proportionality of apportionment.}

Here we discuss the problem in the context of constrained \textit{maximization}, because we find it easier to speak in terms of hill-climbing and possible gains to the figure of merit, but of course constrained maximization of a function $\phi(\bv{x})$ is equivalent to constrained minimization of the additive inverse $-\phi(\bv{x})$ of that same function.

\subsection{Constrained Optimization of Discrete, Additive, Concave Functions}

Suppose we seek a \textit{constrained maximum} of an additive,\footnote{Some sources on operations research or optimization theory instead refer to this property as \textit{separability} rather than \textit{additivity}, but the latter terminology seems more clear.} real-valued, discretely concave (down) function of \textit{integer} arguments, of the form:
\be
\phi( \bv{x} ) = \sum\limits_{j = 1}^{N} f_j(x_j),
\ee
for some choice of scoring functions  $f_j \colon \integersymbol \to \realsymbol$, where
$\bv{x} = (x_1, \dotsc, x_N)^{\text{T}} \in \A(N) \subseteq \integersymbol^N$ is an ordered $N$-tuple of integers taking discrete values inside some \textit{allowed} hyper-rectangle $\A(N) \subset \Z^{N}$ defined by
\be
\label{rectangle1}
-\infty < \lambda_j \le x_j \le u_j  < +\infty \hspace{18pt} \text{ for } j = 1, \dotsc, N,
\ee
where the lower bounds $\lambda_j$ and upper bounds $u_j$ can be taken to be integers,
and where the optimization is further subject to a single linear constraint, namely a requirement that
\be
\label{line1}
\sum\limits_{j = 1}^{N} x_j = X,
\ee
for some choice for the total budget $X \in \integersymbol$ (which must necessarily lie in the interval $\sum\limits_j \lambda_j \le X \le \sum\limits_j u_j$ if there is to be any consistent solution).  Together, the inequality constraints on the individual $x_j$ and the equality constraint\footnote{If the problem with an equality constraint on
$\sum\limits_{j} x_j$ is solvable, then inequality constraints can be accommodated in principle by considering in succession all possible values for this sum, within some allowed range.} on the total sum $\sum_j x_j$ define the \textit{feasible} set of possibilities.  We further assume each function $f_j(x)$ satisfies a discrete \textit{concavity} condition of the form:
\be
f_j(x+1) -  f_j(x) \le f_j(x) -  f_j(x-1) \hspace{18pt} \text{ if } \lambda_j < x < u_j ,
\ee
so that any gains from incrementing $x_j$ are subject to diminishing marginal returns.\footnote{This condition is implied by, but weaker than, the requirement that $\tfrac{\del^2}{\del x^2} f_j(x) \le 0$, but the latter inequality may be simpler to verify when the $f_j(x)$ functions can be defined for real-valued arguments, and are sufficiently smooth.}  Equivalently, this condition says that 
$f(x)$ lies above the arithmetic average of its neighboring values:
\be
f_j(x) \ge \tfrac{1}{2} \bigl[\, f_j(x-1) +  f_j(x+1) \,\bigr] \hspace{18pt} \text{ if } \lambda_j < x < u_j .
\ee
This discrete concavity condition implies that if $f_j(x^*)$ is maximal, then $f_j(x)$ must be non-decreasing in either direction moving away from $x = x^*$.

Such an optimization problem naturally arises when $X$ represents some fixed, discretized resource that is to be allocated amongst $N$ different groups, agents, regions, or categories, based on maximizing some additive utility function $\phi(\bv{x})$ exhibiting diminishing marginal utility, where fractional allocations are not possible or not allowed, and non-trivial lower and/or upper bounds on each share may be imposed.\footnote{In the context of utility theory, we would also naturally require that each $f_j(x)$ be an increasing function of $x$ over its allowed domain (more is better when it comes to utility, just typically at a decreasing marginal rate), but such a monotonicity property is not actually needed for our conclusions here, so is not assumed.}

\subsubsection{Greedy Is Global}

Under these conditions, we claim that the obvious sequential ``greedy'' algorithm, doling out one unit of resource at a time according to whichever $f_j(x_j)$ will be increased the most, finds not just a local but a global (constrained) maximum of $\phi(\bv{x})$.  That is, when the objective function $\phi(\bv{x})$ satisfies the conditions above,  by allocating each unit  in succession, and the constraints are mutually consistent, the following procedure constructs a sequence $\bv{x}(0), \dotsc, \bv{x}(n) = \bv{x}^*$ culminating in an allocation $\bv{x}^*$ corresponding to a constrained global maximum of $\phi(\bv{x})$ subject to $\lambda_j \le x_j \le u_j$, for all $j = 1, \dotsc, N$, and  to $\sum\limits_{j = 1}^{N} x_j = X$:
\begin{enumerate}[(1)]
\item initially (i.e., for stage $t= 0$), assign $x_j(0) = \lambda_j \text{ for each } j = 1, \dotsc, N$, and define $n = X - \sum\limits_j \lambda_j$; 
\item for $t= 1, 2, \dotsc$, let $x_j(t) =  x_j(t-1) + \delta_{jk(t)}$ , where $\delta_{jk}$ is the Kronecker delta, and the index $k = k(t)$ corresponds to that variable $x_k$ for which the change $f_k\bigl(x_k(t-1) +1\bigr) - f_k\bigl(x_k(t-1)\bigr)$ is \textit{maximal} (choosing randomly or by some other prescription in the event of an exact tie), amongst all remaining unsaturated variables $x_j$ for which $x_j(t-1) < u_j $;
\item repeat process (2) until $t = n$.
\end{enumerate}

If $n < 0$, the constraints are mutually inconsistent, and no solution at all will be feasible.  If $n = 0$, there is only one feasible solution satisfying all constraints, namely $\bv{x}^* = \bv{x}(0) = (\lambda_1, \dotsc, \lambda_N)\trans$.  But if $n > 0$, distinct allocations may be feasible, and at each iteration $t = 1, \dotsc, n$ as specified above, we define the corresponding ``greedy gains'' $\Delta \phi_1, \dotsc, \Delta\phi_n$ to the objective function $\phi(\bv{x})$ such that 
\be
\Delta \phi_i = \phi\bigl(\bv{x}(t)\bigr) - \phi\bigl(\bv{x}(t-1)\bigr) = f_k\bigl(x_{k(t)} (t-1) +1\bigr) - f_{k(t)}\bigl(x_{k(t)}(t-1)\bigr).
\ee
For simplicity, we refer to these as ``gains,'' even thought they are not necessarily positive if the $f_j(x)$ are not monotonically increasing functions.  Notice that the gain at any stage depends only on the one variable being incremented, and not any of the other variables.  The algorithm can identify and notify whether any randomization is needed to break a tie.  However, note that even if ties do occur, they can affect the \textit{final} answer $\bv{x}^* = \bv{x}(n)$ only if we in effect run out of units before all groups once tied receive another unit.

Furthermore, it must be the case that gains at subsequent stages of the greedy allocation are subject to diminishing marginal returns:
\be
\Delta \phi_1 \ge  \dotsb  \ge  \Delta \phi_{n}.
\ee
Why?  At the $t$th step, $\Delta \phi_{t}$ can never be larger than its immediate predecessor $\Delta \phi_{t-1}$ at the $(t-1)$th step:  if two successive increments are made to the same variable $x_k$, the later change in $\phi(\bv{x})$ cannot be larger than the preceding one, because by assumption every $f_j(x_k)$ satisfies the discrete concavity condition over its allowed domain.  If instead successive increments are made to distinct variables (say $x_k$, then $x_{k'}$ for $k' \neq k$), then the later gain in $\phi(\bv{x})$ arising from incrementing $x_{k'}$ cannot be larger than the earlier gain due to incrementing $x_k$, or else the later gain would instead have been selected at the earlier stage of the greedy allocation.

To verify that the greedy algorithm works as claimed, and leads to a global constrained optimum $\bv{x}^*$, we can actually prove a slightly stronger result, which will be useful for subsequent developments.  Suppose $\bv{x}'$ is any other feasible allocation.  We will infer that $\phi(\bv{x}') \le \phi(\bv{x}^*)$, by connecting $\bv{x}'$ to $\bv{x}^*$ along a path of feasible allocations generated by \textit{pairwise exchanges} of units of the resource, for which $\phi(\bv{x})$ is non-decreasing while the allocations remain feasible at each point along the path.  (Obviously, unit pairwise swaps automatically preserve the total budget $X$.  To be feasibility-preserving, they additionally must keep each $x_j$ between its allowed upper and lower bound at all times).

Evidently, in any feasible allocation $\bv{x}'$ distinct from $\bv{x}^*$, one or more of the individual variables $x'_j$ in $\bv{x}'$ must be greater than the corresponding variable $x^*_j$ in the greedy optimum $\bv{x}^*$, and to compensate, one or more  $x'_k$ must be less than the corresponding $x^*_k$.  (Otherwise, either the allocations were not actually distinct, or else the constraint fixing the total allocation $X$ could not be satisfied in both allocations).  Thus, the distinct feasible allocation $\bv{x}'$ can be connected to the greedy allocation $\bv{x}^*$ by some minimal number $\ell \le n$ of feasibility-preserving, unit pairwise exchanges $\Delta \bv{x}_{jk}$---that is, swaps transferring one unit at a time, maintaining consistency with all constraints at each step---resulting in a sequence of transitions $\bv{x}' = \bv{x}'(0) \to \bv{x}'(1) \dotsb \to \bv{x}'(\ell) = \bv{x}^*$ made to the independent variables, where at each successive step, one unit is removed from an $x_j$ for which $\lambda_j \le x^*_j <  x'_j \le u_j$ before the swap, and transferred to an $x_k$ for which $\lambda_k \le  x'_k <   x^*_k  \le u_k$ prior to the swap.  The constructed path will be of minimal length (measured in numbers of pairwise swaps), in the sense that only the variables that need to be changed are changed to get from $\bv{x}'$ to $\bv{x}^*$, and each such variable is either incremented or decremented monotonically.  

Moreover, if
\be
\Delta \phi'_t = \phi\bigl(\bv{x}'(t) \bigr) - \phi\bigl( \bv{x}'(t-1)\bigr) \hspace{12pt} \text{ for } t = 1, \dotsc, \ell,
\ee
are the net changes to the objective function $\phi(\bv{x})$ associated with each such pairwise swap, we can choose the sequence of exchanges taking $\bv{x}'$ to $\bv{x}^*$ so that feasibility is always maintained, but also 
\be\label{deltaphi2}
\Delta \phi'_1 \ge \dotsb \ge \Delta \phi'_{\ell} \ge 0.
\ee 
How, and why?  Starting with $\bv{x}'$, we can construct a list of the $\ell$ independent variables (multiplicities included) $x_{j_1}, \dotsc, x_{j_{\ell}}$ to be decremented, sorted in order of decreasing (or at least non-increasing) changes $-\delta \phi'_{-_t}$ to the function $\phi(\bv{x})$ associated with each removal of a unit, such that
\be\label{dp3}
-\delta\phi'_{-_1} \ge \dotsb \ge -\delta\phi'_{-_\ell}.
\ee
Likewise, we can can construct a second list of $\ell$ independent variables (multiplicities included) $x_{k_1}, \dotsc, x_{k_{\ell}}$ to be incremented, sorted in order of decreasing (or at least non-increasing) changes $\delta\phi'_{+_t}$ to the function $\phi(\bv{x})$ corresponding to the successive replacements of the units, such that
\be\label{dp4}
\delta\phi'_{+_1} \ge \dotsb \ge \delta\phi'_{+_\ell}.
\ee
We claim that it is always possible to effect the $\ell$ unit pairwise swaps taking $\bv{x}'$ to $\bv{x}^*$ by choosing paired removals and compensatory replacements from the sorted lists in the order specified, removing a unit from $x_{j_t}$ and adding it to $x_{k_t}$ for $t = 1, \dotsc, \ell$.  This follows because feasible increments or decrements to distinct variables can be effected in either order, whereas the assumed concavity of the $f_j(x)$ implies that multiple decrements to any one variable, or multiple increments to one variable, will lead to decreasing changes to $\phi(\bv{x})$ as indicated.

This particular choice of path will then lead to net changes $\Delta\phi'_t = \delta\phi'_{+_t} - \delta\phi'_{-_t}$ for each swap, increments that must satisfy
\be\label{deltaphi2b}
\Delta \phi'_1 \ge \dotsb \ge \Delta \phi'_{\ell},
\ee 
since both \eqref{dp3} and \eqref{dp4} are satisfied for all $t = 1, \dotsc, \ell$.  We further argue that these changes $\Delta\phi'_t$ must all be nonnegative.  First, it must be the case that
$\delta \phi'_{+_{\ell}}  \in \{ \Delta\phi_1, \dotsc, \Delta \phi_n \}$, because any unit (above the prescribed minimum) belonging to the allocation $\bv{x}^*$ must necessarily have been introduced at some point during the greedy allocation.  It follows that  $\Delta \phi_n \le  \delta \phi'_{{\ell}_+} \le \Delta \phi_1$.
But it also turns out that $\delta \phi'_{-_{\ell}} \le  \Delta \phi_n$:  note that $+\delta \phi'_{-_{\ell}}$ corresponds to a gain in the objective function due to an increment which was in fact never chosen during the greedy allocation, but which would have increased one variable, say $x_j$, to a level beyond the value $x_j^*$ actually obtained in the greedy allocation.  So if $x_j$ happens to be the same variable $x_{k(n)}$ that was incremented at the final ($n$th) stage of the greedy allocation, then concavity ensures that any subsequent additions to $x_j$ would have resulted in gains to $\phi(\bv{x})$ no larger than $\Delta\phi_n$.  Otherwise, if $x_j$ is not the same variable $x_{k(n)}$ incremented at the $n$th stage of the greedy allocation, then it is still the case that $\delta\phi'_{-_{\ell}}$ can be no larger than $\Delta \phi_n$, for otherwise incrementing $x_j$ instead of $x_{k(n)}$ would have resulted in a gain at least as large as $\delta \phi'_{-_{\ell}} > \Delta \phi_n$, and the choice of $x_{k(n)}$ in the greedy allocation would not have been optimal after all.

We may therefore infer that $\delta\phi'_{-_{\ell}} \le \Delta \phi_n \le \delta \phi_{+_{\ell}}$, which then implies  $\Delta \phi'_{\ell} = \delta\phi_{+_{\ell}} - \delta\phi_{-_{\ell}} \ge 0$,  meaning all of the inequalities \eqref{deltaphi2} must hold, as claimed.  So therefore it is possible to connect $\bv{x}'$ to $\bv{x}^*$ along a feasible path of pairwise exchanges for which $\phi(\bv{x})$ never decreases (and furthermore, at an ever non-increasing rate).  In particular, the global optimality of $\bv{x}^*$ amongst all feasible allocations follows.  Do keep in mind, however,  that such a global optimum need not always be unique.\footnote{In the language of operations research, a solution $\bv{x}^*$ that we have called a constrained global optimum would instead be said to be \textit{undominated}.}

\subsubsection{Local is Global: Optimality Under Unit Pairwise Swaps Is Necessary and Sufficient}

For functions of integer arguments, the change $\Delta \phi$ to $\phi(\bv{x})$ under a unit pairwise swap $\bv{x} \to \bv{x} + \Delta \bv{x}_{jk}$  is the closest analog we have to the derivative of a smooth function of real arguments.\footnote{Here again $\Delta \bv{x}_{jk}$ denotes an ordered $N$-tuple with the $j$th component equal to $-1$, the $k$  component equal to $+1$, and the remaining components equal to zero.}

It turns out that non-increase of $\phi(\bv{x})$ under all feasibility-preserving, unit pairwise swaps is both a necessary and sufficient condition for any global constrained optimum of an additive, discretely concave function $\phi(\bv{x})$.  In other words, local optima are global optima, for such functions.

By construction, such swaps preserve feasibility of the solution, and as we have just seen, if $\bv{x}^*$ is an optimal solution, then $\phi(\bv{x}^*) \ge \phi(\bv{x}')$ for all feasible solutions $\bv{x}'$, including those which are connected to $\bv{x}^*$ by just one feasibility-preserving unit pairwise swap: $\bv{x}' = \bv{x}^* + \Delta \bv{x}_{jk}$.

To verify the converse, suppose that $\bv{x}'$ is a feasible allocation for which $\phi(\bv{x}' + \Delta \bv{x}_{jk}) \le \phi(\bv{x}')$ for all feasibility-preserving unit pairwise swaps $\Delta \bv{x}_{jk}$ starting from $\bv{x}'$.   As we have seen, some finite sequence of $\ell \le n$ feasibility-preserving, unit pairwise swaps can always connect $\bv{x}'$ to a greedy optimum $\bv{x}^*$ along a minimal monotonic path, such that the changes to $\phi(\bv{x})$ along the path satisfy \eqref{deltaphi2}.
But by hypothesis, necessarily $\Delta \phi'_1 \le 0$ as well, so
\be
0 \ge \Delta \phi'_1 \ge \dotsb \ge \Delta \phi'_{\ell} \ge 0,
\ee
which is only consistent if
\be
 \Delta \phi'_1 = \dotsb = \Delta \phi'_{\ell} = 0,
\ee
meaning
\be
\phi(\bv{x}') = \dotsb = \phi(\bv{x}^*),
\ee
so $\bv{x}'$ is an optimal solution as well.  Either $\bv{x}'$ coincides with $\bv{x}^*$, or else it is an equally valid solution, as are all points on the minimal monotonic path which connects $\bv{x}'$ to $\bv{x}^*$.

Also notice that the criteria of local optimality of an additive function under all (feasible) pairwise swaps can be translated into inequality conditions on a corresponding family of \textit{ranking} indices.  Under the greedy algorithm, if allocations to $j$ and $k$ are below their allowed upper bounds, then the next unit can be awarded to $j$ rather than to $k$ if $\phi(x_j+1) + \phi(x_k) \ge \phi(x_j) + \phi(x_k+1)$, or equivalently, if $\phi(x_j+1) - \phi(x_j) \ge \phi(x_k+1) - \phi(x_k)$, for all $k \in \{1, \dotsc, N\}$ that have not already reached their allowed upper bounds.

\subsection{Lagrange Multipliers}

For these sorts of constrained optimization problems, a discrete analog of the method of Lagrange multipliers can also be introduced \cite{everett:1963,foster:1964}.  Until noted otherwise, in the following discussion we need \textit{not} even presume the concavity of the functions $f_j(x)$.

Suppose that for some real number $\mu \in \realsymbol$, we define the Lagrangian function
\be
\mathscr{L}(\bv{x},\mu) = \phi(\bv{x}) + \mu \sum\limits_j x_j = \sum\limits_j \psi_j(x_j, \mu),
\ee
where
\be
\psi_j(x, \mu) = f_j(x) + \mu\, x \hspace{12pt} \text{ for } j = 1, \dotsc, N,
\ee
constitute a set of augmented scoring functions.  If, for each $j = 1, \dotsc, N$,  and a fixed, \textit{common} value of the multiplier $\mu$, we can find an integer solution $x_j^*(\mu)$ for 
which $\psi_j(x, \mu)$ is maximal amongst all integers in the interval $\lambda_j \le x \le u_j$, then we claim $\bv{x}^*(\mu) = \bigl( x_1^*(\mu), \dotsc, x^*_N(\mu)  \bigr)\trans$ is a constrained global optimum of $\phi(\bv{x})$ subject to the constraints $\lambda_j \le x_j \le u_j$ for $j = 1, \dotsc, N$ as well as $\sum\limits_j x_j = X(\mu) =  \sum\limits_j x^*_j(\mu)$.

To verify this simple but sometimes useful result, first imagine fixing the value of the Lagrange multiplier $\mu$.   Let $\bv{x}'$ be any  feasible solution for which $\lambda_j \le x'_j \le u_j$ and $\sum\limits_k x'_j = X(\mu) = \sum\limits_j x^*(\mu)$.  Since necessarily $\psi_j\bigl(x'_j, \mu\bigr) \le \psi_j\bigl(x^*(\mu), \mu\bigr)$, by summing over $j$ we find that
\be
\begin{split}
0 &\le \sum\limits_j  \Bigl[ \psi_j\bigl(x^*(\mu), \mu\bigr) - \psi_j\bigl(x'_j, \mu\bigr)  \Bigr] = 
 \sum\limits_j  f_j\bigl(x^*(\mu), \mu\bigr) -  \sum\limits_j  f_j\bigl(x'_j, \mu\bigr)  + \mu \Bigl[ \sum\limits_j  x^*_j(\mu) - \sum\limits_j  x'_j  \Bigr]\\ 
&=  \phi\bigl( \bv{x}^*(\mu) \bigr) - \phi\bigl( \bv{x}' \bigr) + \mu\bigl[ X(\mu) - X(\mu)\bigr] = \phi\bigl( \bv{x}^*(\mu) \bigr) - \phi\bigl( \bv{x}' \bigr) ,
\end{split}
\ee
implying 
\be
\phi\bigl( \bv{x}^*(\mu) \bigr) \ge \phi\bigl( \bv{x}' \bigr),
\ee
and proving our claim.

In a continuum problem, the value of a Lagrange multiplier would tell us how much the value of the constrained optimum changes as the value the constraint is varied.  A somewhat analogous result holds in the discrete case.  Specifically, suppose $\bv{x}^*(\mu_1)$ and $\bv{x}^*(\mu_1)$ are optimal solutions for two different Lagrange multipliers satisfying $\mu_1 < \mu_2$ (strictly), and therefore for possibly two different total budgets $X(\mu_1)$ and $X(\mu_2)$.  Because $\bv{x}^*(\mu_1)$ is optimal for the budget $X(\mu_1)$, it follows that
\be
\phi\bigl( \bv{x}^*(\mu_1) \bigr) + \mu_1 X(\mu_1) \ge \phi\bigl( \bv{x}^*(\mu_2) \bigr) + \mu_1 X(\mu_2),
\ee
and because $\bv{x}^*(\mu_2)$ is optimal for the budget $X(\mu_2)$, it follows that
\be
\phi\bigl( \bv{x}^*(\mu_2) \bigr) + \mu_2 X(\mu_2) \ge \phi\bigl( \bv{x}^*(\mu_1) \bigr) + \mu_2 X(\mu_1).
\ee
Re-arranging,  and remembering that $\mu_1 < \mu_2$ by assumption, we find that
\be
X(\mu_1) \le X(\mu_2),
\ee
and if it is the case that $X(\mu_1) < X(\mu_2)$ strictly, then
\be
\mu_1 \le \frac{ \phi\bigl( \bv{x}^*(\mu_1) \bigr) - \phi\bigl( \bv{x}^*(\mu_2) \bigr)  }{X(\mu_2) - X(\mu_1)} \le \mu_2,
\ee
where at least one of the two inequalities in this last expression must be strict, since $\mu_1 < \mu_2$.

Lagrange multipliers can be quite useful, because they can allow us to decouple the constrained optimization problem involving a collective constraint on $\sum_j x_j$ into a set of simpler single-variable optimization problems with rather trivial (interval) constraints on individual variables, at the price of not knowing ahead of time for what budget $X(\mu)$ the resulting solution will be optimal.

Any optimum of the Lagrangian will lead to a constrained optimum of the original problem for some choice of the budget.  The converse question naturally arises as to whether all constrained optima (under any allowed budget $X$ consistent with lower and upper bound constraints) can be generated in this manner.  In general, the answer is unfortunately negative, as there can be so-called \textit{gaps} in the solution space not covered by allocations generated by a Lagrange multiplier.  

But we can see what additionally would be necessary in order to ensure gaplessness and capture all possible solutions via a Lagrangian: for each $j = 1, \dotsc, N$, every integer in the interval $\lambda_j \le x_j \le  u_j$ is to become optimal for some choice of the Lagrange multiplier, and as $\mu$ is continuously varied over the range leading to feasible solutions, no $x_j^*(\mu)$ need ever jump by more than one unit at a time.

Notice that if every $f_j(x)$ is both bounded and \textit{strictly} concave over its allowed domain, then no gaps will occur.  As $\mu \to -\infty$, $\psi_j(x)$ will become increasingly dominated by the added $+\mu\, x$ penalty term, and the constrained maximum of $\psi_j(x)$ must eventually become $x^* = \lambda_j$.  As $\mu \to +\infty$, the constrained maximum will instead eventually reach $x^* = \lambda_j$.  In between, suppose the maximum of $\psi_j(x)$ is achieved at some feasible $x^{*}-1$ for multiplier $\mu_{} - \epsilon$ for suitably small but non-zero $\epsilon$, but at $x^{*}+1$ for multiplier $\mu_{}+\epsilon$, while $x^*$ is not optimal in either case.
Then
\bsub
\begin{align}
f_j(x^{*}+1) + (\mu_{} + \epsilon)(x^{*}+1) > f_j(x^{*}) +  (\mu_{}+\epsilon)\,x^{*},\\
f_j(x^{*}-1) + (\mu_{} - \epsilon)(x^{*}-1) > f_j(x^{*}) +  (\mu_{}-\epsilon) \,x^{*},
\end{align}
\esub
which after some rearrangements imply
\bsub
\begin{align}
f_j(x^{*}+1)  - f_j(x^{*})  >   -(\mu_{}+\epsilon),\\
f_j(x^{*}-1) - f_j(x^{*})  > +(\mu_{}-\epsilon).
\end{align}
\esub
Adding, we find
\be
f_j(x^{*}+1)  - 2 f_j(x^{*}) +  f_j(x^{*}-1)  >  -2\epsilon,\\
\ee
and considering the limit as $\epsilon \to 0^+$, we may infer $f_j(x^{*}+1)  - 2 f_j(x^{*}) +  f_j(x^{*}-1) \ge 0$, or
\be\label{f_implication}
f_j(x^{*}+1)  -  f_j(x^{*}) \ge f_j(x^{*})  - f_j(x^{*}-1),
\ee
which is impossible if $f_j$ is strictly concave.\footnote{And furthermore, jumps by three or more units also prove impossible, by a similar argument.}  So as $\mu$ is varied continuously over some suitable interval, the optimal solution $x_j^*$ starts at $\lambda_j$, can be taken to jump by \textit{at most} one unit at a time as $\mu$ is continuously increased, and eventually ends up at $x_j^* = u_j$, after a finite number of unit jumps.

The only catch is that as the Lagrange multiplier is increased, it is possible that more than one $x_j^*$ may be inclined to jump at exactly the same value of $\mu$ under independent optimizations (either by coincidence, or more likely, symmetry).  But in this event, one can simply temporarily delay all but one of the increments suggested by the Lagrangian, in order of gains contributed to $\phi(\bv{x})$ (or randomly in the case of an actual tie).  For any delayed variables, concavity of each $\psi_j(x)$ ensures that $x_j^*-1$ is the next best thing to $x_j^*$ amongst all possibilities less than $x_j^*$.  So the Lagrangian algorithm, with a suitable modification in case of predicted simultaneous jumps, can find all possible constrained optima when the $f_j(x)$ are strictly concave and bounded.  Of course, it is in precisely these same cases that the straightforward greedy algorithm will also work, as we have seen in the discussion above.

\subsection{Convexity of Various Objective Functions Suggested for Apportionment}

The simple greedy algorithm introduced above can be used to implement several  of the usual apportionment methods, including not just our recommended entropic scheme, but also the dual entropic, Webster-Willcox, and Huntington-Hill criteria, since each turns out to be equivalent to \textit{constrained minimization}, with respect to $\bv{a}$, of a certain convex objective function $\U(\bv{a}; \bv{q}; R,P,S) = - \phi(\bv{a}; \bv{q}; R,,S)$ that satisfies
\be
\begin{split}
\U(&a_1, \dotsc, a_s,  \dotsc, a_S; q_1, \dotsc, q_s; R,P,S) \le \\
&\tfrac{1}{2} \bigl[   \U(a_1, \dotsc, a_s-1,  \dotsc, a_S; q_1, \dotsc, q_s; R,P,S) 
+\U(a_1, \dotsc, a_s+1, \dotsc, a_S; q_1, \dotsc, q_s; R,P,S)  \bigr]
\end{split}
\ee
for each $s = 1, \dotsc, S$, whenever $a_s > 0$, and for any allowed fixed values of the other seat assignments $a_{s'}$ for $s' \neq s$, given fixed values of $\bv{q}$, $R$, $P$, and $S$.

In many cases, the objective function $\U(\bv{a}; \bv{q}; R,P,S)$, though only ever evaluated at integer-valued seat assignments, is actually a smooth function of the $\bv{a}$ variables, so this discrete convexity condition may be deduced from a continuous one,
\be
\tfrac{\del^2}{\del a_s^2} \U(a_1, \dotsc, a_S; q_1, \dotsc, q_s; R,P) \ge 0 \;\; \text{ for all } s = 1, \dotsc, S,
\ee
which generally is easier to verify directly than the discrete version whenever second derivatives exist.

\subsubsection{Entropic/Identric Mean Method}

If we take as our objective function the Kullback-Leibler divergence, or relative entropy
\be
\K(\bv{a}; \bv{q},R,S) =  \sum\limits_{s=1}^{S} \tfrac{a_s}{R} \log_2 \bigl[ \tfrac{a_s/R}{q_s/R}\bigr] =  \tfrac{1}{R} \sum\limits_{s=1}^{S} a_s \log_2 \bigl[ \tfrac{a_s}{q_s}\bigr],
\ee
then a straightforward calculation reveals after a bit of algebra that
\be
\tfrac{\del^2}{\del a_s^2} \K(\bv{a}; \bv{q},R,S) =   \tfrac{1}{\ln 2} \tfrac{1}{R} \tfrac{1}{a_s} \ge 0,
\ee
from which the non-strict, discrete convexity property follows by standard arguments from analysis.  So for fixed $R$ and $\bv{q}$, $\K$ can be minimized (or $-\K$ maximized) with respect to the $\bv{a}$, by employing the greedy algorithm outlined above.

\subsubsection{Webster-Willcox Method}

If instead we adopt as our objective function the Webster-Willcox ``chi-squared'' objective function,
\be
\U{\stext{W}}(\bv{a}; \bv{q},R,P,S) =  \tfrac{R}{P^2} \sum\limits_{s=1}^{S} \tfrac{(a_s - q_s)^2}{q_s},
\ee
then
\be
\tfrac{\del^2}{\del a_s^2}\U{\stext{W}}(\bv{a}; \bv{q},R,P,S) =   \tfrac{R}{P^2}  \tfrac{2}{q_s} \ge 0,
\ee
which establishes convexity for of $\U{\stext{W}}$.

\subsubsection{Huntington-Hill Method}

For the Huntington-Hill objective function,
\be
\U_{\stext{H}}(\bv{a}; \bv{q},R,P,S) =  \tfrac{P^2}{R^3} \sum\limits_{s=1}^{S} \tfrac{(a_s - q_s)^2}{a_s},
\ee
we find
\be
\tfrac{\del^2}{\del a_s^2}\U{\stext{H}}(\bv{a}; \bv{q},R,P,S) =  \tfrac{P^2}{R^3} \tfrac{2 q_s^2}{a_s^3}  \ge 0,
\ee
from which the needed sort of convexity follows from the mean value theorem.

\subsubsection{Dual Entropic/Logarithmic Mean Method}

Recall that the \textit{dual} Kullback-Leibler divergence is given by swapping the roles of $\bv{a}$ and $\bv{q}$ in the relative entropy:
\be
\tilde{\K}(\bv{a}; \bv{q}, R,S) =  \sum\limits_{s=1}^{S} \tfrac{q_s}{R} \log_2 \bigl[ \tfrac{q_s/R}{a_s/R}\bigr] =  \tfrac{1}{R} \sum\limits_{s=1}^{S} q_s \log_2 \bigl[ \tfrac{q_s}{a_s}\bigr].
\ee
Taking derivatives with respect to any of the $a_s$, we find
\be
\tfrac{\del^2}{\del a_s^2} \tilde{\K}(\bv{a}; \bv{q}, R,S)  =   \tfrac{1}{\ln 2} \tfrac{1}{R} \tfrac{q_s}{a_s^2} \ge 0,
\ee
which establishes convexity.

\subsubsection{Hamilton-Vinton Method}

Recall that the Hamilton-Vinton method can be cast as a constrained minimization of the total squared-error
\be
\U_{\stext{V}}(\bv{a}; \bv{q}, S) = \sum\limits_{s=1}^{S} ( a_s - q_s)^2
\ee
between the apportionments and quotas of all states.

Second derivatives are straightforward to evaluate, resulting in 
\be
\tfrac{\del^2}{\del a_{s}^2} \U_{\stext{V}}(\bv{a}; \bv{q},S) = +2 \ge 0,
\ee
and confirming convexity.

\subsubsection{Conditional Entropic Apportionment, Given Optimal Intra-State Districting}

As a somewhat more complicated case, consider  the conditional relative entropy, assuming precise knowledge of subsequent best-case districting:
\be
\begin{split}
\K_d( \bv{a}; \bv{q}; R,P,S) &=  \log_2 \Dbar -  \tfrac{1}{R}\sum\limits_{s=1}^{S} \sum\limits_{k=1}^{a_s} \log_2 d^*_{sk} \\
&= \log_2 \Dbar -  \tfrac{1}{R}\sum\limits_{s=1}^{S} \bigl\{   (a_s - \eta_s)\log_2 \lfloor \bar{d}_s \rfloor + \eta_s \log_2 \lceil \bar{d}_s \rceil   \bigr\} \\
&= \log_2 \Dbar -  \tfrac{1}{R}\sum\limits_{s=1}^{S} \bigl\{   (a_s - \eta_s)\log_2 \lfloor \bar{d}_s \rfloor + \eta_s \log_s \bigl( \lfloor \bar{d}_s \rfloor + 1 \bigr)  \bigr\},
\end{split}
\ee
where, for each $s = 1, \dotsc, S$, the quantities
\be
d^*_{sk} = 
\begin{cases}
\lceil \bar{d}_s \rceil  &\text{ if } 1 \le k  \le \eta_s \\
\lfloor \bar{d}_s  \rfloor &\text{ if }  \eta_s+1 \le k \le a_s 
\end{cases}
\ee
represent the optimal (i.e., most uniform) district sizes given $\bv{a}$ and $\bv{p}$, and $\eta_s = p_s \mod a_s = p_s - a_s \lfloor \bar{d}_s \rfloor = a_s \bigl( \bar{d}_s - \lfloor \bar{d}_s \rfloor \bigr)$ is the explicit remainder upon dividing $p_s$ by $a_s$.

Discrete convexity is a bit more difficult establish in this situation, because the objective function is not differentiable, nor even continuous.  The trick involves relating the original apportionment problem with populations $\bv{p}$ to an imagined situation with the populations all doubled, to $2\bv{p}$.

Consider for a moment just the sum over the logarithms of the $\bar{d}^*_{sk}$:
\be
\psi(\bv{a}; \bv{p},S) = \sum\limits_{s=1}^{S} \bigl\{   (a_s - \eta_s)\log_2 \lfloor \bar{d}_s \rfloor + \eta_s \log_s \bigl( \lfloor \bar{d}_s \rfloor + 1 \bigr)  \bigr\}.
\ee
From our derivation of optimal choice of district sizes (see Appendix~\ref{district_sizes}), we may infer that for any other choice of (sub-optimal) district sizes $d_{sk}$ consistent with populations $\bv{p}$ and apportionments $\bv{a}$, it must be the case that 
\be
\sum\limits_{s=1}^{S} \sum\limits_{k=1}^{a_s} \log_2 d_{sk} \le \psi(\bv{a}; \bv{p},S).
\ee
Now we notice that if we were to double all state populations and simultaneously double all apportionments, $p_s \to 2 p_s$ and $a_s \to 2 a_s$, then
$\eta_s$ would also double: $\eta_s \to 2 \eta_s$, but $\bar{d}_s = p_s/a_s$ (and hence $\lfloor \bar{d}_s  \rfloor$) would remain unchanged, so that
\be
\psi(2\bv{a}; 2\bv{p},S) = 2 \,\psi(\bv{a}; \bv{p},S).
\ee
But now consider the actual problem of apportioning $2R$ seats to states with populations $2\bv{p} = (2p_1, \dotsc, 2p_S)$.  Under conditional entropic apportionment, the optimal assignments will just be $2\bv{a}$, and the optimal districts will just amount to twofold copies of the optimal districts for the original case of apportioning $R$ seats amongst states with population $\bv{p}$. Therefore we may infer that for any other set of seat assignments for the doubled problem, it follows that
\be
\sum\limits_{s=1}^{S} \sum\limits_{k=1}^{2a_s} \log_2 d_{sk} \le \psi(2\bv{a}; 2\bv{p},S) = 2 \, \psi(\bv{a}; \bv{p},S).
\ee
In particular, one sub-optimal arrangement would be the following: we imagine splitting the (doubled) population of all states into two exactly equal halves, $2\bv{p} = \bv{p}+\bv{p}$.  For all states but one, we still assign the optimal number of seats ($a_{s'}$ to each half) and the corresponding optimal district sizes ($d^*_{s'k}$ within each half).  But for a single state $s$, for which $a_s > 0$, instead of assigning $a_s$ districts to each half, we can imagine assigning $(a_s-1)$ districts to one half, and $(a_s + 1)$ to the other, then choosing best-case district sizes within each half separately.  In this case,
\be
\begin{split}
\sum\limits_{s=1}^{S} \sum\limits_{k=1}^{2a_s} \log_2 d_{sk} &= \psi(a_1, \dotsc, a_{s-1}, a_s -1, a_{s+1}, \dotsc, a_S, \bv{p},S)\\
& + \psi(a_1, \dotsc, a_{s-1}, a_s +1, a_{s+1}, \dotsc, a_S, \bv{p},S),
\end{split}
\ee
corresponds to a sub-optimal arrangement for the doubled-population problem, so
\be
\begin{split}
\psi(a_1, &\dotsc, a_{s-1}, a_s -1, a_{s+1}, \dotsc, a_S, \bv{p},S) \\
&+  \psi(a_1, \dotsc, a_{s-1}, a_s +1, a_{s+1}, \dotsc, a_S, \bv{p},S) \le  2 \, \psi(\bv{a}; \bv{p},S),
\end{split}
\ee
from which the discrete convexity of $\K_d(\bv{a}; \bv{q}; R,P,S) =  \log_2 \tfrac{P}{R} - \tfrac{1}{R} \,\psi(\bv{a},\bv{q})$ follows.

While we do not actually advocate this approach, it is reassuring that it could be accommodated within the same greedy optimization framework, if desired.



\newpage
\section{\uppercase{Optimal Choice of District Sizes}}\label{district_sizes}

Given an actual or proposed apportionment $\bv{a} = (a_1, \dotsc, a_S)$ assigning a whole number of representatives to each state, minimization of the very same relative entropy functional used to allocate representatives (and possibly to choose the overall House size) can also be used to determine optimal sizes for single-member districts within each state.  Of course intuition alone immediately suggests that if the goal is equality of representation, all districts should be chosen to be as similar in size as possible, while remaining exclusive and exhaustive in terms of their membership, while containing a whole number of represented inhabitants of a single state.  But confirming that this natural standard emerges \textit{automatically} via minimization of the conditional relative entropy offers a good sanity check, and, more importantly, establishes that we could use the Kullback-Leibler divergence to \textit{quantify} the extent of inequality in choice of district sizes, for the purposes of evaluating proposed districts drawn up by state legislatures or adjudicating court challenges, in place of more traditional but \textit{ad hoc} measure likes the variance, range, or relative RMS differences.

If a state has only a single representative, then obviously it must in effect have only one district, consisting of the entire represented state population, and \textit{a fortiori} there can be no variation nor inequity between the size of districts within that state alone.

Suppose that a state $s$ has (at least) two districts, one with size $d_{s1}$ and the other with size $d_{s2}$ under some proposed boundaries. Then, conditional on the proposed districting, the contribution to the conditional relative entropy $\K$ just from this pair of districts can be written as
\be
\K = \dotsb - \tfrac{1}{R} \log \bigl( d_{s1} d_{s2}   \bigr) + \dotsb = \dotsb - \tfrac{1}{R} \log\Bigl[ \tfrac{1}{4} (  d_{s1}  +  d_{s2}  )^2 -  \tfrac{1}{4} (  d_{s1}  -  d_{s2}  )^2\Bigr] + \dotsb.
\ee
Because the logarithm is a strictly monotonic function, without violating any constraints on seats, or altering aspects of the apportionment for any other states, or the sizes for \textit{any other} districts within the state in question, we can decrease $\K$ by \textit{minimizing} the product $d_{s1} d_{s2}$ while keeping the total represented population $(d_{s1} + d_{s2})$ within this pair of districts fixed.  Keeping track of the minus signs, we see that this is equivalent to \textit{minimizing} the magnitude of the difference $\abs{ d_{s1} - d_{s2} }$ while fixing the sum $(d_{s1} + d_{s2})$.  Therefore, whatever the choice for $R$ and  any other details of apportionment of these representatives between the various states, $\K$ will be locally minimized when the differences between intra-state district sizes are made as small as possible, namely such that all districts in a given state differ in size by no more than a single represented individual.  In particular: in the $s$th state with $a_s$ single-member districts, the very best arrangements will have $\eta_{s} = p_s \mod a_s$ districts of size $\lceil \bar{d}_s \rceil$ and the remaining $(a_s - \eta_s)$ districts of size $\lfloor \bar{d}_s \rfloor$.

One might of course argue that such fastidiousness in choosing district sizes within states could constitute a case of false precision.  After all, census counts for U.S.\ States are all almost surely subject to some uncertainty at a level significantly higher than $\pm 1$ person, and at best they only provide a once-a-decade snapshot of the represented populations that are inevitably almost immediately out-of-date due to the cumulative effects of migration or other demographic shifts (e.g., inhabitants being born, moving, dying, or naturalizing).

While the Supreme Court acknowledges that exact or maximal equality may obviously be impractical or impossible,\footnote{In oral arguments for \textit{Evenwel vs.\ Abbott} (2016) Justice Breyer noted that the ``Constitution does not demand mathematical perfection,'' although this was in the context of redistricting for state-level legislatures.} current standards for U.S.\ Congressional districting are actually quite strict, with equality of represented populations required\footnote{Articulated in \textit{Wesberry vs.\ Sanders}, in 1964 \cite{westbury:1964}.} ``as nearly as is practicable,'' interpreted to mean that each state must make good-faith efforts to draw districts with almost exactly the same number of people in each district within the state, and where any appreciable deviations from the ``ideal'' district-size $\bar{d}_s$ must be specifically justified by consistent state policy.  In practice, policies that cause even a $1\%$ spread from largest to smallest district may be regarded as un-Constitutional unless a compelling justification for the variation can be mustered.

\newpage

So the entropic approach conforms nicely to common sense and longstanding Supreme Court jurisprudence involving districting, and indeed provides a more fundamental measure of inequity relative to ideal district sizes, than does either the standard deviation of district sizes, or the percentage difference between the largest and smallest districts in the state.  We again stress that the near-equality of \textit{intra-state} district sizes does not arise from any \textit{additional} requirement or desideratum, whether \textit{ad hoc} or not, but rather from optimization, with respect to the intra-state district sizes $d_{sk}$ for each $s$, of the very same Kullback-Leibler divergence  whose minimization with respect to the $a_s$ yielded the congressional apportionment itself. Only at this stage, the minimization of $\K$ is to be performed with respect to the district sizes $d_{sk}$, conditional on the chosen $a_s$.

 
\newpage
\section{\uppercase{Best Case District-Conditioning versus Average-District Conditioning}}\label{district_conditioning}

We have tried to argue in the main text why we should assume average district sizes rather than best case district sizes.  
But in any event, for the case of U.S.\ Congressional apportionment, any numerical differences between using averaged intra-state district sizes and best-case integral district sizes are expected to be small, and very unlikely to change the optimal outcome.  This is demonstrated here.

First, we estimate the typical magnitude of the entropy $\tfrac{1}{R} \K = \sum\limits_j a_j \log \tfrac{a_j}{q_j}$ under optimal apportionment.  Though entropic apportionment is not a quota method (in the absence of strong constraints), the $a_j$ are expected to be reasonably close to the $q_s$ on average, so we can expand $\K$ in a Taylor series in the apportionments $a_s$ centered on the quotas $q_s$.  The zeroth-order and first-order terms cancel, so the leading-order contribution becomes:
\be
\K =  \tfrac{1}{R} \sum\limits_s  a_s \log \tfrac{a_s}{q_s} \sim  \tfrac{1}{2 \ln 2}  \tfrac{1}{R} \sum\limits_s \tfrac{(a_s - q_s)^2}{q_s} + \dotsb ,
\ee
assuming the logarithms are taken base-$2$.  Assuming $a_s \sim O(\bar{a}) \sim O\bigl( \tfrac{R}{S} \bigr)$ and $\abs{a_s - q_s} \sim O\bigl( \tfrac{1}{2} \bigr)$, we find
\be
\K \sim O\bigl(\tfrac{3}{4} \bigr) \, \tfrac{1}{R} \,O\bigl( S \tfrac{ (1/2)^2}{\bar{a}} \bigr) \sim O\bigl(\tfrac{1}{4} \tfrac{S^2}{R^2} \bigr), 
\ee
which for current parameters pertinent to U.S.\ Congressional apportionment, suggests $\K \sim 3 \cdot 10^{-3}$.
 
If we assume districting would be both predictable and best-case under a proposed apportionment, 
the district sizes and identities constitute additional constraints which would increase the relative entropy, by an amount
\be
\begin{split}
\Delta \K &= \Bigl\{ \log \Dbar - \tfrac{1}{R} \sum\limits_s\sum\limits_k \log d_{ks} \Bigr\} - \Bigl\{ \log \Dbar - \tfrac{1}{R} \sum\limits_s a_s \log \bar{d}_s \Bigr\} \\
&= \tfrac{1}{R} \sum\limits_s a_s \log \bar{d}_s - \tfrac{1}{R} \sum\limits_s \bigl[ (a_s - \eta_s) \log \bigl( \lfloor \bar{d}_s \rfloor \bigr) + \eta_s \log\bigl( \lceil \bar{d}_s \rceil \bigr)  \bigr],
\end{split}
\ee
where $\eta_s = p_s \mod a_s$.  In this expression, we can then replace $\eta_s \log (\lceil \bar{d}_s \rceil)$ with $\eta_s \log \bigl( \lfloor \bar{d}_s \rfloor + 1 \bigr)$ because
$\lceil \bar{d}_s \rceil = \lfloor \bar{d}_s \rfloor + 1$ unless $\lceil \bar{d}_s \rceil = \lfloor \bar{d}_s \rfloor = \bar{d}_s$, which holds occurs if and only if $\eta_s = 0$, and because $p_s = a_s \bar{d}_s  = a_s \lfloor \bar{d}_s \rfloor + \eta_s$, we can also substitute $\lfloor \bar{d}_s \rfloor = \bar{d}_s - \tfrac{\eta_s}{a_s}$:
\be
\begin{split}
\Delta \K &= \tfrac{1}{R} \sum\limits_s a_s \log \bar{d}_s - \tfrac{1}{R} \sum\limits_s \bigl[ (a_s - \eta_s) \log \bigl( \lfloor \bar{d}_s \rfloor \bigr) + \eta_s \log\bigl( \lfloor \bar{d}_s \rfloor + 1 \bigr)  \bigr] \\
&= \tfrac{1}{R} \sum\limits_s a_s \log \bar{d}_s - \tfrac{1}{R} \sum\limits_s \bigl[ (a_s - \eta_s) \log \bigl( \bar{d}_s - \tfrac{\eta_s}{a_s}  \rfloor \bigr) + \eta_s \log\bigl( \lfloor \bar{d}_s +1 - \tfrac{\eta_s}{a_s} \bigr)  \bigr].
\end{split}
\ee
In the case of U.S.\ Congressional apportionment, the average district sizes all satisfy $\bar{d}_s = p_s/a_s \gg 1$, while it is always the case that $0 \le \tfrac{\eta_s}{a_s} < 1$, so we can now expand the logarithms in a Taylor series centered on $\bar{d}_s $:
\be
\log_2( \bar{d}_s + \epsilon) = \log_2( \bar{d}_s )  + \tfrac{1}{\ln 2} \tfrac{\!\epsilon}{\bar{d}_s} - \tfrac{1}{2 \ln 2} \tfrac{\epsilon^2}{\bar{d}_s^2} + \dotsb,
\ee
assuming the logarithms are here taken base-$2$.  The zeroth-order terms will obviously cancel.  Somewhat less obviously, the first-order terms also cancel, resulting in leading contributions from the second-order terms:
\be
\Delta \K \approx  \tfrac{1}{2 \ln 2} \tfrac{1}{R} \sum\limits_s \tfrac{1}{\bar{d}_s^2} \, \eta_s \bigl( 1 - \tfrac{\eta_s}{a_s} \bigr) \ge 0.
\ee
Approximating $\bar{d}_s \sim O(\Dbar) \sim O\big(\tfrac{P}{R} \bigr)$, $a_s \sim O(\bar{a}) \sim O\bigl( \tfrac{R}{S} \bigr) > 1$,  $\eta_s \sim O\bigl( \tfrac{\bar{a} - 1}{2}  \bigr)$, so that $\eta_s ( 1 - \eta_s/a_s) \sim O\bigl( \tfrac{\bar{a}^2 - 1}{4 \bar{a}} \bigr)$,  we can estimate
\be
\Delta \K \approx O(1)  \,  O\bigl( 1- \tfrac{S^2}{R^2} \bigr) \, O\big(\tfrac{R^2}{P^2} \bigr). 
\ee
For recent U.S.\ demographic parameters, this comes out to $\Delta \K \sim 2 \cdot 10^{-12}$.  So the relative entropy based on average-district sizes and that based on best-case exact district sizes are expected to differ by a few parts in $10^{12}$. 
In comparison, the population of even the largest states can be specified to only a few parts in $10^8$, suggesting that we should typically not expect such small differences in the relative entropies to have an observable effect in terms of the apportionments.

Actually, to get a sense of whether the difference between the district-averaged and district-conditioned version of the relative entropy should matter, a better comparison would look at a typical difference in the relative entropy between the best and second best apportionments.  Because of the convexity of the entropy, we need look only at the very last seat awarded during the greedy allocation.  Supposing we shift this seat from the most deserving state (say state $1$ without loss of generality) to the second most deserving state (call it state $2$), the relative entropy would be increased by an amount
\be
\begin{split}
\delta \K &= \tfrac{1}{R} \bigl\{   (a_2+1) \log \tfrac{a_2+1}{q_2 } +  (a_1-1) \log \tfrac{a_1-1}{q_1} -  a_2 \log \tfrac{a_2}{q_2}   -  a_1 \log \tfrac{a_1}{q_1}  \bigr\}  \\
&=    \tfrac{1}{R} \bigl\{ a_2 \log \tfrac{a_2+1}{a_2} + a_1 \log \tfrac{a_1-1}{a_1}  + \log \bigl[\tfrac{a_2+1}{a_1-1}\tfrac{q_1}{q_2} \bigr]\bigr\}.
\end{split}
\ee
At the optimum, the argument of each of the logarithms on the right hand side  is not expected to be too different from unity, so we can Taylor expand, leading to an estimate of $\delta \K \sim O\bigl(\tfrac{1}{R}\bigr)$.  But we should also account for the fact that we are looking at the difference between the best and second-best values of the relative entropy, i.e., the \textit{smallest} of the entropic penalties that would be incurred if we were to shift the last-awarded seat to one of the other $(S-1)$ states.  Thinking about order statistics, we might guess that the minimal such increment is likely to be smaller by a factor of something like $O\bigl(\tfrac{1}{S} \bigr)$ than the \textit{typical} increment, so we shall predict $\delta \K \sim O\bigl(\tfrac{1}{R \,S}\bigr)$.  For parameters pertinent to the current U.S.\ Congress, this evaluates to $\delta\K \sim 5 \cdot 10^{-5}$, smaller than $\K$ itself but still several orders of magnitude larger than $\Delta \K$, suggesting that changes on the scale of the latter would tend to be unlikely to change the actual seat assignments, although of course it would be \textit{possible} for such differences to matter.

In the end, however, we reiterate that this idea is not recommended as a matter of principle, because the required information is very unlikely to be available, and it is dubious whether it ought to be used for apportionment even if it were available.

%
%
%

\newpage
\section{\uppercase{Additional Tables and Examples}}\label{tables}

\mbox{}

\begin{table}[h!]
\begin{small}
\begin{tabular}{ccccccccc}
\hline
State &  Population & Quota & Adams (SD) & Dean (HM) & Huntington (EP) & Webster (MF) & Jefferson (GD) & Entropic  \\
\hline\hline
A & 9061 & 9.061 & 9 & 9 & 9 & 9 & 10 & 9 \\
B & 7179 & 7.179 & 7 & 7 & 7 & 8 & 7   & 8 \\
C & 5259 & 5.259 & 5 & 5 & 6 & 5 & 5   & 5 \\
D & 3319 & 3.319 & 3 & 4 & 3 & 3 & 3   & 3 \\
E & 1182 & 1.182 & 2  & 1 & 1 & 1 & 1  & 1\\
\hline 
TOTAL & 26\,000 & 26 & 26 & 26 & 26 & 26 & 26 & 26 \\
\hline
Rel.\ entropy &   &   &  0.01435 & 0.004953  &  0.004606  &  0.004582 & 0.004796 &  0.004582 \\
\hline
\hline
\end{tabular}
\end{small}
\caption{Comparison of entropic apportionment to divisor methods for a hypothetical apportionment problem  constructed by Balinski and Young \protect\cite{balinski_young:1975} using $P = 26\hspace{0.95pt}000$, $S = 5$,  and $R = 26$, for which the five traditional divisor methods all differ.  Here the entropic method agrees with the Webster Method (Major Fractions), but not the Huntington Method (Equal Proportions), although it does turn out that $R = 26$ as well as $R = 27$ are particularly discordant choices for the total number of seats given this population distribution.}\label{Table:all_different}
\end{table}

\begin{table}[h!]
\begin{small}
\begin{tabular}{ccccccc}
\hline
State &  Population &  Quota & Dean &  Huntington & Webster & Entropic  \\
\hline\hline
 A & 729 & 7.29 &  7 & 7 &  8  & 8  \\
 B & 534 & 5.34 &  5 &  6 &  5 &  5 \\
 C & 337 & 3.37 &  4 &  3 &  3  & 3  \\
 \hline
 TOTAL & 1600 & 16 & 16 & 16 & 16 & 16 \\
\hline
Relative Entropy & & & 0.00653 & 0.00596 & 0.00154 &0.00154 \\ 
\hline
\end{tabular}
\end{small}
\caption{Apportionments for a $P = 1\hspace{0.95pt}600$, $S = 3$, $R = 16$ test case of Huntington \protect\cite{huntington:1928}, illustrating that the Dean, Huntington, and Webster methods can all differ, while all remaining on quota.  The entropic method agrees with the Webster Method in this case.}\label{table:huntington_cmp_abc}
\end{table}

\begin{table}[h!]
\begin{small}
\begin{tabular}{ccccc}
\hline
State  &  Population &  Quota &  Hamilton &  Entropic  \\
\hline\hline
 $J$ & 987 & 9.87   &  10  & 9  \\
 $K$ & 157 & 1.57  &   2 &  2   \\
 $L$ & 156 & 1.56  &   1  & 2  \\
\hline
TOTAL & 1300 & 13 & 13 & 13 \\
\hline
Relative Entropy & & & 0.0189  & 0.0167 \\ 
\hline
\end{tabular}
\end{small}
\caption{Apportionments for a $P = 1\hspace{0.95pt}300$, $S = 3$, $R = 13$  test case of Huntington \protect\cite{huntington:1928}, illustrating that the size of the fractional remainders of the quotas do not determine the apportionments in the Huntington, Webster, or entropic schemes, all of which agree here.}\label{table:not_quota_remainders}
\end{table}

\newpage

\begin{table}[ht!]
\begin{small}
\begin{tabular}{crrccc}
\hline
State &  Population & \hspace{6pt} Quota & \hspace{6pt} Huntington & Webster & Entropic \\
\hline\hline
 1 & 60272 & 61.477 & 68 & 70 & 70 \\
 2 & 1226 & 1.251  & 1 & 1 & 1 \\
 3 & 1227 & 1.252  & 1 & 1 & 1 \\
 4 & 1228 & 1.253 & 1 & 1 & 1 \\
 5 & 1229 & 1.254  & 1 & 1 & 1 \\
 6 & 1230 & 1.255 & 1 & 1 & 1 \\
 7 & 1231 & 1.256 & 1 & 1 & 1 \\
 8 & 1232 & 1.257 & 1 & 1 & 1 \\
 9 & 1233 & 1.258 & 1 & 1 & 1 \\
 10 & 1234 & 1.259 & 1 & 1 & 1 \\
 11 & 1235 & 1.260 & 1 & 1 & 1 \\
 12 & 1236 & 1.261 & 1 & 1 & 1 \\
 13 & 1237 & 1.262 & 1 & 1 & 1 \\
 14 & 1238 & 1.263 & 1 & 1 & 1 \\
 15 & 1239 & 1.264 & 1 & 1 & 1 \\
 16 & 1240 & 1.265 & 1 & 1 & 1 \\
 17 & 1241 & 1.266 & 1 & 1 & 1 \\
 18 & 1242 & 1.267 & 1 & 1 & 1 \\
 19 & 1243 & 1.268 & 1 & 1 & 1 \\
 20 & 1244 & 1.269 & 1 & 1 & 1 \\
 21 & 1245 & 1.270 & 1 & 1 & 1 \\
 22 & 1246 & 1.271 & 1 & 1 & 1 \\
 23 & 1247 & 1.272 & 1 & 1 & 1 \\
 24 & 1248 & 1.273 & 1 & 1 & 1 \\
 25 & 1249 & 1.274 & 1 & 1 & 1 \\
 26 & 1250 & 1.275 & 1 & 1 & 1 \\
 27 & 1251 & 1.276 & 1 & 1 & 1 \\
 28 & 1252 & 1.277 & 1 & 1 & 1 \\
 29 & 1253 & 1.278 & 1 & 1 & 1 \\
 30 & 1254 & 1.279 & 1 & 1 & 1 \\
 31 & 1255 & 1.280 & 1 & 1 & 1 \\
 32 & 1256 & 1.281 & 2 & 1 & 1 \\
 33 & 1257 & 1.282 & 2 & 1 & 1 \\
 \hline
 TOTAL & 100000 & 102.0 & 102 & 102 & 102 \\
\hline
\end{tabular}
\end{small}
\caption{Apportionments for a $P = 100
\,000$, $S = 33$, $R = 102$ test case of Balinski and Young \protect\cite{balinski_young:1975}, for which the Huntington Method, Webster Method, and  entropic method all violate upper quota. The latter two methods happen to agree, while the former awards extra seats to two states whose populations only very slightly exceed that of thirty other states.}\label{table:above_quota}
\end{table}

\begin{table}[ht!]
\begin{small}
\begin{tabular}{crrccc}
\hline
State &  Population & \hspace{6pt} Quota & \hspace{6pt} Huntington & Webster & Entropic \\
\hline\hline
 1 & 68010 & 69.370 & 60 & 64 & 63 \\
 2 & 1590 & 1.622 & 1 & 1 & 1 \\
 3 & 1591 & 1.623 & 1 & 1 & 1 \\
 4 & 1592 & 1.624 & 2 & 1 & 1 \\
 5 & 1593 & 1.625 & 2 & 1 & 1 \\
 6 & 1594 & 1.626 & 2 & 1 & 1 \\
 7 & 1595 & 1.627 & 2 & 1 & 2 \\
 8 & 1596 & 1.628 & 2 & 2 & 2 \\
 9 & 1597 & 1.629 & 2 & 2 & 2 \\
 10 & 1598 & 1.630 & 2 & 2 & 2 \\
 11 & 1599 & 1.631 & 2 & 2 & 2 \\
 12 & 1600 & 1.632 & 2 & 2 & 2 \\
 13 & 1601 & 1.633 & 2 & 2 & 2 \\
 14 & 1602 & 1.634 & 2 & 2 & 2 \\
 15 & 1603 & 1.635 & 2 & 2 & 2 \\
 16 & 1604 & 1.636 & 2 & 2 & 2 \\
 17 & 1605 & 1.637 & 2 & 2 & 2 \\
 18 & 1606 & 1.638 & 2 & 2 & 2 \\
 19 & 1607 & 1.639 & 2 & 2 & 2 \\
 20 & 1608 & 1.640 & 2 & 2 & 2 \\
 21 & 1609 & 1.641 & 2 & 2 & 2 \\
\hline
 TOTAL & 100000 & 98.00 & 98 & 98 & 98 \\
\hline
\end{tabular}
\end{small}
\caption{Apportionments for a $P = 100
\,000$, $S = 21$, $R = 98$ test case of Balinski and Young \protect\cite{balinski_young:1975}, for which the Huntington Method, Webster Method, and  entropic method all violate lower quota. Note that all three apportionments disagree for this artificial population distribution.  The entropic method tends to ``split the difference'' in terms of the seats awarded to the large state, and the level at which it begins allotting a second seat to the smaller states.}\label{table:below_quota}
\end{table}


\begin{table}[h!]
\begin{small}
\begin{tabular}{ccccccc}
\hline
Scenario $a$: \\
  &   &   &  Entropic   & Sub-Optimal  \\
 State  &  Population &  Quota &  Apportionment   & Apportionment  \\
\hline\hline
 $X$ & 731 & 7.31 &   8  &   7 \\
 $Y$ & 535 & 5.35 &   5  &  6  \\
 $Z$ & 334 & 3.34  &  3  &  3  \\
 \hline
 TOTAL & 1600 & 16 & 16 & 16  \\
\hline
Relative Entropy & & & 0.00552  & 0.00564 \\ 
\hline\\
\end{tabular}
\vspace{20pt}
\begin{tabular}{ccccccc}
\hline
Scenario $b$: \\
  &   &   &  Entropic   & Sub-Optimal  \\
 State  &  Population &  Quota &  Apportionment   & Apportionment  \\
\hline\hline
 $X$ & 729 & 7.29 &   7  &   8 \\
 $Y$ & 535 & 5.35 &   6  &  5  \\
 $Z$ & 336 & 3.36  &  3  &  3  \\
\hline
TOTAL & 1600 & 16 & 16 & 16  \\
\hline
Relative Entropy & & & 0.00576  & 0.00588 \\ 
\hline \\
\end{tabular}
\end{small}
\caption{Apportionments for $P = 1\,600$, $S = 3$, $R = 16$ test cases of Huntington \protect\cite{huntington:1928}, showing that one state's quota by itself does not determine its optimal apportionment---in both cases the population and quota for state $Y$ are the same, but its optimal apportionment differs under the two scenarios, as the number of seats received may depend on small shifts in population between other states.  Here the Huntington apportionment, Webster apportionment, and optimal entropic apportionment all coincide for both scenarios.  The sub-optimal apportionment refers to the optimal apportionment from the other scenario.}\label{table:quota_not_enough_a}
\end{table}

\begin{table}[h!]
\begin{small}
\begin{tabular}{cccc}
\hline
State &  Population &  Quota &  Apportionment  \\
\hline\hline
 $A$ & 1536 & 15.36 &  15 \\
 $B$ & 1535 & 15.35 &  15 \\
$C$ & 1534 & 15.34 &  15 \\
$D$ & 1533 & 15.33 &  15 \\
$E$ & 1532 & 15.32 &  15 \\
$F$ & 1530 & 15.30 &  15 \\
$G$ & 162 & \phantom{0}1.62 &  2 \\
$H$ & 161 & \phantom{0}1.61 &  2 \\
$I$ & 160 & \phantom{0}1.60 &  2 \\
$J$ & 159 & \phantom{0}1.59 &  2 \\
$K$ & 158 & \phantom{0}1.58 &  2 \\
 \hline
 TOTAL & 10000 & 100 & 100   \\
\hline\\
\end{tabular}
\vspace{20pt}
\begin{tabular}{ccccccc}
\hline
Group & Population &  Group-Level Quota &  Total Apportionment  \\
\hline\hline
 $ABCDEF$ & 9200 & 92.00 &   90 \\
 $GHIJK$ & 800 & \phantom{0}8.00 &  10  \\
 \hline
TOTAL & 10000 & 100 & 100  \\
\hline
\end{tabular}
\end{small}
\caption{Apportionments for a $P = 10\hspace{0.95pt}000$, $S = 11$, $R = 100$ test case of Huntington \protect\cite{huntington:1928}, showing that if states are all on-quota in the usual (state-level) sense, then groups of states may depart from group-level quota.}\label{table:aggregate_quota}
\end{table}

\newpage

\begin{table}[ht!]
\begin{small}
\begin{tabular}{crrccc}
\hline
State &  Population & Quota & Huntington & Webster & Entropic \\
\hline\hline
 \text{Alabama} & 3893888 & 7.498 & 7 & 7 & 7 \\
 \text{Alaska} & 401851 & 0.774 & 1 & 1 & 1 \\
 \text{Arizona} & 2718215 & 5.234 & 5 & 5 & 5 \\
 \text{Arkansas} & 2286435 & 4.403 & 4 & 4 & 4 \\
 \text{California} & 23667902 & 45.574 & 45 & 45 & 45 \\
 \text{Colorado} & 2889964 & 5.565 & 6 & 6 & 6 \\
 \text{Connecticut} & 3107576 & 5.984 & 6 & 6 & 6 \\
 \text{Delaware} & 594338 & 1.144 & 1 & 1 & 1 \\
 \text{Florida} & 9746324 & 18.767 & 19 & 19 & 19 \\
 \text{Georgia} & 5463105 & 10.520 & 10 & 10 & 10 \\
 \text{Hawaii} & 964691 & 1.858 & 2 & 2 & 2 \\
 \text{Idaho} & 943935 & 1.818 & 2 & 2 & 2 \\
 \text{Illinois} & 11426518 & 22.003 & 22 & 22 & 22 \\
 \text{Indiana} & 5490224 & 10.572 & 10* & 11 & 10* \\
 \text{Iowa} & 2913808 & 5.611 & 6 & 6 & 6 \\
 \text{Kansas} & 2363679 & 4.551 & 5 & 5 & 5 \\
 \text{Kentucky} & 3660777 & 7.049 & 7 & 7 & 7 \\
 \text{Louisiana} & 4205900 & 8.099 & 8 & 8 & 8 \\
 \text{Maine} & 1124660 & 2.166 & 2 & 2 & 2 \\
 \text{Maryland} & 4216975 & 8.120 & 8 & 8 & 8 \\
 \text{Massachusetts} & 5737037 & 11.047 & 11 & 11 & 11 \\
 \text{Michigan} & 9262078 & 17.835 & 18 & 18 & 18 \\
 \text{Minnesota} & 4075970 & 7.849 & 8 & 8 & 8 \\
 \text{Mississippi} & 2520638 & 4.854 & 5 & 5 & 5 \\
 \text{Missouri} & 4916686 & 9.467 & 9 & 9 & 9 \\
 \text{Montana} & 786690 & 1.515 & 2 & 2 & 2 \\
 \text{Nebraska} & 1569825 & 3.023 & 3 & 3 & 3 \\
 \text{Nevada} & 800493 & 1.541 & 2 & 2 & 2 \\
 \text{New Hampshire} & 920610 & 1.773 & 2 & 2 & 2 \\
 \text{New Jersey} & 7364823 & 14.182 & 14 & 14 & 14 \\
 \text{New Mexico} & 1302894 & 2.509 & 3* & 2 & 3* \\
 \text{New York} & 17558072 & 33.809 & 34 & 34 & 34 \\
 \text{North Carolina} & 5881766 & 11.326 & 11 & 11 & 11 \\
 \text{North Dakota} & 652717 & 1.257 & 1 & 1 & 1 \\
 \text{Ohio} & 10797630 & 20.792 & 21 & 21 & 21 \\
 \text{Oklahoma} & 3025290 & 5.825 & 6 & 6 & 6 \\
 \text{Oregon} & 2633105 & 5.070 & 5 & 5 & 5 \\
 \text{Pennsylvania} & 11863895 & 22.845 & 23 & 23 & 23 \\
 \text{Rhode Island} & 947154 & 1.824 & 2 & 2 & 2 \\
 \text{South Carolina} & 3121820 & 6.011 & 6 & 6 & 6 \\
 \text{South Dakota} & 690768 & 1.330 & 1 & 1 & 1 \\
 \text{Tennessee} & 4591120 & 8.841 & 9 & 9 & 9 \\
 \text{Texas} & 14229191 & 27.399 & 27 & 27 & 27 \\
 \text{Utah} & 1461037 & 2.813 & 3 & 3 & 3 \\
 \text{Vermont} & 511456 & 0.985 & 1 & 1 & 1 \\
 \text{Virginia} & 5346818 & 10.296 & 10 & 10 & 10 \\
 \text{Washington} & 4132156 & 7.957 & 8 & 8 & 8 \\
 \text{West Virginia} & 1949644 & 3.754 & 4 & 4 & 4 \\
 \text{Wisconsin} & 4705767 & 9.061 & 9 & 9 & 9 \\
 \text{Wyoming} & 469557 & 0.904 & 1 & 1 & 1 \\
\hline
\end{tabular}
\end{small}
\caption{U.S.~Congressional apportionments of $R = 435$ seats following the 1980 census were assigned via the Huntington-Hill Method.  The entropic method assigns the same seats as the Huntington-Hill Method, differing from the Webster Method for Indiana and New Mexico.}\label{Table:1980}
\end{table}

\begin{table}[ht!]
\begin{small}
\begin{tabular}{crrccc}
\hline
State &  Population & Quota & Huntington & Webster & Entropic \\
\hline\hline
 \text{Alabama} & 4062608 & 7.097 & 7 & 7 & 7 \\
 \text{Alaska} & 551947 & 0.964 & 1 & 1 & 1 \\
 \text{Arizona} & 3677985 & 6.425 & 6 & 6 & 6 \\
 \text{Arkansas} & 2362239 & 4.126 & 4 & 4 & 4 \\
 \text{California} & 29839250 & 52.124 & 52 & 52 & 52 \\
 \text{Colorado} & 3307912 & 5.778 & 6 & 6 & 6 \\
 \text{Connecticut} & 3295669 & 5.757 & 6 & 6 & 6 \\
 \text{Delaware} & 668696 & 1.168 & 1 & 1 & 1 \\
 \text{Florida} & 13003362 & 22.715  & 23 & 23 & 23 \\
 \text{Georgia} & 6508419 & 11.369  & 11 & 11 & 11 \\
 \text{Hawaii} & 1115274 & 1.948  & 2 & 2 & 2 \\
 \text{Idaho} & 1011986 & 1.768  & 2 & 2 & 2 \\
 \text{Illinois} & 11466682 & 20.030  & 20 & 20 & 20 \\
 \text{Indiana} & 5564228 & 9.720  & 10 & 10 & 10 \\
 \text{Iowa} & 2787424 & 4.869  & 5 & 5 & 5 \\
 \text{Kansas} & 2485600 & 4.342  & 4 & 4 & 4 \\
 \text{Kentucky} & 3698969 & 6.461  & 6 & 6 & 6 \\
 \text{Louisiana} & 4238216 & 7.403  & 7 & 7 & 7 \\
 \text{Maine} & 1233223 & 2.154  & 2 & 2 & 2 \\
 \text{Maryland} & 4798622 & 8.382  & 8 & 8 & 8 \\
 \text{Massachusetts} & 6029051 & 10.532 & 10* & 11 & 10* \\
 \text{Michigan} & 9328784 & 16.296  & 16 & 16 & 16 \\
 \text{Minnesota} & 4387029 & 7.663  & 8 & 8 & 8 \\
 \text{Mississippi} & 2586443 & 4.518  & 5 & 5 & 5 \\
 \text{Missouri} & 5137804 & 8.975 & 9 & 9 & 9 \\
 \text{Montana} & 803655 & 1.404  & 1 & 1 & 1 \\
 \text{Nebraska} & 1584617 & 2.768  & 3 & 3 & 3 \\
 \text{Nevada} & 1206152 & 2.107  & 2 & 2 & 2 \\
 \text{New Hampshire} & 1113915 & 1.946  & 2 & 2 & 2 \\
 \text{New Jersey} & 7748634 & 13.536  & 13 & 13 & 13 \\
 \text{New Mexico} & 1521779 & 2.658  & 3 & 3 & 3 \\
 \text{New York} & 18044505 & 31.521  & 31 & 31 & 31 \\
 \text{North Carolina} & 6657630 & 11.630  & 12 & 12 & 12 \\
 \text{North Dakota} & 641364 & 1.120  & 1 & 1 & 1 \\
 \text{Ohio} & 10887325 & 19.018  & 19 & 19 & 19 \\
 \text{Oklahoma} & 3157604 & 5.516  & 6* & 5 & 6* \\
 \text{Oregon} & 2853733 & 4.985  & 5 & 5 & 5 \\
 \text{Pennsylvania} & 11924710 & 20.830  & 21 & 21 & 21 \\
 \text{Rhode Island} & 1005984 & 1.757  & 2 & 2 & 2 \\
 \text{South Carolina} & 3505707 & 6.124  & 6 & 6 & 6 \\
 \text{South Dakota} & 699999 & 1.224  & 1 & 1 & 1 \\
 \text{Tennessee} & 4896641 & 8.554  & 9 & 9 & 9 \\
 \text{Texas} & 17059805 & 29.801 & 30 & 30 & 30 \\
 \text{Utah} & 1727784 & 3.018  & 3 & 3 & 3 \\
 \text{Vermont} & 564964 & 0.987  & 1 & 1 & 1 \\
 \text{Virginia} & 6216568 & 10.859  & 11 & 11 & 11 \\
 \text{Washington} & 4887941 & 8.538  & 9 & 9 & 9 \\
 \text{West Virginia} & 1801625 & 3.147  & 3 & 3 & 3 \\
 \text{Wisconsin} & 4906745 & 8.571  & 9 & 9 & 9 \\
 \text{Wyoming} & 455975 & 0.797 & 1 & 1 & 1 \\
 \hline
\end{tabular}
\end{small}
\caption{U.S.~Congressional apportionments of $R  = 435$ seats following the 1990 census were assigned via the Huntington-Hill Method.  The entropic method assigns the same seats as the Huntington-Hill Method, differing from the Webster Method for Massachusetts and Oklahoma.}\label{Table:1990}
\end{table}

\begin{table}[ht!]
\begin{small}
\begin{tabular}{crrccc}
\hline
State &  Population & Quota & Huntington & Webster & Entropic \\
\hline\hline
\text{Alabama} & 4461130 & 6.896 & 7 & 7 & 7 \\
 \text{Alaska} & 628933 & 0.972  & 1 & 1 & 1 \\
 \text{Arizona} & 5140683 & 7.946 & 8 & 8 & 8 \\
 \text{Arkansas} & 2679733 & 4.142  & 4 & 4 & 4 \\
 \text{California} & 33930798 & 52.447  & 53 & 53 & 53 \\
 \text{Colorado} & 4311882 & 6.665  & 7 & 7 & 7 \\
 \text{Connecticut} & 3409535 & 5.270  & 5 & 5 & 5 \\
 \text{Delaware} & 785068 & 1.213  & 1 & 1 & 1 \\
 \text{Florida} & 16028890 & 24.776 & 25 & 25 & 25 \\
 \text{Georgia} & 8206975 & 12.686 & 13 & 13 & 13 \\
 \text{Hawaii} & 1216642 & 1.881  & 2 & 2 & 2 \\
 \text{Idaho} & 1297274 & 2.005  & 2 & 2 & 2 \\
 \text{Illinois} & 12439042 & 19.227  & 19 & 19 & 19 \\
 \text{Indiana} & 6090782 & 9.415  & 9 & 9 & 9 \\
 \text{Iowa} & 2931923 & 4.532  & 5 & 5 & 5 \\
 \text{Kansas} & 2693824 & 4.164  & 4 & 4 & 4 \\
 \text{Kentucky} & 4049431 & 6.259  & 6 & 6 & 6 \\
 \text{Louisiana} & 4480271 & 6.925  & 7 & 7 & 7 \\
 \text{Maine} & 1277731 & 1.975 & 2 & 2 & 2 \\
 \text{Maryland} & 5307886 & 8.204  & 8 & 8 & 8 \\
 \text{Massachusetts} & 6355568 & 9.824  & 10 & 10 & 10 \\
 \text{Michigan} & 9955829 & 15.389  & 15 & 15 & 15 \\
 \text{Minnesota} & 4925670 & 7.614  & 8 & 8 & 8 \\
 \text{Mississippi} & 2852927 & 4.410 & 4 & 4 & 4 \\
 \text{Missouri} & 5606260 & 8.666  & 9 & 9 & 9 \\
 \text{Montana} & 905316 & 1.399  & 1 & 1 & 1 \\
 \text{Nebraska} & 1715369 & 2.651  & 3 & 3 & 3 \\
 \text{Nevada} & 2002032 & 3.095  & 3 & 3 & 3 \\
 \text{New Hampshire} & 1238415 & 1.914  & 2 & 2 & 2 \\
 \text{New Jersey} & 8424354 & 13.021  & 13 & 13 & 13 \\
 \text{New Mexico} & 1823821 & 2.819  & 3 & 3 & 3 \\
 \text{New York} & 19004973 & 29.376  & 29 & 29 & 29 \\
 \text{North Carolina} & 8067673 & 12.470  & 13 & 13 & 13 \\
 \text{North Dakota} & 643756 & 0.995  & 1 & 1 & 1 \\
 \text{Ohio} & 11374540 & 17.582  & 18 & 18 & 18 \\
 \text{Oklahoma} & 3458819 & 5.346  & 5 & 5 & 5 \\
 \text{Oregon} & 3428543 & 5.230  & 5 & 5 & 5 \\
 \text{Pennsylvania} & 12300670 & 19.013  & 19 & 19 & 19 \\
 \text{Rhode Island} & 1049662 & 1.622  & 2 & 2 & 2 \\
 \text{South Carolina} & 4025061 & 6.222  & 6 & 6 & 6 \\
 \text{South Dakota} & 756874 & 1.170  & 1 & 1 & 1 \\
 \text{Tennessee} & 5700037 & 8.811  & 9 & 9 & 9 \\
 \text{Texas} & 20903994 & 32.312  & 32 & 32 & 32 \\
 \text{Utah} & 2236714 & 3.457  & 3 & 3 & 3 \\
 \text{Vermont} & 609890 & 0.943  & 1 & 1 & 1 \\
 \text{Virginia} & 7100702 & 10.976  & 11 & 11 & 11 \\
 \text{Washington} & 5908684 & 9.133  & 9 & 9 & 9 \\
 \text{West Virginia} & 1813077 & 2.802  & 3 & 3 & 3 \\
 \text{Wisconsin} & 5371210 & 8.302  & 8 & 8 & 8 \\
 \text{Wyoming} & 495304 & 0.766  & 1 & 1 & 1 \\
 \hline
 \end{tabular}
 \end{small}
\caption{U.S.~Congressional apportionments of $R = 435$ seats following the 2000 census were assigned via the Huntington-Hill Method.  Both the entropic method and the Webster Method lead to the same apportionment as the Huntington-Hill Method.}\label{Table:2000}
\end{table}

\begin{table}[ht!]
\begin{small}
\begin{tabular}{crrccc}
\hline
State &  Population & Quota & Huntington & Webster & Entropic \\
\hline\hline
\text{Alabama} & 4802982 & 6.75747 & 7 & 7 & 7 \\
 \text{Alaska} & 721523 & 1.01513 & 1 & 1 & 1 \\
 \text{Arizona} & 6412700 & 9.02223 & 9 & 9 & 9 \\
 \text{Arkansas} & 2926229 & 4.117 & 4 & 4 & 4 \\
 \text{California} & 37341989 & 52.5376 & 53 & 53 & 53 \\
 \text{Colorado} & 5044930 & 7.09787 & 7 & 7 & 7 \\
 \text{Connecticut} & 3581628 & 5.03911 & 5 & 5 & 5 \\
 \text{Delaware} & 900877 & 1.26747 & 1 & 1 & 1 \\
 \text{Florida} & 18900773 & 26.5921 & 27 & 27 & 27 \\
 \text{Georgia} & 9727566 & 13.686 & 14 & 14 & 14 \\
 \text{Hawaii} & 1366862 & 1.92308 & 2 & 2 & 2 \\
 \text{Idaho} & 1573499 & 2.21381 & 2 & 2 & 2 \\
 \text{Illinois} & 12864380 & 18.0993 & 18 & 18 & 18 \\
 \text{Indiana} & 6501582 & 9.14728 & 9 & 9 & 9 \\
 \text{Iowa} & 3053787 & 4.29647 & 4 & 4 & 4 \\
 \text{Kansas} & 2863813 & 4.02919 & 4 & 4 & 4 \\
 \text{Kentucky} & 4350606 & 6.12101 & 6 & 6 & 6 \\
 \text{Louisiana} & 4553962 & 6.40711 & 6 & 6 & 6 \\
 \text{Maine} & 1333074 & 1.87554 & 2 & 2 & 2 \\
 \text{Maryland} & 5789929 & 8.14603 & 8 & 8 & 8 \\
 \text{Massachusetts} & 6559644 & 9.22897 & 9 & 9 & 9 \\
 \text{Michigan} & 9911626 & 13.945 & 14 & 14 & 14 \\
 \text{Minnesota} & 5314879 & 7.47767 & 8 & 8 & 8 \\
 \text{Mississippi} & 2978240 & 4.19018 & 4 & 4 & 4 \\
 \text{Missouri} & 6011478 & 8.45774 & 8 & 8 & 8 \\
 \text{Montana} & 994416 & 1.39908 & 1 & 1 & 1 \\
 \text{Nebraska} & 1831825 & 2.57725 & 3 & 3 & 3 \\
 \text{Nevada} & 2709432 & 3.81199 & 4 & 4 & 4 \\
 \text{New Hampshire} & 1321445 & 1.85918 & 2 & 2 & 2 \\
 \text{New Jersey} & 8807501 & 12.3916 & 12 & 12 & 12 \\
 \text{New Mexico} & 2067273 & 2.90851 & 3 & 3 & 3 \\
 \text{New York} & 19421055 & 27.3241 & 27 & 27 & 27 \\
 \text{North Carolina} & 9565781 & 13.4584 & 13* & 14 & 13* \\
 \text{North Dakota} & 675905 & 0.950952 & 1 & 1 & 1 \\
 \text{Ohio} & 11568495 & 16.2761 & 16 & 16 & 16 \\
 \text{Oklahoma} & 3764882 & 5.29693 & 5 & 5 & 5 \\
 \text{Oregon} & 3848606 & 5.41473 & 5 & 5 & 5 \\
 \text{Pennsylvania} & 12734905 & 17.9171 & 18 & 18 & 18 \\
 \text{Rhode Island} & 1055247 & 1.48466 & 2* & 1 & 2* \\
 \text{South Carolina} & 4645975 & 6.53657 & 7 & 7 & 7 \\
 \text{South Dakota} & 819761 & 1.15335 & 1 & 1 & 1 \\
 \text{Tennessee} & 6375431 & 8.9698 & 9 & 9 & 9 \\
 \text{Texas} & 25268418 & 35.5509 & 36 & 36 & 36 \\
 \text{Utah} & 2770765 & 3.89828 & 4 & 4 & 4 \\
 \text{Vermont} & 630337 & 0.886841 & 1 & 1 & 1 \\
 \text{Virginia} & 8037736 & 11.3085 & 11 & 11 & 11 \\
 \text{Washington} & 6753369 & 9.50153 & 10 & 10 & 10 \\
 \text{West Virginia} & 1859815 & 2.61663 & 3 & 3 & 3 \\
 \text{Wisconsin} & 5698230 & 8.01702 & 8 & 8 & 8 \\
 \text{Wyoming} & 568300 & 0.799559 & 1 & 1 & 1 \\
\hline
\end{tabular}
\end{small}
\caption{U.S.~Congressional apportionments of $R = 435$ seats following the 2010 census were assigned via the Huntington-Hill Method.  The entropic method assigns the same seats as the Huntington-Hill Method, differing from the Webster Method for North Carolina and Rhode Island.}\label{Table:2010}
\end{table}


\begin{table}[ht!]
\begin{small}
\begin{tabular}{crrcccrr}
\hline
State & Population & Exact Quota  & Huntington & Entropic & Difference & $\K_{\stext{H}}$ contribution & $\K_{\stext{E}}$ Contribution  \\
\hline\hline
\text{Alabama}  & $5030053$ & $6.608328$ & $7$ & $7$ & $0$ & $0.00133675$ & $0.00133675$ \\
 \text{Alaska}  & $736081$ & $0.967040$ & $1$ & $1$ & $0$ & $0.00011115$ & $0.00011115$ \\
 \text{Arizona}  & $7158923$ & $9.405171$ & $9$ & $9$ & $0$ & $-0.00131440$ & $-0.00131440$ \\
 \text{Arkansas}  & $3013756$ & $3.959379$ & $4$ & $4$ & $0$ & $0.00013541$ & $0.00013541$ \\
 \text{California}  & $39576757$ & $51.994717$ & $52$ & $52$ & $0$ & $0.00001752$ & $0.00001752$ \\
 \text{Colorado}  & $5782171$ & $7.596437$ & $8$ & $8$ & $0$ & $0.00137337$ & $0.00137337$ \\
 \text{Connecticut}  & $3608298$ & $4.740470$ & $5$ & $5$ & $0$ & $0.00088388$ & $0.00088388$ \\
 \text{Delaware}  & $990837$ & $1.301731$ & $1$ & $1$ & $0$ & $-0.00087455$ & $-0.00087455$ \\
 \text{Florida}  & $21570527$ & $28.338690$ & $28$ & $28$ & $0$ & $-0.00111654$ & $-0.00111654$ \\
 \text{Georgia}  & $10725274$ & $14.090533$ & $14$ & $14$ & $0$ & $-0.00029929$ & $-0.00029929$ \\
 \text{Hawaii}  & $1460137$ & $1.918283$ & $2$ & $2$ & $0$ & $0.00027671$ & $0.00027671$ \\
 \text{Idaho}  & $1841377$ & $2.419144$ & $2$ & $2$ & $0$ & $-0.00126205$ & $-0.00126205$ \\
 \text{Illinois}  & $12822739$ & $16.846117$ & $17$ & $17$ & $0$ & $0.00051268$ & $0.00051268$ \\
 \text{Indiana}  & $6790280$ & $8.920859$ & $9$ & $9$ & $0$ & $0.00026363$ & $0.00026363$ \\
 \text{Iowa}  & $3192406$ & $4.194084$ & $4$ & $4$ & $0$ & $-0.00062856$ & $-0.00062856$ \\
 \text{Kansas}  & $2940865$ & $3.863617$ & $4$ & $4$ & $0$ & $0.00046021$ & $0.00046021$ \\
 \text{Kentucky}  & $4509342$ & $5.924234$ & $6$ & $6$ & $0$ & $0.00025288$ & $0.00025288$ \\
 \text{Louisiana}  & $4661468$ & $6.124092$ & $6$ & $6$ & $0$ & $-0.00040736$ & $-0.00040736$ \\
 \text{Maine}  & $1363582$ & $1.791432$ & $2$ & $2$ & $0$ & $0.00073051$ & $0.00073051$ \\
 \text{Maryland}  & $6185278$ & $8.126027$ & $8$ & $8$ & $0$ & $-0.00041471$ & $-0.00041471$ \\
 \text{Massachusetts}  & $7033469$ & $9.240354$ & $9$ & $9$ & $0$ & $-0.00078668$ & $-0.00078668$ \\
 \text{Michigan}  & $10084442$ & $13.248627$ & $13$ & $13$ & $0$ & $-0.00081680$ & $-0.00081680$ \\
 \text{Minnesota}  & $5709752$ & $7.501295$ & $8$ & $8$ & $0$ & $0.00170778$ & $0.00170778$ \\
 \text{Mississippi}  & $2963914$ & $3.893898$ & $4$ & $4$ & $0$ & $0.00035664$ & $0.00035664$ \\
 \text{Missouri}  & $6160281$ & $8.093186$ & $8$ & $8$ & $0$ & $-0.00030727$ & $-0.00030727$ \\
 \textbf{Montana}  & $1085407$ & $1.425974$ & $\mathbf{2}$ & $\mathbf{1}$ & $-1$ & $0.00224392$ & $-0.00117689$ \\
 \text{Nebraska}  & $1963333$ & $2.579366$ & $3$ & $3$ & $0$ & $0.00150308$ & $0.00150308$ \\
 \text{Nevada}  & $3108462$ & $4.083801$ & $4$ & $4$ & $0$ & $-0.00027506$ & $-0.00027506$ \\
 \text{New Hampshire}  & $1379089$ & $1.811804$ & $2$ & $2$ & $0$ & $0.00065551$ & $0.00065551$ \\
 \text{New Jersey}  & $9294493$ & $12.210817$ & $12$ & $12$ & $0$ & $-0.00069311$ & $-0.00069311$ \\
 \text{New Mexico}  & $2120220$ & $2.785479$ & $3$ & $3$ & $0$ & $0.00073819$ & $0.00073819$ \\
 \textbf{\textit{New York}}  & $20215751$ & $26.558827$ & $\mathbf{26}$ & $\mathbf{27}$ & $+1$ & $-0.00183374$ & $0.00147525$ \\
 \text{North Carolina}  & $10453948$ & $13.734073$ & $14$ & $14$ & $0$ & $0.00089044$ & $0.00089044$ \\
 \text{North Dakota}  & $779702$ & $1.024348$ & $1$ & $1$ & $0$ & $-0.00007978$ & $-0.00007978$ \\
 \textbf{\textit{Ohio}}  & $11808848$ & $15.514099$ & $\mathbf{15}$ & $\mathbf{16}$ & $+1$ & $-0.00167646$ & $0.00163649$ \\
 \text{Oklahoma}  & $3963516$ & $5.207144$ & $5$ & $5$ & $0$ & $-0.00067315$ & $-0.00067315$ \\
 \text{Oregon}  & $4241500$ & $5.572351$ & $6$ & $6$ & $0$ & $0.00147140$ & $0.00147140$ \\
 \text{Pennsylvania}  & $13011844$ & $17.094557$ & $17$ & $17$ & $0$ & $-0.00031273$ & $-0.00031273$ \\
 \textbf{Rhode Island}  & $1098163$ & $1.442733$ & $\mathbf{2}$ & $\mathbf{1}$ & $-1$ & $0.00216642$ & $-0.00121564$ \\
 \text{South Carolina}  & $5124712$ & $6.732688$ & $7$ & $7$ & $0$ & $0.00090392$ & $0.00090392$ \\
 \text{South Dakota}  & $887770$ & $1.166325$ & $1$ & $1$ & $0$ & $-0.00051027$ & $-0.00051027$ \\
 \text{Tennessee}  & $6916897$ & $9.087205$ & $9$ & $9$ & $0$ & $-0.00028783$ & $-0.00028783$ \\
 \text{Texas}  & $29183290$ & $38.340102$ & $38$ & $38$ & $0$ & $-0.00112294$ & $-0.00112294$ \\
 \text{Utah}  & $3275252$ & $4.302925$ & $4$ & $4$ & $0$ & $-0.00096844$ & $-0.00096844$ \\
 \text{Vermont}  & $643503$ & $0.845414$ & $1$ & $1$ & $0$ & $0.00055694$ & $0.00055694$ \\
 \text{Virginia}  & $8654542$ & $11.370069$ & $11$ & $11$ & $0$ & $-0.00120716$ & $-0.00120716$ \\
 \text{Washington}  & $7715946$ & $10.136971$ & $10$ & $10$ & $0$ & $-0.00045119$ & $-0.00045119$ \\
 \text{West Virginia}  & $1795045$ & $2.358274$ & $2$ & $2$ & $0$ & $-0.00109302$ & $-0.00109302$ \\
 \text{Wisconsin}  & $5897473$ & $7.747917$ & $8$ & $8$ & $0$ & $0.00084950$ & $0.00084950$ \\
 \text{Wyoming}  & $577719$ & $0.758989$ & $1$ & $1$ & $0$ & $0.00091459$ & $0.00091459$ \\
 \hline
 TOTAL &  $331108434$ & $435.000000$ &  $435$ & $435$  & $0$  & $0.00189995$ &  $0.00171902$ \\
\hline
\end{tabular}
\end{small}
\caption{\textit{Proposed} U.S.~Congressional apportionments of $R = 435$ seats following the 2020 Census have been calculated via the Huntington-Hill Method of Equal Proportions. The Huntington and entropic methods disagree, with the entropic method shifting one seat each from Montana and Rhode Island to New York and Ohio. State-by-state  contributions to the discrimination informations  are also listed for both apportionments.}\label{Table:2020}
\end{table}

\newpage

\begin{table}[ht!]
\begin{scriptsize}
\begin{tabular}{ccrc | llll | llll}
\hline
         &  Counted                  & ``Exact'' & Baseline  & \hspace{6pt} (a) $0.6\%$ & \textit{undercount} &  &   & \hspace{6pt} (b) $1.2\%$ & \textit{undercount} &  &  \\
State & Population & Quota & Seats  & Avg.\ Shift & RMS Shift & $P$(loss) & $P$(gain)  & Avg.\ Shift & RMS Shift & $P$(loss) & $P$(gain)  \\
\hline\hline
 \text{Alabama}  & $5030053$ & $6.60833$ & $7$ & $\phantom{+}0$ & $0$ & $0.00002$ & $0.00002$ & $-0.00718$ & $0.08473$ & $0.00720$ & $0.00002$ \\
 \text{Alaska}  & $736081$ & $0.96704$ & $1$ & $\phantom{+}0$ & $0$ & $0.00002$ & $0.00002$ & $\phantom{+}0$ & $0$ & $0.00002$ & $0.00002$ \\
 \text{Arizona}  & $7158923$ & $9.40517$ & $9$ & $\phantom{+}0.09780$ & $0.31273$ & $0.00002$ & $0.09782$ & $\phantom{+}0.18530$ & $0.43046$ & $0.00002$ & $0.18530$ \\
 \text{Arkansas}  & $3013756$ & $3.95938$ & $4$ & $\phantom{+}0$ & $0$ & $0.00002$ & $0.00002$ & $\phantom{+}0$ & $0$ & $0.00002$ & $0.00002$ \\
 \text{California}  & $39576757$ & $51.99472$ & $52$ & $\phantom{+}0.09558$ & $0.32302$ & $0.00172$ & $0.09468$ & $\phantom{+}0.10910$ & $0.62496$ & $0.10218$ & $0.17884$ \\
 \text{Colorado}  & $5782171$ & $7.59644$ & $8$ & $-0.00012$ & $0.01095$ & $0.00014$ & $0.00002$ & $-0.03926$ & $0.19814$ & $0.03928$ & $0.00002$ \\
 \text{Connecticut}  & $3608298$ & $4.74047$ & $5$ & $\phantom{+}0$ & $0$ & $0.00002$ & $0.00002$ & $\phantom{+}0$ & $0$ & $0.00002$ & $0.00002$ \\
 \text{Delaware}  & $990837$ & $1.30173$ & $1$ & $\phantom{+}0$ & $0$ & $0.00002$ & $0.00002$ & $\phantom{+}0$ & $0$ & $0.00002$ & $0.00002$ \\
 \text{Florida}  & $21570527$ & $28.33869$ & $28$ & $\phantom{+}0.19282$ & $0.43966$ & $0.00002$ & $0.19259$ & $\phantom{+}0.27420$ & $0.54593$ & $0.00004$ & $0.26279$ \\
 \text{Georgia}  & $10725274$ & $14.09053$ & $14$ & $\phantom{+}0.00400$ & $0.06325$ & $0.00002$ & $0.00402$ & $\phantom{+}0.03662$ & $0.19147$ & $0.00002$ & $0.03662$ \\
 \text{Hawaii}  & $1460137$ & $1.91828$ & $2$ & $\phantom{+}0$ & $0$ & $0.00002$ & $0.00002$ & $\phantom{+}0$ & $0$ & $0.00002$ & $0.00002$ \\
 \text{Idaho}  & $1841377$ & $2.41914$ & $2$ & $\phantom{+}0.00526$ & $0.07253$ & $0.00002$ & $0.00528$ & $\phantom{+}0.04302$ & $0.20741$ & $0.00002$ & $0.04304$ \\
 \text{Illinois}  & $12822739$ & $16.84612$ & $17$ & $\phantom{+}0.00070$ & $0.02646$ & $0.00002$ & $0.00072$ & $\phantom{+}0.01404$ & $0.12884$ & $0.00122$ & $0.01518$ \\
 \text{Indiana}  & $6790280$ & $8.92086$ & $9$ & $\phantom{+}0$ & $0$ & $0.00002$ & $0.00002$ & $\phantom{+}0.00218$ & $0.04669$ & $0.00002$ & $0.00220$ \\
 \text{Iowa}  & $3192406$ & $4.19408$ & $4$ & $\phantom{+}0$ & $0$ & $0.00002$ & $0.00002$ & $\phantom{+}0.00088$ & $0.02966$ & $0.00002$ & $0.00090$ \\
 \text{Kansas}  & $2940865$ & $3.86362$ & $4$ & $\phantom{+}0$ & $0$ & $0.00002$ & $0.00002$ & $\phantom{+}0$ & $0$ & $0.00002$ & $0.00002$ \\
 \text{Kentucky}  & $4509342$ & $5.92423$ & $6$ & $\phantom{+}0$ & $0$ & $0.00002$ & $0.00002$ & $\phantom{+}0.00012$ & $0.01095$ & $0.00002$ & $0.00014$ \\
 \text{Louisiana}  & $4661468$ & $6.12409$ & $6$ & $\phantom{+}0.00002$ & $0.00447$ & $0.00002$ & $0.00004$ & $\phantom{+}0.00258$ & $0.05079$ & $0.00002$ & $0.00260$ \\
 \text{Maine}  & $1363582$ & $1.79143$ & $2$ & $\phantom{+}0$ & $0$ & $0.00002$ & $0.00002$ & $\phantom{+}0$ & $0$ & $0.00002$ & $0.00002$ \\
 \text{Maryland}  & $6185278$ & $8.12603$ & $8$ & $\phantom{+}0.00028$ & $0.01673$ & $0.00002$ & $0.00030$ & $\phantom{+}0.00882$ & $0.09391$ & $0.00002$ & $0.00884$ \\
 \text{Massachusetts}  & $7033469$ & $9.24035$ & $9$ & $\phantom{+}0.00512$ & $0.07155$ & $0.00002$ & $0.00514$ & $\phantom{+}0.04064$ & $0.20159$ & $0.00002$ & $0.04066$ \\
 \text{Michigan}  & $10084442$ & $13.24863$ & $13$ & $\phantom{+}0.02070$ & $0.14387$ & $0.00002$ & $0.02072$ & $\phantom{+}0.08112$ & $0.28531$ & $0.00002$ & $0.08099$ \\
 \text{Minnesota}  & $5709752$ & $7.50130$ & $8$ & $-0.31186$ & $0.55844$ & $0.31187$ & $0.00002$ & $-0.52912$ & $0.72741$ & $0.52912$ & $0.00002$ \\
 \text{Mississippi}  & $2963914$ & $3.89390$ & $4$ & $\phantom{+}0$ & $0$ & $0.00002$ & $0.00002$ & $\phantom{+}0$ & $0$ & $0.00002$ & $0.00002$ \\
 \text{Missouri}  & $6160281$ & $8.09319$ & $8$ & $\phantom{+}0.00012$ & $0.01095$ & $0.00002$ & $0.00014$ & $\phantom{+}0.00616$ & $0.07849$ & $0.00002$ & $0.00618$ \\
 \text{Montana}  & $1085407$ & $1.42597$ & $1$ & $\phantom{+}0.00234$ & $0.04837$ & $0.00002$ & $0.00236$ & $\phantom{+}0.02764$ & $0.16625$ & $0.00002$ & $0.02766$ \\
 \text{Nebraska}  & $1963333$ & $2.57937$ & $3$ & $\phantom{+}0$ & $0$ & $0.00002$ & $0.00002$ & $\phantom{+}0$ & $0$ & $0.00002$ & $0.00002$ \\
 \text{Nevada}  & $3108462$ & $4.08380$ & $4$ & $\phantom{+}0$ & $0$ & $0.00002$ & $0.00002$ & $\phantom{+}0.00010$ & $0.01000$ & $0.00002$ & $0.00012$ \\
 \text{New Hampshire}  & $1379089$ & $1.81180$ & $2$ & $\phantom{+}0$ & $0$ & $0.00002$ & $0.00002$ & $\phantom{+}0$ & $0$ & $0.00002$ & $0.00002$ \\
 \text{New Jersey}  & $9294493$ & $12.21082$ & $12$ & $\phantom{+}0.01002$ & $0.10010$ & $0.00002$ & $0.01004$ & $\phantom{+}0.05522$ & $0.23524$ & $0.00002$ & $0.05517$ \\
 \text{New Mexico}  & $2120220$ & $2.78548$ & $3$ & $\phantom{+}0$ & $0$ & $0.00002$ & $0.00002$ & $\phantom{+}0$ & $0$ & $0.00002$ & $0.00002$ \\
 \text{New York}  & $20215751$ & $26.55883$ & $27$ & $-0.18028$ & $0.42708$ & $0.18135$ & $0.00108$ & $-0.43636$ & $0.68984$ & $0.45558$ & $0.01870$ \\
 \text{North Carolina}  & $10453948$ & $13.73407$ & $14$ & $\phantom{+}0.00002$ & $0.00447$ & $0.00002$ & $0.00004$ & $-0.00362$ & $0.10909$ & $0.00778$ & $0.00416$ \\
 \text{North Dakota}  & $779702$ & $1.02435$ & $1$ & $\phantom{+}0$ & $0$ & $0.00002$ & $0.00002$ & $\phantom{+}0$ & $0$ & $0.00002$ & $0.00002$ \\
 \text{Ohio}  & $11808848$ & $15.51410$ & $16$ & $-0.29694$ & $0.54492$ & $0.29695$ & $0.00002$ & $-0.51784$ & $0.72227$ & $0.51976$ & $0.00194$ \\
 \text{Oklahoma}  & $3963516$ & $5.20714$ & $5$ & $\phantom{+}0.00010$ & $0.01000$ & $0.00002$ & $0.00012$ & $\phantom{+}0.00470$ & $0.06856$ & $0.00002$ & $0.00472$ \\
 \text{Oregon}  & $4241500$ & $5.57235$ & $6$ & $\phantom{+}0$ & $0$ & $0.00002$ & $0.00002$ & $-0.02668$ & $0.16334$ & $0.02670$ & $0.00002$ \\
 \text{Pennsylvania}  & $13011844$ & $17.09456$ & $17$ & $\phantom{+}0.00888$ & $0.09423$ & $0.00002$ & $0.00890$ & $\phantom{+}0.05690$ & $0.24004$ & $0.00002$ & $0.05655$ \\
 \text{Rhode Island}  & $1098163$ & $1.44273$ & $1$ & $\phantom{+}0.01652$ & $0.12853$ & $0.00002$ & $0.01654$ & $\phantom{+}0.07600$ & $0.27568$ & $0.00002$ & $0.07601$ \\
 \text{South Carolina}  & $5124712$ & $6.73269$ & $7$ & $\phantom{+}0$ & $0$ & $0.00002$ & $0.00002$ & $\phantom{+}0$ & $0$ & $0.00002$ & $0.00002$ \\
 \text{South Dakota}  & $887770$ & $1.16632$ & $1$ & $\phantom{+}0$ & $0$ & $0.00002$ & $0.00002$ & $\phantom{+}0$ & $0$ & $0.00002$ & $0.00002$ \\
 \text{Tennessee}  & $6916897$ & $9.08720$ & $9$ & $\phantom{+}0.00014$ & $0.01183$ & $0.00002$ & $0.00016$ & $\phantom{+}0.00862$ & $0.09284$ & $0.00002$ & $0.00864$ \\
 \text{Texas}  & $29183290$ & $38.34010$ & $38$ & $\phantom{+}0.25102$ & $0.50551$ & $0.00002$ & $0.24877$ & $\phantom{+}0.32940$ & $0.62655$ & $0.00044$ & $0.30145$ \\
 \text{Utah}  & $3275252$ & $4.30292$ & $4$ & $\phantom{+}0.00036$ & $0.01897$ & $0.00002$ & $0.00038$ & $\phantom{+}0.01042$ & $0.10208$ & $0.00002$ & $0.01044$ \\
 \text{Vermont}  & $643503$ & $0.84541$ & $1$ & $\phantom{+}0$ & $0$ & $0.00002$ & $0.00002$ & $\phantom{+}0$ & $0$ & $0.00002$ & $0.00002$ \\
 \text{Virginia}  & $8654542$ & $11.37007$ & $11$ & $\phantom{+}0.07610$ & $0.27586$ & $0.00002$ & $0.07612$ & $\phantom{+}0.16076$ & $0.40140$ & $0.00002$ & $0.16058$ \\
 \text{Washington}  & $7715946$ & $10.13697$ & $10$ & $\phantom{+}0.00118$ & $0.03435$ & $0.00002$ & $0.00120$ & $\phantom{+}0.02024$ & $0.14227$ & $0.00002$ & $0.02026$ \\
 \text{West Virginia}  & $1795045$ & $2.35827$ & $2$ & $\phantom{+}0.00012$ & $0.01095$ & $0.00002$ & $0.00014$ & $\phantom{+}0.00510$ & $0.07141$ & $0.00002$ & $0.00512$ \\
 \text{Wisconsin}  & $5897473$ & $7.74792$ & $8$ & $\phantom{+}0$ & $0$ & $0.00002$ & $0.00002$ & $\phantom{+}0.00018$ & $0.01342$ & $0.00002$ & $0.00020$ \\
 \text{Wyoming}  & $577719$ & $0.75899$ & $1$ & $\phantom{+}0$ & $0$ & $0.00002$ & $0.00002$ & $\phantom{+}0$ & $0$ & $0.00002$ & $0.00002$ \\
\hline
 TOTAL &  $331108434$ & $435.00000$  &  $435$ &  $\phantom{+}0.00000$ & & $0.64618$  & $0.64618$ & $\phantom{+}0.00000$ & & $0.93914$ & $0.93914$   \\
\hline
\end{tabular}
\end{scriptsize}
\caption{Sensitivity of 2020 Congressional apportionment to undercounts, based on Monte Carlo sampling and entropic apportionment.  Case (a) assumes an average undercount rate of $0.6\%$ of each State's population, which is the approximate level estimated for the previous 2010 U.S.\ Census, while Case (b) assumes an average rate twice as high, $1.2\%$, anticipating that 2020 posed additional technical and political challenges.  The House size was fixed throughout at the prescribed value of $R = 435$, and all apportionments were calculated via the entropic method. The baseline apportionment reflects the officially reported populations for apportionment.  Possible undercounts were (somewhat arbitrarily) assumed to be drawn from independent exponential distributions for each State, and a sample of $n_{\stext{MC}} = 50\hspace{0.95pt} 000$ random Census results were simulated in each case, based on the assumed undercount rates and the best guess of the true populations given the actual Census data.  The average shifts and root-mean-square (RMS) shifts in allocated seats per state have been measured relative to the baseline entropic apportionment.  Probabilities reflect estimates for losing or gaining at least one seat.}\label{Table:2020_sensitivity_long}
\end{table}


%
%




\end{document}